\documentclass[12pt]{report}

\usepackage{UCLAPhDStyle/uclaphd-2-cochairs}

\usepackage{graphicx}

\begin{document}

\Author{Roy Gerrit Hemker}
\Title{Particle-In-Cell Modeling \\
       of Plasma-Based Accelerators \\
       in Two and Three Dimensions}
\DegreeDate{2000}
\Major{Physics}

\Copyright


%
%


\CommitteeMember{Steve Cowley}
\CommitteeMember{Chandrasekhar J. Joshi}

\CommitteeCoChair{John M. Dawson}
\CommitteeChair{Warren B. Mori}

\Dedication{
  \vspace*{1in}
  \begin{center}
    DEDICATION \\
    \vspace{\baselineskip}
    To my family and my friends
  \end{center}
}

\SingleSpaceLists
\ListOfFigures
\LOSDots


%

\Acknowledgements{First of all, I want to express my thanks and appreciation
 to Prof. Warren Mori. His strong support and his guidance throughout my
 research were essential to the success of this work. I would not have
 been able to complete this project without it. 
 \\ \\
 I would like to thank Prof. John Dawson and Prof. Chan Joshi for their
 support and active interest in my research. In their conversations with 
 me they have inspired me with their broad view of the field of plasma
 physics which helped me greatly to see my own research in perspective.
 My special thanks goes to Dr. Viktor Decyk. Much of the work in this
 dissertation builds on the foundation of his ideas. His suggestions and
 advice have been invaluable.
 \\ \\
 I am grateful to Seung Lee for her help with the cylindrically-symmetric
 algorityhms, Dr. Frank Tsung for his help with the 3D current 
 deposition scheme, and Brian Duda for his collaboration on the long 
 wavelength hosing. My discussions with Dr. Kuo-Cheng Tzeng and
 Prof. Tom Katsoules helped me with many parts of my research and 
 contributed significantly to the success of this work.
 \\ \\
 Last but not least I would like to thank the staff, researchers, and students
 who I had the pleasure of working with for making UCLA such a friendly and
 welcoming place.
 }

\Vita{
  \VitaItem{1993}              {M.S., Physics     \\
                                California State University, Northridge}
  \VitaItem{2000}              {Ph.D. in Physics \\
                                University of California, Los Angeles }
}


\Publication{R.G.~Hemker, W.B.~Mori, S.~Lee, T.~Katsouleas,
             ``Three-dimensional wake structures from asymmetric drive 
             beams'',
             in preparation.}

\Publication{F.S.~Tsung, R.G.~Hemker, C.~Ren, L.O.~Silva, W.B.~Mori,
             T.~Katsouleas,
             ``Generation of a single-cycle laser pulse via photon 
              deceleration'',
             submitted to Phys. Rev. Lett.}

\Publication{R.G.~Hemker, W.B.~Mori, S.~Lee, T.~Katsouleas,
             ``Dynamic Effects in Plasma Wakefield Excitation'',
             submitted to Phys. Rev. Spec. Top. - Acc.\&Beams.}

\Publication{S.~Lee, T.~Katsouleas, R.G.~Hemker, W.B.~Mori, E.~Esarey,
             C.~Schroeder,
             ``Simulations of a meter long plasma wakefield accelerator'',
             accepted by in Phys. Rev. E.}

\Publication{B.J.~Duda, R.G.~Hemker, K.C.~Tzeng, W.B.~Mori,
             ``A long-wavelength hosing instability in laser-plasma
              interactions'',
             Phys. Rev. Lett., vol. 83, 1978(1999). }

\Publication{R.G.~Hemker, F.S.~Tsung, V.K.~Decyk, W.B.~Mori, S.~Lee,
             T.~Katsouleas,
             ``Development of a Parallel Code for Modeling Plasma
              Based Accelerators'',
             Proceedings of the 1999 Particle Accelerator Conference.}

\Publication{R.G.~Hemker, K.-C.~Tzeng, W.B.~Mori, C.E.~Clayton, T.~Katsouleas,
             ``Computer simulations of cathodeless, high-brightness
              electron-beam production by multiple laser beams in plasmas'',
             Phys. Rev. E, vol. 57, 5920(1998). }

\Publication{R.G.~Hemker, K.C.~Tzeng, W.B.~Mori, C.E.~Clayton, T.~Katsouleas,
             ``Cathodeless, high-brightness electron-beam production
             by multiple laser beams in plasmas'',
             Proceedings of the 1997 Particle Accelerator Conference,
             vol. 3, p.2870-2 (1998).}

\Publication{N.~Kioussis, H.~Watanabe, R.G.~Hemker, W.~Gourdin, A.~Gonis,
             P.E.~Johnson,
             "Effect of boron and hydrogen impurities on the
              electronic structure of Ni3Al",
              Mater. Res. Soc. Proc., vol. 319, 363(1994).}



\Abstract{Particle-In-Cell Modeling of Plasma-Based Accelerators
          in Two and Three Dimensions}
{
    In this dissertation, a fully object-oriented, fully relativistic,
    multidimensional Particle-In-Cell code was developed and applied
    to answer key questions in plasma-based accelerator research. The
    simulations increase the understanding of the processes in laser
    plasma and beam-plasma interaction, allow for comparison with
    experiments, and motivate the development of theoretical models.
    \\ \\
    The simulations support the idea that the injection of electrons in 
    a plasma wave by using
    a transversely propagating laser pulse is possible. The beam parameters
    of the injected electrons found in the simulations compare
    reasonably with beams produced by conventional methods and therefore
    laser injection is an interesting concept for future plasma-based
    accelerators.
    Simulations of long laser pulses, such as the ones used in self-modulated
    laser wakefield acceleration, predict the existence of a hosing
    instability with a wavelength longer than the plasma wavelength.
    It is found that this effect might increase the emittance
    of electron beams produced by this acceleration method.
    \\ \\
    Simulations of the optical guiding of a laser wakefield driver in a parabolic
    plasma channel support the idea that electrons can be accelerated over
    distances much longer than the Rayleigh length in a channel.
    Simulations of plasma wakefield acceleration in the nonlinear blowout regime
    give a detailed picture of of the highly nonlinear processes involved. 
    Using OSIRIS, we have also been able to perform full scale simulations
    of the E-157 experiment at the Stanford Linear Accelerator Center.
    These simulations have aided the experimentalists and they have 
    assisted in the development of a theoretical model that is able to reproduce
    some important aspects of the full PIC simulations. 
}

\MakePreliminaryPages

\chapter{Introduction}
  \label{chap:intro}

\section{Introduction}
  \label{sect:intro:intro} 

One of the most important practical problems in high energy physics
has always been how to increase the energy of the beams that are used in
lepton collision experiments. This is also true today but the problem is more
fundamental than it was in the past because the conventional
design of accelerators is constrained by some fundamental physical and
practical limits. The maximum acceleration gradient that can be achieved
using RF-waveguides in existing facilities, e.g., SLAC,
is of the order of about 25MeV/m.
This technology is limited because the material used to
build the waveguides will be destroyed by tunneling ionization at 
field strength near 100MeV/m using existing RF frequencies. This means
that in order to get an energy gain of 50GeV an electron would have to be
accelerated over 2000m with a gradient of 25MeV/m.
The two conventional accelerator designs to achieve this are linear
accelerators or accelerator
rings which accelerate a particle by sending it repeatedly through
the same RF-waveguide for acceleration. For both these designs
achieving higher energies means that they have to get bigger in size.

Current linear accelerators for electrons and positrons are a few
miles long and current
$e^{+}e^{-}$ rings accelerators require diameters of the order of 10 miles.
Their energies are a couple of 10GeVs. It seems unlikely that accelerators
of significantly higher energy and therefore size are going to be
build using this conventional technology. What is required is a
significant increase in the magnitude of the accelerating field
and plasma-based acceleration does offer an answer to this problem.

The remainder of this chapter will first introduce the different basic
ideas for plasma-based acceleration and experiments using them. It will then
briefly examine the different fields that are of importance to this research
and then consider the role that computer simulations can play in physics
research and in particular in the research on plasma-based accelerators.
Finally, it will briefly explain the usefulness of advanced programing
concepts for computer simulations.

\section{Plasma-Based Accelerator Concepts}
  \label{sect:intro:accelerators} 

\begin{figure}
   \begin{center}
      \epsfig{ file=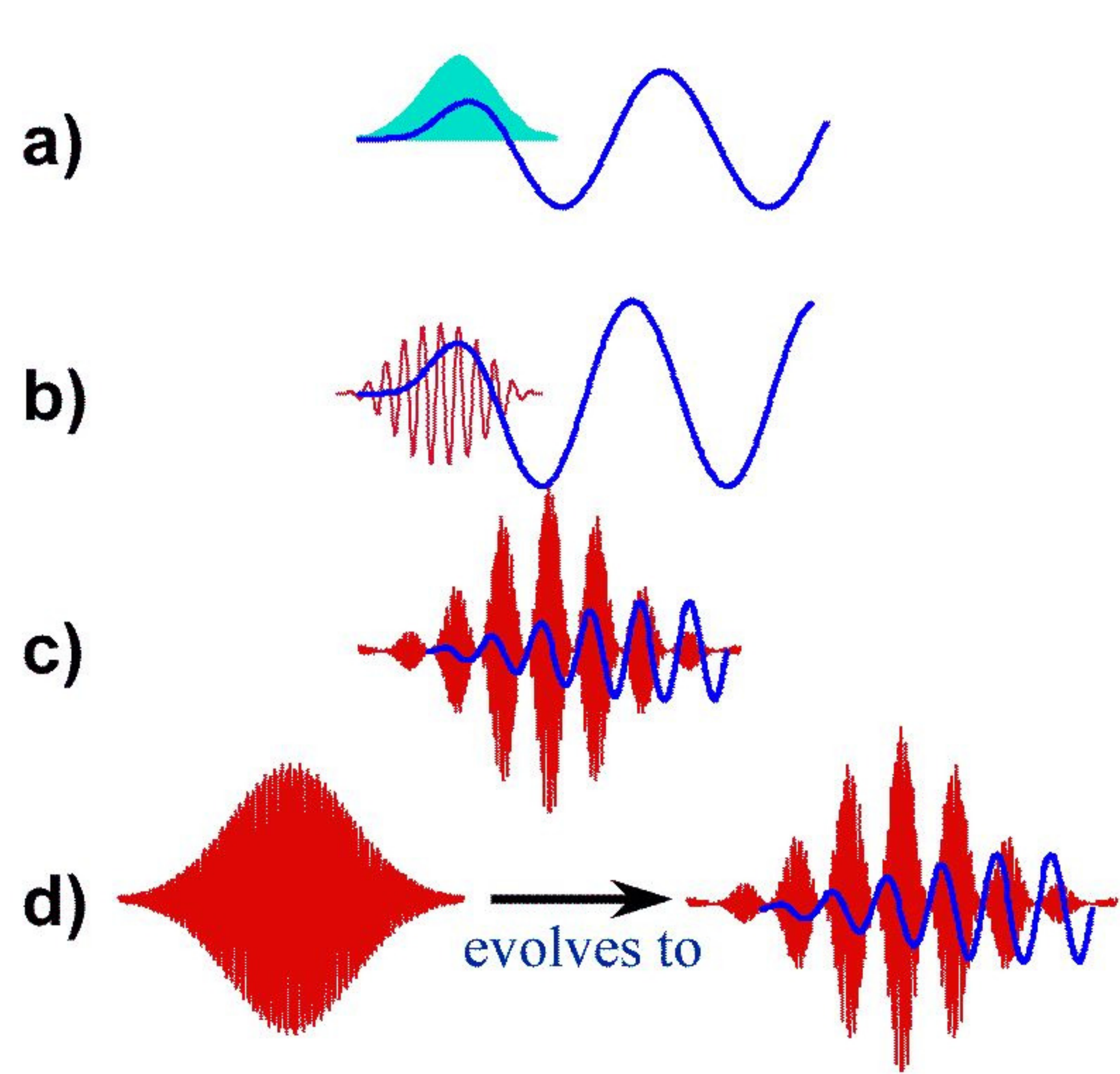, width=3.5in }
      \caption{Plasma-based accelerator concepts: a) Plasma Wakefield
               Accelerator (PWFA) b) Laser Wakefield Accelerator (LWFA)
               c) Plasma Beatwave Accelerator (PBWA) d) Self-modulated
               Laser Wakefield Accelerator (SMLWFA)}
      \label{fig:intro:acc-schemes}
   \end{center}
\end{figure}

Several ways of using a plasma for particle acceleration have been
suggested \cite{Tajima79b,Joshi84,ChenP85}.
Fig.~\ref{fig:intro:acc-schemes} shows four
different concepts. Fig.~\ref{fig:intro:acc-schemes}~a) shows a plasma
wakefield accelerator (PWFA) \cite{ChenP85}. This concept uses an
electron bunch moving through a plasma to create a wake. The charge of
the bunch electrons will push the plasma electrons out of the path of
the bunch. These displaced electrons will then oscillate back after
the bunch has passed through setting up a plasma oscillation. The
phase velocity of the plasma oscillation is the velocity of
the electron bunch that created the wake. For highly relativistic
electrons this is a velocity very close to speed of light. Since
plasma space-charge waves have an electric field component parallel to their
propagation direction a particle placed in this wake with an energy above a
certain threshold energy \cite{Williams90,Katsouleas85,Mora92a,Mora92b,Esarey95}
can stay in phase with this accelerating field for long distances and gain
significant amounts of energy.  In order for the plasma oscillation to have
a large amplitude and therefore a large accelerating field the length of the
driving bunch has to be of the order of the plasma wavelength.

The other concepts shown in Fig.~\ref{fig:intro:acc-schemes}
share the idea of generating a wakefield in a plasma and only differ in the
driver used to generate this wakefield. In Fig.~\ref{fig:intro:acc-schemes}~b),
the laser wakefield accelerator (LWFA) concept is shown\cite{Tajima79b}.
This concept relies
on the ponderomotive force of a short laser pulse to set up a plasma oscillation.
In Fig.~\ref{fig:intro:acc-schemes}~c), the plasma beatwave accelerator (PBWA)
concept is shown\cite{Tajima79b,Joshi84}. In this concept two
laser pulses which are long by comparison with the plasma wavelength 
and which have frequencies that differ by the plasma frequency are used
to excite the wake. The beatwave resulting from
these two laser pulses can be viewed as a series of successive pulses each of
which has a length of one plasma wavelength. Each of these pulses then
contributes to setting up a plasma wave in the same way as in concept b).
In  Fig.~\ref{fig:intro:acc-schemes}~d) the so called self-modulated
laser wakefield accelerator (SMLFWA) concept is shown
\cite{Esarey90,Mori94,Andreev94}. 
In this concept a long single frequency pulse
first generates frequency components shifted by the plasma frequency due
to Raman forward scattering. The Raman-scattered light beats with the original
frequency generating a beatwave and therefore a plasma wave results 
similarly as in the PBWA concept.

\begin{figure}
   \begin{center}
      \epsfig{ file=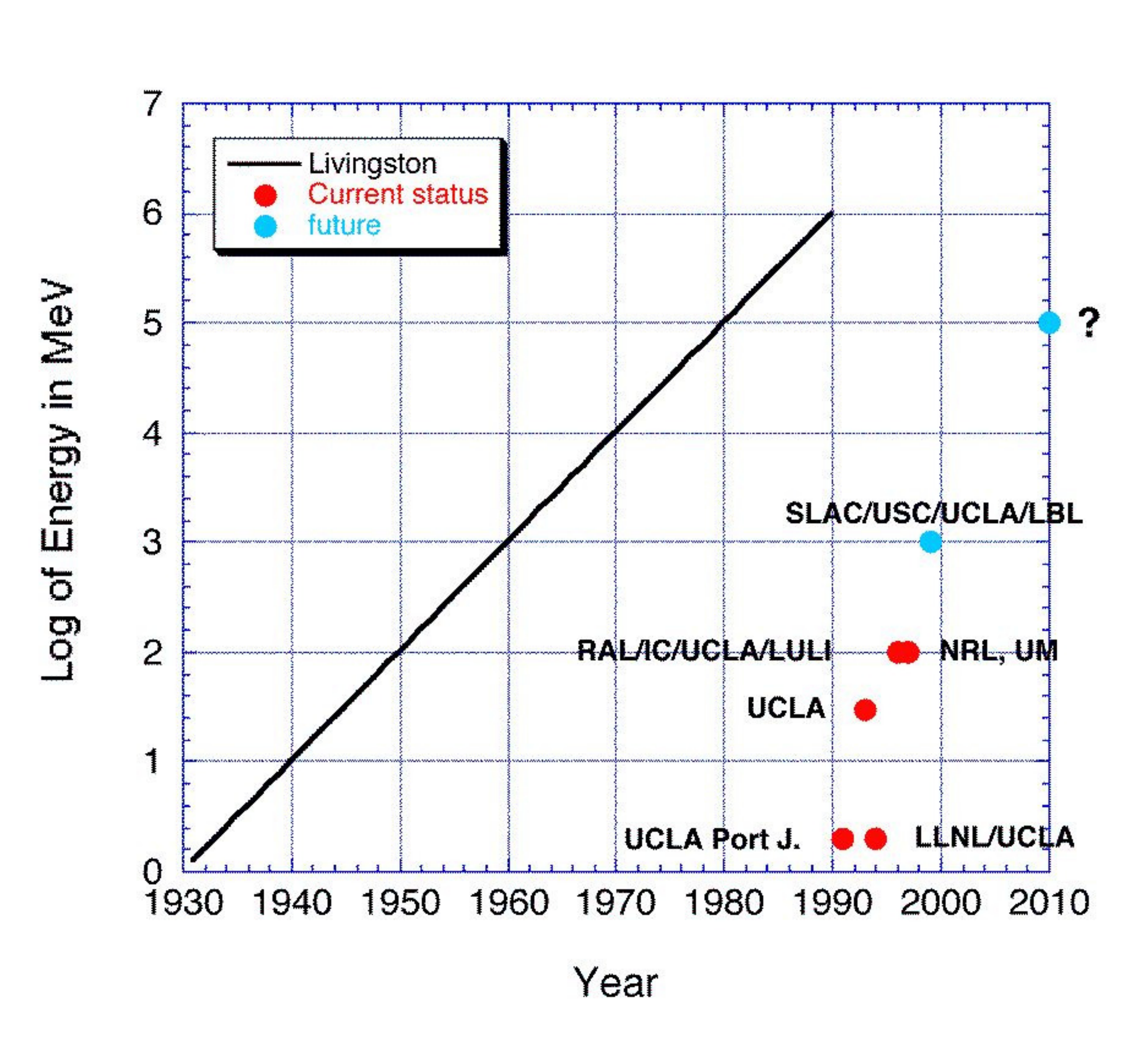, width=3.5in }
      \caption{Comparison of the increase of energy in plasma-based
               accelerations against the Livingston curve
               (Courtesy of T. Katsouleas)}
      \label{fig:intro:acc-increase}
   \end{center}
\end{figure}

Using plasma wakes generated by particle beam or laser
pulse drivers for particle acceleration has been a topic of research
since the idea was first
published \cite{Tajima79b}. However, it has only been during the last 5 to 10 years
that significant experimental progress has been made \cite{Esarey96,Leemans99}.
Fig.~\ref{fig:intro:acc-increase} shows the energy gains
measured in plasma-based acceleration experiments as a function
of time. The solid black line indicates the past and projected increase
of conventional accelerator technology (the Livingston curve) \cite{Livingston},
the red data points
show results of past plasma-based accelerators \cite{Leemans99}, and the blue
data points are the expected results of current and future experiments with
plasma-based accelerators \cite{Assmann97}. The data points for plasma-based
acceleration experiments do indeed suggest a faster increase of the 
output energy than is to be expected for conventional accelerators.
Most of the past experiments were
based on the PBWA and SMLWFA concept since the laser and plasma
parameters for these concepts were easier to realize experimentally.
The rapid increase in a laser power and the simultaneous
shortening of laser pulse length \cite{Strickland85,Tzeng98b} now make LWFA
experiments possible. Several such experiments are being conducted at
laboratories around the world.

In contrast to this the most recent
experiment indicated in Fig.~\ref{fig:intro:acc-increase} is based
on the PWFA concept. The data point labeled with ``SLAC/USC/UCLA/LBL''
refers to the E-157 experiment conducted at the Stanford Linear
Accelerator Center (SLAC) as a collaboration of research groups between
SLAC, the University of Southern California (USC), the University of
California at Los Angeles (UCLA), and the Lawrence Berkeley Laboratory
(LBL)\cite{Assmann97}. The goal of this experiment is to use the high quality
electron beam generated by the Stanford linear accelerator as a driver for
the generation of a wakefield in a plasma. The expected accelerating field
for this experiment is up to about 1GeV/m, which is about one order of magnitude
larger than what can be achieved with conventional technology. It is worth noting
that a significant breakthrough that has made this experiment possible was the
construction of a plasma source that is able to generate a uniform plasma 
over 1m distances \cite{Muggli99}. All previous experiments for plasma-based
acceleration were limited to a couple of mm for acceleration. Because of the
increased length of the acceleration distance in this experiment, energy gains
of up to 1GeV are expected.

Modeling the E-157 experiment has served as a motivation for
much of the research and code development presented in this dissertation
and it will be described in more detail in chapter \ref{chap:pwfa}.

\section{Research Areas Relevant to Plasma-Based Accelerators}
  \label{sect:intro:research-fields}

\begin{figure}
   \begin{center}
      \epsfig{ file=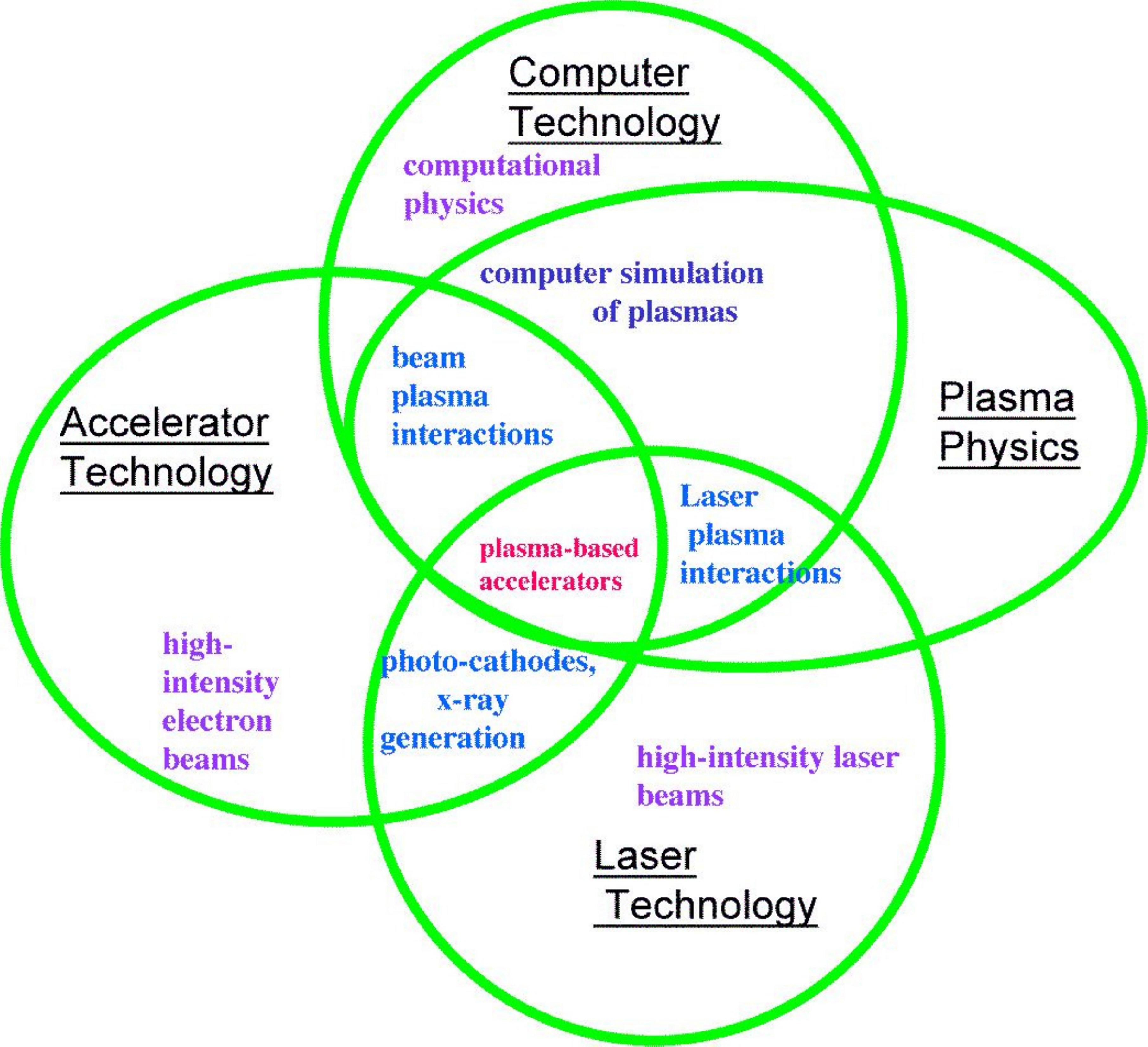, width=3.5in }
      \caption{Multidisciplinary areas which have contributed to plasma-based
               accelerator research}
      \label{fig:intro:expertise}
   \end{center}
\end{figure}

Computer simulations play a role in many areas of physics and the topic of this
dissertation is at the intersection of several of these research areas.
Fig.~\ref{fig:intro:expertise} shows schematically the different relevant fields
in physics that are of importance to plasma-based accelerator concepts and how
they are connected to each other and computer technology. The importance of computer
technology to any research involving simulations is obvious and will be examined
further below. The other three fields of importance are laser technology, accelerator
physics, which includes charged-beam dynamics, and plasma physics. 

There are three key research topics that need to be studied and understood for
the successful development of plasma-based accelerators. The first one is the
evolution of the drive-beam as it generates the plasma wakefield . The 
second is the excitation of the wakefield by a given drive beam. The third is
evolution of a trailing bunch of particles which is loaded into the wakefield.
Each of these topics has been investigated separately,
but to get an  integrated understanding of an eventual accelerator they have
to be answered in a self-consistent
manner since the drive beam as well as the accelerated particle beam are interacting
with the plasma wake, i.e., are modifying it and are being modified by it. So far this kind of
self-consistent understanding could only be gained by using computer simulations.
The importance of the different research areas indicated in
Fig.~\ref{fig:intro:expertise} is straight forward to understand from these
questions. The initial qualities of a laser drive-beam are a problem of laser
technology. The initial qualities of a particle drive-beam and the evolution of
an accelerated particle bunch are problems also dealt with in accelerator physics
and the evolution of either type of drive-beam is a question of laser-plasma or
beam-plasma physics.

\section{The Role of Simulations in Plasma and Accelerator Research}
  \label{sect:intro:simulations}
  
The process of science is mainly a process of comparing the predictions
of theories and hypotheses with the results of actual experiments. If
a prediction does not agree with a measured result then either the theory that
gave rise to the prediction, or the manner in which the prediction was 
inferred from the theory has to be modified. In this way scientific
theories become better and better models of the facts they are trying
to explain. Computer simulations like the ones presented in this dissertation
come into this picture as a method of bridging the gap between theory and
experiment. They are a way to derive predictions from complex and 
integrated theoretical models which can be compared to experimental
results or they are a means to test theories directly.

Problems in many areas of physics today
are systems with many degrees of freedom, as for example in plasmas 
physics. Often the equations that determine the evolution of each degree
of freedom over time are very well
established, but to track and understand the simultaneous evolution of more
than a few variables is beyond the capacity of the human mind. Computer
simulations allow us to do these things. First, they make it possible to derive
results from basic theories that can be compared to experimental results.
This is particularly important for areas where no simplified analytical
model exists at all. Secondly, the insight into the physical processes
gained by evaluating the simulation results can in some cases lead to
the development of simplified analytical models that are tractable.
All these consideration apply directly to the case of
plasma physics where research involves a very large number of particles
or degrees of freedom.

The view of simulations outlined above is one that follows directly from the 
standard methodology of science and it should always be kept in mind
when using computer simulations. For the case of Particle-In-Cell simulations
a different viewpoint, but one which is not in contradiction but is an
extension of the one above, is useful as well. Particle-in-cell (PIC) codes
used for the research presented in this dissertation make 
no assumptions in physics except the validity of classical physics. 
That is, the full set of Maxwell's equations and the relativistic equations
of motion for individual particles is self-consistently evolved.
Within the validity of classical physics (quantum effects are ignored) and within
the limits of numerical accuracy, PIC-simulations of plasmas give exact and
very detailed
information on the processes within a plasma. They could therefore be considered
as numerical experiments, that provide a third kind of methodology to the scientific
method on an equal footing with experiment and theory. As such the value of
simulations lies in providing us with extremely detailed and accurate information
about a simulated problem to an extent that far exceeds the possibilities of either
theory or experiment.

\section{The Case for the Use of Object-Oriented Simulation Codes}
  \label{sect:intro:advanced-codes}

The topic of this dissertation in not only the physics learned by conducting
simulations but also the development of a completely new kind of PIC code that
uses modern state-of-the-art software methods. Developing a new code like this
is the equivalent of the development of a new kind of sophisticated 
experimental laboratory (apparatus and diagnostic techniques)
or the development of a new kind of analytical approach to a theoretical
problem. 
In order to be reproducible not only the results of scientific work but also
the methods that were applied need to be well documented. In case of a 
new method this is particularly important in order to make it possible for 
others to apply the same method to other problems. 
The development of a new simulation code requires therefore that the new
algorithms used in the code as well as sufficient instructions on how to
actually use the code should be documented.
This kind of documentation for the newly developed code OSIRIS (Object-oriented
Simulation Rapid Implementation System) will be part of this dissertation.

\begin{figure}
   \begin{center}
      \epsfig{ file=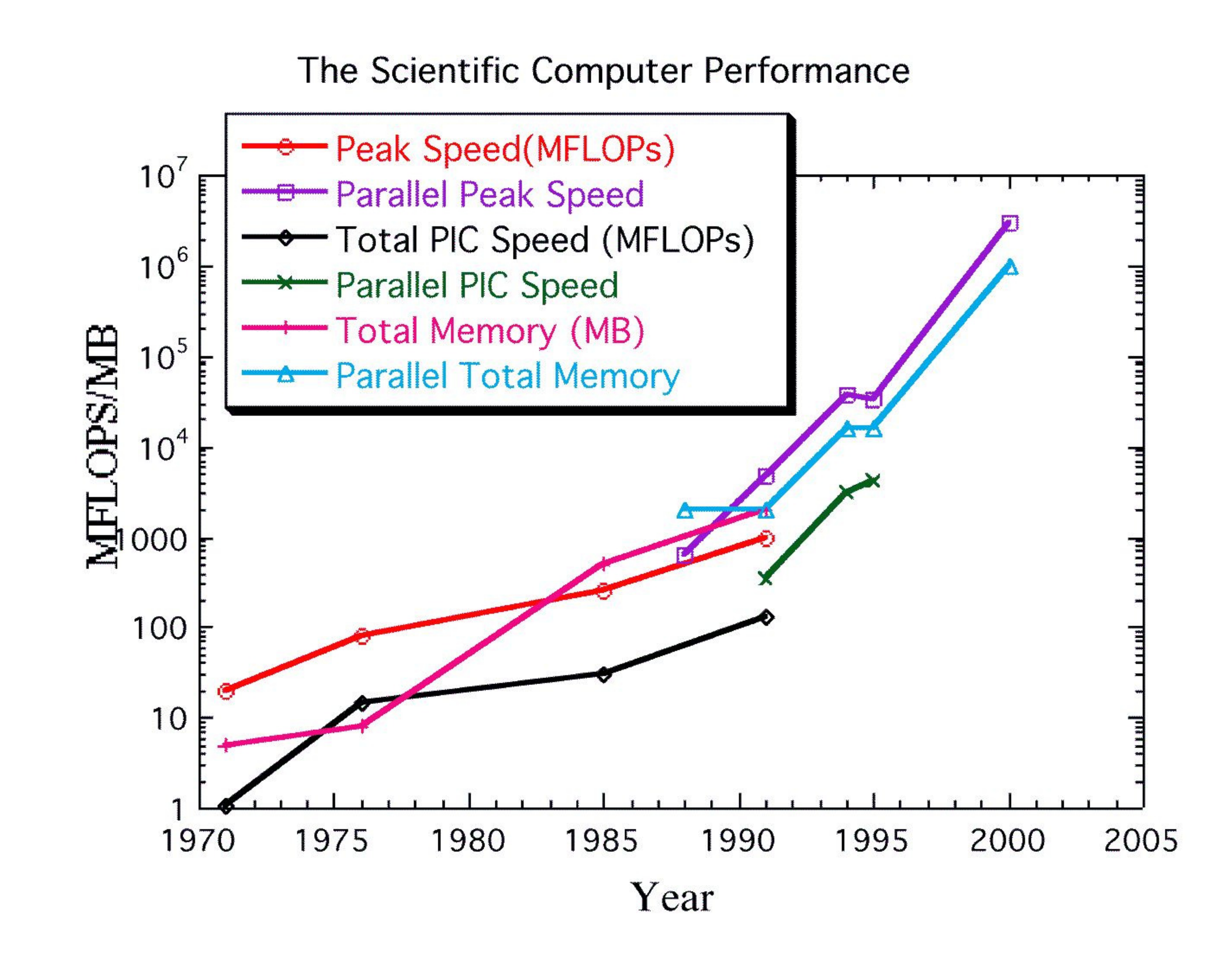, width=4.0in }
      \caption{Increase in computing speed and memory for different
               supercomputer architectures}
      \label{fig:intro:comp-increase}
   \end{center}
\end{figure}

In order to see the necessity of a new approach to PIC-simulations, the
possibilities opened up by the rapid increase in available computing power have
to be understood. Fig.~\ref{fig:intro:comp-increase} shows the advances made
in computing speed over time and it indicates an exponential increase in the
computing speed as well as in the available memory.  These advances in
computational speed and memory now make it possible to do full scale 2D and
3D PIC simulations of laser and beam plasma interactions. However, the
increased complexity of these codes and interactions make it necessary to
apply modern programming approaches like an object oriented programing style
to the development of codes.

An important fact to note is that the growth in computing power over the
last ten years has largely been due to the use of massively parallel computers
which now have hundreds of processors. In order to take full advantage of
this development it has become necessary to use more
complex simulation codes. The increased complexity of codes arises for two
reasons. One reason is that the realistic simulation of a problem requires
a larger number of more complex algorithms interacting with each other than
the simulation of a rather simple model system. For example,
initializing an arbitrary laser or particle beam in 3D is a much more difficult
problem than doing the same in 1D or 2D. The other reason that simulation
codes are becoming more complex is that the computer systems are more complex.
Questions a code developer has to consider include things like memory management,
operating systems calls, threads, and message passing. As a result the
performance obtained from a system can dramatically differ depending on
the code strategy. Parallelized codes that have to handle the problems of
parallel communication and parallel I/O are an example of this. A way to deal
with this increased complexity is to use an object oriented programming style,
which divides the code and data structures into independent classes of objects.
This programming style maximizes code reusability and reliability.

The goal of the code development program, that is part of the research presented
in this dissertation, was to create a code that breaks up the large problem of a
simulation into a set of essentially independent smaller problems that can be
solved separately from each other. Object oriented programming achieves this by
handling different aspects of the problem in different modules (classes) that
communicate through well-defined interfaces. The programming language we chose
for this purpose was Fortran 90, mainly because it allowed us to more easily
integrate already available Fortran algorithms into the new OSIRIS-framework.
As a result of the intensive code development effort OSIRIS now contains
algorithms for 1D, 2D, and 3D simulations in Cartesian coordinates and for
2D simulations in cylindrically symmetric coordinates. For all of these algorithms
the code is fully relativistic and presently uses a charge-conserving current
deposition algorithm. It allows for a moving simulation window and arbitrary
domain decomposition for any number of dimensions. This large number of
algorithms in one code was only possible due to the object-oriented style of the
code. It makes the code a useful tool for many different research problems with
the possibility to be extended much further by adding new modules.

There are past and ongoing efforts by other plasma simulation research
groups to take advantage of object-oriented programming. Forslund 
\cite{Forslund90} introduced a parallel object-oriented PIC code written in C++
that ran in parallel on a network of workstations. Haney \cite{Haney91} used 
a hybrid C++/Fortran77 code for tokamak modeling. Reynders \cite{Reynders94}
seems to have continued Forslund's work and developed an object-oriented 
particle simulation library. An ongoing effort to develop an object-oriented
library for scientific programming including plasma simulations is
the C++ based POOMA project \cite{Wilson96}. Verboncoeur \cite{Verboncoeur95} 
developed the object-oriented 2D parallel code OOPIC using C++.

The use of Fortran90 for object-oriented codes in plasma physics has been
investigated by Norton \cite{Norton96} and Decyk\cite{Decyk98a,Decyk98b}.
Many of the code development results presented in this dissertation build
on the results of their research. Qiang \cite{Qiang99} has used Fortran90
to develop an object-oriented code for electrostatic simulations of beam 
dynamics in linear accelerators.
The contribution made in this dissertation is a code that combines already
existing and new algorithms in a way that leads to significantly improved
qualities with regard to operating and extending the code.

\section{Overview}

This chapter has tried to explain the motivations that led to the research results
presented in this dissertation. The remainder of this dissertation is structured
as follows. First a brief review of basic physics and computer algorithms will
be given. This will be followed by a presentation of the implementation details of
the new code OSIRIS with particular emphasis on newly developed algorithms. The
remaining chapters will then present research results for laser injection of
electrons into a plasma based accelerator, long wave length hosing of lasers in
plasmas, laser wakefield acceleration in a parabolic plasma channel, and
plasma wakefield excitation and acceleration in the blowout regime.

\chapter{Review of Plasma-Based Accelerator Physics}
  \label{chap:review-physics}
  
This chapter will review the basic physics that governs the
behavior of plasma-based wakefield accelerators in general and
their drivers, lasers and particle beams. The equations and symbols
introduced in this chapter will be used throughout this dissertation.

\section{Single Particle Dynamics in a Wakefield}
  \label{sect:review-physics:single-particle} 

We will start with reviewing the behavior of a single particle in a
given wakefield. Consider an electron being accelerated in a plasma
wave of the form

\begin{equation}
  \label{equ:review-physics:harmonic-potential}
  \phi=\phi_{0}\left(1-x_{2}^{2}/w_{p}^{2}\right) 
  \sin\left[k_{p}\left(x_{1}-v_{\phi}t\right)\right]\
\end{equation}

\noindent{where} $v_{\phi}$ is the phase velocity of the wave and 
$w_{p}$ is a parameter describing the width of the plasma wave. This 
potential describes the behavior of particles close to the center of 
a typical plasma wave. We assume $v_{\phi}\cong c$, i.e., 
relativistic plasma waves. The subscripts 1 and 2 refer to directions 
parallel and perpendicular, respectively, to the plasma wave's 
direction of propagation. The equations of motion for an individual 
electron are

\begin{equation}
  \label{equ:review-physics:p1-evol}
  \frac{d}{dt}p_{1}=-eE_{1}=e\phi_{0}k_{p}\left(1-x_{2}^{2}/w_{p}^{2}\right)
  \cos\left[k_{p}\left(x_{1}-v_{\phi}t\right)\right] \\
\end{equation}

\begin{equation}
  \label{equ:review-physics:p2-evol}
  \frac{d}{dt}p_{2}=-eE_{2}=-2e\phi_{0}\frac{x_{2}}{w_{p}^{2}}
  \sin\left[k_{p}\left(x_{1}-v_{\phi}t\right)\right]
\end{equation}

\noindent{The} acceleration of single electrons in these fields has 
been studied extensively \cite{Williams90,Mora92a,Mora92b,Fedele86}. An 
injected electron accelerated along the axis, $x_{2}=0$, will be 
trapped if its injection energy (the initial kinetic energy) exceeds 
the trapping threshold\cite{Williams90,Mora92a,Mora92b,Esarey95}. 

\begin{equation}
  \label{equ:review-physics:threshold}
  W_{i}\approx mc^{2}\left(\gamma_{\phi}^{2}\left\{\bar{\phi}_{0}
  +1/\gamma_{\phi}-\beta_{\phi}\left[\left(\bar{\phi}_{0}
  +2/\gamma_{0}\right)\bar{\phi}_{0}\right]^{1/2}\right\}-1\right) 
\end{equation}
\begin{center}
  with  $\bar{\phi}_{0}=e\phi_{0}/\left(mc^{2}\right)$
\end{center}

\noindent{which} reduces to 
$\frac{1}{2}\left[\bar{\phi}_{0}+\left(1/\bar{\phi}_{0}\right)\right]-1$ 
as $\gamma_{\phi}\rightarrow\infty$. Here $\beta_{\Phi} = v_{\Phi}/c$
and $\gamma_{\Phi} \: = \: 1/\sqrt{\,1\, -\, \left(v_{\Phi}/c\right)^{2}\,}$.

Once trapped an electron is accelerated and its speed eventually 
exceeds the phase velocity of the wave. The acceleration process 
ceases after the electron outruns the wave and encounters decelerating 
forces. If $x_{2}=0$, then the maximum energy gain is 
\cite{Tajima79b,Williams90,Mora92a,Mora92b,Esarey95}.

\begin{equation}
  \label{equ:review-physics:energy-gain}
  W_{f}-W_{i}\equiv \Delta W\cong 
  2\gamma_{\phi}\left[1+\eta\bar{\phi}_{0}\gamma_{\phi}\right]mc^{2}
\end{equation}

\noindent{where} $\eta$ is 2 if the particle slips through a full 
$\pi$ phase of the accelerating bucket. $\eta$ usually has a value 
smaller than 2 depending on certain conditions explained below. 
$\Delta W$ is approximately 
$2\eta\bar{\phi}_{0}\gamma_{\phi}^{2}mc^{2}$ if 
$\bar{\phi}_{0}\gamma_{\phi}\gg 1$. The dephasing distance can be 
estimated by calculating the distance it takes for the electron 
moving at the speed of light, c, to move forward a half wavelength in 
a wave moving at $v_{\phi}\cong c$. This gives 
\cite{Williams90,Mora92a,Mora92b,Esarey95}

\begin{equation}
  \label{equ:review-physics:dephasing}
  L_{dp}=\frac{1}{2}\eta\gamma_{\phi}^{2}\lambda_{p}=\eta\pi\gamma_{\phi}^{2}c/\omega_{p}
\end{equation}

An electron which is not on the axis, $x_{2}\neq 0$, will also feel 
transverse, or so called defocusing/focusing fields, as given by 
Eq.~(\ref{equ:review-physics:p2-evol}). Electrons in the defocusing phase of 
the wave accelerate away from the axis and are eventually lost 
\cite{Williams90,Mora92a,Mora92b,Fedele86}. Electrons in the focusing phase 
execute betatron oscillations (in $x_{2}$) as they accelerate along 
$x_{1}$ so only electrons which reside in both focusing and 
accelerating fields are accelerated to the dephasing limit 
\cite{Williams90,Mora92a,Mora92b,Fedele86}. These fields are $\pi/2$ out of phase and 
therefore only a quarter of a plasma wave wavelength can be used for 
acceleration. This reduces the maximum energy gain and the dephasing 
length given above by roughly a factor of 2 [i.e., $\eta$=1 in
Eq.~(\ref{equ:review-physics:energy-gain})]. 
In finite-width plasma waves additional second order focusing terms 
may extend the range of phases which have both focusing and 
accelerating forces\cite{Gorbunov96,Mori87b}\footnote{The total dc
focusing force is $3/2$ times larger than given in Ref. \cite{Gorbunov96},
because of an additional electrostatic field.}. In this case we have
$1 < \eta < 2$.

\section{Laser Beams}
  \label{sect:review-physics:laser-beams}  

A laser beam in vacuum can be described as a Hermite Gaussian beam. This is an solution to
the paraxial wave equation which is an approximation to Maxwell's equations\cite{Milonni}.
The lowest order Hermite Gaussian beam propagating in $z$ is given by:

\begin{equation}
  \label{equ:review-physics:gaussian-beam}
  E \left( x, y, z, t \right) = A
  \times \frac{e^{-i\Phi\left(z\right)}}{\sqrt{1+\left( \frac{z}{z_{R}}\right)^{2}}}
  \times e^{i \frac{k \left( x^{2} + y^{2} \right) }{2 R\left(z\right)}}
  \times e^{- \frac{  \left( x^{2} + y^{2} \right) }{  w\left(z\right)^{2}}}
  \times e^{i \left( k z - \omega t \right) }
\end{equation}

\begin{figure}
   \begin{center}
      \epsfig{ file=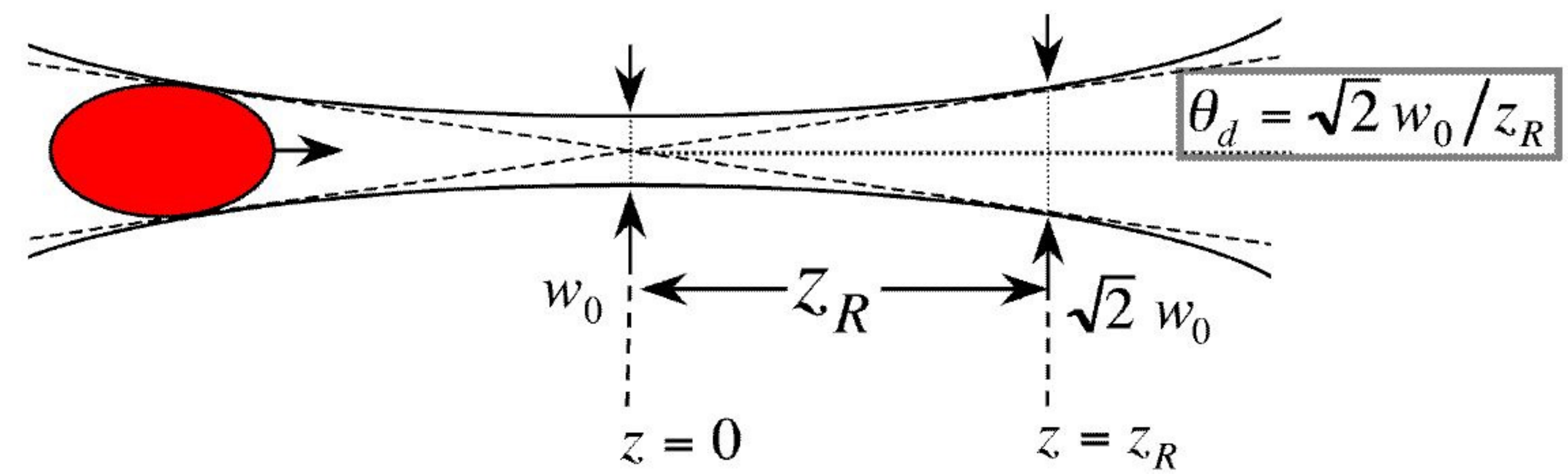, width=5.8in }
      \caption{The figure illustrates the evolution of the spot size
               of a Gaussian beam during its propagation through a vacuum.
               After propagating away from the focal plane at $z=0$, where
               the beam has its minimum spot size $w_{0}$, to a distance of
               one Rayleigh length $z_{R}$ the spot size has increased to
               $\sqrt{2}w_{0}$} 
      \label{fig:review-physics:gaussian-beam}
   \end{center}
\end{figure}

\noindent{Here} we use $\Phi\left(z\right) = \arctan{\frac{z}{z_{R}}}$ and
$R\left(z\right) = z + \frac{z_{R}^{2}}{z}$. The spotsize $w$ of the laser beam is given by

\begin{equation}
  \label{equ:review-physics:spot-size}
  w\left(z\right) = w_{0} \sqrt{1+\left( \frac{z}{z_{R}}\right)^{2}}  
\end{equation}

where the Rayleigh length $z_{R}$ is defined as
$z_{R}= \frac{\pi}{\lambda} w_{0}^{2} = \frac{1}{2} k w_{0}^{2}$.
This solution shows that the evolution of the spotsize of a laser beam is characterized
by two parameters; it's
wavenumber $k$ and the spotsize $w_{0}$ in the focal plane where the beam is narrowest.
Eq.~(\ref{equ:review-physics:spot-size}) shows that the Rayleigh length is the distance
from the focal plane at which the spotsize is $\sqrt{2}$ times the spotsize  $w_{0}$
in the focal plane and
Fig.~\ref{fig:review-physics:gaussian-beam} illustrates the physical meaning of
$z_{R}$ and $w_{0}$. 

The evolution of the spotsize of an approximately Gaussian beam in 
either a uniform plasma or a plasma channel is given
by the envelope equation for the evolution of the laser spotsize.
This equation can be derived by a variety of methods, e.g., the source
dependent expansion\cite{Sprangle87b} or the variational principle 
techniques\cite{Duda2000}.
For a plasma with a parabolic density profile

\begin{equation}
  \label{equ:review-physics:channel-profile}
  n\left(r\right) = n_{0} + \Delta n \; \frac{r^{2}}{r_{0}^{2}}.
\end{equation}

\noindent{the}  envelope equation for the normalized laser spotsize
$W\left( z \right)=w\left( z \right)/w_{0}$
of a laser beam is \cite{Esarey96,Leemans96}

\begin{equation}
  \label{equ:review-physics:laser-envelope}
  \frac{d^{2}W }{dz^{2}} = \frac{I}{z_{R}^{2} W^{3}}
  \left[
  1 - \frac{P}{P_{c}} - \frac{\Delta n}{\Delta n_{c}} W^{4}
  \right] 
\end{equation}


\noindent{where} $\Delta n$ and $r_{0}$ characterize the channel and we set
$r_{0}=w_{0}$ without loss of generality.
The three terms in the bracket are due to I) diffraction, II) relativistic self-focusing,
and III) the external focusing forces (e.g., the plasma channel), respectively.
$P/P_{c} = a^{2} w_{0}^{2}$/32 is the laser power normalized to the critical
power for relativistic self-focusing,
$P_{c} \cong 17 GW \times \left( \frac{\omega}{\omega_{p}} \right)^{2}$
\cite{Max74,Sprangle87a,Mori88},
and $\Delta n_{c}=\left( \pi r_{e} w_{0}^{2} \right)^{-1}$
with $r_{e}= e^{2}/ \left( m_{e}c^{2} \right)$ (the classical electron radius).

The evolution of the spotsize given by Eq.~(\ref{equ:review-physics:spot-size})
can be recovered as the solution of Eq.~(\ref{equ:review-physics:laser-envelope})
if only the diffraction term on the right side of the equation is kept.
For small laser power, Eq.~(\ref{equ:review-physics:laser-envelope})
has focusing solutions for normalized
spotsizes $W< \left( \, \Delta n \, / \, \Delta n_{c} \, \right)^{1/4}$
if $\Delta n \geq \Delta n_{c}$. 
It also has a stable stationary solution
with $W = \left( \, \Delta n \, / \, \Delta n_{c} \, \right)^{1/4}$
if $\Delta n \geq \Delta n_{c}$.
This stationary solution is the matched beam solution for a parabolic density
channel. In the absence of a density channel, there are focusing 
solutions if $P/P_{c}>1$ and a stable stationary solution for $P/P_{c}=1$.
However, it is now well known, that a laser pulse with a pulse 
length $ \stackrel{<}{{\scriptstyle \sim}} \pi c/\omega_{p}$, as is the 
case for LWFA drivers, does not relativistically 
self-focus\cite{Sprangle90a,Sprangle90b}. Therefore
to optically guide a LWFA driver it is necessary to use a density 
channel \cite{Leemans96,Antonsen92}.

\section{Charged Particle Beams}
  \label{sect:review-physics:particle-beams}

The evolution of the spotsize of an accelerated particle beam is 
determined
by its energy, number of particles, spotsize, and normalized 
emittance $\varepsilon_{n}$ where $\varepsilon_{n}$ is a measure of 
the area of the beam in transverse phase space. For a relativistic beam
(i.e., $\gamma\gg 1$), this area is 
given by the product of the beam's transverse spot size, $\sigma$, 
angular divergence, $\theta = \Delta p_{2}/p_{1}$, and energy, 
$\gamma \simeq p_{1}/mc$; therefore
$\varepsilon_{n} =\pi\gamma\theta\sigma \simeq \frac{\Delta p_{2}}{mc}\sigma$,
and it is conserved under ideal conditions.
This can be derived by showing that
Eq.~(\ref{equ:review-physics:p2-evol}) has the adiabatic invariant
$p_{2}x_{2}$ for each individual particle.

The envelope equation \cite{Lawson88} describes the evolution
of the beam's spotsize.

\begin{equation}
  \label{equ:review-physics:beam-envelope}
  \frac{d^{2}}{dx_{1}^{2}}\;\sigma 
  +\frac{1}{\gamma}\frac{d\gamma}{dx_{1}}\frac{d\sigma}{dx_{1}}
  -\left(\frac{\varepsilon_{n}}{\pi}\right)^{2}\frac{1}{\gamma^{2}\sigma^{3}}
  \left[1+\frac{2\pi^{2}}{\gamma}\left(\frac{\sigma}{\varepsilon_{n}}\right)^{2}
  \frac{I}{I_{A}}-\frac{\gamma\omega_{B}^{2}\sigma^{4}}{c^{2}}\left(
  \frac{\pi}{\varepsilon_{n}}\right)^{2}\right]=0
\end{equation}

\noindent{Here} I is the beam's current, $I_{a}\equiv mc^{3}/e$, 
is the Alfven current, and 
$\omega_{B}^{2}=2\left|\bar{\phi}_{0}\right|c^{2}/w_{p}^{2}$ is 
the betatron frequency for the potential given by
Eq.~(\ref{equ:review-physics:harmonic-potential}). The 
three terms in the bracket are due to I) diffraction, II) self space 
charge, and III) the external focusing forces (i.e., of the plasma 
wave), respectively.

The parameter characterizing the ratio of the space charge term to the 
diffraction term in the beam envelope is given by:

\begin{equation}
  \label{equ:review-physics:space-charge-parameter}
  \rho 
  =\frac{2\pi^{2}}{\gamma}\left(\frac{\sigma^{2}}{\varepsilon_{n}^{2}}\right)
  \frac{I}{I_{A}}
\end{equation}

If the effects of space charge can be neglected, then the equilibrium 
state of a matched beam ($\sigma$ doesn't change during the 
acceleration) can be obtained by balancing the two remaining force 
terms. These two terms are the one arising from the diffraction and 
the transverse external force term. The external force term can be 
related to the amplitude $E_{10}$ of the accelerating electric field 
of the plasma wave, which is a quantity we observe in our simulations, 
i.e., $\phi_{0}=-E_{10}/k_{p}$. The resulting condition for a 
matched beam is:

\begin{equation}
  \label{equ:review-physics:matched-beam}
  \frac{1}{4\pi^{2}\gamma}\frac{mc\omega_{p}}{eE_{10}}
  \left(\frac{\varepsilon_{n}}{\sigma}\right)^{2}\left(\frac{w_{L}}{\sigma}
  \right)^{2}=1
\end{equation}

\noindent{Here,} we also replace $w_{p}$ with $w_{L}/\sqrt{2}$, 
where $w_{L}$ is the laser spot size because the transverse profile 
of the longitudinal field of the plasma wave is proportional to the 
transverse profile of the laser intensity $E_{10}\propto E_{L}^{2}$, 
since the ponderomotive force of the laser pulses causes the plasma 
wake \cite{Fedele86,Sprangle88}. If the expression on the left 
side of the equation is larger than unity, the focusing forces 
dominate diffraction.

The evolution of the particle beam spotsize is the same as the 
evolution of the laser spotsize given by Eq.~(\ref{equ:review-physics:spot-size})
if all terms except the diffraction term on the right side of 
Eq.~(\ref{equ:review-physics:beam-envelope}) can be neglected and 
the assumptions $\frac{1}{\gamma} , \: \frac{d\gamma}{dt} \: \ll \: 1$ 
hold. In this case Eq.~(\ref{equ:review-physics:beam-envelope}) can be
rewritten as

\begin{equation}
  \label{equ:review-physics:beam-envelope-diffraction-only}
  \frac{d^{2}}{dx_{1}^{2}}\;\bar{\sigma}
  =\frac{1}{\left( \frac{\pi}{\varepsilon} \sigma_{0}^{2} \right)^{2}}
   \frac{1}{\bar{\sigma}^{3}}
\end{equation}

\noindent{where} $\sigma_{0}$ is the minimum spotsize, 
$\bar{\sigma}=\sigma/\sigma_{0}$ the normalized spotsize, and 
$\varepsilon = \varepsilon_{n}/\gamma$ the emittance. If we define
the quantity $\beta^{*} \equiv  \frac{\pi}{\varepsilon} \sigma_{0}^{2}$ and 
compare Eq.~(\ref{equ:review-physics:beam-envelope-diffraction-only})
with Eq.~(\ref{equ:review-physics:laser-envelope}) then it is clear
that $\beta^{*}$ of a particle beam corresponds to the Rayleigh length
$z_{R}$ of a laser and that the emittance $\varepsilon$ of a particle
beam corresponds to the wavelength of a laser. Since
Eq.~(\ref{equ:review-physics:spot-size}) is the solution of 
Eq.~(\ref{equ:review-physics:laser-envelope}) if only the diffraction
term is kept we can use Eq.~(\ref{equ:review-physics:spot-size}) also
to describe the evolution of a particle beam as long as the approximations
mentioned above hold.

\section{Wakefield Generation}
  \label{sect:review-physics:wake-generation}

The plasma wave wakes can be generated via beatwave, laser wakefield, 
Raman forward scattering or plasma wakefield excitation. For the 
purposes of this dissertation we only review wakefield excitation
by short laser or particle beams.
We begin with laser wakefield excitation. Two quantities that are helpful
to define first when discussing the generation of wakefields are the normalized
scalar potential 

\begin{equation}
  \label{equ:review-physics:normalized-scalar}
 \bar{\Phi} = e \Phi / \left( m_{e} c^{2} \right)
\end{equation}

\noindent{and} the normalized vector potential

\begin{equation}
  \label{equ:review-physics:normalized-vector}
 \vec{a} = e \vec{A} /  \left( m_{e} c^{2} \right)
\end{equation}

\noindent{The} maximum electric field that a plasma wave in a cold plasma
can support is determined by the amplitude at which the wave breaks 
\cite{Akhiezer56,Mori89}.
It is given by:

\begin{equation}
  \label{equ:review-physics:wake-breaking}
  E_{WB} \; = \; \sqrt{2} \; \left(\,\gamma_{p}\,-\,1\,\right)^{1/2} \; E_{0} 
  \left[ V/cm\right]
\end{equation}

\noindent{with} $\gamma_{p}= 1/\sqrt{1-\left(v_{\Phi}/c\right)^{2}}$ 
where $v_{\Phi}$ is the phase velocity of the plasma wave. $E_{0}$ is
here given by 

\begin{equation}
  \label{equ:review-physics:wake-breaking-non-relativistic}
  E_{0} = m_{e} \, c \, \omega_{p} / e
  \simeq 0.96 \: \sqrt{ \frac{n_{0}}{ \left[ cm^{-3} \right] } } 
  \; \left[ V/cm\right]
\end{equation}

The excitation of a plasma wake by a non-evolving circularly polarized
laser pulse with cylindrically-symmetric envelope can be
solved analytically \cite{Sprangle88}.
For a laser with a Gaussian profile of width $L$ in propagation direction
and with a peak normalized vector potential $a_{0}$ in a plasma with plasma
wavenumber $k_{p}$ the resulting amplitude of the excited wake is \cite{Esarey96}

\begin{equation}
  \label{equ:review-physics:wake-laser-gaussian1}
  E_{max} = E_{0} \left( \sqrt{\pi} a_{0}^{2} / 2 \right) k_{p} L
  e^{ - \frac{k_{p}^{2} \: L^{2}}{4}}
\end{equation}

\noindent{For} the optimal length $ L = \lambda_{p} / \left( \pi \sqrt{2} \right)$
this becomes

\begin{equation}
  \label{equ:review-physics:wake-laser-gaussian2}
  E_{max} = E_{0} \: a_{0}^{2} \left( \frac{\pi}{2e} \right)^{1/2}
  \simeq  E_{0} \: 0.76 \: a_{0}^{2}
\end{equation}

\noindent{The} normalized vector potential $a_{0}$ is related to the laser intensity $I$ by

\begin{equation}
  \label{equ:review-physics:laser-a0(i)}
  a_{0} = \left( 2 \: e^{2}\lambda_{L}^{2} I/
          \left( \pi m_{e}^{2} c^{5}  \right) \right)^{1/2}
\end{equation}

\noindent{$\lambda_{L}$} is here the laser wavelength. The total laser power
of a Gaussian beam with the spotsize $w_{0}$ is related to the intensity by

\begin{equation}
  \label{equ:review-physics:laser-intensity}
  I = 2 P / \left( \pi w_{0}^{2} \right)
\end{equation}

\noindent{This} can be combined to give a direct relationship
between power and normalized vector potential.

\begin{equation}
  \label{equ:review-physics:laser-power}
  P \simeq 21.5 \left[ GW \right] \left( a_{0} \, w_{0} / \lambda_{L} \right)^{2}
\end{equation}

\noindent{Substituting} this into 
Eq.~(\ref{equ:review-physics:wake-laser-gaussian2})
and assuming vacuum diffraction gives an estimate
for the maximum diffraction limited energy gain of

\begin{equation}
  \label{equ:review-physics:energy-gain-diffraction}
  \Delta W_{max,\, diff}
  \cong \int_{-z_{R}}^{z_{R}} \: e \, E_{max}\left( z \right) \: dz  
  \cong  e \, E_{max}\left( 0 \right) \: 2 z_{R}
  \cong 1.4\cdot 10^{3} \: mc^{2} \: 
  \frac{\omega_{p}}{\omega_{L}} \: \frac{P}{\left[TW\right]}
\end{equation}

Next we review plasma wakefield excitation. An expression for the wakefield
amplitude of a symmetric Gaussian electron bunch can be obtained from 2D linear
theory\cite{Katsouleas89}.

\begin{equation}
  \label{equ:review-physics:wake-particle-linear-large-rb}
   e\, E_{wake} \; = \; \sqrt{ \frac{n_{p}}{cm^{-3}}}
   \; \frac{eV}{cm} \times \frac{n_{b}}{n_{p}} \times 
   \frac{k_{p} \: \sigma_{z} \;
   e^{-k_{z}^{2}\sigma_{p}^{2}/2}}{1+\frac{1}{k_{p}^{2} \, \sigma_{r}^{2}}}
\end{equation}

\noindent{Here} $n_{p}$ is the plasma density, $n_{b}$ is the beam density, $k_{p}$ 
the plasma wave vector, $\sigma_{z}$ the width of the Gaussian in 
propagation direction, and $\sigma_{r}$ the width of the Gaussian 
perpendicular to the propagation direction.

In the blowout regime of a PWFA, most of the electron driver
bunch will propagate through the positively charged ion column created
by the blowout at the head of the beam\cite{Rosenzweig91}.
For highly relativistic beams, i.e., those for which
$\frac{1}{\gamma} , \: \frac{d\gamma}{dt} \: \ll \: 1$,
Eq.~(\ref{equ:review-physics:beam-envelope}) can be reduced to

\begin{equation}
  \label{equ:review-physics:beam-envelope-in-ion-column}
  \frac{d^{2}}{dx_{1}^{2}}\;\sigma \; + \; k_{B}^{2}\sigma \; = \; 0
\end{equation}

\noindent{where} $k_{B}^{2}$, is now due to the ion column. We can 
find the correct $k_{B}$  from Gauss' law applied to a uniformly
charged ion column using a cylindrical surface around the axis.
The radial electric field due to the ions is determined by

\begin{equation}
  \label{equ:review-physics:gauss-ion-column-1}
   2 \, \pi \, r \,E_{r} = \; 4\, \pi \,  e \, n_{p} \, \pi \, r^{2}
\end{equation}

\noindent{For} this case the force on the beam electrons is

\begin{equation}
  \label{equ:review-physics:gauss-ion-column-2}
  F_{r} = \left(-e\right) E_{r} \simeq \gamma \, m \, \frac{d^{2}}{dt^{2}}r
  = - \frac{4 \pi e^{2} n_{p}}{2} \: r^{2}
  = - \, m \, \omega_{B}^{2} \: r^{2}
\end{equation}

\noindent{with} $\omega_{B}=\omega_{p}/\sqrt{2}$ and therefore

\begin{equation}
  \label{equ:review-physics:beam-k-beta-in-ion-column}
   k_{B}^{2} = \frac{1}{c^{2}}\frac{\omega_{B}^{2}}{\gamma}
             =  \frac{4 \pi  e^{2} n_{p} }{2 \gamma m c^{2}}
\end{equation}

\noindent{gives} the $k_{B}$ due to an ion column in
Eq.~(\ref{equ:review-physics:beam-envelope-in-ion-column}).


  



\chapter{Review of Basic Particle-In-Cell Algorithms}
  \label{chap:review-codes}

There are many variations of the the Particle-In-Cell or PIC
method\cite{BirdsallLangdon}. This chapter will review the
general idea of the 
PIC method and then look at the
specific
algorithms implemented in the simulations codes used for
this dissertation, PEGASUS \cite{Tzeng96} and OSIRIS. Wherever possible
a short and concise review of the actual equations implemented
is given; otherwise a short description without the actual equations
of the implemented method is given. The 2D Cartesian algorithms described in this chapter are
common to PEGASUS and OSIRIS with the exception of one of the
current deposition schemes. The 3D Cartesian and 2D 
cylindrically-symmetric algorithms are only part of OSIRIS.

\section{The PIC-Method}
  \label{sect:review-codes:pic}

The basic equations governing the behavior of a plasma are
well known. Each particle moves according to
the Lorentz force exerted on it by the electromagnetic field
at its position.

\begin{equation}
  \label{equ:review-codes:lorentz-force}
  \vec{F} = q \left(\vec{E}+\frac{\vec{v}}{c}\times\vec{B}\right)
\end{equation}

The field in turn evolves according to Maxwell's Equations
with the sources given by the particles of the plasma.

\begin{equation}
  \label{equ:review-codes:maxwell-1}
  \vec{\nabla}\cdot\vec{E} = 4\pi\rho
\end{equation}

\begin{equation}
  \label{equ:review-codes:maxwell-2}
  \vec{\nabla}\times\vec{B} = \frac{1}{c} \frac{\partial\vec{E}}{\partial t}
                       + \frac{4\pi}{c}\vec{j}
\end{equation}

\begin{equation}
  \label{equ:review-codes:maxwell-3}
  -\vec{\nabla}\times\vec{E} = \frac{1}{c} \frac{\partial\vec{B}}{\partial t}
\end{equation}

\begin{equation}
  \label{equ:review-codes:maxwell-4}
   \vec{\nabla}\cdot\vec{B} = 0
\end{equation}

\begin{equation}
  \label{equ:review-codes:maxwell-jay}
   \vec{j}\left( \vec{x} \right) =
  \sum_{i=1}^{n} q_{i}v_{i}\delta \left(\vec{x}-\vec{x_{i}}\right)
\end{equation}

\begin{equation}
  \label{equ:review-codes:maxwell-rho}
  \rho \left( \vec{x} \right) =
  \sum_{i=1}^{n} q_{i}\delta\left(\vec{x}-\vec{x_{i}} \right)
\end{equation}

Together these equations perfectly describe a plasma and
in principle completely predict its behavior within the
limits of classical physics; but actually solving these
equations for a large collection of particles is computationally
challenging. 

\begin{figure}
   \begin{center}
      \epsfig{ file=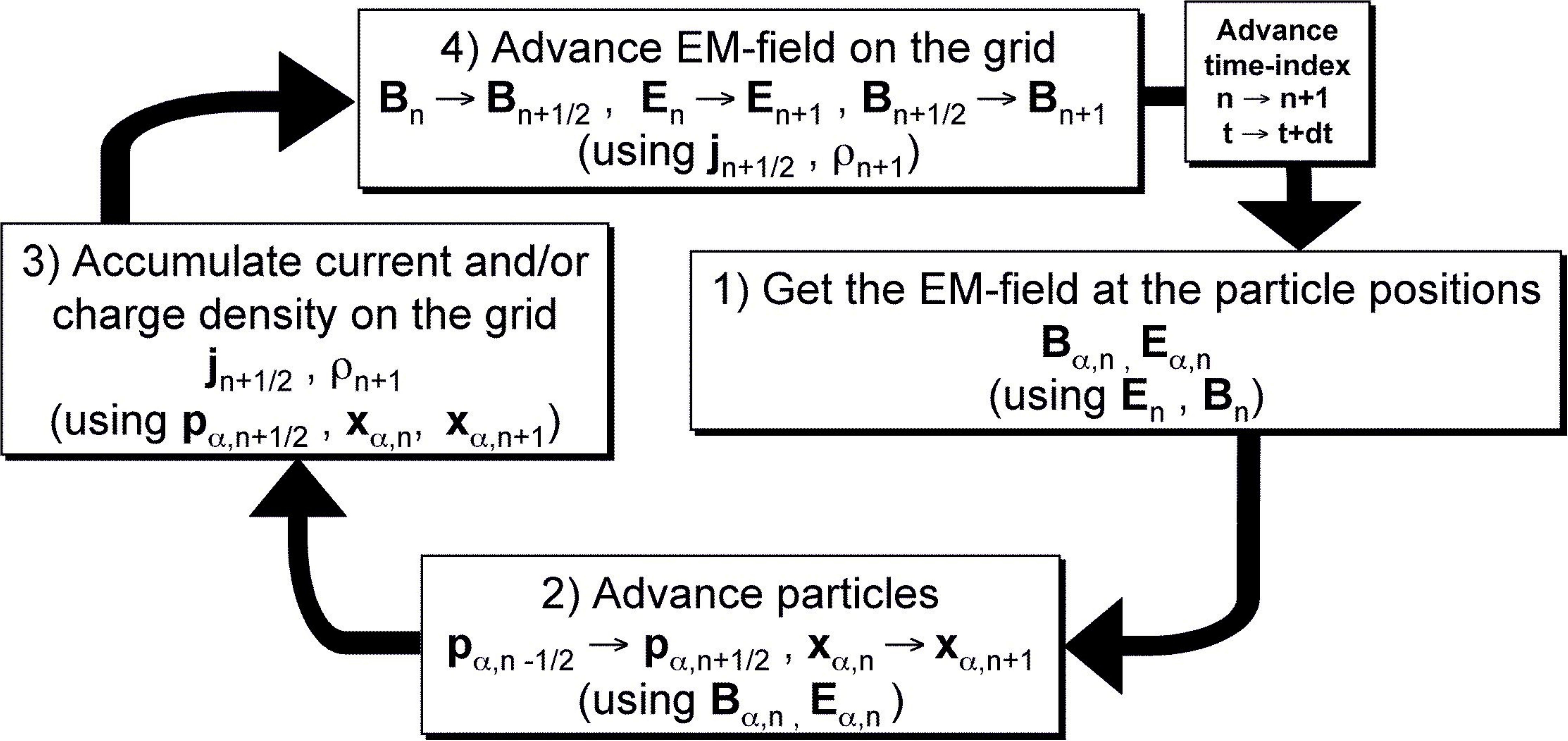, width=5.8in }
      \caption{The basic loop for PIC simulations. Time is increased
               in steps of $\Delta t$ so that $t = t_{o} + n \times \Delta t$}
      \label{fig:review-codes:pic-loop}
   \end{center}
\end{figure}

One way to solve these equations is the PIC
method \cite{BirdsallLangdon,Dawson83}. It breaks up the problem
into four distinct steps.
Fig.~\ref{fig:review-codes:pic-loop} shows these steps.
Given an initial configuration of particles with certain positions and
momenta, and electromagnetic field values known on a staggered grid that
is defined throughout the simulation space, a PIC-code first calculates
the fields at the particle positions by interpolating the fields on the
grid to the particle positions. The dimensions of the grid cells are chosen
to resolve the minimum wavelength of interest for the simulated problem.
The code then uses these fields and the particle information to calculate
the new positions and new momenta of the particles after a suitably chosen
timestep, dt. The updated position and momentum data are then used to find
the sources of the electromagnetic field, i.e., the current and the charge
density are deposited onto the grid. In the final step of the loop, 
the sources are used to advance the electromagnetic fields in time by
a timestep, dt, via Maxwell's equations.

In the following sections of this chapter we will review some of the
details for the numerical algorithms of this loop for the 2D cartesian,
3D cartesian, and 2D cylindrically-symmetric simulations that are
possible with OSIRIS.

\section{PIC Algorithms for 2D-Cartesian Simulations}
  \label{sect:review-codes:2d-slab}

\begin{figure}
   \begin{center}
      \epsfig{file=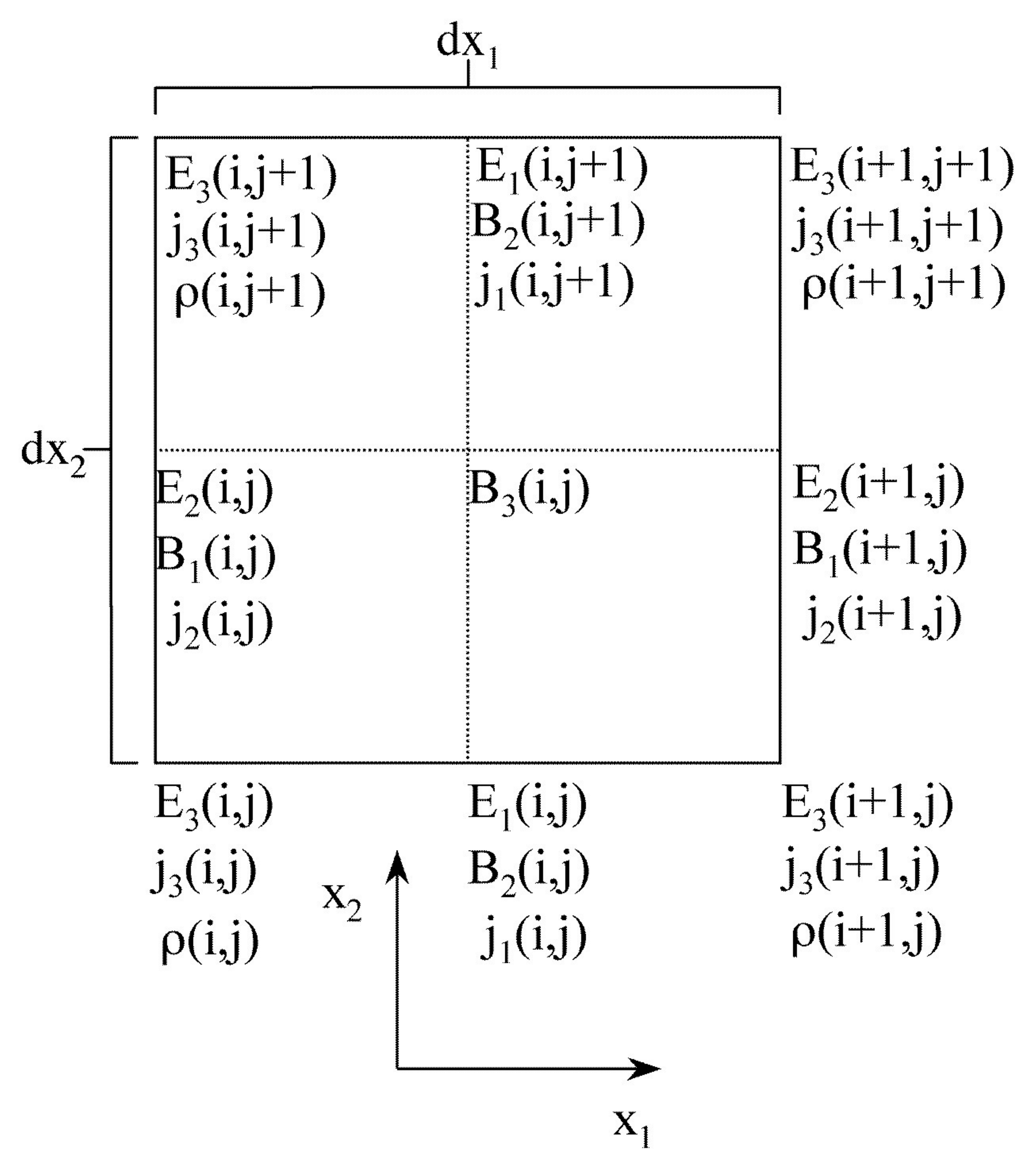, width=3.0in}
      \caption{The grid for a 2D PIC simulation. The staggered
               spacing of the $E$, $B$, and $j$ components,
               and of $\rho$ allows for 
               a higher precision of the calculations}
      \label{fig:review-codes:grid-2d}
   \end{center}
\end{figure}

Fig.~\ref{fig:review-codes:grid-2d} shows where the different quantities
are located on staggered grids in 2D simulations. The staggered grids 
are used since they increase the numerical accuracy \cite{BirdsallLangdon}.
The field solve in OSIRIS works in three different steps
to advance the fields at a grid point with the indices $i1$ and $i2$
by a timestep, $dt$, from a time index $n$ to a time index $n+1$.
It starts with the $\vec{E}$ and $\vec{B}$ fields fields at time $n$
and the current density $\vec{j}$ at the centered time $n+\frac{1}{2}$. 

The first step is to advance $\vec{B}^{n}$ by only half a timestep,
$dt/2$, to $\vec{B}^{n+\frac{1}{2}}$, using $\vec{E}^{n}$, through 
Faraday's law:

\begin{eqnarray}
       B_{1,i1,i2}^{n+\frac{1}{2}}
     = B_{1,i1,i2}^{n}
     - c \frac{dt}{2} \times
       \frac{E_{3,i1,i2+1}^{n}-E_{3,i1,i2}^{n}}{dx_{2}}
       \nonumber \\
       \label{equ:review-code:2d-field-solve-b-part1}
       B_{2,i1,i2}^{n+\frac{1}{2}}
     = B_{2,i1,i2}^{n}
     + c \frac{dt}{2} \times
       \frac{E_{3,i1+1,i2}^{n}-E_{3,i1,i2}^{n}}{dx_{1}}
       \\
       B_{3,i1,i2}^{n+\frac{1}{2}}
     = B_{3,i1,i2}^{n}
     - c \frac{dt}{2} \times
       \frac{E_{2,i1+1,i2}^{n}-E_{2,i1,i2}^{n}}{dx_{1}} 
       \nonumber \\
     + c \frac{dt}{2} \times
       \frac{E_{1,i1,i2+1}^{n}-E_{1,i1,i2}^{n}}{dx_{2}}
       \nonumber
\end{eqnarray}

The next step is to advance $\vec{E}^{n}$ by a full timestep, $dt$,
to $\vec{E}^{n+1}$, using $\vec{B}^{n+\frac{1}{2}}$ and
$\vec{j}^{n+\frac{1}{2}}$, through Ampere's law:

\begin{eqnarray}
       E_{1,i1,i2}^{n+1}
     = E_{1,i1,i2}^{n}
     - 4 \pi dt \times j_{1,i1,i2}^{n+\frac{1}{2}}
       \nonumber \\
     + c \: dt \times \frac{B_{3,i1, i2 }^{n+\frac{1}{2}}
                      -B_{3,i1,i2-1}^{n+\frac{1}{2}}}{dx_{2}}
       \nonumber \\
       \label{equ:review-code:2d-field-solve-e}
       E_{2,i1,i2}^{n+1}
     = E_{2,i1,i2}^{n}
     - 4 \pi dt \times j_{2,i1,i2}^{n+\frac{1}{2}}
       \\
     - c \: dt \times \frac{B_{3, i1 ,i2}^{n+\frac{1}{2}}
                      -B_{3,i1-1,i2}^{n+\frac{1}{2}}}{dx_{1}}
       \nonumber \\
       E_{3,i1,i2}^{n+1}
     = E_{3,i1,i2}^{n}
     - 4 \pi dt \times j_{1,i1,i2}^{n+\frac{1}{2}}
       \nonumber \\
     + c \: dt \times \frac{B_{2, i1 , i2 }^{n+\frac{1}{2}}
                      -B_{2,i1-1, i2 }^{n+\frac{1}{2}}}{dx_{1}}
       \nonumber \\
     - c \: dt \times \frac{B_{1, i1 , i2 }^{n+\frac{1}{2}}
                      -B_{1, i1 ,i2-1}^{n+\frac{1}{2}}}{dx_{2}}
       \nonumber
\end{eqnarray}

The final step is to again advance $\vec{B}$ by another half a 
timestep, $dt/2$,
from $\vec{B}^{n+\frac{1}{2}}$ to $\vec{B}^{n+1}$, using $\vec{E}^{n+1}$,
through Faraday's law:

\begin{eqnarray}
       B_{1,i1,i2}^{n+1}
     = B_{1,i1,i2}^{n+\frac{1}{2}}
     - c \frac{dt}{2} \times
       \frac{E_{3,i1,i2+1}^{n+1}-E_{3,i1,i2}^{n+1}}{dx_{2}}
       \nonumber \\
       \label{equ:review-code:2d-field-solve-b-part2}
       B_{2,i1,i2}^{n+1}
     = B_{2,i1,i2}^{n+\frac{1}{2}}
     + c \frac{dt}{2} \times
       \frac{E_{3,i1+1,i2}^{n+1}-E_{3,i1,i2}^{n+1}}{dx_{1}}
       \\
       B_{3,i1,i2}^{n+1}
     = B_{3,i1,i2}^{n+\frac{1}{2}}
     - c \frac{dt}{2} \times
       \frac{E_{2,i1+1,i2}^{n+1}-E_{2,i1,i2}^{n+1}}{dx_{1}} 
       \nonumber \\
     + c \frac{dt}{2} \times
       \frac{E_{1,i1,i2+1}^{n+1}-E_{1,i1,i2}^{n+1}}{dx_{2}}
       \nonumber
\end{eqnarray}

These equations above can be derived in a straight forward manner
from the differenced form of Maxwell's equation if the translational
invariance in $x_{3}$ is used to remove the $dx_{3}^{-1}$ terms in
the full equations.
The first and the second part of advancing $\vec{B}$ have
the same form and only the arguments differ. This is therefore implemented 
in the codes by calling the same subroutine once with $\vec{B}^{n}$ 
and $\vec{E}^{n}$ as arguments and then again later with
$\vec{B}^{n+\frac{1}{2}}$ and $\vec{E}^{n+1}$.
The benefit of splitting up the advancement of $\vec{B}$ is that
$\vec{E}$ and $\vec{B}$ after the advancement are both known at the
same time index $n$.
This keeps the particle push and the field solve time centered.
An equivalent implementation would be to set
$\vec{B}^{n+\frac{1}{2}}\;=\;
\frac{1}{2}\:\left( \vec{B}^{n+1} + \vec{B}^{n} \right)$,
but the above implementation requires less memory since $\vec{B}^{n}$ 
and $\vec{B}^{n+1}$ are not needed simultaneously.

The fields are interpolated to a particle position by weighting
each field component linearly from the particle's nearest four 
grid points for which each component of $\vec{E}$ or $\vec{B}$
is known. For example, if a particle is in the grid cell
$i1$ in $x_{1}$ and $i2$ in $x_{2}$ at a position
$\left( \varepsilon_{1} dx_{1} ,\varepsilon_{2} dx_{2} \right)$
within that grid cell then the field component $E_{3}$ at the
particle position is given by:

\begin{eqnarray}
  \label{equ:review-code:2d-field-weighting-e3}
  E_{3}  =  &   (1-\varepsilon_{1})   & (
           \:   (1-\varepsilon_{2})   \: E_{3, i1 , i2 }
           \: + \: \varepsilon_{2}    \: E_{3, i1 ,i2+1} \: )
        \\
         +  &      \varepsilon_{1}    & (
           \:   (1-\varepsilon_{2})   \: E_{3,i1+1, i2 }
           \: + \: \varepsilon_{2}    \: E_{3,i1+1,i2+1} \:)
        \nonumber
\end{eqnarray}
  
The fields at the particle's position are then used to update the momentum
of the particle. This happens in several steps in order to increase
the accuracy of the momentum push. A derivation of this method can
be found in the literature \cite{BirdsallLangdon}. This so called
Boris push can be summarized as,

\begin{eqnarray}
  \vec{p}\,'    & = & \vec{p}\,^{n-\frac{1}{2}}
                + q \: \frac{dt}{2} \: \vec{E}^{n} \nonumber \\                
  \label{equ:review-code:momentum-update}
  \vec{p}\,''   & = & \vec{p}\,' + q \: \frac{dt}{2} 
                    \: \vec{p}\,'  \times \vec{B}^{n}
                    \: \frac{1}{\sqrt{1 + \vec{p}\,'^{2}}}  \\                     
  \vec{p}\,'''  & = & \vec{p}\,' +  q \: dt
                    \: \vec{p}\,'' \times \vec{B}^{n}   
                    \: \frac{1}{\sqrt{1 + \vec{p}\,'^{2}}}
                    \: \frac{1}{ 1 + \left(\vec{B}^{n}\right)^{2}}  \nonumber \\
  \vec{p}\,^{n+\frac{1}{2}} & = & \vec{p}\,'''  +  q \frac{dt}{2} \vec{E}^{n} \nonumber  
\end{eqnarray}

\noindent{where} $\vec{p}^{n-\frac{1}{2}}$ and $\vec{p}^{n+\frac{1}{2}}$
are the momenta of the particle before and after the push.
The updated momentum, $\vec{p}^{n+\frac{1}{2}}$, is then used to update the particle position
according to:
         
\begin{eqnarray}               
  \label{equ:review-code:2d-position-update}
  \vec{x}\,^{n+1} & = & \vec{x}\,^{n} 
                  \; + \; q \: dt \:
                     \frac{\vec{p}\,^{n+\frac{1}{2}}}
                         {\sqrt{1 + \left(\vec{p}\,^{n+\frac{1}{2}}\right)^{2}}} 
\end{eqnarray}

\noindent{where} only the components in the simulation plane are 
updated.

For 2D simulations OSIRIS provides two different current deposition
algorithms\cite{Morse70,Villasenor92,Eastwood90} 
both of which use the old position $\vec{x}\,^{n}$ and the new position 
$\vec{x}\,^{n+1}$
of each particle to calculate the current on the grid. Both of these 
current deposition schemes have in common that they rigorously obey
the continuity equation:

\begin{equation}               
  \label{equ:review-code:continuity}
  \vec{\nabla} \cdot \vec{j} + \frac{\partial \rho}{\partial t} = 0
\end{equation}

\noindent{However}, this leaves the possibility of adding an arbitrary curl
to the current. Therefore, both of these charge conserving
algorithms give the same value for $\vec{\nabla} \cdot \vec{j}$
but sometimes give different values for $\vec{\nabla} \times \vec{j}$.

In both methods a particle is viewed as a finite size particle which 
contributes a charge density $\rho$ to the nearest grids using a 
weighting function.
In OSIRIS, the weighting 
functions differ between the methods but this is not the fundamental
difference between them. Both methods are based on the idea that the
contribution of a particle to the charge density on the grid before and 
after the push can be used to infer the current that has to be 
assigned to the grid. Due to these common
ideas both methods satisfy Eq.~(\ref{equ:review-code:continuity}). If a
particle stays within a cell during a timestep then both methods give
the same answer.
The difference between the methods lies in the assumed paths for a particle 
when it crosses a cell boundary.

\begin{figure}
   \begin{center}
      \epsfig{file=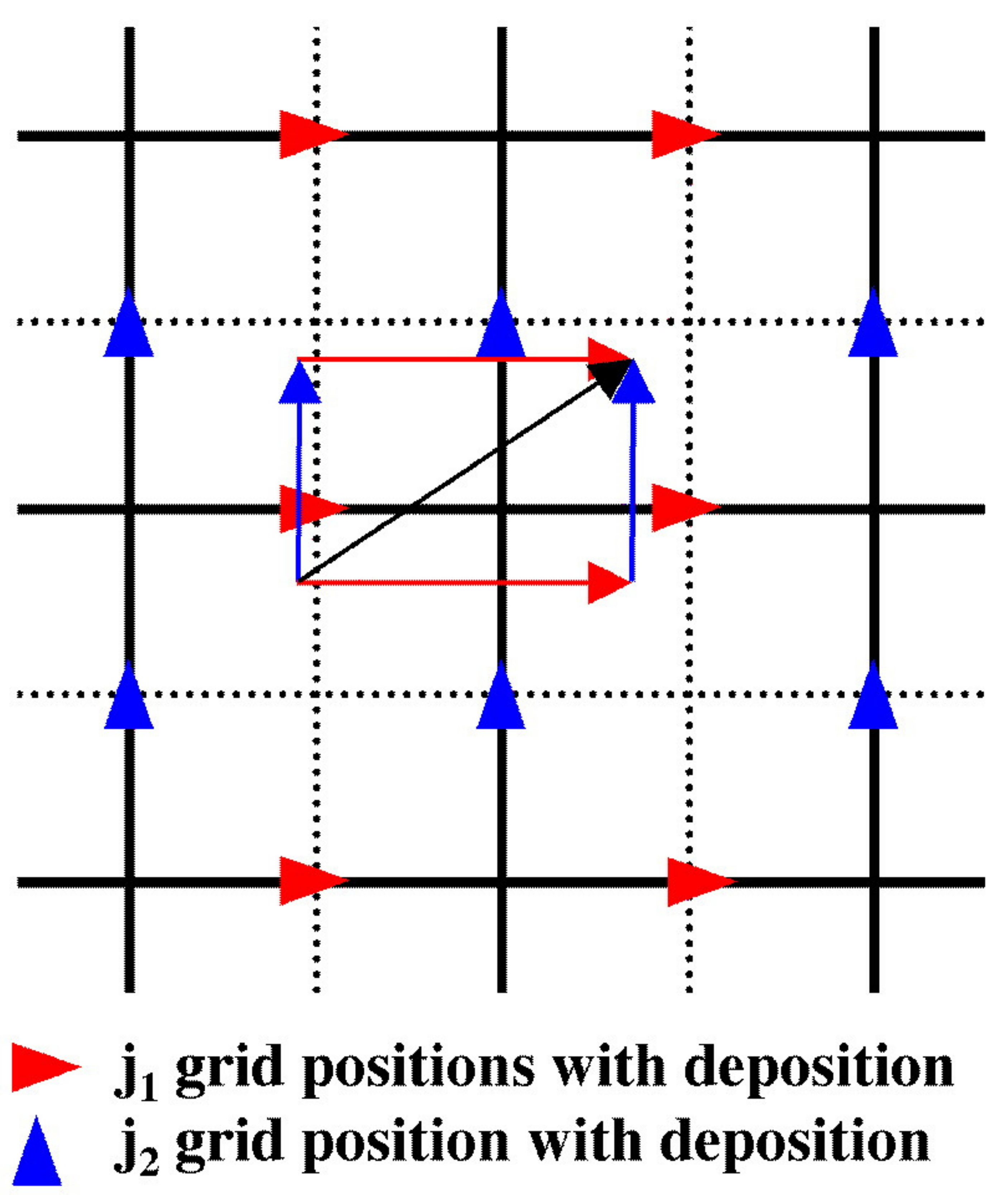, width=3.0in}
      \caption{The figure shows an example of the paths over which
               the ISIS method averages using the particle position
               before and after the particle push. It also shows where
               current is deposited on the grid for the case of this example.}
      \label{fig:review-codes:depos-isis}
   \end{center}
\end{figure}

\begin{figure}
   \begin{center}
      \epsfig{file=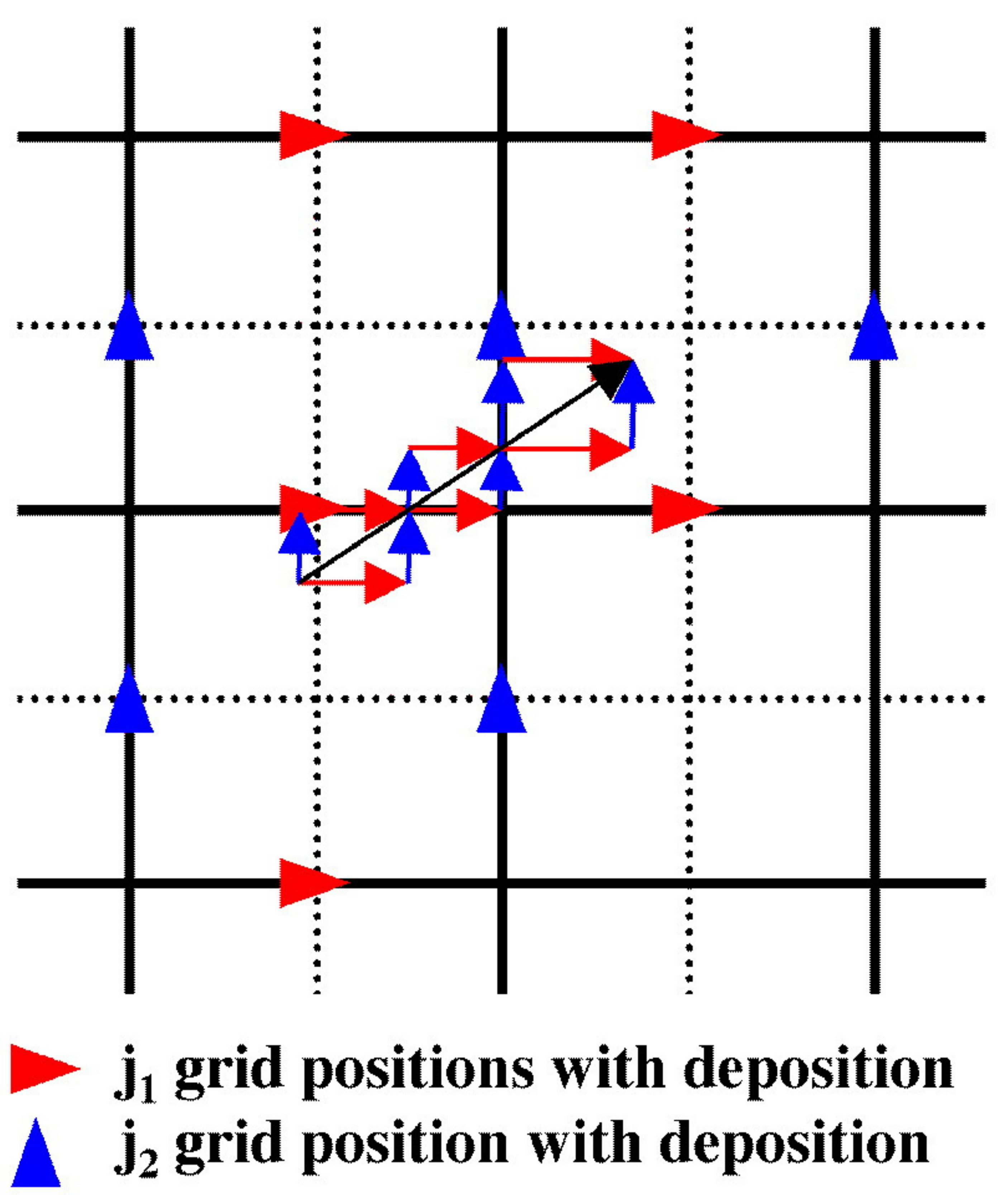, width=3.0in}
      \caption{The figure shows an example of the paths over which
               the TRISTAN method averages using the particle position
               before and after the particle push. It also shows where
               current is deposited on the grid for the case of this example.}
      \label{fig:review-codes:depos-tristan}
   \end{center}
\end{figure}

An example for the paths that the first method, which
is also implemented in PEGASUS, assumes is illustrated in
Fig.~\ref{fig:review-codes:depos-isis}. This scheme was taken 
directly from ISIS and it is similar in spirit to appendix A of 
Ref.\cite{Morse70}. Fig.~\ref{fig:review-codes:depos-tristan} on
the other hand gives an example for the paths that the virtual particle
method\cite{Villasenor92,Eastwood90} assumes for 
the same actual particle motion as shown in
Fig.~\ref{fig:review-codes:depos-isis}. This method is identical to
that of Villasenor and Buneman and it is the one used in the well known
TRISTAN code\cite{Villasenor92,Buneman92}.

The differences between the two methods, which we will refer to as ISIS or 
TRISTAN methods, can be understood by noting 
that in two dimensions any straight line trajectory can be decomposed 
into two orthogonal moves. If we define
$\Delta x_{1}\equiv x_{1}^{n+1} - x_{1}^{n}$ and
$\Delta x_{2}\equiv x_{2}^{n+1} - x_{2}^{n}$
as the changes in each 2D coordinate during a push then the trajectory
can be viewed as
$\hat{x}_{1} \Delta x_{1} + \hat{x}_{2} \Delta x_{2}$.
However, if this motion is used to determine the current, there is an 
ambiguity between  letting the particle first move in $\hat{x}_{1}$ 
and then in $\hat{x}_{2}$ or vice versa. The difference in the paths 
is an overall current loop, so there is an ambiguity in the curl. In 
the ISIS algorithm the current deposited is that from the average of 
the two moves. In the TRISTAN (or virtual particle) method the same 
procedure is used as long as the particle stays within a cell.
However, when it doesn't, as in the case shown in
Fig.~\ref{fig:review-codes:depos-isis} and
Fig.~\ref{fig:review-codes:depos-tristan},
the complete straight line trajectory is broken up into two straight line paths
which connect at the cell face boundaries, and each separate straight 
line is then broken up as the average of the two types of orthogonal moves 
as described earlier. The TRISTAN method is more
accurate since it approximates the path of a particle more closely,
but it is also computationally more expensive.

\section{3D-Cartesian and 2D-Cylindrically-Symmetric Algorithms}
  \label{sect:review-codes:3D-2D-cyl} 

\begin{figure}
   \begin{center}
      \epsfig{file=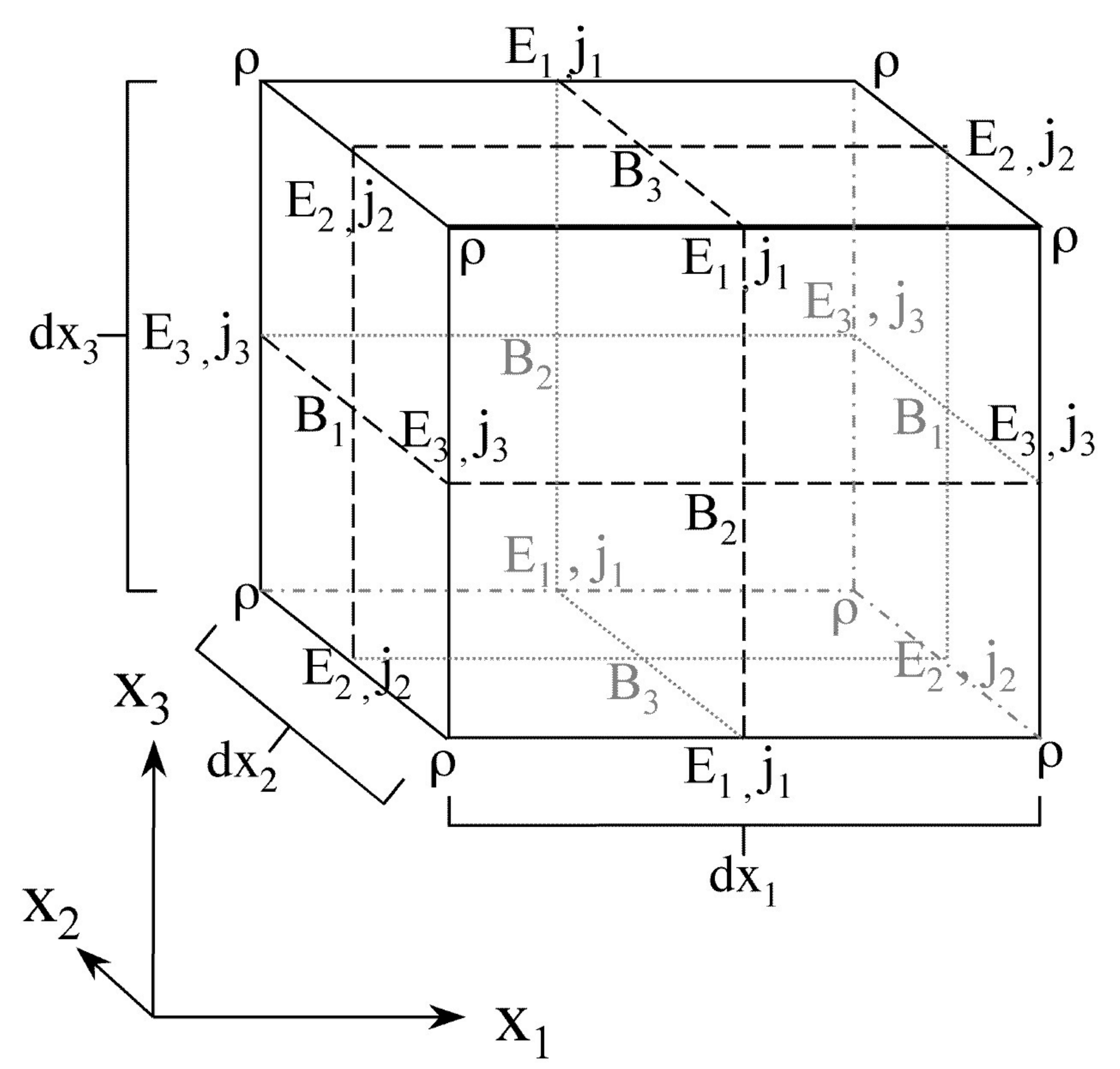, width=3.0in }
      \caption{The grid for a 3D PIC simulation. The staggered
               placing of the $E$, $B$, and $j$ components,
               and of $\rho$ allows for 
               a higher precision of the calculations}
      \label{fig:review-codes:grid-3d}
   \end{center}
\end{figure}

The field solve for 3D Cartesian simulations is a simple extension to the one
used for 2D Cartesian simulations. In Fig.~\ref{fig:review-codes:grid-3d} 
we show how the different quantities are staggered on a 3D grid. 
The main difference between the 2D and 3D setup is that there are
additional terms in the calculation of $E_{1}$, $E_{2}$, $B_{1}$,
and $B_{2}$, because derivatives with respect to $x_{3}$ are not
assumed to be zero. Some other straightforward changes arise from the
additional staggering of some grid quantities in $x_{3}$. This
can also be seen in Fig.~\ref{fig:review-codes:grid-3d}.

%
%
%
%
%

\begin{figure}
   \begin{center}
      \epsfig{file=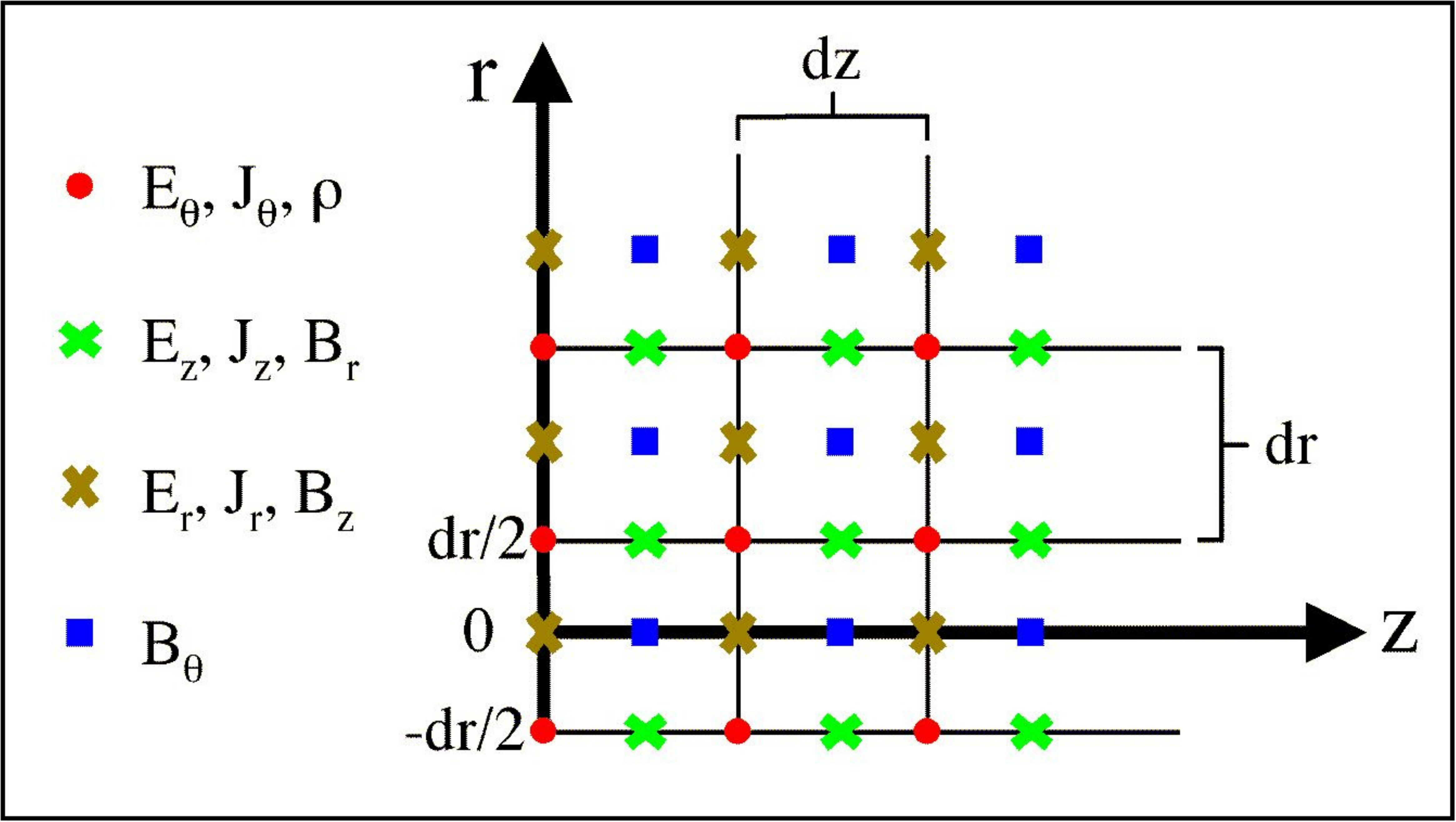, width=5.0in}
      \caption{The grid for a 2D cylindrically-symmetric PIC simulation.
               The axis has been placed through the middle of the first
               grid cell in order to avoid having to calculate $j_{z}$
               on for $r=0$.}
      \label{fig:review-codes:grid-2d-cyl}
   \end{center}
\end{figure}

On the other hand the 2D cylindrically symmetric field solve algorithm
is not a straightforward extension of the 2D Cartesian case.
In Fig.~\ref{fig:review-codes:grid-2d-cyl} the grid for the
2D cylindrically symmetric case is shown. This grid can be derived from
the 2D Cartesian grid by making the substitutions:

$dx_{1} \rightarrow dz$, $dx_{2} \rightarrow dr$ \linebreak
$B_{1} \rightarrow B_{z}$, $B_{2} \rightarrow B_{r}$,
$B_{3} \rightarrow B_{\Theta}$ \linebreak
$E_{1} \rightarrow E_{z}$, $E_{2} \rightarrow E_{r}$,
$E_{3} \rightarrow E_{\Theta}$ \linebreak
$j_{1} \rightarrow j_{z}$, $j_{2} \rightarrow j_{r}$,
$j_{3} \rightarrow j_{\Theta}$ \linebreak

\noindent{In} the field solve for 2D cylindrically-symmetric simulations
the way $E_{z}$ and $B_{z}$ are calculated differs
significantly from the way $E_{1}$ and $B_{1}$ are calculated for
2D Cartesian simulations.
For the other components of
the fields it is possible to derive the correct equations for the
2D cylindrically-symmetric case by applying the substitutions above.
The equations for $z$-components have an explicit dependency on $r$
in cylindrical coordinates and the modified equations for the
$x_{1}$-components of 
Eq.~(\ref{equ:review-code:2d-field-solve-b-part1}),
Eq.~(\ref{equ:review-code:2d-field-solve-e}), and
Eq.~(\ref{equ:review-code:2d-field-solve-b-part2}) are: 

\begin{eqnarray}
       B_{z,i1,i2}^{n+\frac{1}{2}} & = & B_{z,i1,i2}^{n}
     - c \frac{dt}{2} \times \frac{1}{r_{i2+\frac{1}{2}}}
       \frac{ r_{i2+1} E_{\Theta,i1,i2+1}^{n}
            - r_{ i2 } E_{\Theta,i1, i2 }^{n}}{dr}
       \nonumber \\
       \label{equ:review-code:2d-field-solve-cyl}
       E_{z,i1,i2}^{n+1} & = & E_{z,i1,i2}^{n}
     - 4 \pi dt \times j_{z,i1,i2}^{n+\frac{1}{2}}
       \\
 & &  + c \: dt \times \frac{1}{r_{i2}}
       \frac{ r_{i2+\frac{1}{2}} B_{\Theta, i1 ,i2}^{n+\frac{1}{2}}
            - r_{i2-\frac{1}{2}} B_{\Theta,i1-1,i2}^{n+\frac{1}{2}}}{dr}
       \nonumber \\
       B_{z,i1,i2}^{n+1} & = & B_{z,i1,i2}^{n+\frac{1}{2}}
     - c \frac{dt}{2} \times \frac{1}{r_{i2+\frac{1}{2}}}
       \frac{ r_{i2+1} E_{\Theta,i1,i2+1}^{n+1}
            - r_{ i2 } E_{\Theta,i1, i2 }^{n+1}}{dr} \nonumber
\end{eqnarray}

Here the following conventions are used:
$i1$ is the grid index for $z$ and $i2$ is the grid index for $r$.
The grid cells on axis in Fig.~\ref{fig:review-codes:grid-2d-cyl}
have the index $i2=1$ and $r_{i2} = dr \: (i2-\frac{3}{2})$.

The calculation of the field right on the $z$-axis is also not straightforward.
In OSIRIS the $r=0$ axis is not the lower boundary of the simulations.
The simulation space extends to $r=-dr/2$ as shown in
Fig.~\ref{fig:review-codes:grid-2d-cyl}.
This is done in order to avoid having to calculate $E_{z}$
on axis. It is easier to calculate $B_{z}$ since it does not require 
interpolating the current component $j_{z}$ to the $r=0$ axis. It is
interesting to note that we  initially used the $r=0$ as the lower boundary.
This led to substantial short-wavelength ($\lambda\sim dz$) noise near the
$r=0$ axis. Based on work by Seung Lee 
the present algorithm for the cylindrically-symmetric field solve was
implemented into OSIRIS.

For the chosen staggered grid the axial boundary conditions also differ
for the vector field components parallel and perpendicular to the axis.
In particular for the perpendicular components we have:

\noindent{if} $f\left(z,r\right)=B_{r}$, $B_{\Theta}$,
$E_{r}$, $E_{\Theta}$, $j_{r}$, $j_{\Theta}$ then 
$f(z,r) = -f(z,-r) \rightarrow f(z,0) = 0$

\noindent{While} for the parallel components (and scalars) we have:

\noindent{if} $f(z,r) = B_{z}$, $E_{z}$, $j_{z}$, $\rho$ then
$f(z,r) = f(z,-r) \rightarrow $ no restrictions for $f(z,0)$

Here $f(z,r)$ stands for a generic field component used in the
simulation. With these boundary conditions and
the standard difference equations for the fields, we can find all the required
quantities in the grid cell on axis except $B_{z}$. We use the 
integral form of Faraday's law to find $B_{z}$ on axis by integrating
$\oint\vec{E}\cdot d\vec{l}$ around a loop half a grid cell off axis. The change
of $B_{z}$ on axis is then given by: 

\begin{equation}               
  \label{equ:review-code:2d-field-solve-cyl-b1-axis}
  B_{z,i1,1}^{n+\frac{1}{2}} = B_{z,i1,1}^{n}
  - \;4 \: c \: \frac{dt}{2} \times \frac{ E_{\Theta,i1,2}^{n}}{dr}
\end{equation}

\noindent{Note} that again this advance of $B_{z}$ has to be applied a second
time - after $\vec{E}$ has been advanced - in order to obtain $B_{z,i1,1}^{n+1}$.

The weighting of the $\vec{E}$ and $\vec{B}$-fields to the particle positions
works the same in 2D cylindrically-symmetric simulation as in 2D Cartesian and
the method is straightforward to extend to a 3D algorithm. Extending 
the push is even simpler because the 2D and 3D Cartesian, and 2D cylindrically symmetric
simulations in OSIRIS all use the same subroutine for the momentum
update. The algorithm was described above, but it is worth commenting 
on the 2D cylindrically-symmetric algorithm. The reason why the same algorithm can
still be used in
the 2D cylindrically-symmetric coordinates is that the
2D cylindrically-symmetric algorithms keep track of
$\vec{p} = \left(p_{z}, p_{r}, p_{\Theta} \right)$. This is the momentum vector
represented in the local Cartesian coordinate system at any given point
$\left[ \vec{e_{z}}, \vec{e_{r}}, \vec{e_{\Theta}}\right]$ and 
therefore it can
be updated by a momentum update in Cartesian coordinates.

\begin{figure}
   \begin{center}
      \epsfig{file=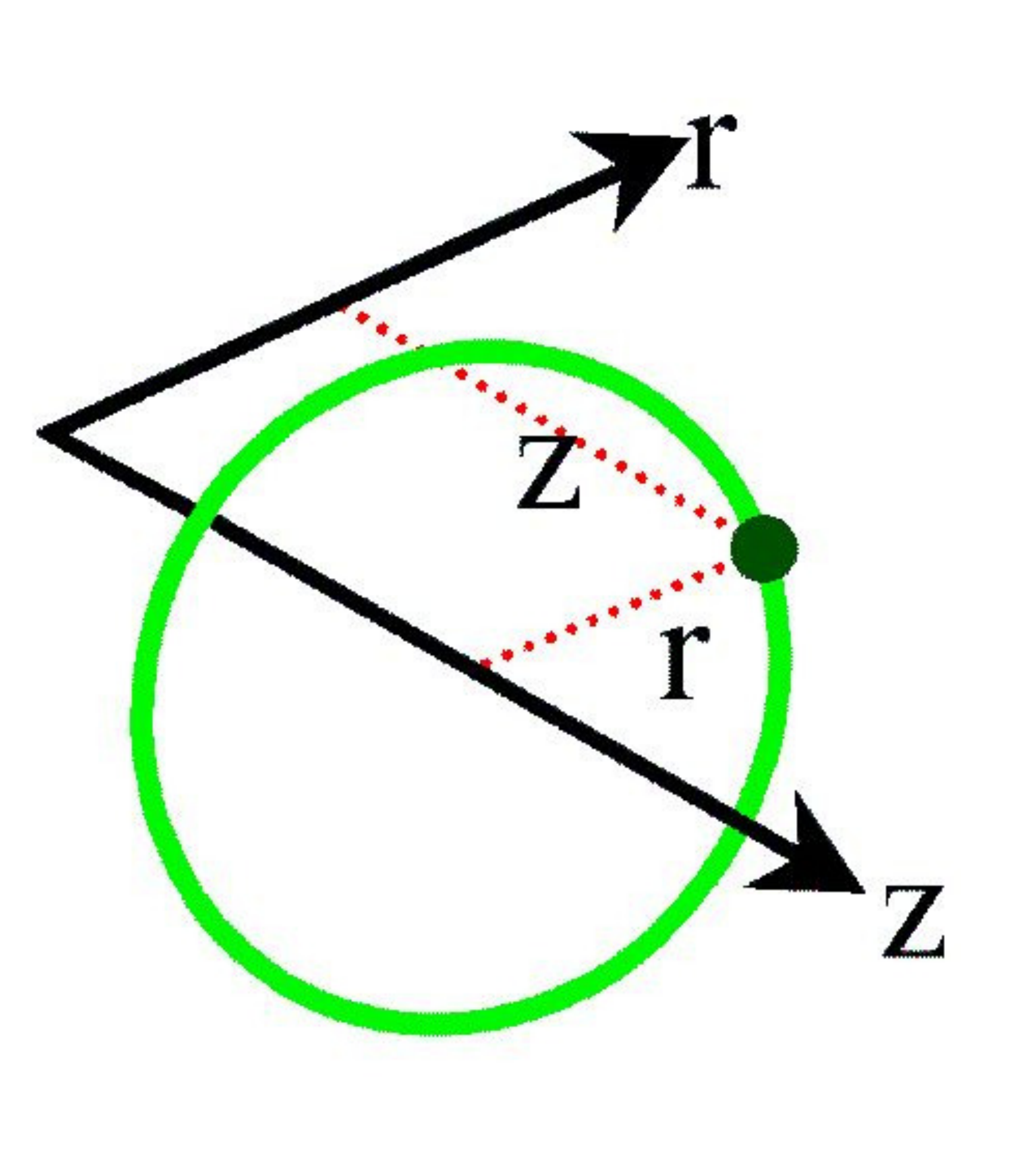, width=3.0in}
      \caption{The charge represented by a simulation particle in
               the $r$-$z$-plane is a ring}
      \label{fig:review-codes:ring-charge}
   \end{center}
\end{figure}

\begin{figure}
   \begin{center}
      \epsfig{file=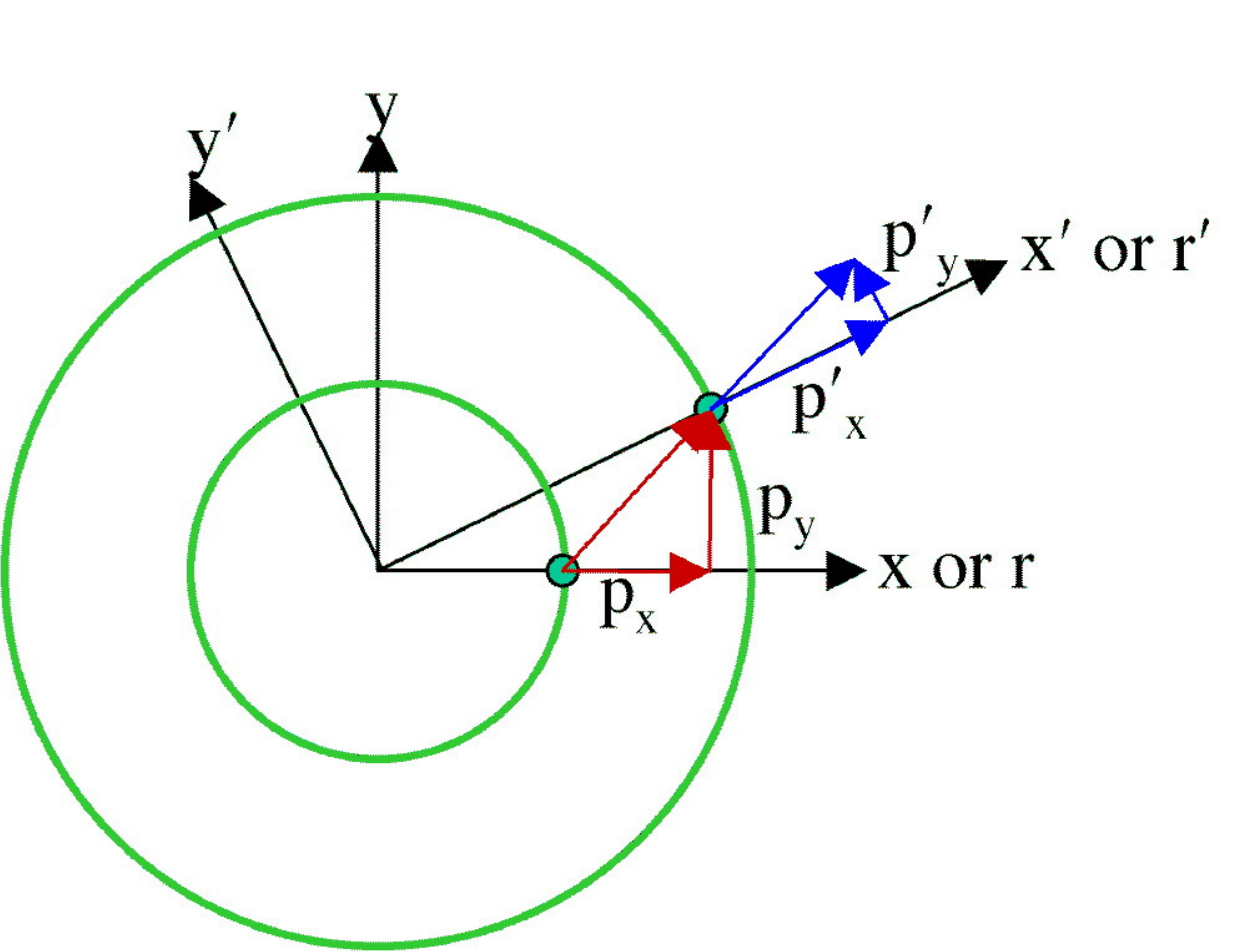, width=3.0in}
      \caption{The position update for a particle in 2D cylindrically-symmetric
               coordinates. The use of a ``pseudo'' 3D push followed by a
               rotation also requires an update of the momentum.}
      \label{fig:review-codes:2d-cyl-push}
   \end{center}
\end{figure}

The position update for 3D Cartesian Coordinates is a straightforward extension
from the 2D Cartesian algorithm and does not need to be described any further.
However, the position update for 2D cylindrically symmetric is more complicated and
requires an understanding of what is actually represented by a simulation
particle in a 2D cylindrically-symmetric simulation.
As shown in Fig.~\ref{fig:review-codes:ring-charge}, a simulation particle
in this type of simulation represents a ring of charge.
Combining this with the fact that the momentum for the particle is known in the
local Cartesian coordinates of the particle, leads to a method of position
update. The particle position is first updated in a 3D Cartesian space
in a way that is identical to the one in 3D Cartesian simulations. Afterwards
all $\vec{x}$ and $\vec{p}$ quantities are transformed into a rotated
frame of reference that eliminates the change in position in the third dimension
\cite{BirdsallLangdon}.
This is illustrated in Fig.~\ref{fig:review-codes:2d-cyl-push} which 
shows the Cartesian position update in the $x$-$y$-plane with successive rotation from
$r$ to $r'$. The $x$ and $y$ coordinates in this figure are used to describe the
plane transverse to the $z$ axis. For the momentum, the substitutions
$p_{r} \rightarrow p_{x}$, and
$p_{\Theta} \rightarrow p_{y}$ are used. It should be noted that there 
is an additional
change of the momentum vector caused by the rotation of the coordinate system.
The next paragraph will give a more detailed explanation of the 2D cylindrically-symmetric
position update.

Once the new momenta are calculated each particle is pushed to its
new positions as follows. 
First, the a ``pseudo'' 3D push is done to get the new temporary position vector
in a 3D Cartesian representation,
  
\begin{eqnarray}               
  \label{equ:review-code:2d-cyl-position-update-1}
  \vec{x}_{3D,new}^{n+1} & = &
  \left(\vec{x}_{2D}^{n}, 0\right) 
  + q \cdot dt \frac{\vec{p}^{n+\frac{1}{2}}}
  {\sqrt{1 + \left(\vec{p}^{n+\frac{1}{2}}\right)^{2}}}
\end{eqnarray}

\noindent{Secondly}, the new position vector for the particle in 2D cylindrical
coordinates is obtained in the rotated coordinate system:

\begin{eqnarray} 
  \label{equ:review-code:2d-cyl-position-update-2}
  x_{1,2D}^{n+1} & = & x_{1,3D,new} \\ 
  x_{2,2D}^{n+1} & = &
  \sqrt{x_{2,3D,new}^{2}+x_{3,3D,new}^{2}} = r^{n+1} \nonumber
\end{eqnarray} 

\noindent{Finally} the momentum vector in the new rotated coordinate system is
calculated: \linebreak
(using $r^{n}=x_{2}^{n}$)

\begin{eqnarray}  
  \label{equ:review-code:2d-cyl-position-update-3}
  p_{2,new}^{n+\frac{1}{2}} & = &
  \left( p_{2}^{n+\frac{1}{2}} x_{2,3D,new}
  + p_{3}^{n+\frac{1}{2}} x_{3,3D,new} \right) / r^{n+1} \\
  p_{3,new}^{n+\frac{1}{2}} & = & p_{3}^{n+\frac{1}{2}} r^{n} / r^{n+1} \nonumber
\end{eqnarray}     

There is a subtle detail about the calculation of $p_{3,new}^{n+\frac{1}{2}}$
that is easy to miss. The transformation of the components of $\vec{p}$ 
suggested in Fig.~\ref{fig:review-codes:2d-cyl-push} is correct for a
point particle. However, the code assumes ``ring particles'', so for a 
uniform ring any contribution that the component $p_{2}$ in the old coordinate 
system will make to the new $p_{3,new}$ after the rotation has to 
cancel out for reasons of symmetry. The equation for 
$p_{3,new}^{n+\frac{1}{2}}$
above  takes this fact into account, because it is derived from the 
conservation of angular momentum instead of using the straight forward
transformation of the momentum of a point particle.

The current deposition scheme used for 3D simulations in OSIRIS is a simple 
extension of the 2D TRISTAN deposition method to three 
dimensions and it is well described in the
literature\cite{Villasenor92,Eastwood90}.
The deposition algorithms and subroutines used for the 2D 
cylindrically-symmetric simulations are the same as for the 2D cartesian
simulations. This is true for ISIS method as well as for the TRISTAN method
and it is possible because the quantity which is deposited 
is not the current density  but the current due to the particles.
The difference between the current deposition schemes in the 2D 
Cartesian and 2D cylindrically-symmetric operating modes of OSIRIS
is simply in calculating the current density, $\vec{j}$ from the 
current, $\vec{I}$. 
In the cylindrical geometry the volume of a grid cell depends on its
radial position. For a grid cell with a distance r from the axis we
get $j_{ r  } \; = \; I_{ r  }/V\left(r\right)$, 
    $j_{\Phi} \; = \; I_{\Phi}/V\left(r\right)$, and
    $j_{ z  } \; = \; I_{ z  }/V\left(r\right)$ with
$V\left(r\right) = dr\:r\,d\Phi\:dz$.


\chapter{The Implementation of the Object-Oriented Code OSIRIS}
  \label{chap:os-algorithms}

In this chapter, we describe the strategy used to develop OSIRIS 
and the important features in object oriented structure of code.

\section{Development Strategy and Code Design}
  \label{sect:os-algorithms:basic-design}
  
The code OSIRIS is written in the programing language Fortran90\cite{MetcalfReid}
and is implemented using an object-oriented style of problem solving
\cite{Rumbaugh}. It is the first fully object oriented, multi-dimensional,
electromagnetic PIC code written in Fortran90 that is also being used to
undertake large scale production runs. As a result, it has been used to
gain valuable new insights into physics problems. Several of these will
be discussed in the subsequent chapters.

The central goal of developing OSIRIS was to get a code that would 
support multiple algorithms for multi-dimensional PIC simulations in
a distributed computing environment by using multi-dimensional domain
decomposition. An additional, essential requirement for OSIRIS was the
implementation of the dynamic simulation space concept, which is explained
below. In order to achieve these goals 
the development of OSIRIS followed a step by step process that allowed
each step to be verified by comparing the code's results to previous
results. This was done as follows:

\begin{enumerate}

\item An Object oriented single-node 2D PIC code based on
the numerical algorithms of ISIS/Pegasus \cite{Tzeng96} was developed.

\item The dynamic simulation space algorithm was implemented.

\item Parallelization was implemented.

\item OSIRIS was ported to several different architectures.

\item A 2D cylindrically-symmetric algorithm was implemented.

\item A fully 3D algorithm was implemented

\item New algorithms and physics packages are continually incorporated.

\end{enumerate}

Even though the development of OSIRIS had these distinct
stages there were a number of general principles that were
used in designing the objects and algorithms of the code throughout
the development. These principles were motivated by the eventual
goal of the code development which was explained above.
The general principles we used were:

\begin{itemize}

\item All real physical quantities should have a corresponding
object in the code and distinct physical processes should have a
corresponding application of a method in the main loop of the code.
Following this principle makes the physics being modeled in the code
clear and therfore easier to modify and extend.

\item There should be, as far as possible, a distinction between the
physical objects of the code, e.g., the particle object,
and the numerical objects of the code, e.g., the
grid object. Physical objects encapsulate information about physical
aspects of the simulation. Numerical objects encapsulate 
information about the numerical algorithms being used. This isolation
of numerical and physical aspects allows one to change one of them without
having to change the other one.

\item All information required frequently throughout the simulation
should be declared as a variable in the main program. This gives clarity
about which information is available at any point in time, which 
means that unintended changes of variable values are less likely to
happen. This increases the safety and reliability of the code.

\item The input file of the code should define as far as possible only
the global physical problem to be simulated. Node specific information
should be avoided as far as possible. In this way the user of the code
can focus on defining the physical problem and does not have to be
concerned with parallelization issues.

\item As far as possible all classes and objects should refer to a 
single node and should not be affected by parallelization issues. This
is realized by treating all communication of physical objects
between nodes as boundary conditions for the physical object
on each node. This strategy simplifies incorporation of new 
algorithms into the code.

\item The code should be written in such a way that is largely independent
from the dimensionality or the coordinate system used in order to allow for
polymorphism\cite{Decyk98b}. This way much of the code can be reused when incorporating 
a new algorithm with a different dimensionality or with a different
coordinate system.

\item Objects and methods should be designed to allow for easy incorporation
of old but fast Fortran77 style algorithms as subroutines. This should
make it easy to incorporate algorithms from one of the many
Fortran77 legacy codes.

\item All system dependent parts of the code should be encapsulated in 
as few modules as possible. This ensures easy portability of the code 
to other systems.

\end{itemize}

In addition to these general principles there are a couple of conventions
used in OSIRIS. These conventions are listed here
for the benefit of people who are interested in understanding or modifying
OSIRIS for future research. With few exceptions these conventions are applied
throughout most of the code.

\begin{itemize}

\item Subroutines/Methods modifying specific data are in the same module/class as
the type-/object-definition of those data. This is a general principle of
object-oriented programming. It is not maintained in this code for certain
utility modules which do not contain any type/object definition but supporting
subroutines that are used by more than one other module/class. Another
exception is the VDF-class described below. It provides direct access 
to the arrays of different dimensionality it contains in order to allow
for polymorphism in the code.

\item With the exceptions of dummy arguments with the pointer attribute all
dummy arguments in subroutines are declared with an intent. (It is not
possible to declare an intent for dummy arguments with the pointer attribute.)


\item In the argument list of any subroutine first the arguments
with the ``intent(out)'' attribute, then the arguments
with the ``intent(inout)'' attribute, and finally the arguments
with the ``intent(in)'' attribute are listed.

\item In the declaration part of any subroutine
the order in which dummy arguments are declared corresponds to their order
in the subroutine call argument list. Exceptions are commented on in the code.

\item The names of modules have the structure m\_$<$rest~of~the~name$>$.
The names of types have the structure t\_$<$rest~of~the~name$>$.
If a type is defined in a module then $<$rest~of~the~name$>$ is
the same for the type and the module.

\item The names of all compile-time parameters have the form
p\_$<$rest~of~the~name$>$.

\end{itemize}

\section{High Level Description}
  \label{sect:os-algorithms:high-level-description}

\begin{figure}
   \begin{center}
      \epsfig{file=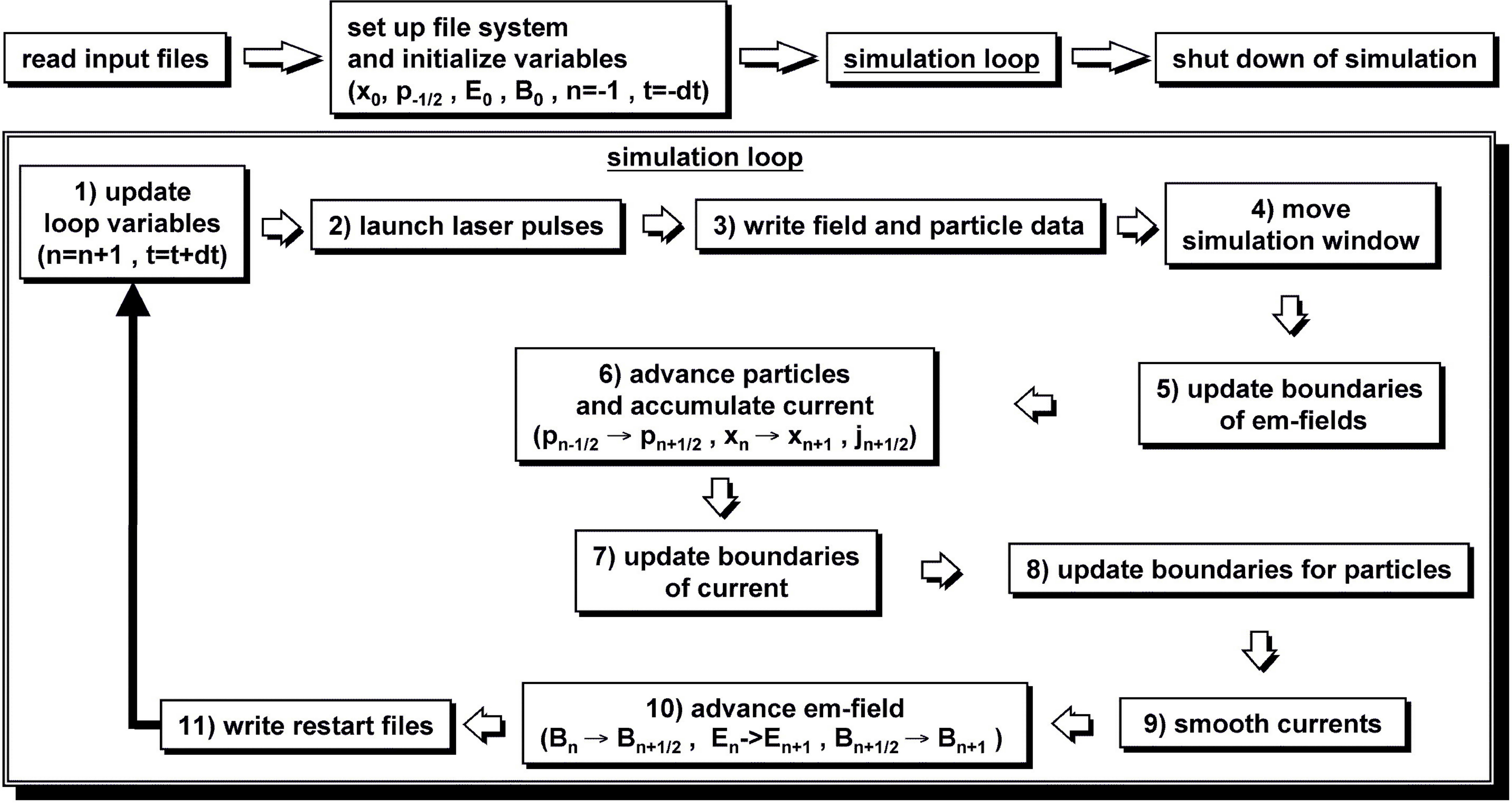, width=5.8in }
      \caption{The flow control diagram of OSIRIS. It follows the general 
               structure of Fig.~\ref{fig:review-codes:pic-loop} but shows
               differences that arise from the specific details of the
               implementation of OSIRIS.}
      \label{fig:os-algorithms:flow-chart}
   \end{center}
\end{figure}

\begin{figure}
   \begin{center}
      \epsfig{file=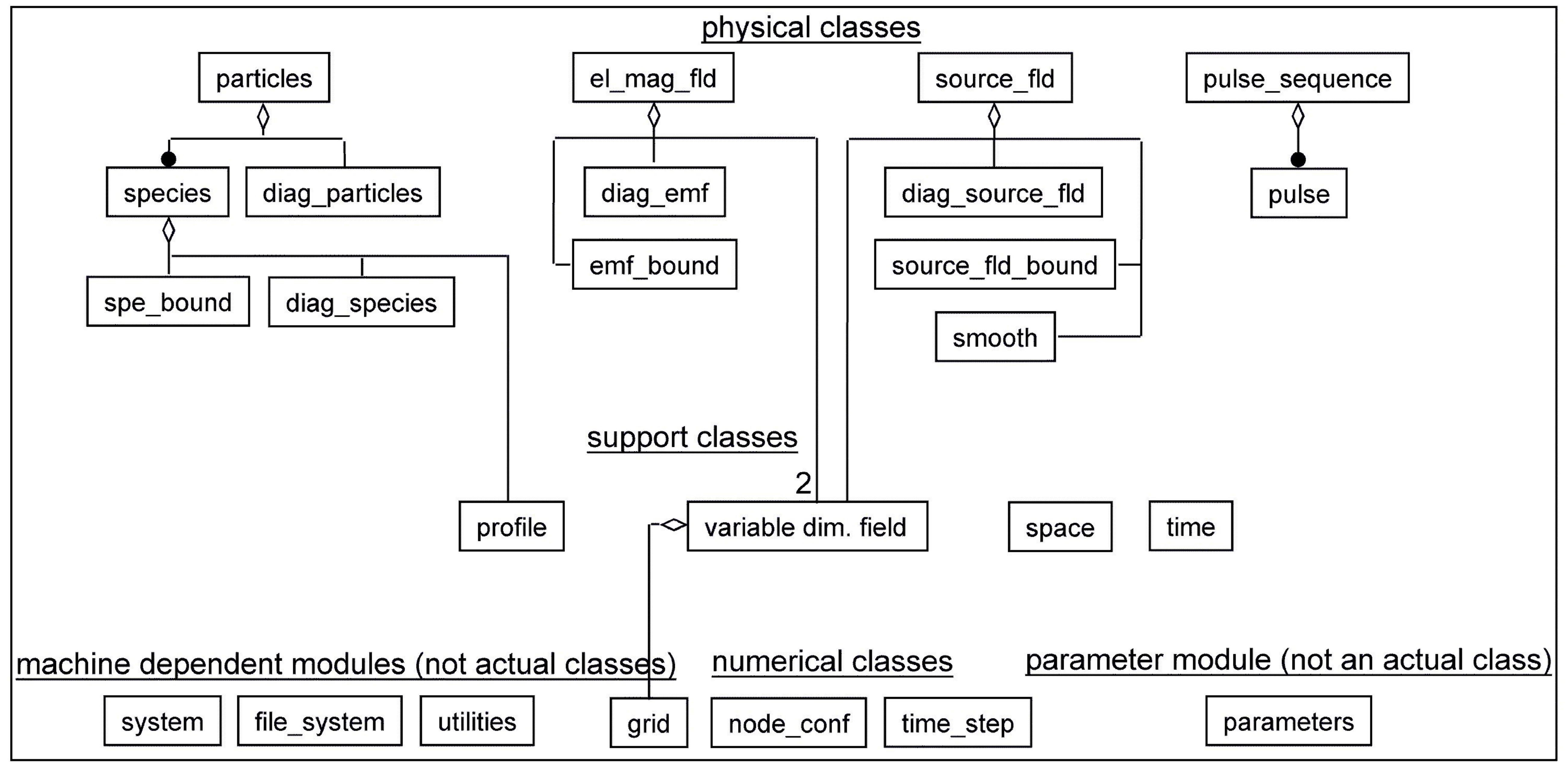, width=5.8in }
      \caption{The class hierarchy of OSIRIS. The figure shows most of the
               classes and modules used but omits some for the goal of 
               clarity.}
      \label{fig:os-algorithms:hierarchy}
   \end{center}
\end{figure}

Fig.~\ref{fig:os-algorithms:flow-chart} and Fig.~\ref{fig:os-algorithms:hierarchy}
show the flow chart and the class hierarchy of OSIRIS. Together, these two
figures give a high level description of the code. It is worth noting, that in the
main loop of OSIRIS the distinct physical and other (diagnostic, restart, etc.)
operations correspond to a specific step in the loop. It differs from the loop shown
in
Fig.~\ref{fig:review-codes:pic-loop} by the fact that step one to three of
Fig.~\ref{fig:review-codes:pic-loop} are all part of step six of
Fig.~\ref{fig:os-algorithms:flow-chart}. For computational efficiency all
these steps are best taken care of in a combined algorithm on this level 
of the code. The different subroutines that are called
for these different steps are called on a lower level of the code.

Fig.~\ref{fig:os-algorithms:hierarchy} is using the Object Modeling
Technique Notation (OMT)\cite{Rumbaugh} to describe the class hierarchy of OSIRIS.
The top level of the class hierarchy shows four different classes, particles,
electromagnetic fields, source fields, and the laser pulse sequence. The first three
correspond to distinctly different physical quantities. The laser pulse
sequence actually does not belong on this level and should in future
versions of the code become a sub-object of the electromagnetic field.
 
Each of the physical objects contains a sub-object that describes its boundary
conditions and another sub-object that describes the data diagnostic 
for the
object. In the case of the particle object, which is composed of an arbitrary
number of species objects, the boundary conditions are actually defined for
each single species object separately. A species object contains all information
required for one particular particle population, e.g., the actual data for each
single particle, the initial density of the species and the temperature.

The electromagnetic field object contains the information for electric and magnetic
fields, and the source field objects contain the information for the source terms in
Maxwell's equation, the current and the charge density. The distinction 
between the electromagnetic and source fields is made because 
particle information is needed to be deposited onto the grid to 
calculate the source fields. This requires an object with different properties
than the electromagnetic field object. In addition to the field data both 
classes
of field objects also contain sub-classes that define the boundary conditions and
the diagnostics for the fields.

\section{Variable Dimension Field Objects}
  \label{sect:os-algorithms:variable-dimension-field-objects}

For storage of the actual field information the electromagnetic field 
objects as
well as the source field objects have variable-dimension-field objects (VDF objects)
as sub-objects.
An object of this type contains the field and grid information for a scalar or
vector field in a 1D, 2D, or 3D simulations space. The use of these polymorph
objects \cite{Decyk98b}
makes it possible to avoid the explicit use of the dimensionality of a
simulation in most of the code. The dimensionality only becomes important when
the actual data within one of the arrays in a VDF object need to be used. For
this case the VDF class
provides functions to inquire about the dimensionality and other properties
of the VDF-object. It also provides pointer-valued functions that allow 
direct access to the actual arrays.
Fig.~\ref{fig:os-algorithms:vdf-type} shows the actual type definition
of the vdf-class. Note that the class methods make sure that only one of the
pointers is used for a given VDF object.
The grid component of the VDF object contains the information about 
the computational grid that the field is defined on.
Polymorph objects like the VDF-objects were suggested by Decyk et al. 
as a way to allow Fortran90 codes to simulate certain uses of
C++ templates \cite{Decyk98b}.

\begin{figure}
   \begin{center}
      \small{
      \begin{verbatim}

        type :: t_vdf

!         allow access to type components only to module procedures
          private

!         variable to indicate status of vdf-object
          logical :: associated

!         pointer to field data for field 1D space f(j,i1)
          real(p_k_rdoub), dimension(:,:), pointer :: f1

!         pointer to field data for field 2D space f(j,i1,i2)
          real(p_k_rdoub), dimension(:,:,:), pointer :: f2

!         pointer to field data for field 3D space f(j,i1,i2,i3)
          real(p_k_rdoub), dimension(:,:,:,:), pointer :: f3

!         grid information for this object on the local node
          type( t_grid ), pointer :: grid

        end type t_vdf

      \end{verbatim}
      }
      \caption{The definition of the VDF type in OSIRIS}
      \label{fig:os-algorithms:vdf-type}
   \end{center}
\end{figure}

\section{Global and Local Objects}
  \label{sect:os-algorithms:global-and-local-objects}  

Data in OSIRIS can be classified into two different categories,
global data and local data. Global data are referring to the whole
simulation. Local data are referring to the part of the simulation running
on one specific node. In OSIRIS most classes refer only to local data. The exceptions
are the node-configuration class, the space class, and the grid class.
A node-configuration object always contains global as well as local
data. The space class and the grid class have global as well as local
instances. However, each space or grid object contains either global or
local data but not both. The
global space and global grid describe the space and the grid of the
whole simulation. The local space and the local grids describe the 
space and the grid on the node that a process is running on.
All other classes of OSIRIS contain only local data.
This section will describe how the global and local objects of a 
simulations are initialized at startup and how this relates to the
one object per node strategy that is used in OSIRIS. 

A simulation starts with initializing all necessary variables of the code
with the correct initial values. This happens in OSIRIS by first reading
in from the input file for each object separately the information the 
object requires from that file. The input file contains,
with the exception of the node-configuration information, only
information about the global physical problem to be simulated.
The node configuration object contains the information on how many
computing nodes the simulation will run on and on how the whole
simulated space is decomposed to the different nodes.
Therefore, after reading in the input file information, the code first 
fully initializes the node-configuration object, then a space object
that describes the global simulation space and then a grid object that
describes the global simulation grid. A grid object contains the 
information about a computational grid that is necessary to define
fields in a space. The next step uses the global space object and the
node-configuration object to generate a space object that describes the
local space. A local grid object is generated in a similar fashion
(from the global grid and the node-configuration object). All other
objects of the simulation are then
initialized as local objects for each specific node using the 
information from the local space and local grid objects to adapt as 
much as necessary the
information read in from the input file (which is global information)
to the local node.

\begin{figure}
   \begin{center}
      \epsfig{file=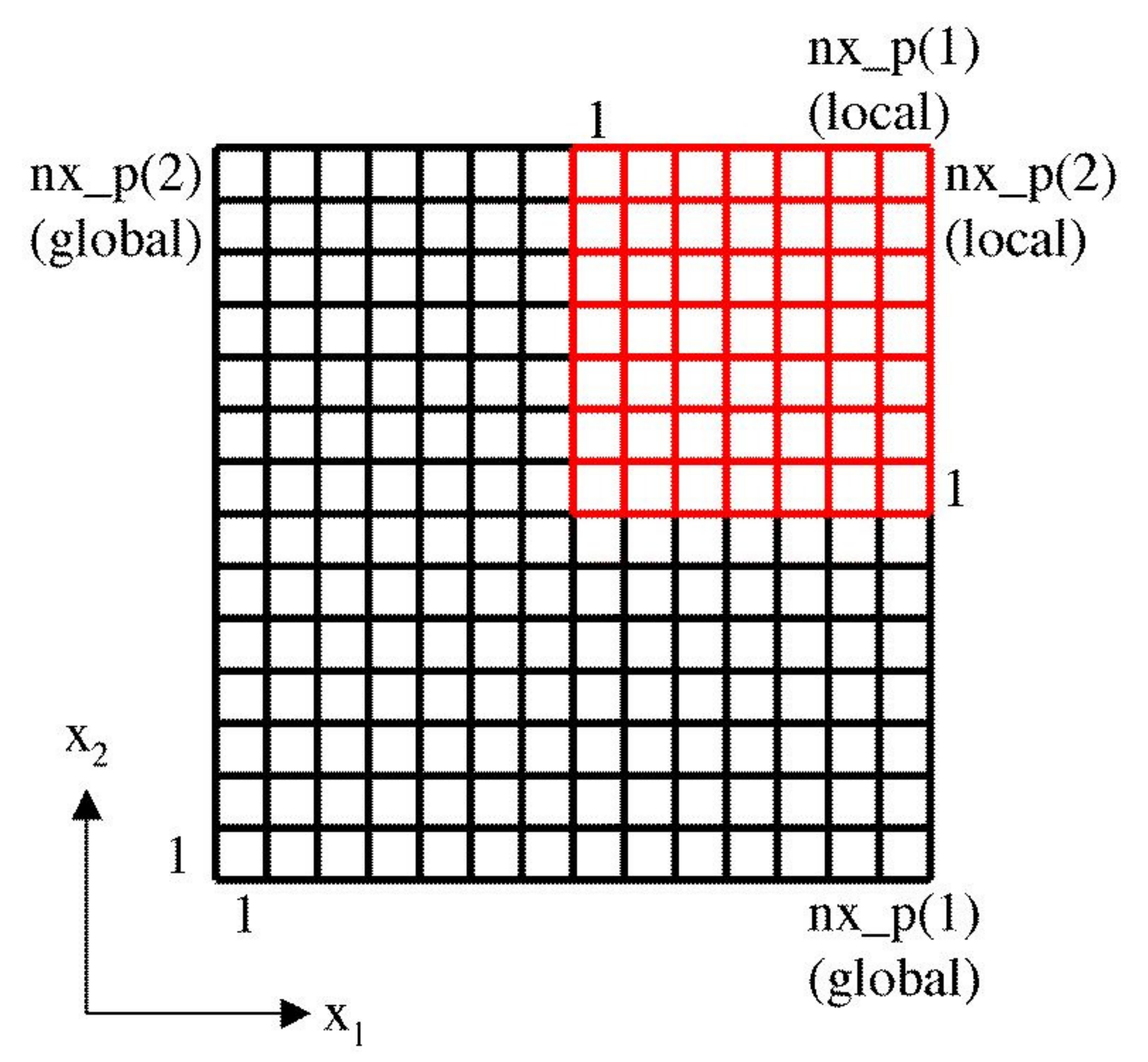, width=3.0in }
      \caption{The code running on each node has several instances of a grid
               object. The global grid object describes the the grid of
               the whole simulation (black) and how it has moved. The local 
               grid objects that are part of each VDF object contain the
               same information for the domain assigned to a specific
               node (red).}
      \label{fig:os-algorithms:grids}
   \end{center}
\end{figure}

Fig.~\ref{fig:os-algorithms:grids} shows the global and the local grid
which would be used for a 2D-simulationon on $2 \times 2$ nodes. The global
grid has indices 
from $1$ to $nx\_p(1)(global)$ in the $x_{1}$-direction and from  $1$ to
$nx\_p(2)(global)$ in the $x_{2}$-direction. The local grid has indices
from $1$ to $nx\_p(1)(local)$ in the $x_{1}$-direction and from  $1$ to
$nx\_p(2)(local)$ in the $x_{2}$-direction. For the currently 
implemented algorithm all local grids have the same size in a given
direction if the global grid can be divided up evenly over the number
of nodes in this direction.
If the global grid can not be divided up evenly over the number
of nodes then a certain number of nodes will have one grid cell less 
than the other nodes.
 
To make the generation of the local space and grid possible, each node
is assigned a certain position in a
regular grid of nodes. This node-grid is 3D for 3D-simulation, and of  
lower dimensionality for 1D- and 2D-simulations. The assignment is done
by placing the unique task (or process) IDs for all nodes in a 1D array
which has a size given by the total number of nodes used. The task IDs are
provided to the program by the message-passing library that is used. 
The current implementation of OSIRIS uses MPI
\cite{MPI-specification,Gropp97}for message-passing.
The position in the array is then used as an ID number (AID)
to access the task ID when necessary.  If the number of processors in 
each direction is given by $nx(1)$, $nx(2)$, $nx(3)$ then the total number
of processors is $nx(1)$~$\times$~$nx(2)$~$\times$~$nx(3)$. The Array ID
for the node at the node grid position n(1), n(2), n(3) is then given by:

\begin{equation}
  \label{equ:os-algorithms:aid}
  AID = \left(n\left(1\right)-1\right)
      + \left(n\left(2\right)-1\right) \times nx\left(1\right)
      + \left(n\left(3\right)-1\right) \times nx\left(1\right)
                                       \times nx\left(2\right)
      + 1
\end{equation}

\begin{figure}
   \begin{center}
      \epsfig{file=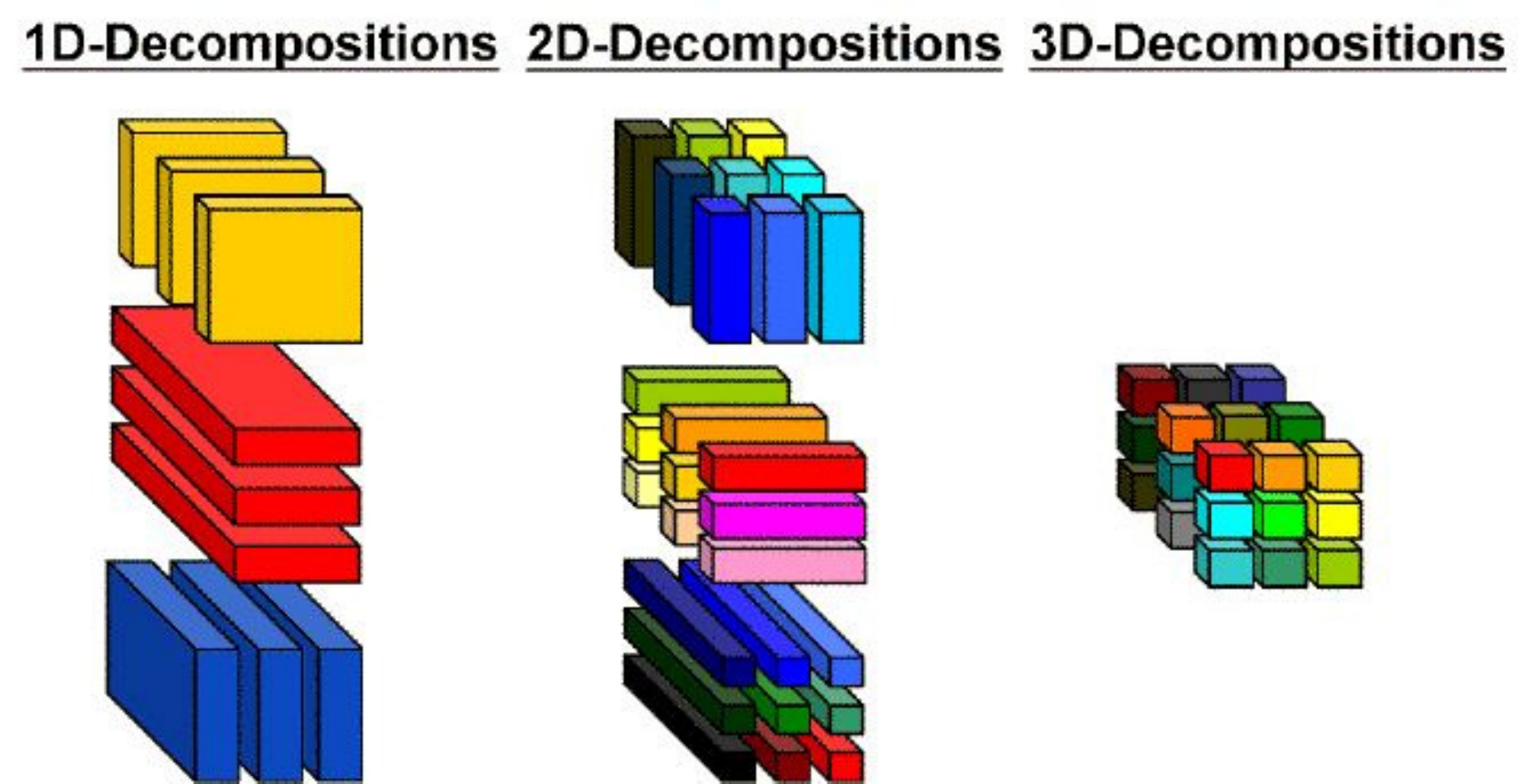, width=5.8in }
      \caption{The possible decompositions in a 3D simulation.}
      \label{fig:os-algorithms:decompositions}
   \end{center}
\end{figure}

\noindent{This} can be inverted to give a unique node-grid position
for a given array ID. Fig.~\ref{fig:os-algorithms:decompositions} shows
the different type of decompositions that this node-assignment
algorithm allows for 3D simulations. The parallel efficiency of these
decomposition has been investigated previously by Lyster et al. for an 
electrostatic PIC code\cite{Lyster95}.
 
\begin{figure}
   \begin{center}
      \epsfig{file=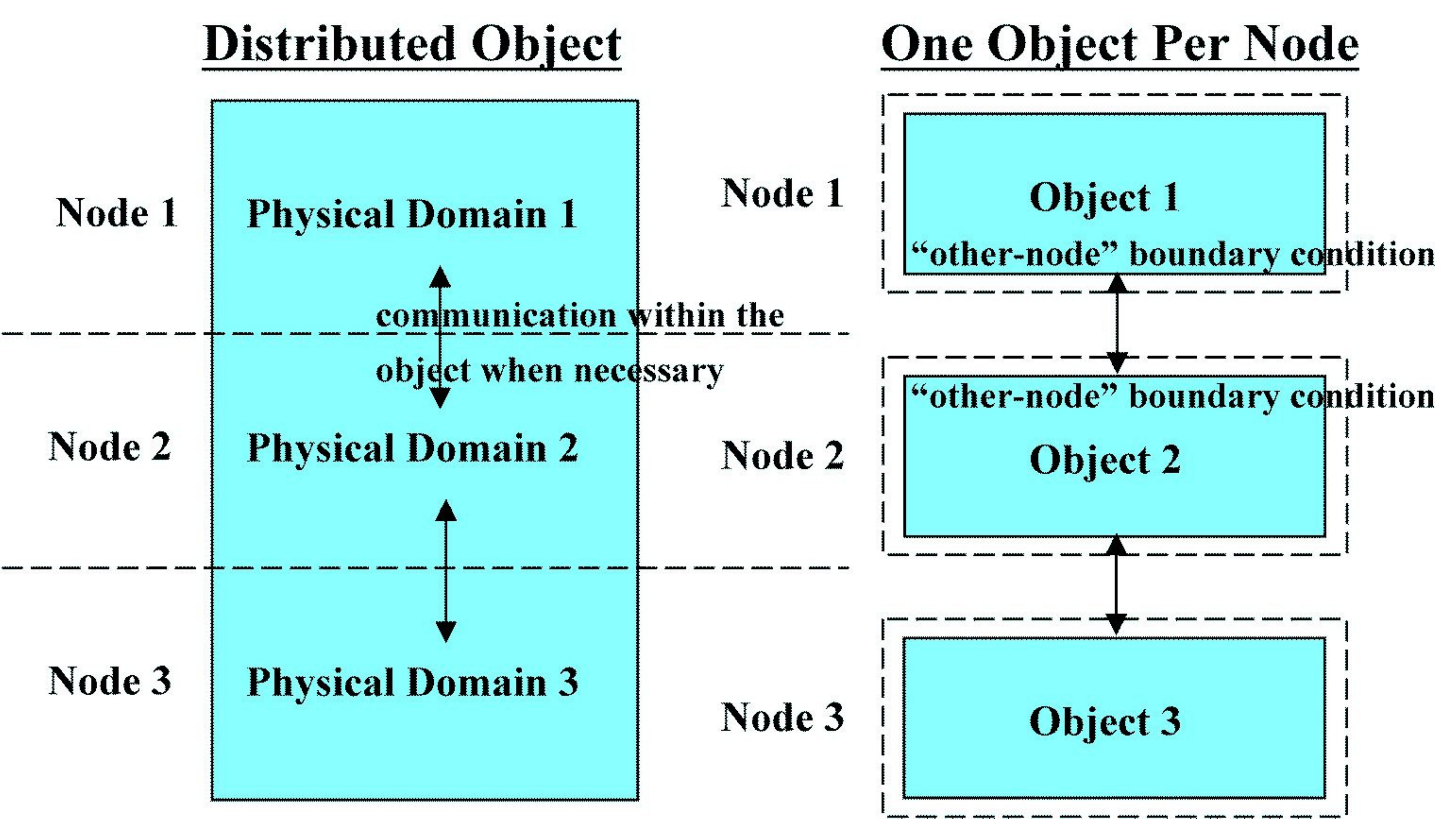, width=5.0in }
      \caption{Two different concepts for objects on a parallel computer.
               In a parallel
               object communication is done by the methods of a class whenever
               information from another node is needed within that method. The
               concept of one object-per-node treats boundaries with other nodes
               as one particular kind of boundary condition.}
      \label{fig:os-algorithms:object-per-node}
   \end{center}
\end{figure}

The separation of global and local data described above together with the
node-configuration object make it possible to implement a one-object-per-node strategy,
since all information within the physical objects is local and all information
required for communication with other nodes is contained
in the node-configuration object.
When a physical object, which contains only local data, needs to exchange
information with a neighboring node at a given boundary it only has to specify the
information and the boundary and hand this information to the 
node-configuration object. This process is implemented as one particular boundary 
condition. The node-configuration object then manages the details of which
nodes send or receive information.
The one-object-per-node strategy implemented in OSIRIS is
illustrated in Fig.~\ref{fig:os-algorithms:object-per-node}. Since 
this strategy makes it much easier to extend OSIRIS with new algorithms it 
was an important goal of the code development. The design of the space
and grid objects and of the setup was done in the way described above 
because of the one-object-per-node strategy.
 
\section{Dynamic Simulation Spaces}
  \label{sect:os-algorithms:dynamic-simulation-spaces}

Once the global and local objects have been defined, they have to be
consistently maintained throughout the run. This includes consistency
with each other, with the respective global objects, and with the
objects on other nodes. The implementation of this is not straightforward
since in OSIRIS moving boundaries are possible with a special case being the
motion of boundaries required for a moving simulation window\cite{Tzeng96}.
There are two types
of moving boundaries in OSIRIS; one that moves outward from the
simulation window with the speed of light and one that moves
inward with the speed of light. An outward moving boundary 
requires us to extend the simulation window by one grid cell whenever
enough time has passed for the boundary to have moved outward by at 
least that far. An inward moving
boundary requires us to shorten the simulation window by one grid cell
whenever enough time has passed for the boundary to have moved inward
by at least that far.
If $dx$ is the grid cell size and $dt$ is the timestep size then we
know that the Courant condition \cite{BirdsallLangdon} for electromagnetic
PIC codes requires that $c \, dt \: < \: dx$. Therefore a moving boundary
can not move by one grid cell at every timestep. It is only moved by $dx$
when the mismatch between the location where the boundary actually is and
the location where the boundary should be becomes larger than $dx$.
At timesteps where the boundary is not moved by a grid cell no special
boundary algorithms are needed. 

\begin{figure}
   \begin{center}
      \epsfig{file=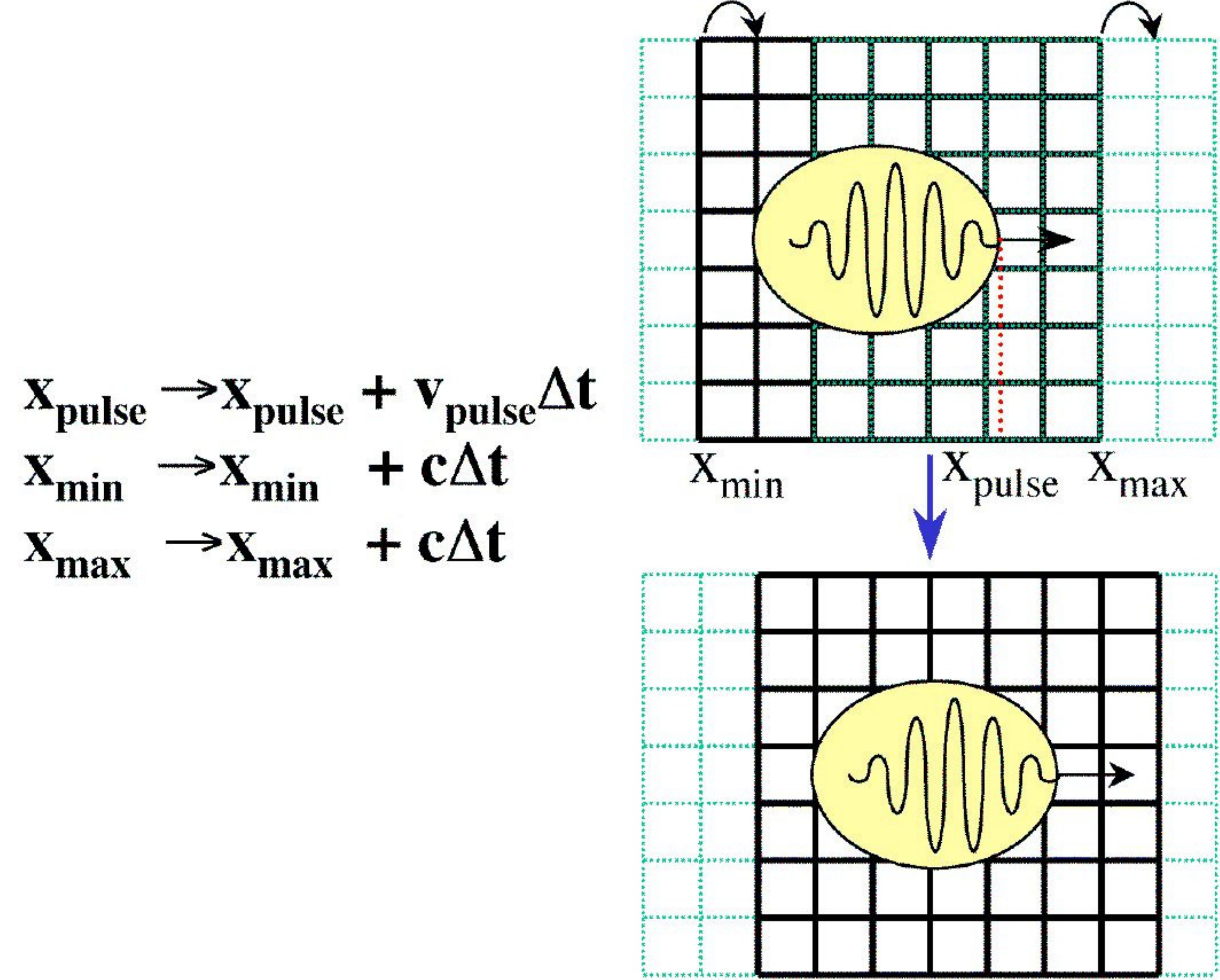, width=5.8in }
      \caption{For a dynamic space the boundaries can move inward or outward
               from there current position. A moving window is the special
               case of the front boundary moving outward and the back boundary
               moving inward.}
      \label{fig:os-algorithms:dynamic-space}
   \end{center}
\end{figure}

All objects, the
space objects (global and local), the grid objects (global and local),
and all the physical objects, have
to be modified according to this change of the space. The fact that
a boundary is moving outward with the speed of light makes it
possible to simply initialize the plasma in the new space as a
thermal plasma with zero electromagnetic field since it is in an
area that can not have been affected in any way by the the interior
of the simulation. The boundary condition for a boundary moving inward
with the speed of light can be implemented in a similar way only that
simulation space has to be removed and the plasma in that space has 
to be discarded. Again the fact that the boundary is moving at the
speed of light makes this boundary condition simple since none of the
discarded space can have any further affect on the remaining simulation
space. With these kind of moving
boundaries the simulation space becomes a dynamic window that moves 
in space and can change in size as it follows the physics of interest.
A dynamic window is very useful for plasma-based accelerator 
simulations. In such cases it is useful for the front boundary of the
simulation window to move outward with the speed of light and the
back boundary to move inward with the speed of light.
Fig.~\ref{fig:os-algorithms:dynamic-space} illustrates the 
implementation of a moving window for laser-plasma physics using the
dynamic space concept. Another application of dynamic space
is that of an expanding or collapsing space. This is not fully
implemented yet, but planed for a future version of the code.

\begin{figure}
   \begin{center}
      \epsfig{file=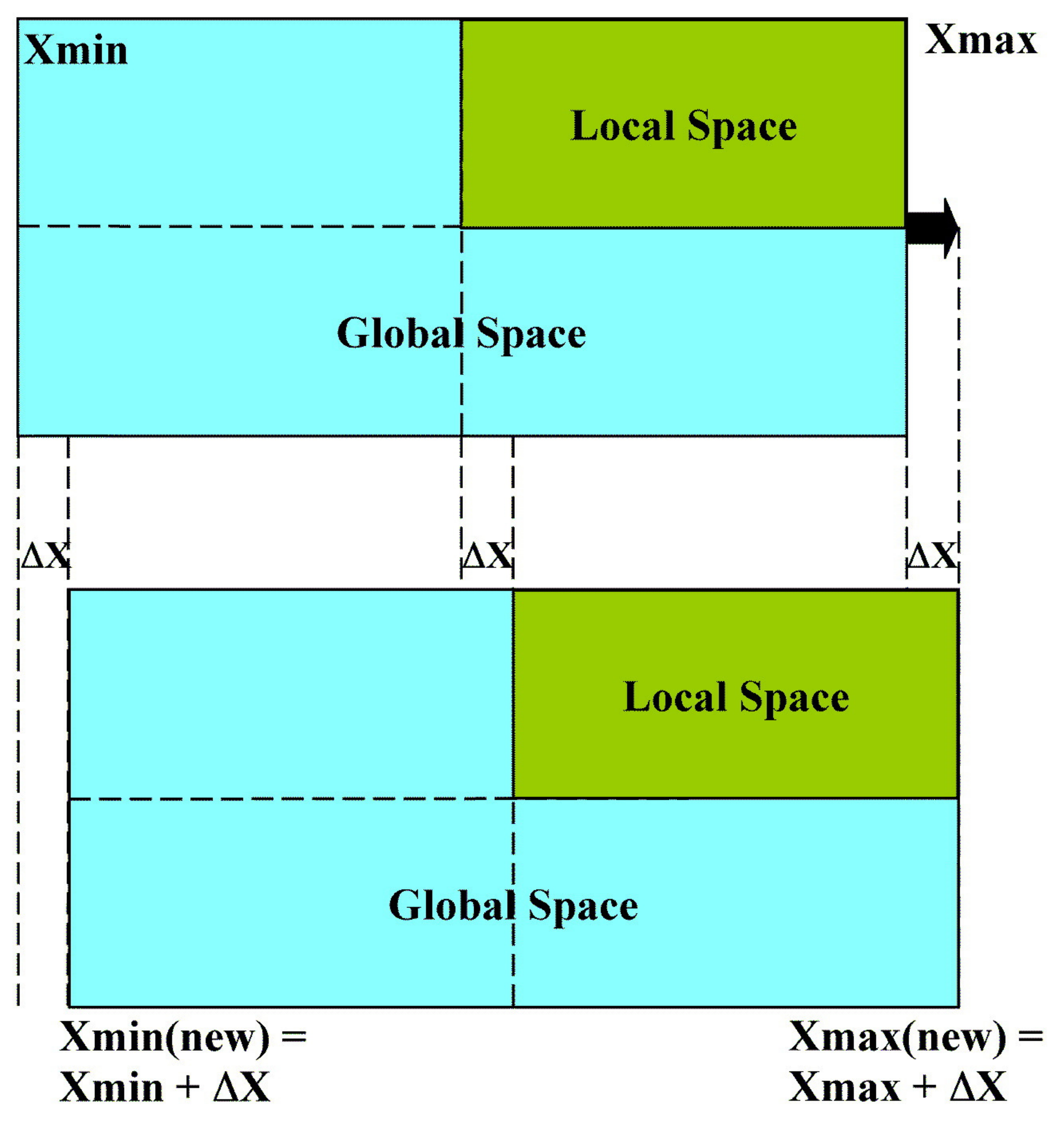, width=4.0in }
      \caption{The code running on each node has two instances of a space
               object. The global space object describes the space of
               the whole simulation and its motion. The local space object
               contains the same information for the domain assigned to the
               specific node. The figure shows the update of the space boundaries
               explicitly only for the global object but for the local space the
               same updating of Xmin and Xmax is done.}
      \label{fig:os-algorithms:spaces}
   \end{center}
\end{figure}

Fig.~\ref{fig:os-algorithms:spaces} illustrates the relationship
between the global and the local space object for a moving window.
The global, as well as the local, space contain the information about
the lower and upper boundaries of the simulation window in each
direction. In general these boundaries are different for the global
and local space on a given node unless the code is running on only one
node. In the case of a single-node simulation the global and the local
space describe the same space (the same is true for the global and
local grid). If the window of the simulation moves then the boundary
information of both spaces has to be updated each time the boundary
moves. Actually, the information that describes the motion of the
simulation window is stored as part of the global space object.
If the global simulation window moves at a certain timestep then
the local space object will move accordingly.
Note that even though in Fig.~\ref{fig:os-algorithms:spaces} only the
update of the values Xmin and Xmax of the global space boundary is
explicitly shown, this is done for the local space boundaries as well
(as the actual picture in
Fig.~\ref{fig:os-algorithms:spaces} also shows). The next section of 
this chapter will give a more detailed description of the implementation
of moving boundaries between nodes.
 
After the space objects have been moved, the motion of the 
global grid is determined according to the motion of the global space.
All other objects use information from the local space, which has been 
updated using the global space object, to decide 
whether and how to move each boundary. This is motivated by
the strategy to use local-node information wherever possible.

\begin{figure}
   \begin{center}
      \epsfig{file=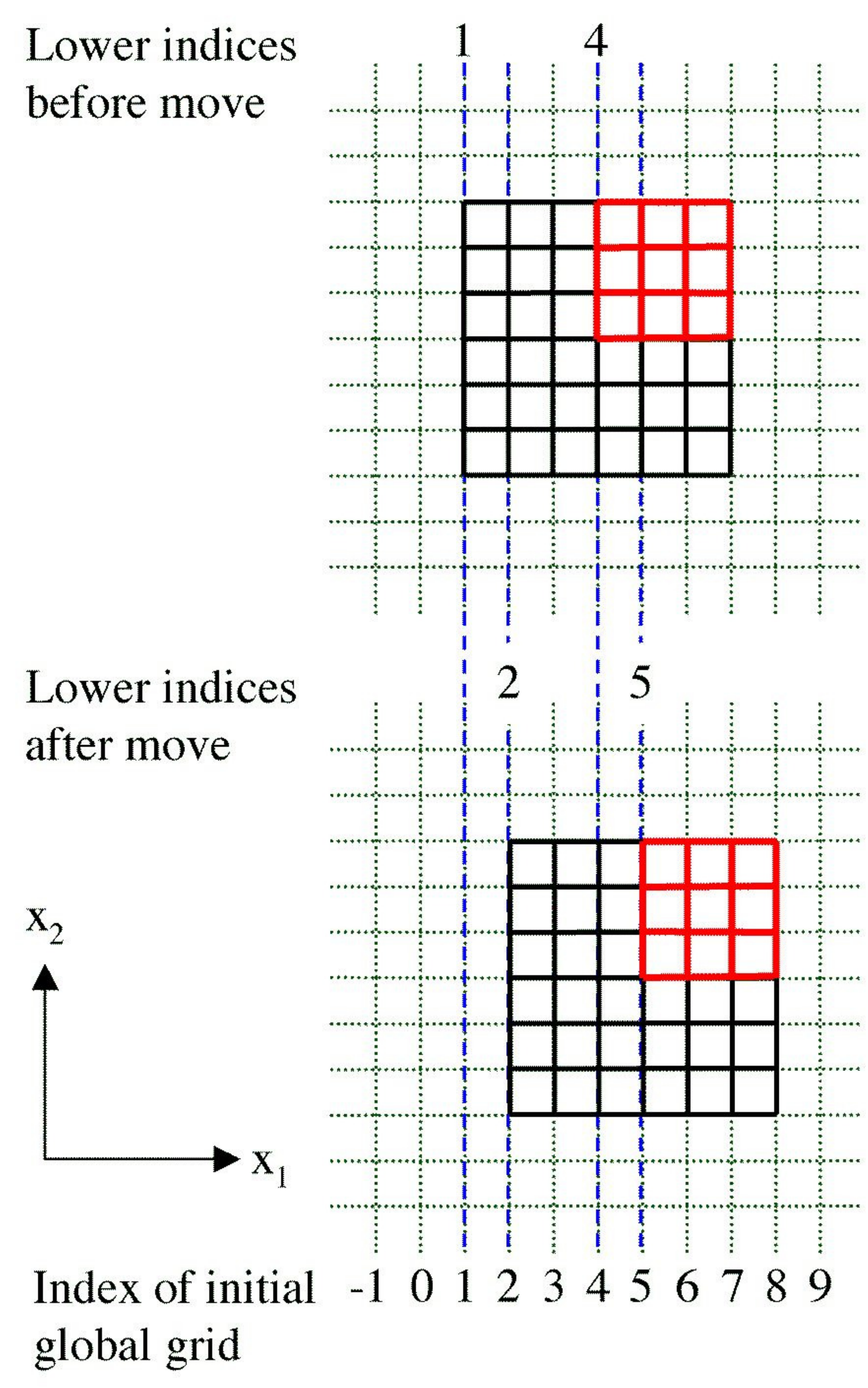, height=7.0in }
      \caption{The figure shows how the motion of the lower boundary of
               the global and local grid is followed. The upper boundaries
               are tracked in the same way.}
      \label{fig:os-algorithms:grid-motion}
   \end{center}
\end{figure}

There is a second data structure in the code that keeps track of the
motion of objects and boundaries in space. Each grid object keeps
track of the motion of its boundaries as it moves with respect to
the initial global grid (note that the global grids on all nodes
are identical to each other at all times). In this way the grid
objects on each node can keep track of their size and motion in space.
The global grid keeps track of its size and motion with regard to
its own initial state. This information is in a certain sense
redundant with some of the information stored in the space objects
but the redundancy is justified by
providing easy access to this information without having to 
translate from the continuous space boundary description
to the discrete grid cell description of it every single time 
it is needed. It therefore simplifies a lot of algorithms.
Fig.~ \ref{fig:os-algorithms:grid-motion} illustrates the way
the  global and local grid objects keep track of the boundary
motion for their lower boundaries. They do so by simply adjusting
the variable that describes the boundary position with respect to
the initial grid each time the boundary is moved. The upper boundary
is followed in the same way and together these two numbers also
describe the size of the grid. Please note that following the positions 
of the grid boundaries is completely independent from the grid indices provided 
by grid objects to the algorithms of the code. The grid indices 
start for a given grid always with 1 at the lower boundary and go up 
to the maximum number of grid points in a given direction for this grid.

\section{The Motion of Internal Boundaries}
  \label{sect:os-algorithms:motion-of-internal-boundaries}

In the previous section we explained that the information in the local
space object is used to steer the boundary movement of all objects except
of the global space and the global grid.
This decouples the motion of the boundary of a node
from the motion of the global boundary. In the case where a node has a boundary
that is also a global, external boundary of the whole simulation the
result has to be the same, but in the case where the boundary is an internal
boundary to another node the signal to move the boundary will
have to make use of different algorithms than the ones described above
which are just applicable to the external boundaries of the 
simulation.

The additional complications that need to be addressed when moving
an internal boundary between nodes, and that are outlined
below, are independent from the reasons why the boundary is moved.
Before describing details of how internal boundary motion is handled
it is therefore worth noting that although we have implemented moving boundaries
as part of the motion of the whole simulation space, this ability 
makes it possible to reassign a piece of simulation space
from one node to its neighboring node for the purpose of dynamic load
balancing. While dynamic load balancing is currently not
implemented in OSIRIS, it could easily be done by adding some sort
of algorithm, probably as part of the node-configuration object, that
modifies the boundary motion of the local space object after
the boundary motion of this object due to the global window motion
has been done, but before it is used to steer the motion of other 
objects.

\begin{figure}
   \begin{center}
      \epsfig{file=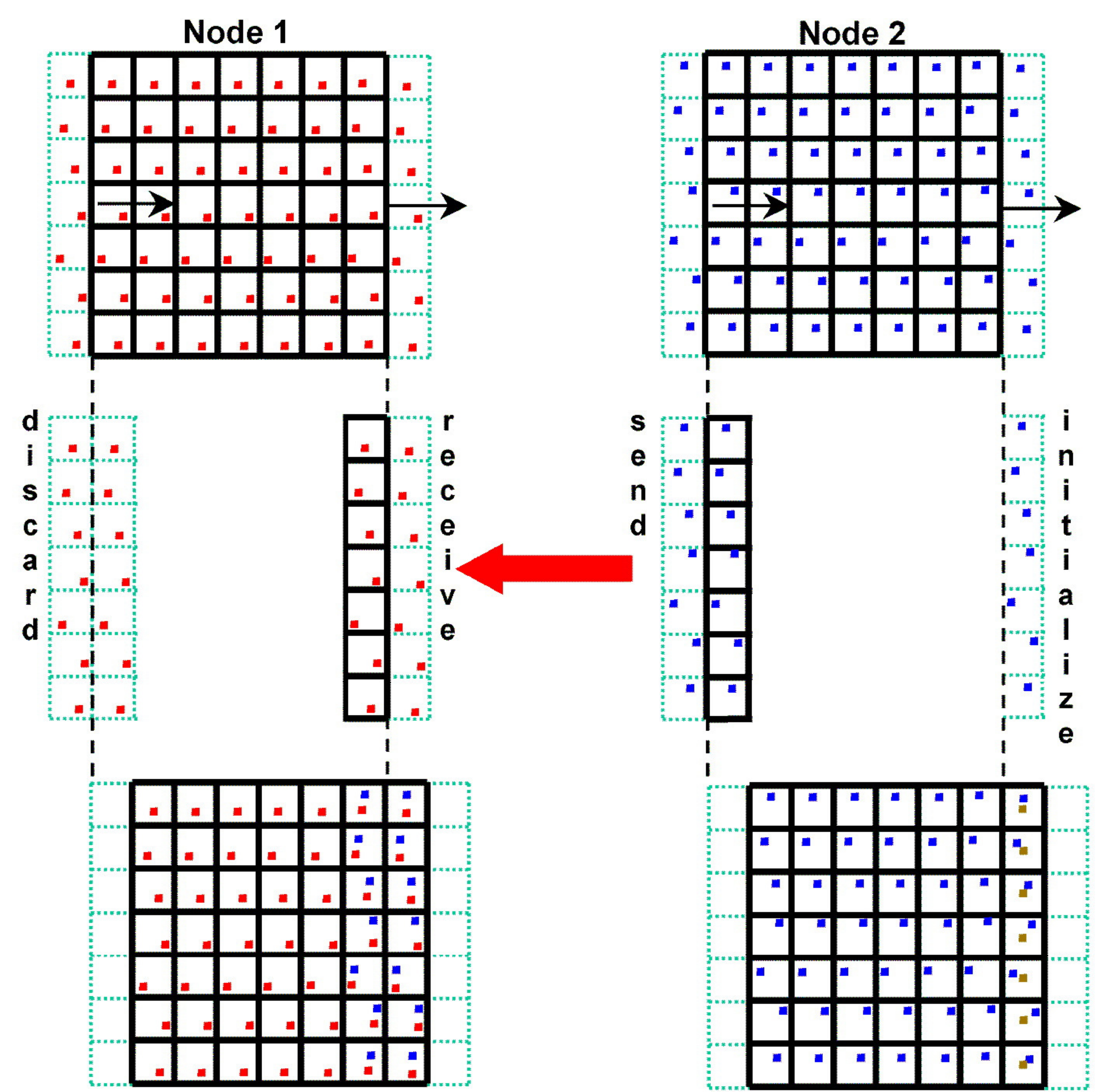, width=5.8in }
      \caption{If a boundary between nodes moves then the boundary condition
               handling this case has to move the necessary data. This figure shows
               the motion of particles between nodes. The red, blue, and yellow
               colors for the particles in the figure are used to distinguish the
               different groups of particles. The red and blue particles are
               originally on the left and right node respectively. The yellow
               particles are newly initialized particles.}
      \label{fig:os-algorithms:node-bnd-par}
   \end{center}
\end{figure}

We next describe as an example to illustrate a moving internal 
boundary how the particle boundary conditions are handled in
the moving window frame.
Fig.~\ref{fig:os-algorithms:node-bnd-par} shows some of the
details of the process of moving an internal boundary in a moving
window simulation for a (particle-)species object. The first
row of the figure shows the space and the grid of a 2-node
simulation moving to the right. The black cells indicate the
physical space of the simulations while the green cells correspond to
the guard cells. If the simulation were not in a moving window then
the following would have to be done to take care of the particle
boundary conditions. The left boundary of the left node (node 1) and the
right boundary of the right node (node 2) would have to treated according
to some other global, external boundary condition defined there.
The particles in the right guard cells of the node 1 would have
to be sent to the node 2 and placed at the corresponding positions
within the physical space of the node 2. The left side of node 2 would
be treated similarly.

For a moving window this process is more complicated. First, consider 
what happens at the external, global boundaries of the simulation window, .i.e.,
the left boundary of node 1 and the right boundary of node 2 in
Fig.~\ref{fig:os-algorithms:node-bnd-par}.
The particles in the guard cells as well as in the first column of
cells on the left side of the node 1 have to be discarded. The particles
in the guard cells on the right side of node 2 are kept
and new particles are initialized according to the plasma that
needs to be simulated in the new area. At the internal boundary, a 
boundary that is physically moving to the right, the right node is
sending particle information and the left node is receiving
particle information due to the motion of the boundary.
No particles are send from the left node to the right node.
This asymmetry in the message passing will also be true
for the electromagnetic fields. In the current version of the code
the source fields are recalculated from scratch at every timestep
after the whole system has been moved already. Therefore, the source fields do not
need to be explicitly moved since they are calculated from the 
particles, which were already moved. For this reason the source field
message passing is always
symmetric. The third row of Fig.~\ref{fig:os-algorithms:node-bnd-par}
shows the final particle distribution after the motion of the boundary
and the passing of particles. Note that all the guard cells are now free of
particles as they need to be.

\section{Multi-Dimensional Issues}
  \label{sect:os-algorithms:multi-dimensional-issues}

So far all the discussions of algorithms and the figures illustrating
them ignored the problems arising from simulations with more than one
spatial dimension. This is justified because extending all algorithms
described above is straightforward when all boundaries of a given node,
whether they are external or internal, are handled one dimension 
after another. The suggestion to exchange messages between nodes one
direction at a time and therefore have multiple exchanges of messages for
a multi-dimensional domain decomposition has been made before\cite{Lyster95}.

\begin{figure}
   \begin{center}
      \epsfig{file=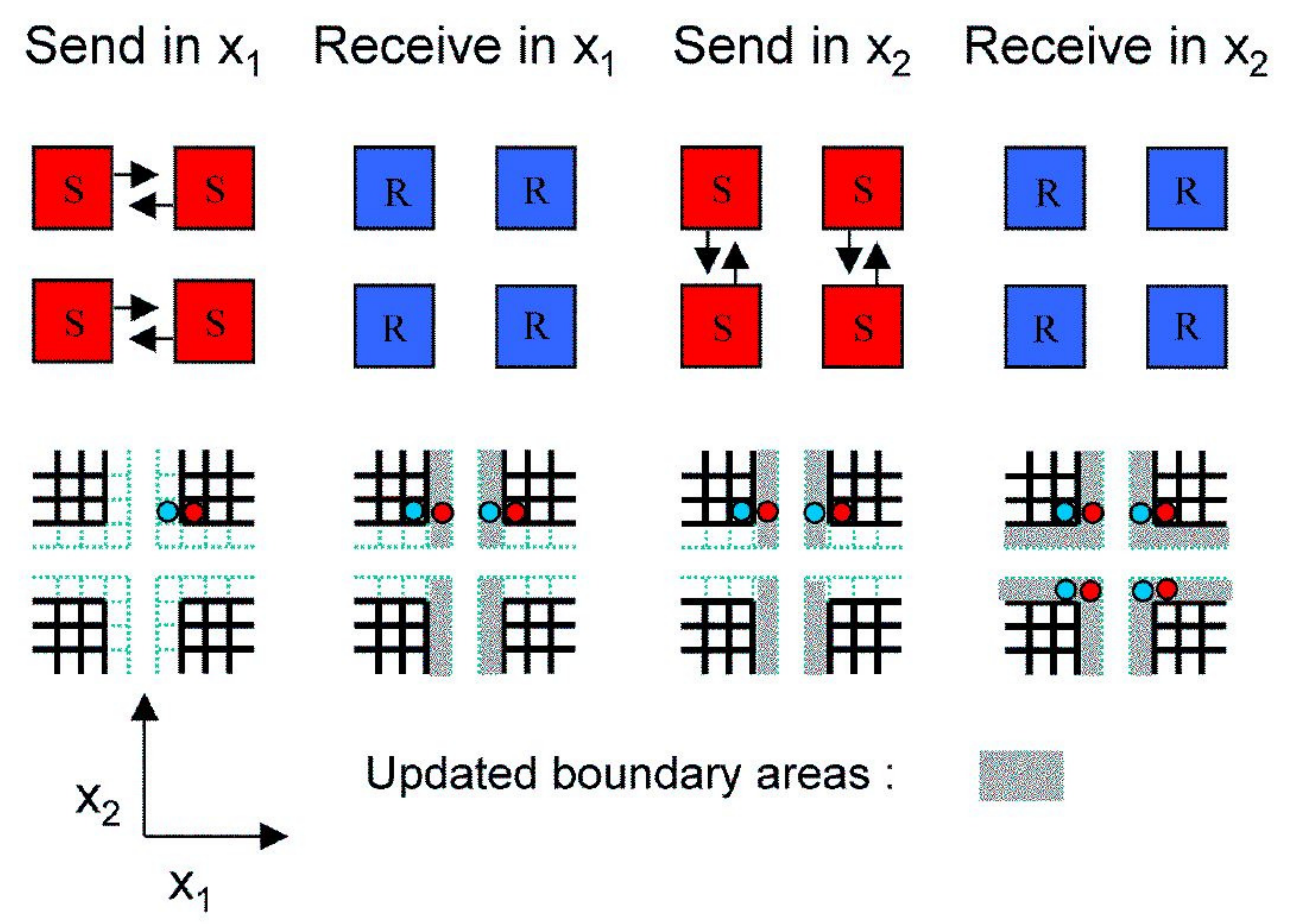, width=5.0in}
      \caption{The communication pattern of OSIRIS for 2D domain
               decomposition. The communications and boundary handling
               takes place for one direction at a time}
      \label{fig:os-algorithms:communications}
   \end{center}
\end{figure}

The idea is illustrated for the case of four nodes with
internal boundaries between them in Fig.~\ref{fig:os-algorithms:communications}.
The figure shows the communication patterns between the nodes.
First each node takes care of its boundaries in $x_{1}$ (by first 
sending and then receiving information) and then it does the same thing 
in $x_{2}$. This approach can obviously be extended to a third
dimension as well.

To our knowledge the novel aspect of this idea as it is implemented in OSIRIS is
that message passing at internal boundaries and boundary conditions 
at external boundaries are all handled in the same way since message
passing is treated as one particular kind of boundary condition. 
The approach of taking care of all
boundary conditions in a given direction first before taking care of
the boundary conditions in another direction also works for cases where
internal node-to-node boundaries in one direction are followed by external
boundary conditions in another. For example, a node in a 3D 
simulation could have internal boundaries in $x_{1}$ which it takes care
of by sending and receiving messages and then have external, conducting
boundaries in $x_{2}$ which can be applied after the $x_{1}$ boundary conditions
have been taken care of. Finally if the $x_{3}$ boundaries are internal
boundaries again it will exchange information but this time with its
neighbors in the $x_{3}$ direction.






\section{Conclusion}
  \label{sect:os-algorithms:conclusion}

This chapter described the most important object-oriented strategies and
parallel algorithms of OSIRIS. Using object-oriented programing
in Fortran 90 made it possible to combine several features and algorithms,
some of which are novel. The most important ones are:

\begin{itemize}

\item The implementation of multi-dimensional domain decomposition.

\item The implementation of the one-object-per node strategy.

\item The implementation of a dynamic space algorithm using moving 
      boundaries.

\item The implementation of information exchanges between nodes as a boundary 
      condition.

\item The handling of boundaries one direction at a time including 
      node-to-node boundaries.

\item The encapsulation of the dimensionality and the coordinate system
      of a simulation by using polymorphic objects.

\end{itemize}

This code is operational and has been used to model a variety of
problems for the first time.
In addition the structure of its objects and the codes
modularity make it easily extendable so that in the future new 
algorithms for increasing the efficiency of the code (e.g., dynamic load
balancing, ponderomotive guiding center description for lasers)
or for including new physics (e.g., ionization) can be integrated.

\chapter{Electron Beam Production Using Multiple Laser Beams in Plasmas}
  \label{chap:injection}

\section{Introduction}
  \label{sect:injection:introduction}

\begin{figure}
   \begin{center}
      \epsfig{ file=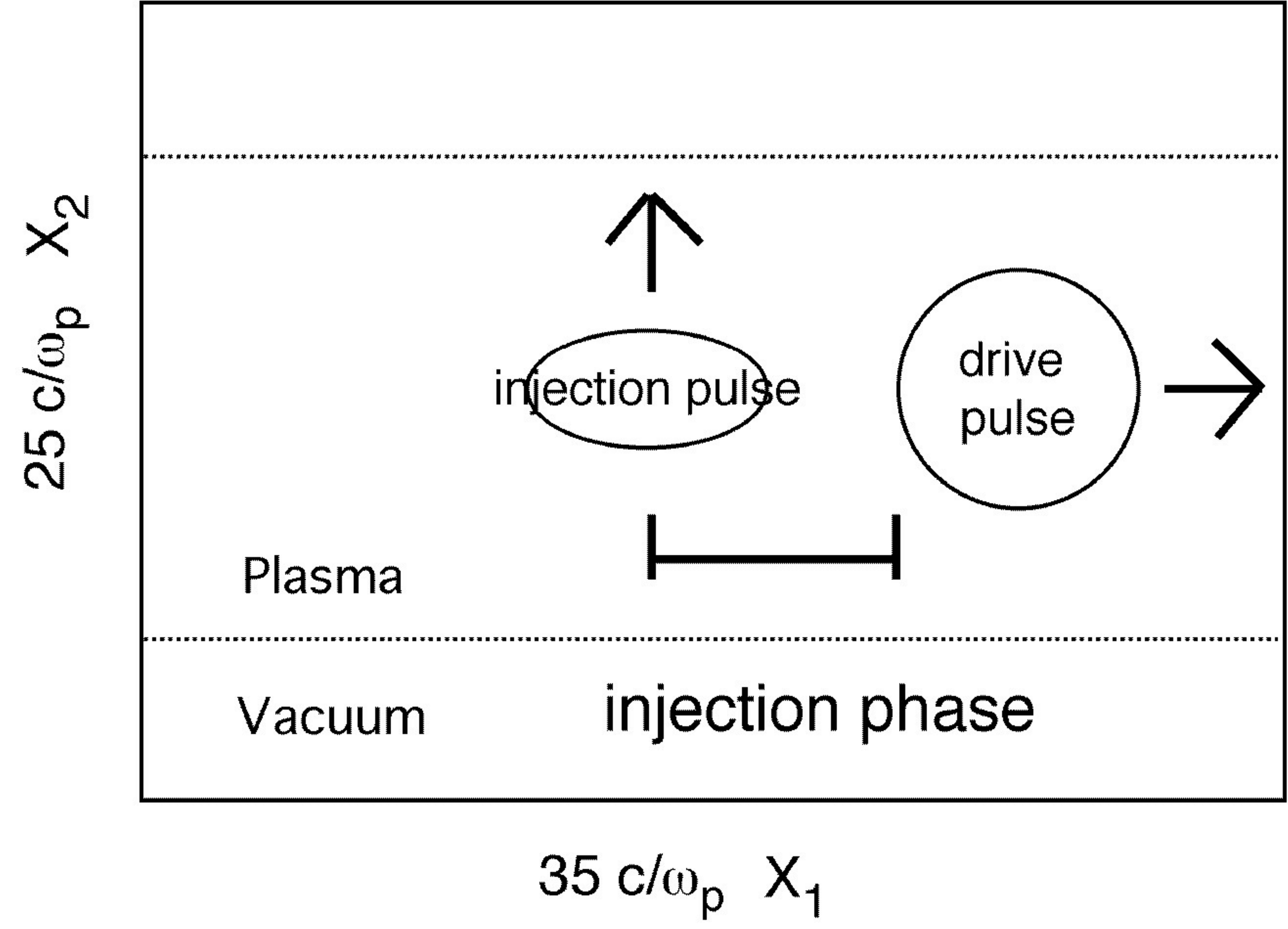, width=4.0in }
      \caption{Geometry of the cathodeless injector concept.
               The injection phase of the injection pulse is defined
               by the distance between the trailing edge of the drive
               pulse and the center of the injection pulse when it
               crosses the drive pulse.}
      \label{fig:injection:scheme}
   \end{center}
\end{figure}

D. Umstadter et al. \cite{Umstadter96a} proposed the use of 
two orthogonal laser pulses in a plasma to trap and accelerate an 
ultra-short bunch of electrons. As envisioned, the first (or drive) 
pulse creates a plasma wave which is below its self-trapping or 
wavebreaking threshold. The transverse ponderomotive force of the 
second (or injection) pulse was argued to give electrons an extra kick 
forward in the wake direction, enabling them to be trapped and 
accelerated in the wake of the drive pulse. This geometry is 
illustrated in Fig.~\ref{fig:injection:scheme}. Such a cathodeless
injector (or perhaps more 
correctly, a plasma cathode) is of interest for a wide variety of 
applications including an injector for future linear accelerator 
technologies with short wavelength accelerating structures, a source 
of short pulses of light or x-rays, or a source of electron bursts for 
pulsed radiology and ultra fast pump-probe chemistry \cite{Ogata96}.

For plasma accelerator applications in particular, the scheme 
naturally overcomes problems of synchronizing the injector with the 
accelerator. Moreover, the rapid acceleration of the bunch in the 
plasma (order of 10-100 GeV/m) \cite{Tajima79b,Joshi84,ChenP85,Esarey96}
minimizes the effect of space charge that would be severe for such dense 
beams ($10^{14}-10^{18}cm^{-3}$) produced from a conventional 
thermionic photocathode \cite{Fraser87}.

The original analysis of Ref. \cite{Umstadter96a} used single particle theory 
and estimates based on one-dimensional (1D) PIC simulations. 
This chapter contains the results from a detailed 2D and 3D PIC 
simulation analysis of this concept. We find that our results support 
the feasibility of such a cathodeless injection scheme, but that in 
the regime studied here the physical mechanism for the trapping is 
different from the one originally suggested.  Furthermore, we
show that the number of particles, emittance, and energy spread can all 
depend sensitively on the laser parameters and the injection phase. 
Depending on the applications, these results place constraints on the 
allowable shot to shot jitter of the injection laser. Last, based on 
the new insight into the trapping mechanism, we put forth additional 
geometries, e.g., co- and counterpropagating pulses, as well as 
related injection schemes.

\section{Acceptance of a Plasma Wave}
  \label{sect:injection:acceptance}

Before considering the simulation results we present here a calculation
of the acceptance\cite{Reiser} of a plasma wave. The acceptance of a plasma
wave is an estimate for the upper limit of the emittance of a beam
in a plasma wave since it is defined as the maximum transverse phase space
volume that can be accelerated by an accelerating system. 
For a plasma wave the acceptance can be approximately calculated by 
assuming a transverse potential profile that is responsible for the 
focusing forces of the plasma wave. For a given transverse 
potential, $\phi_{2}\equiv\phi_{0}(1-x_{2}^{2}/w_{p}^{2})$, we can 
find the maximum transverse momentum $p_{2}$ that a particle can 
have at a given transverse position $x_{2}$ before the particle can 
escape the potential well. Since the plasma wave as well as the 
particle both move with almost the same velocity, c, the potential 
function $\phi_{2}$ will change slowly and we will neglect that 
change here.

We start with the condition that an electron is trapped transversely 
in the plasma wave's potential well, i.e., that the transverse 
kinetic energy has to be smaller than the energy needed to escape the 
transverse potential $\left|E_{k,2}\right|<\left|E_{p,2}\right|$.

\begin{displaymath}
  \sqrt{p_{2}^{2}c^{2}+p_{1}^{2}c^{2}+m^{2}c^{4}}
  -\sqrt{p_{1}^{2}c^{2}+m^{2}c^{4}}<-e\phi_{2}\;\;\;\;\;\; 
  (\phi_{2}\leq 0)
\end{displaymath} 

\noindent{This} can be solved, giving an inequality for the $p_{2}$ 
of a trapped electron.

\begin{displaymath}
  \left|p_{2}\right|c<\sqrt{\left(-e\phi_{2}
  +\sqrt{p_{1}^{2}c^{2}+m^{2}c^{4}}\right)^{2}-m^{2}c^{4}-p_{1}^{2}c^{2}}
\end{displaymath}

\noindent{Rearranging} terms gives the following result:

\begin{equation}
  \label{equ:injection:p2max}
  \left|p_{2}\right|<mc\sqrt{-\frac{2e\phi_{2}\gamma_{1}}{mc^{2}}}
  \sqrt{\, 1 \, + \, \frac{-e\phi_{2}}{2mc^{2}} \: \frac{1}{\gamma_{1}}}\equiv 
  p_{2,max}\left(x_{2}\right)
\end{equation}

\noindent{where} $\gamma_{1}^{2}=1+(p_{1}/mc)^{2}$.

For linear waves $\bar{\phi}_{2}=e\phi_{2}/(mc^{2})\leq 1/2$; so to 
lowest order the second square root term can be approximated as unity. 
We use Eq.~(\ref {equ:injection:p2max}) to calculate the normalized acceptance 
\cite{Reiser}. 

\begin{equation}
  \label{equ:injection:acceptance1}
  A_{n}=2\int_{-\infty}^{\infty}\frac{p_{2.max}}{mc}dx_{2}
  =2\int_{-\infty}^{\infty}\frac{\sqrt{-2me\phi_{2}\gamma_{1}}}{mc}dx_{2}
\end{equation}

\noindent{Assuming} the potential given in
Eq.~(\ref{equ:review-physics:harmonic-potential}), and 
replacing $w_{p}$ with $w_{L}/\sqrt{2}$, we get an approximate 
result for $A_{n}$ by replacing the integration limits with 
$w_{L}/\sqrt{2}$ and $-w_{L}/\sqrt{2}$:

\begin{eqnarray}
  \nonumber
  A_{n}
  & = & 2\sqrt{ \, 2 \, m \, e\phi_{0}\cos\left(\alpha\right)\gamma_{1}}\frac{1}{mc}
     \int^{w_{L}/\sqrt{2}}_{-w_{L}/\sqrt{2}}
     \left(1-2\frac{x_{2}^{2}}{w_{L}^{2}}\right)^{1/2}dx_{2}
  \\ \nonumber
  \\
  \label{equ:injection:acceptance2}
  & = & 2 \pi w_{L} \sqrt{ \gamma_{1} \bar{\phi}_{0} } 
     \sqrt{ \cos \left( \alpha \right) }
\end{eqnarray}

\noindent{where} $\alpha$ is the phase of the electron in the wave 
with respect to the potential maximum. If we assume $\gamma_{1}$ is 
of the order of the trapping threshold then 
$\bar{\phi}_{0}\gamma_{1}=O(1)$, so the $\varepsilon_{n}$ for any 
cathodeless injection scheme is bounded by $\varepsilon<2 \, w_{L} \, \pi$. 
If the trapping of a particle bunch by a plasma wave doesn't take 
place at the maximum of the potential then $\cos (\alpha)$ is smaller 
than 1 and the emittance of the beam can be expected to be smaller 
than this upper bound. Note that if Eq.~(\ref{equ:review-physics:matched-beam})
is solved for 
$\varepsilon_{n}$ then it results in 
$\varepsilon_{n}=2\pi\sqrt{\gamma 
eE_{10}/(mc\omega_{p})}(\sigma/w_{L})^{2}\sigma$. Using 
$\gamma\approx\gamma_{1}$, 
$\bar{\phi}_{0}= k_{p}^{-1}\: e \, E_{10}/(mc^{2})$, and $\sigma=w_{L}$ 
leads to $\varepsilon_{n}=2\pi 
w_{L}\sqrt{\gamma_{1}\bar{\phi}_{0}}$. This means that the 
acceptance is the emittance for a matched beam.

\section{Simulation Parameters}
  \label{sect:injection:parameters}  
  
The simulations were done with Pegasus \cite{Tzeng96} on a single node
with the moving window to follow the laser pulse for extended periods of time. 
Fig.~\ref{fig:injection:scheme} shows the basic set up of the simulations.
The following parameters are 
valid for most of the simulations results presented below unless 
stated differently. The simulation box has a size of 35 
$c/\omega_{p}$ in the $x_{1}$ direction and 25 $c/\omega_{p}$ in 
the $x_{2}$ direction and the simulations run for a time of 105 
$\omega_{p}^{-1}$. The simulations use a 700 x 500 grid, a timestep, 
dt=0.035 $\omega_{p}^{-1}$, and four particles per cell.

In the beginning of the simulation, as the drive pulse enters into
the cold plasma in the $x_{1}$ direction, it creates a plasma wave
in its wake. At a later time the injection pulse is launched in a
vacuum region at the side of the box and propagates in the $x_{2}$ 
direction crossing the path of the drive pulse.
The frequency ratio 
$\omega_{0}/\omega_{p}$ between the laser frequency and the plasma 
frequency is 5 for both pulses, and both have their polarization in 
the plane of the simulation. (This means the drive pulse has mainly 
an $E_{2}$ component and the injection pulse mainly an $E_{1}$ 
component). We adopt the notation of Ref. \cite{Umstadter96a}, where the 
normalized vector potential for the drive pulse is a $\equiv 
eA_{y}/mc^{2}=1$ and for the injection pulse is $b\equiv 
eA_{x}/mc^{2}=2$, unless stated otherwise. We observed in the 
simulation that the plasma wave amplitude caused by $a=1$ is about 
$\bar{\phi}_{0}=0.45$. The transverse profile for each laser is given 
by a Gaussian with a spot size of $3 c/\omega_{p}$. The temporal 
profile has a symmetric rise and fall of the form $f(x)-10\cdot 
x^{3}-15\cdot x^{4}+6\cdot x^{5}$ with $0\leq x=\tau/\tau_{L}\leq 
1$. The value of $\tau_{L}$ is $\pi c/\omega_{p}$ for the drive 
pulse and $\frac{1}{2} \pi c/\omega_{p}$ for the injection pulse; thus the 
simulations have fewer laser cycles than in typical experiments. We 
define the injection phase $\psi$ to be the distance between the back 
of the drive pulse and the center of the injection pulse as it 
crosses the axis. This is shown in Fig.~\ref{fig:injection:scheme}. 

In order to convert the simulation results to physical units, we 
assume a plasma density of $10^{16}cm^{-3}$. If not stated 
differently all quantities are given in normalized Gaussian units with 
the plasma frequency equal to 1. The number of accelerated electrons 
is estimated from the simulations as follows:

\begin{equation}
  \label{equ:injection:particle-number}
  N=\frac{\# \; of \; trapped \; simulation \; particles}{\# \; of \; particles 
  \; per \; cell}\cdot n\cdot dx_{1}\cdot dx_{2}\cdot \Delta x_{3}\times 
  ((mc^{2})/(4\pi e^{2}n))^{3/2}
\end{equation}

Here n is the electron density in $cm^{-3}, dx_{1}$ and $dx_{2}$ 
are the cell sizes in the $x_{1}$ and $x_{2}$ direction, and $\Delta 
x_{3}$ is an assumed extension in the $x_{3}$ direction. $dx_{1}, 
dx_{2}$ and $\Delta x_{3}$ are in normalized units. We assume 
$\Delta x_{3}$ to be equal to $\Delta x_{2}$, the width of the group 
of accelerated particles in $x_{2}$. The normalized emittance is 
calculated as:

\begin{equation}
  \label{equ:injection:emittance}
  \varepsilon_{n}=\gamma\times\frac{\Delta p_{2}}{p_{1}}\cdot \Delta 
  x_{2}\times \left(\left(mc^{2}\right)/\left(4\pi 
  e^{2}n\right)\right)^{1/2}
\end{equation}

\noindent{with} $\gamma=\sqrt{1+\vec{p}\,^{2}}\approx p_{1}$.
Here, $p_{1}$ is the average longitudinal momentum and $\Delta 
p_{2}$ and $\Delta x_{2}$ are the width of the distributions of 
$p_{2}$ and $x_{2}$ for the group of accelerated particles. It 
should be noted that the number of electrons as well as the 
normalized emittance both scale with $n^{-1/2}$. All quantities 
including the energy spread are calculated after the final timestep of 
the calculation, i.e., after a propagation distance of $105 
c/\omega_{p}$ (particles are trapped, $\gamma>\gamma_{\phi}$, 
between 50 and 60 $c/\omega_{p}$-see 
Fig.~\ref{fig:injection:p1-of-t-2d}).
The values of $\Delta 
x_{2}$ and $\Delta p_{2}$ are defined to be the standard deviations 
of the particles bunches for these quantities. The energies of the 
trapped particles are around 10 MeV, which is of the order of the 
theoretical energy gain $\Delta W=16 MeV$
(Eq.~(\ref{equ:review-physics:energy-gain}) with $\eta =1$, 
$\bar{\phi}_{0}=0.45$, and $\gamma_{\phi}=5$) that would be obtained over 
the dephasing distance of $80 c/\omega_{p}$
[see Eq.~(\ref{equ:review-physics:dephasing})]. The trapping 
threshold for these simulations is 0.05 MeV
[see Eq.~(\ref{equ:review-physics:threshold})]. The simulation 
show that trapped particles close to the maximum accelerating 
gradient, which is consistent with the result above.

\section{2D Simulation Results}
  \label{sect:injection:2d-results} 
  
\begin{figure}
   \begin{center}
      \epsfig{ file=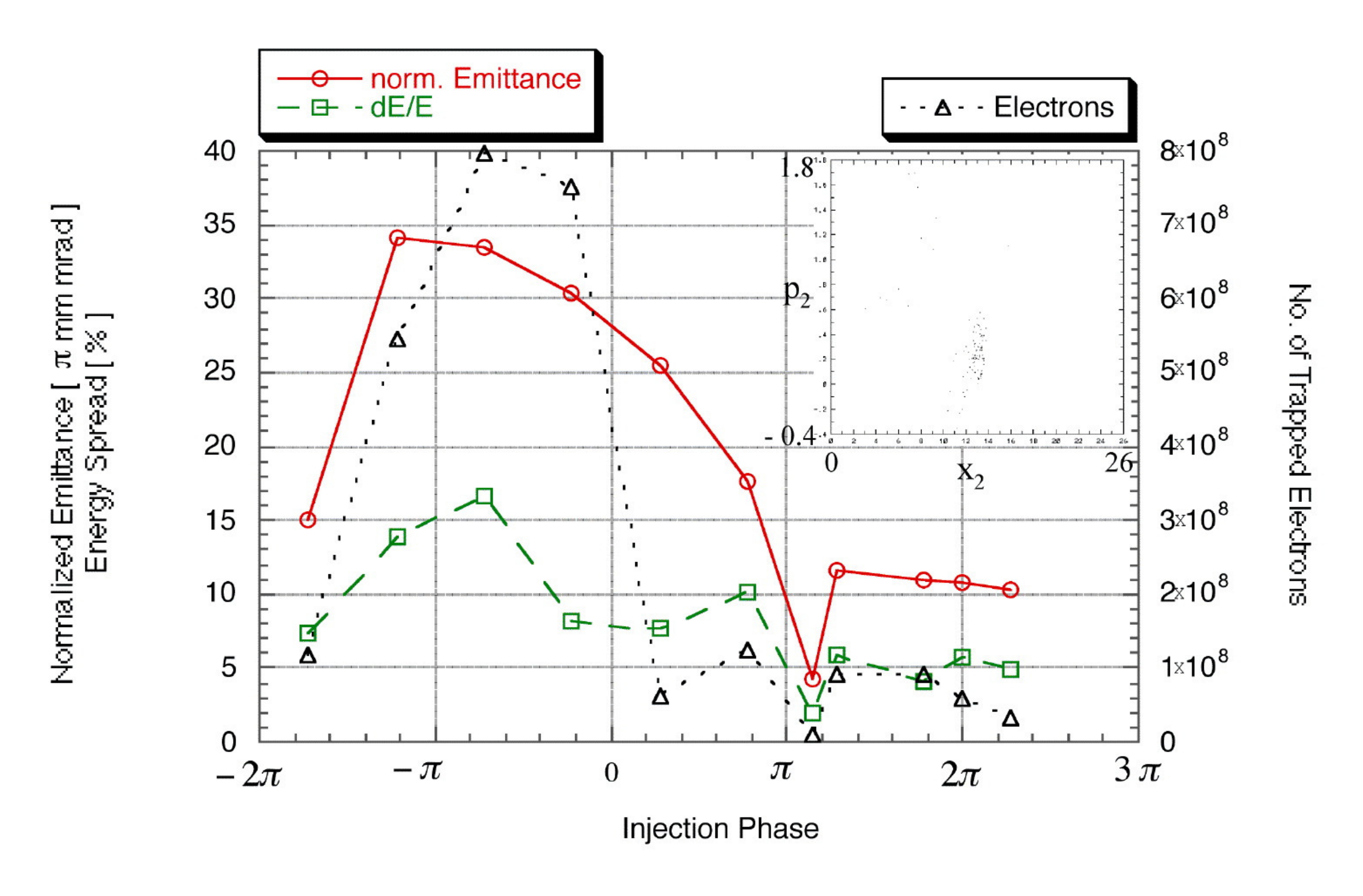, width=5.8in }
      \caption{The number of trapped electrons, the normalized emittance,
               and the energy spread of the trapped particles as a function
               of the injection phase. The injection amplitude $b$ is $2.0$ and
               the drive amplitude $a$ is $1.0$. The connecting lines between the
               data points have been added to make it easier to distinguish
               the different data. The inset shows the raw data for the
               transverse phase space of the trapped particles that is used
               to calculated the emittance for the simulation at 
               $\psi=1.8\pi$.}
      \label{fig:injection:phase}
   \end{center}
\end{figure}

\begin{figure}
   \begin{center}
      \epsfig{ file=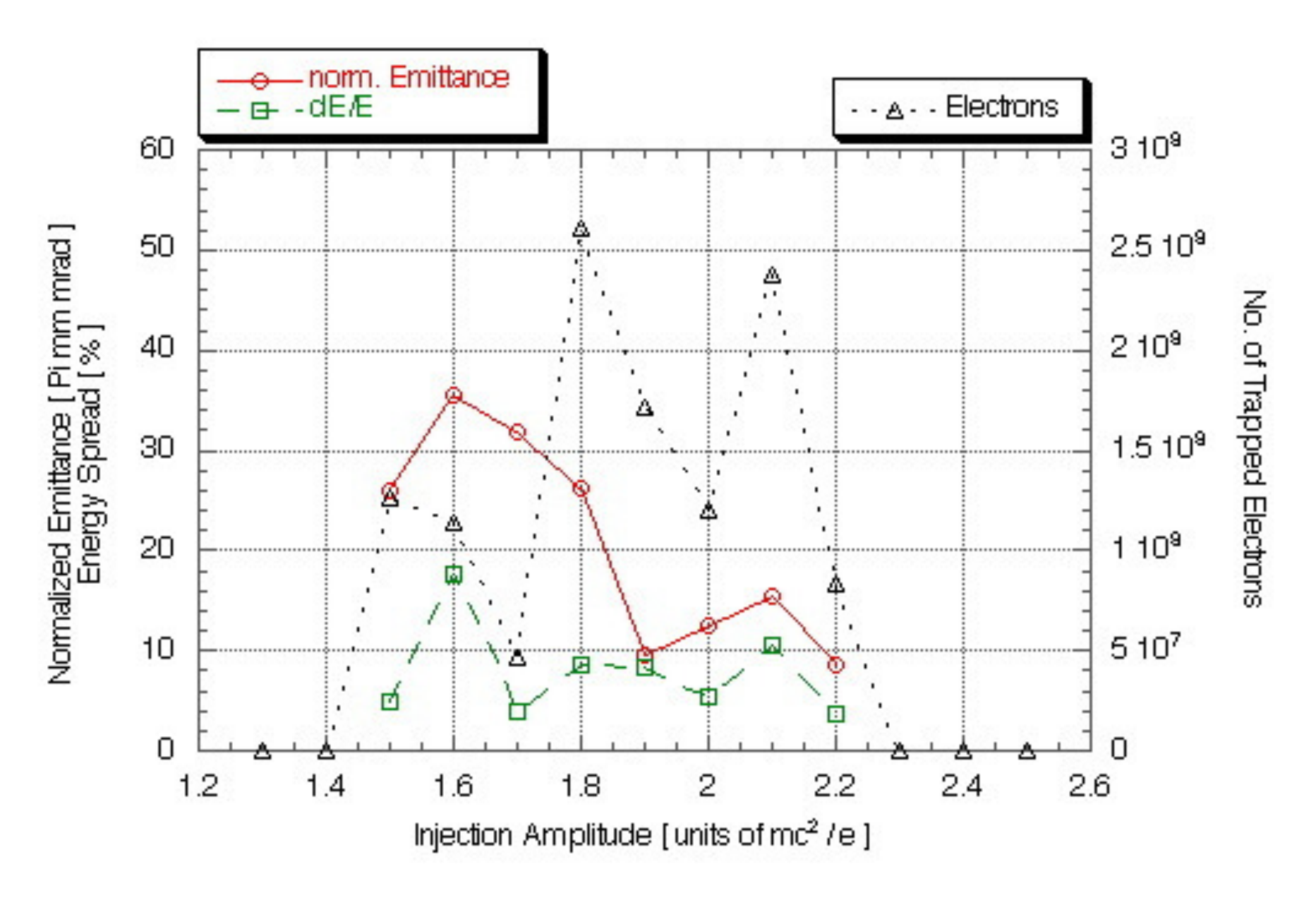, width=5.8in }
      \caption{The number of trapped electrons, the normalized emittance, and
               the energy spread of the trapped particles as a function of
               the injection amplitude. The injection phase $\psi$ is $1.3\pi$.
               All other parameters are the same as the ones used in the
               simulations of Fig.~\ref{fig:injection:phase}. The connecting
               lines between the data points have been added to make it easier
               to distinguish the different data.}
      \label{fig:injection:amplitude}
   \end{center}
\end{figure}

The engineering results of the simulations can be summarized in
Fig.~\ref{fig:injection:phase} and Fig.~\ref{fig:injection:amplitude}.
In Fig.~\ref{fig:injection:phase} we plot the number of trapped
electrons, the emittance,  and the energy spread as a function of
the injection phase for a fixed value of the injection amplitude, b=2.0.
In Fig.~\ref{fig:injection:amplitude}, we plot the same quantities as
in Fig.~\ref{fig:injection:phase} but as a function of the injection 
amplitude for a fixed value of the injection phase, $\psi = 1.3\pi$. 
All other parameters have the values given before. Note that negative 
values for $\psi$ mean that the center of the injection pulse crosses 
the $x_{2}$ axis before the end of the drive pulse.

The most notable feature of Fig.~\ref{fig:injection:phase} is the large
variation of the three beam quantities as a function of $\psi$ and
especially the strong difference in the number of particles and their emittance 
between positive injection phases larger and smaller than $\pi$. The 
direct overlap of the injection pulse with the drive pulse (i.e. 
injection phase smaller than $\pi$) clearly yields the largest number 
of trapped particles. The maximum number of trapped electrons 
corresponds to $8 \times 10^{8}$ at a plasma density of 
$10^{16}cm^{-3}$ (or to $6 \times 10^{7}$ at a density of 
$10^{19}cm^{-3}$. Note that 100\%\ beamloading \cite{Katsouleas87}
corresponds to $N=5 \times 
10^{5}\bar{\phi}_{0}\sqrt{n_{0}cm^{3}}Acm^{-2}\approx 8\times 
10^{9}$ for $n_{0}=10^{16}cm^{-3}$ where we use a laser beam 
cross section of $A=\pi w_{p}^{2}$ with 
$w_{p}=w_{L}/\sqrt{2}=\frac{3}{\sqrt{2}}c/\omega_{p}$. Therefore, 
there is $\leq$ 10\%\ beamloading for negative and $\leq$1\%\ 
positive injection phases.

The number of particles decreases by an order of magnitude for 
injection phases larger than $\pi$. The normalized emittance on the 
other hand is better for injection phases larger than $\pi$, with the 
smallest normalized value of $3\pi$ mm mrad in a $10^{16}cm^{-3}$ 
density plasma (or 0.1$\pi$ mm mrad at $10^{19}cm^{-3}$). Note from 
Eq.~(\ref{equ:injection:acceptance2}) 
that the acceptance for the plasma wave places an upper 
bound on the emittance of $2w_{L}\pi =300 \pi$ mm-mrad for 
$n_{0}=10^{16}cm^{-3}$. However since the particles are getting 
trapped at a phase close to the maximum accelerating phase of the 
plasma wave (i.e., close to a zero for the focusing field), the $\cos$ 
term in Eq.~(\ref{equ:injection:acceptance2})
is small. We therefore expect the emittance to 
be smaller than this upper limit. For injection phases smaller than 
$\pi$ the emittance increases by a factor of five. The emittance 
therefore seems to grow with the number of particles. Although this 
is suggestive of some sort of space charge degradation, we will show 
later that space charge is not important. Instead, we believe that the 
relatively larger emittance and number of particles at smaller $\psi$ 
are both due to a stochastic interaction between the plasma and the 
overlapping laser fields.

The energy spread of the accelerated bunch also varies widely; it is 
between 2\% and 17\% at a beam energy of 10 MeV and we expect the 
energy spread $\Delta E/E$ to scale as $1/\gamma$ for simulations with 
larger dephasing energies (i.e., larger values of 
$\omega_{0}/\omega_{p}$), since $\Delta E/E \propto \Delta E/\gamma$ 
and $\Delta E$ is not expected to change significantly. There is an 
interesting difference between the behavior of the energy spread and 
the number of particles on the one hand and of the emittance on the 
other hand for the larger injection phases in Fig.~\ref{fig:injection:phase}.
The energy spread, and to some extent the number of particles fluctuate as a 
function of $\psi$. The emittance remains almost constant which 
suggests that it is determined by qualities of the accelerating 
plasma wave and not by details of the injection process like the 
injection phase.

Although the simulations with b=2.0 produce similar numbers of 
particles at $\psi =1.3\pi$ or $1.8\pi$ as can be seen from
Fig.~\ref{fig:injection:phase}, for b=1.8 the number of particles changes
from a $10^{8}$ at $\psi=1.3\pi$ (see Fig.~\ref{fig:injection:amplitude})
to nearly zero at $\psi=1.8\pi$ (data not shown in figures). This indicates
that the results of the simulations are quite sensitive to b and $\psi$ so
that the curve found in Fig.~\ref{fig:injection:phase} for the injection phase
dependence at injection amplitudes of 2.0 is not readily applicable to other
values of this parameter but merely indicates the magnitudes of 
various quantities that can be obtained.

The value of $\psi=1.3 \pi$ is used for the simulations of
Fig.~\ref{fig:injection:amplitude} 
since it seems to have close to an optimal injection phase judging 
from the data of Fig.~\ref{fig:injection:phase}. As a function of the
injection amplitude  the normalized emittance and the energy spread do
not seem to show any systematic behavior on the scale that is resolved by
the simulations. The values of the energy spread vary between 4\%\ and 
18\%\ while the values for the emittance are between 10 $\pi$ mm mrad 
and 40 $\pi$ mm mrad. Depending on the application, these variations 
will place a limit on the tolerable shot to shot laser jitter.

The number of trapped electrons on the other hand seems to show a 
systematic behavior. What should be expected is that the number of 
trapped particles first rises with increasing injection amplitude and 
then falls off. This is recognizable in the figure even though the 
curve is quite noisy. The decrease with an increased amplitude causes 
an increase in transverse momentum, $P_{2}$, that is transferred to 
the particles by the injection pulse. At a certain value, the 
transverse momentum becomes large enough to prevent the trapping of 
the particles.

We may use Fig.~\ref{fig:injection:phase} and
Eq.~(\ref{equ:injection:particle-number}) and 
Eq.~(\ref{equ:injection:emittance}) to
obtain an interesting scaling law for the brightness of the plasma cathode 
injector. The normalized brightness can be defined by 
$B_{n}=I/\varepsilon_{n}^{2}$ \cite{Reiser} for axially 
symmetric beams. For the average current of a bunch we find $I \propto 
N/T_{p}=N/\omega_{p}$, where $T_{p}$ is the plasma wave period. As 
noted earlier N scales with $n^{-1/2}$ while $\omega_{p}$ scales as 
$n^{1/2}$. Therefore, the product $N\omega_{p}$ does not depend on 
the density as the simulation results are scaled to different 
densities for fixes values of $\omega_{0}/\omega_{p}$, a, and b. For 
example at $\psi =1.8 \pi$, we get $I_{max}$=220 Amps and 
$\varepsilon_{n} =11\pi mm \;
mrad\left[\left(10^{16}cm^{-3}\right)/n\right]^{1/2}$ which scales 
as $n^{-1/2}$. Combining these results predicts a brightness of 
$B_{n}=1.8 \cdot 10^{7} \times n/\left(10^{16}cm^{-3}\right)\times 
Amps/cm^{2}$ which scales linearly with density.

The insensitivity of the beam current to the plasma density should 
also hold if $\omega_{p}$, a, and b are changed. This can be argued 
as follows. The beam current can be written as $I=en_{b}c\times 
\sigma^{2}\pi$ where $n_{b}$ is the beam density. If we normalize 
$n_{b}$ with respect to the plasma density $n_{0}$ and the spot 
size $\sigma$ with respect to $c/\omega_{p}$ we find that

\begin{equation}
  \label{equ:injection:current}
  I = e n_{0} c \left( \pi c^{2} / \omega_{p}^{2} \right) \frac{n_{b}}{n_{0}}
      \left( \sigma c / \omega_{p} \right)^{2}
    = \frac{I_{A}}{4} \frac{n_{b}}{n_{0}}
      \left( \sigma c / \omega_{p} \right)^{2}
\end{equation}

\noindent{This} expression for I is insensitive to the plasma density 
for various laser parameters, if the normalized beam density and the 
spot size are relatively insensitive to the plasma density. We expect 
that the ratio $n_{b}/n_{0}$ is not a strongly varying function of 
$\gamma_{\phi}=\omega_{0}\omega_{p}$, since the trapping threshold 
asymptotes for large $\gamma_{\phi}$ 
[see Eq.~(\ref{equ:review-physics:threshold})]. Note also that 
since $n_{b}/n_{0}$ is typically less than 1 and $\sigma$ is 
typically $c/\omega_{p}$ or less, this shows that the current is 
typically some fraction of the Alfven current.

It is interesting that despite their high brightness and density, the 
bunches are not space charge dominated. From the discussion above 
$I/I_{A}$ is of the order of $10^{-2}$, while 
($\sigma/\varepsilon_{n})^{2}$ is typically of order unity. Thus, 
using Eq.~(\ref{equ:review-physics:space-charge-parameter}) we
find that $\rho\ll 1$ at all times in the 
plasma and the beam is emittance dominated. We note that once the 
bunch leaves the plasma and expands in free space it can rapidly 
become space charge dominated. For beams generated by the cathodeless 
injection scheme this typically occurs in a distance of the order of 
1 cm $ \times \left[ \left( 10^{16} cm^{-3} \right) / n \right]^{1/2}$. 
Since the effects of space charge can be neglected, it is possible to 
apply Eq.~(\ref{equ:review-physics:matched-beam}),
the condition for matched beams. For the 
simulation parameters the left side of
Eq.~(\ref{equ:review-physics:matched-beam}) has values 
between 2 and 3, which means the external force term is larger than 
the diffraction term. For the beam emittances in the simulations, we 
also note that $\varepsilon_{n}$ is between 0.01 and 0.12 times the 
plasma wave acceptance that was calculated above. The numbers for the 
matched beam condition and the emittance to acceptance ratio indicate 
that once the electrons are ``injected'' they are well within the 
parameters of stable acceleration for the plasma wave.

To achieve high energies in the LWFA the laser pulse must propagate 
through many diffraction or Rayleigh lengths of plasma. One way to 
guide a pulse is to use a parabolic density channel 
\cite{Sprangle92a,Durfee93}. Therefore the cathodeless injection scheme may 
need to work in plasma channels. We have carried out a simulation in 
which the drive pulse propagated down a channel and the injection 
pulse propagated across the channel. The channel had a width of $3.25 
c/\omega_{p}$ and the density was decreased by 40\% in the middle of 
the channel. In the simulation the number of trapped particles as well 
as the emittance of the particle bunch are reduced to about 20\% from 
their values in the uniform plasma case. We also note that for all 
results presented in this chapter so far the initial plasma is cold. We
have done simulations with a 1 KeV plasma and the number of electrons as
well as the emittance decrease to about 40\% of the cold plasma values.

\begin{figure}
   \begin{center}
      \epsfig{ file=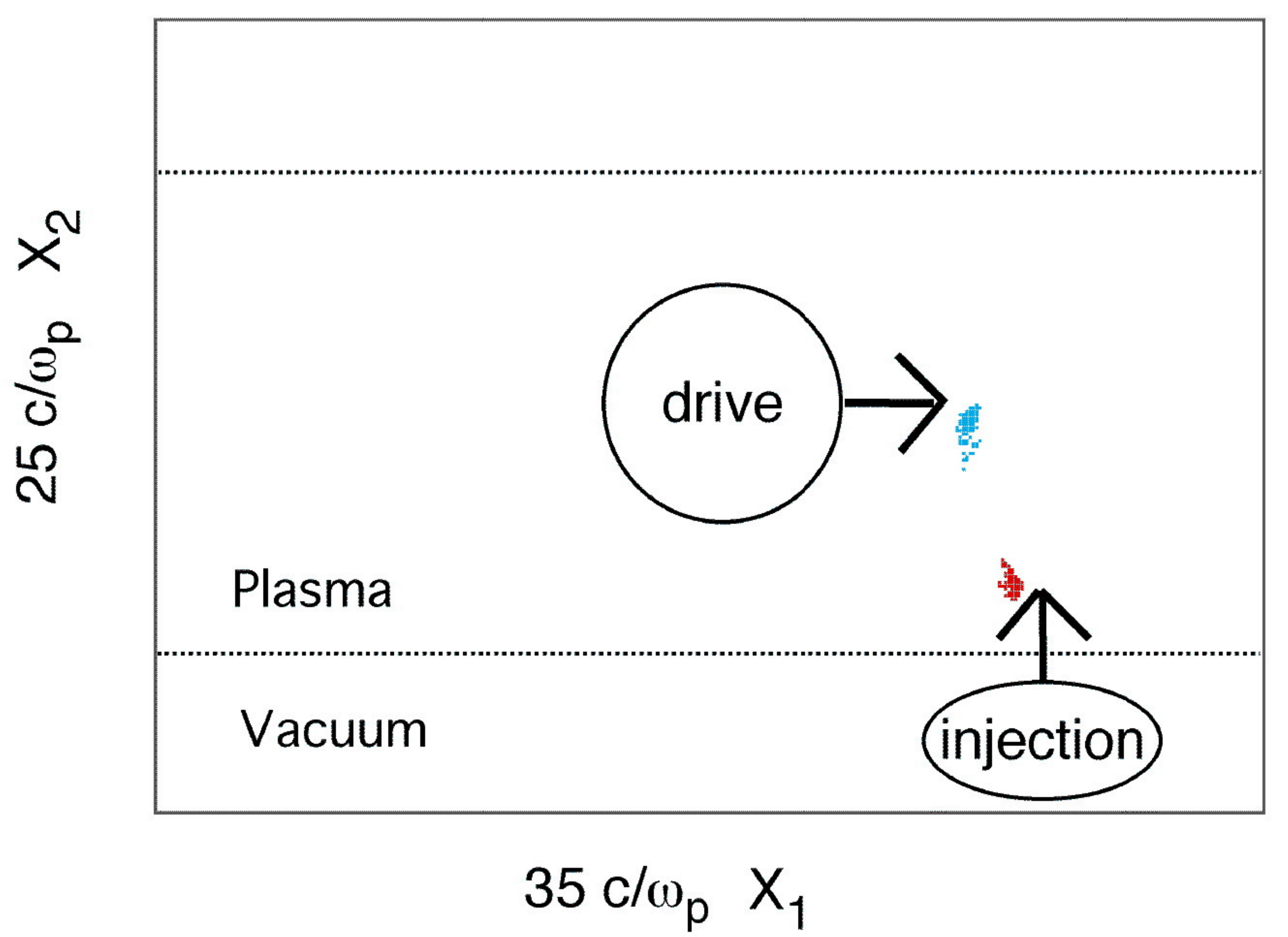, width=4.0in }
      \caption{The figure shows the initial position of trapped particles
               for two different simulations. The red particles come from
               a simulation with $\psi=1.8\pi$ and $b=2.0$. The blue
               particles come from a simulation with with $\psi=1.3\pi$
               and $b=1.8$. The position of the drive pulse in the figure
               is illustrative and does not match $\psi=1.3\pi$ or
               $\psi=1.8\pi$.}
      \label{fig:injection:par.origin}
   \end{center}
\end{figure}

Insight into the mechanism of trapping can be gained by studying the 
original location and the trajectories of the trapped particles. In
Fig.~\ref{fig:injection:par.origin}, we plot the original $(x_{1}, x_{2})$
positions for all the trapped particles from two simulations. The red points
are for $\psi=1.8\pi$ and b=2.0, while the blue points are for a $\psi=1.3\pi$ 
and b=1.8. There are several important points to be noticed. The first 
is that for both cases the particles are to the left of the injection 
pulse. Therefore, these particles feel a transverse ponderomotive 
force to the left not to the right as was presumed in Ref.~\cite{Umstadter96a}.
We have verified this by rerunning the simulations without the 
drive pulse to see only the effect of the injection pulse.

\begin{figure}
   \begin{center}
      \epsfig{ file=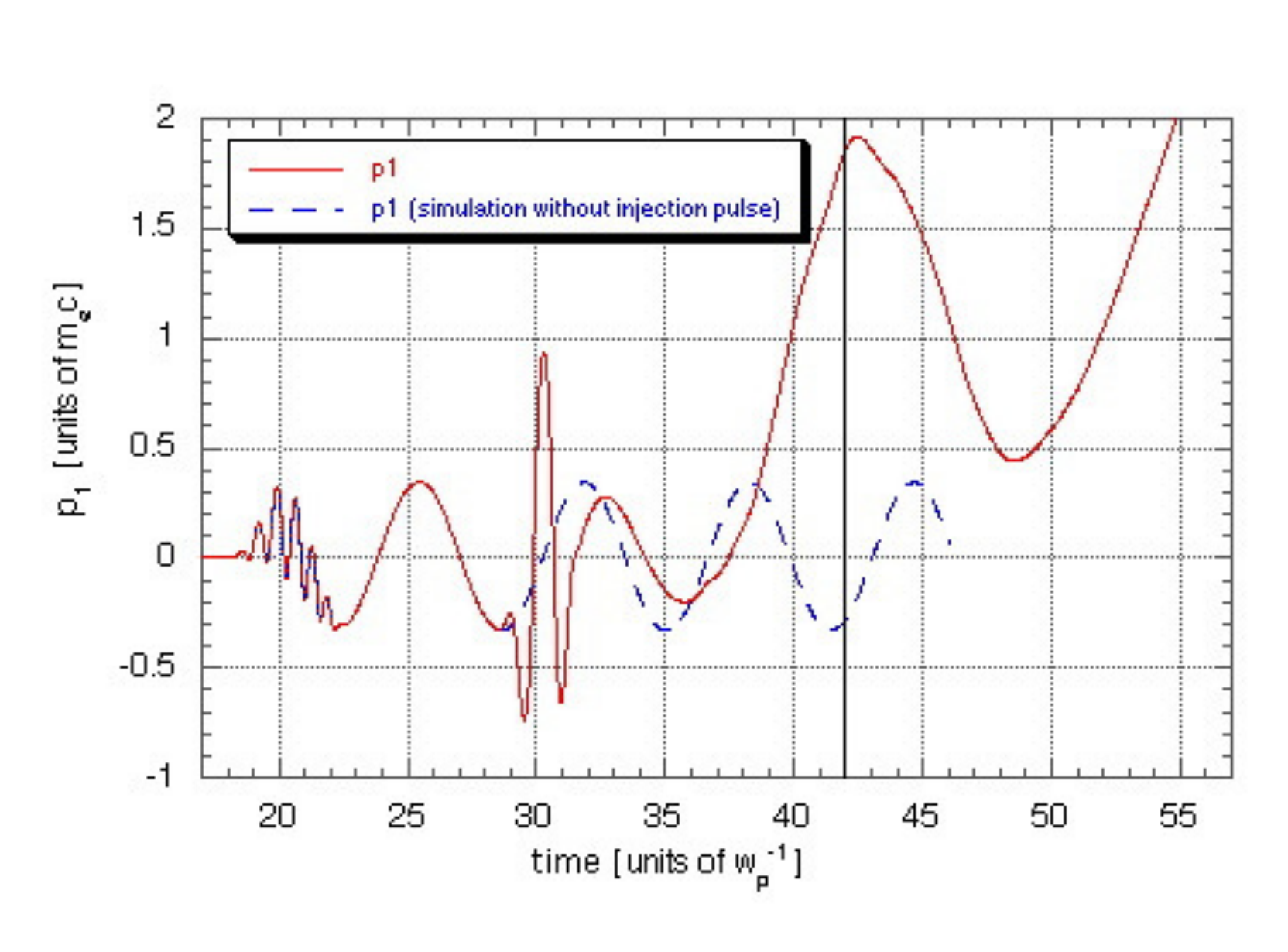, width=5.8in }
      \caption{$p_{1}$ of a test particle as a function of time. The
               two curves are the results from simulations with (solid)
               and without (dashed) an injection pulse. $\psi=1.3\pi$ and
               $b=1.8$ for the simulation with an injection pulse. The
               initial position of the test particle is given by offsets
               of $-2.3$ in $x_{1}$ and $-0.1$ in $x_{2}$ relative to the
               intersection of the pulses (see Fig.~\ref{fig:injection:par.origin}).
               The vertical line in the figure indicates the time $t=42.0$
               at which the electric fields are given in
               Fig.~\ref{fig:injection:e1field}.}
      \label{fig:injection:p1-of-t-2d}
   \end{center}
\end{figure}

To gain a deeper understanding of the process, we follow the momentum 
of a single, typical, trapped particle as function of time in the 2D 
simulation. We consider a particle for the case of $\psi=1.3\pi$ and 
b=1.8. The data are shown in Fig.~\ref{fig:injection:p1-of-t-2d}. The initial
momentum is zero since the simulation uses a cold plasma. We show here the results
for only one particle, but the curves are very similar for other trapped 
particle phase space trajectories in this simulation.

The solid curve in Fig.~\ref{fig:injection:p1-of-t-2d} shows the longitudinal
momentum of the particle; the dotted curve shows $p_{1}$ for the same particle
in a simulation where the injection pulse is not launched. As expected, the 
same particle simply oscillates in the wake of the drive pulse. In the 
full simulation we can see that the injection pulse has completely 
passed by the test particle at about the time t=31.7. Although the 
injection pulse has an impact on the particle, the really large 
changes occur at a time when the injection pulse has already left the 
area of the test particle. This indicates that the trapped particle 
gets the extra momentum needed for getting trapped from the 
interaction of the two plasma wake fields created by these pulses 
rather than from any effect directly related to the laser pulses 
(since those have already left the area of the particle). Note that 
the trapped particle goes through one full oscillation (accelerating, 
decelerating, and accelerating again) before it is trapped. This 
feature that the particles get trapped in a multi-step process 
(acceleration-deceleration-acceleration) caused by the interaction of 
the wake fields is not unique to this particular simulation. Other 
simulations with different values for $\psi$ and b showed the same 
process.

\begin{figure}
   \begin{center}
      \epsfig{ file=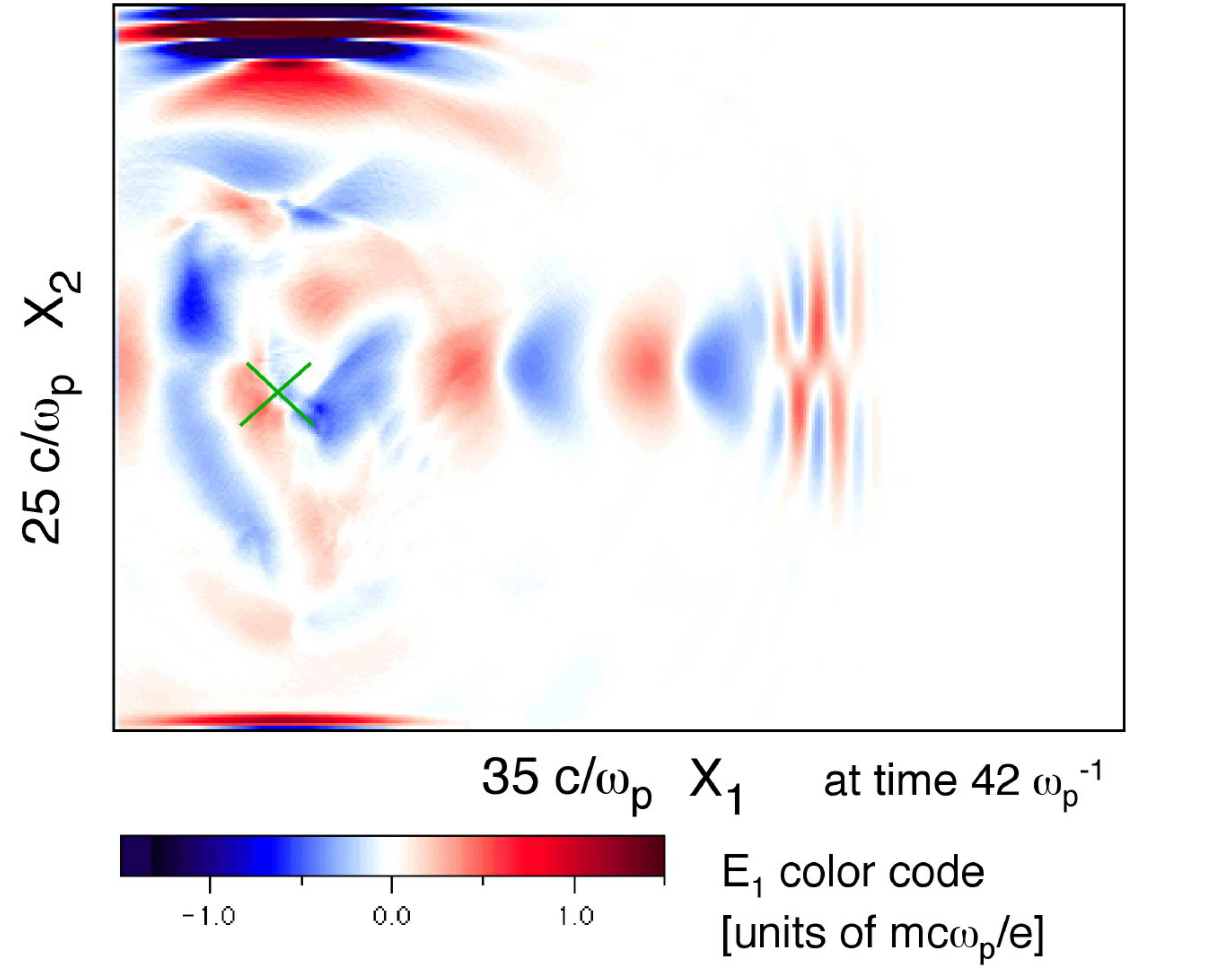, width=5.8in }
      \caption{The $E_{1}$ field at the time $t=42.0$ for $\psi=1.3\pi$ and
               $b=1.8$. The cross indicates the position of the test particle
               shown in Fig.~\ref{fig:injection:p1-of-t-2d}}
      \label{fig:injection:e1field}
   \end{center}
\end{figure}

Fig.~\ref{fig:injection:e1field} shows the $E_{1}$ field at t=42.0. The blue
areas accelerate, while the red areas decelerate electrons with respect to
the $x_{1}$ direction. The green cross marks the position of the test particle 
shown in Fig.~\ref{fig:injection:p1-of-t-2d} at that time. The position of the
particle in this picture is consistent with the development of $p_{1}$ in
Fig.~\ref{fig:injection:p1-of-t-2d}. The  particle is at the edge of the
accelerating area and will slip back into the decelerating area. The field
magnitude of the decelerating area is clearly smaller than that of the
accelerating area. The spatial structure of the $E_{1}$ field seen in this
figure can qualitatively be understood as mainly a superposition of the 
longitudinal field of the drive pulse wake and the transverse field of 
the injection pulse wake.

\section{3D Simulation Results}
  \label{sect:injection:3d-results}
  
A key motivation for developing OSIRIS was to be able to routinely carry
out 3D simulations. In this section we give one example of the usefulness
of 3D simulations. In particular, we use them to check the validity of 
the 2D simulation results presented earlier in this chapter.
The 3D simulation results
presented below are for the same parameters as
the 2D simulation with an injection phase of $\psi=1.3\pi$ and a
normalized vector potential of $b=1.8$ for the injection pulse. The 
2D result for the electron number, the emittance, and the energy spread 
were summarized in Fig.~\ref{fig:injection:amplitude}.
In order to decrease the computational cost of the simulation the
numerical parameters are slightly modified in the 3D simulation.
For the 3D simulation the simulation box has a size of
35.09~$c/\omega_{p}$ in the $x_{1}$ direction and
25.13~$c/\omega_{p}$ in the $x_{2}$ and $x_{3}$ directions.
The simulation uses a $400 \times 280 \times 280 \simeq 31 \times 10^{6}$
grid, a timestep dt=0.0513 $\omega_{p}^{-1}$,
and four particles per cell. The total number of particles in the 
simulation is $\sim 45$ million. The injection laser for the 3D simulation
propagated in the $x_{2}$-direction as it did in the 2D simulations.

\begin{figure}
   \begin{center}
      \epsfig{ file=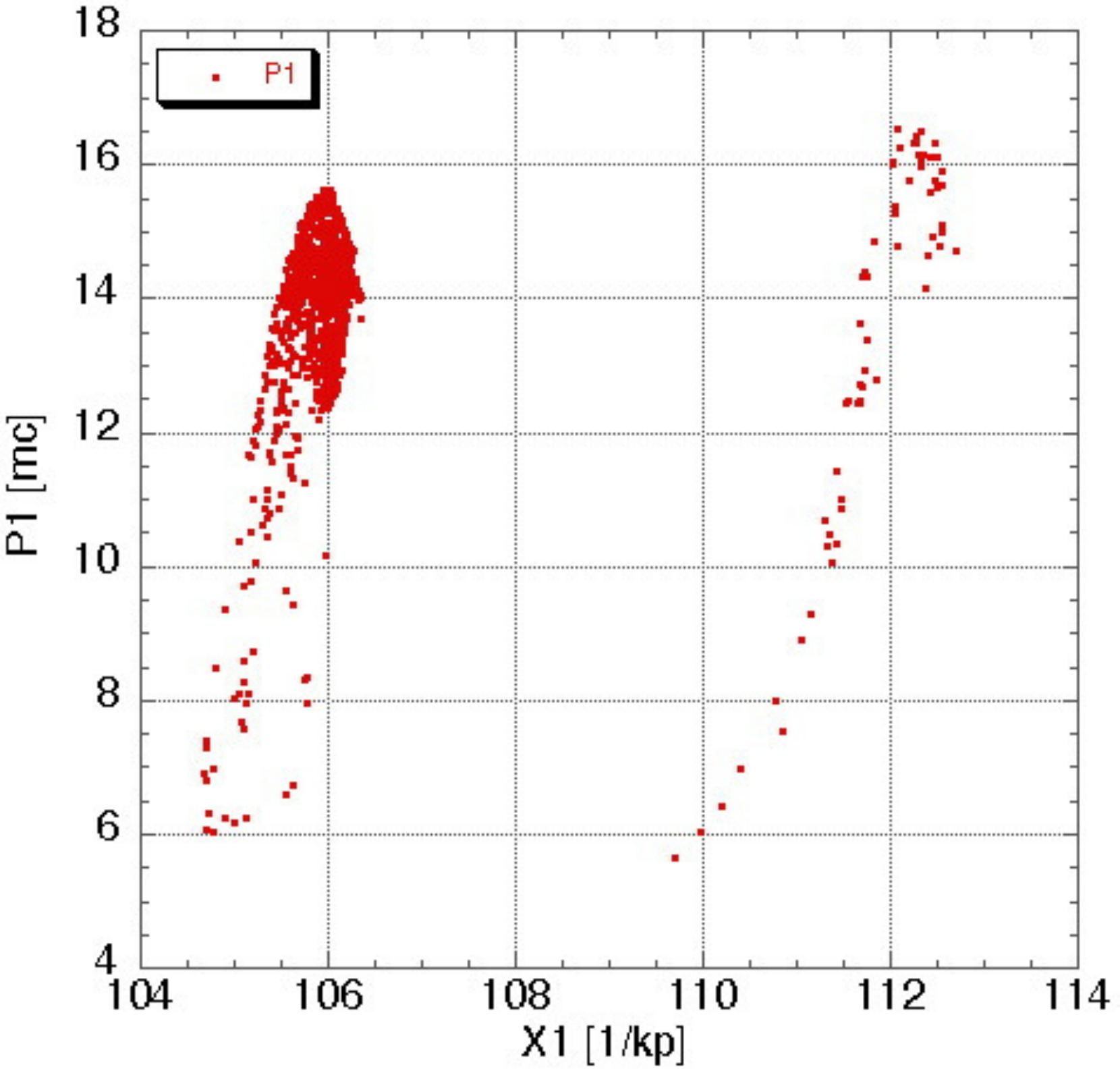, width=3.0in }
      \caption{The longitudinal phase space $x_{1}$-$p_{1}$ of the
               injected particles at the end of the 3D simulation.}
      \label{fig:injection:3d-energy}
   \end{center}
\end{figure}

\begin{figure}
   \begin{center}
      \epsfig{ file=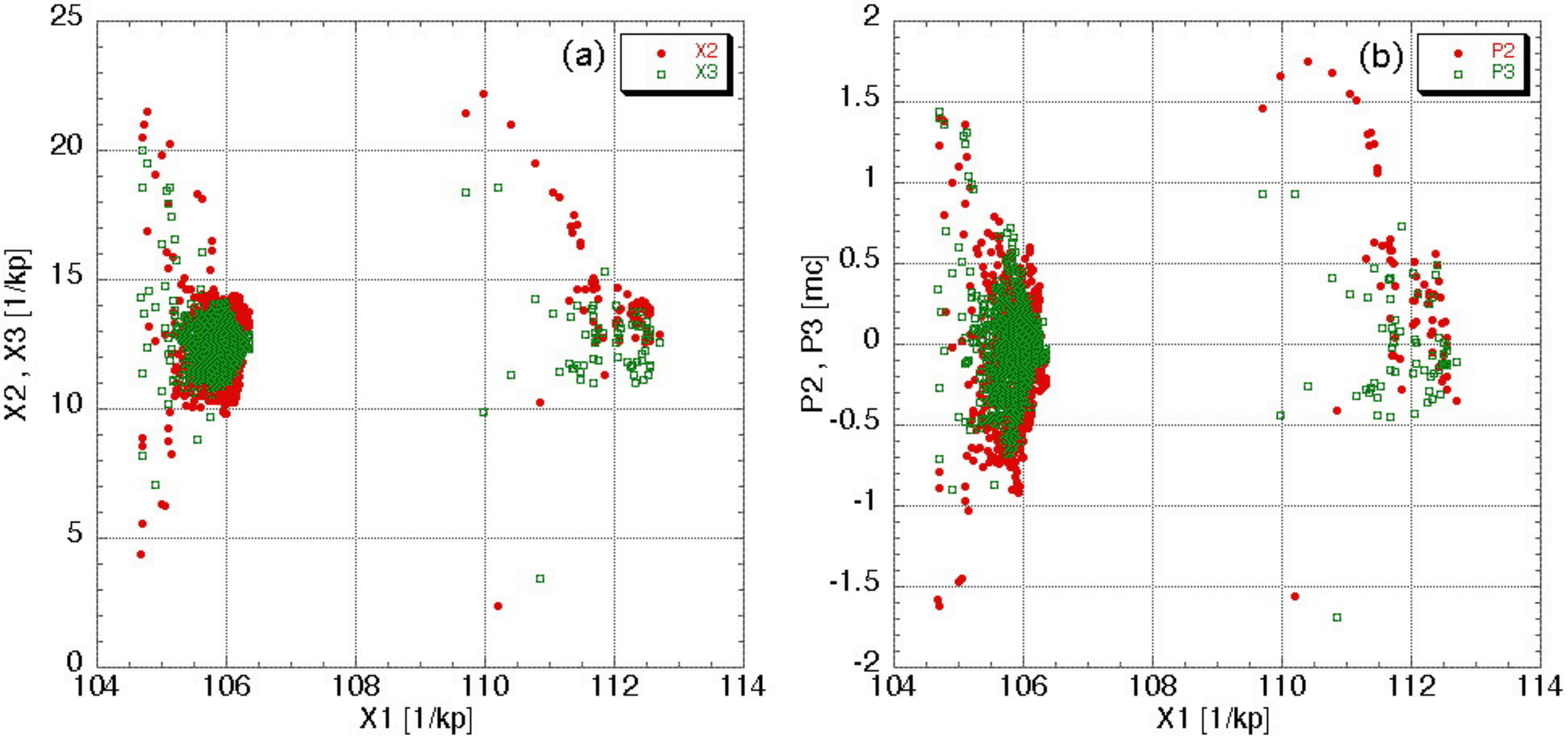, width=5.8in }
      \caption{The transverse phase space data at the end of the
               3D simulation. (a) shows the $x_{1}$-$x_{2}$ and the 
               $x_{1}$-$x_{3}$ distribution of the injected particles.
               (b) shows the $x_{1}$-$p_{2}$ and the $x_{1}$-$p_{3}$
               distribution of the injected particles particles.}
      \label{fig:injection:3d-transvers}
   \end{center}
\end{figure}

The results of the 3D simulation regarding the injected
electrons are summarized in Tab.~\ref{tab:injection:3d-2d-results}.
Tab.~\ref{tab:injection:3d-2d-results} gives the mean values
and widths $\bar{x}_{i}$, $\bar{p}_{i}$, $\sigma_{x_{i}}$, and
$\sigma_{p_{i}}$ as well as the number of the injected electrons after
the final timestep for the
two distinct particle bunches that got injected during the simulation.
Various phase space plots for the particle data are plotted in 
Fig.~\ref{fig:injection:3d-energy} and
Fig.~\ref{fig:injection:3d-transvers}. (Note that in these phase space 
plots the axes that are not shown are axes that have been integrated 
over, e.g., in Fig.~\ref{fig:injection:3d-energy} $x_{2}$, 
$x_{3}$, $p_{2}$, and $p_{3}$ have been integrated over.)

\begin{table}

  \begin{singlespace}\begin{center}
  \begin{tabular}{|l|r|r|r|r|} \hline \hline
   & 3D - Bucket 1 & 3D - Bucket 2 & 2D - Bucket 2 \\ \hline \hline
    
    $\bar{x}_{1}$    & 111.84              & 105.88   &    -      \\ 
    $\sigma_{x_{1}}$ &   0.66              &   0.27   &   0.4     \\ \hline 
    $\bar{x}_{2}$    &   1.79              &  -0.59   &    -      \\ 
    $\sigma_{x_{2}}$ &   2.72              &   1.45   &   1.0     \\ \hline  
    $\bar{x}_{3}$    &  -0.02              &  -0.11   &    -      \\ 
    $\sigma_{x_{3}}$ &   1.85              &   0.97   &    -      \\ \hline  
    $\bar{p}_{1}$    &  13.40              &  13.71   &  16.5     \\ 
    $\sigma_{p_{1}}$ &   2.98              &   1.45   &   0.7     \\ \hline  
    $\bar{p}_{2}$    &   0.39              &  -0.11   &    -      \\ 
    $\sigma_{p_{2}}$ &   0.60              &   0.35   &   0.8     \\ \hline  
    $\bar{p}_{3}$    &  -0.02              &  -0.03   &    -      \\ 
    $\sigma_{p_{3}}$ &   0.38              &   0.25   &    -      \\ \hline  
    N & $1.7 \times 10^{7}$ & $2.6 \times 10^{8}$ &  $2.6 \times 10^{8}$ \\
    \hline \hline 
  \end{tabular}
  \end{center}\end{singlespace}

  \caption{The mean values
           and widths $\bar{x}_{i}$, $\bar{p}_{i}$, $\sigma_{x_{i}}$, and
           $\sigma_{p_{i}}$ as well as the number of the injected electrons
           after the final timestep for the two distinct particle bunches
           that got injected in the 3D simulation as well as for the 
           bunch in the second bucket of the corresponding 2D simulation.
           Some values for the 2D simulation were not available.}
  \label{tab:injection:3d-2d-results}
\end{table}

Fig.~\ref{fig:injection:3d-energy} shows the longitudinal phase space
$x_{1}$-$p_{1}$ of the injected particles at the end of the simulation. The 
figure shows clearly that electrons are injected into two distinct 
buckets of the plasma 
wake but most of the injection takes place in the later bucket.
This later bucket is the same that accelerated electrons in the 2D
simulation. Using Eq.~(\ref{equ:injection:particle-number}) to calculate
the number of trapped electrons gives $1.7 \times 10^{7}$ electrons
for the first electron bunch and $2.6 \times 10^{8}$ electrons for
the second electron bunch. A comparison with the 2D simulation result
shows that the number of trapped electrons in the second bunch for the
2D and 3D simulations is the same within less than$\%5$. That the first bunch is not seen 
in the 2D simulation can be understood when considering the number
of simulation particles involved in the 2D and 3D simulations. The first
and second groups of electrons are represented by $\sim 50$ and $\sim 
1000$, simulation particles in the 3D simulation. The
number of particles used in the 2D simulation to represent the bunch
which corresponds to the second bunch in the 3D simulation is $\sim 
100$. This means that the smaller bunch in the 3D simulation is 
probably too
small to be correctly resolved in the 2D simulation. For this reason
only the results for the second, larger group of accelerated electrons
will be considered when comparing the 2D and 3D simulations.

The final energy in the 2D simulation is $16.5 mc^{2}$. This is more 
than the final energy in the 3D simulation which is $13.7 mc^{2}$. The
difference is probably due to the fact that the laser pulse
is diffracting more quickly in 3D than in 2D. As a result the laser intensity
decreases more quickly and therefore the wakefield is smaller in the 
3D simulation.
[see Eq.~(\ref{equ:review-physics:wake-laser-gaussian1})].
The relative energy spread is larger in the
3D simulation than in the 2D simulation. The 3D simulation shows a
$\Delta E/E = 21\%$ while the 2D simulation has an energy spread of
about $9\%$.


In Fig.~\ref{fig:injection:3d-transvers} the various transverse phase 
spaces of the injected electrons are shown. As before there are two 
bunches but only the properties of the larger group (i.e. later bunch)
will be discussed here.
In Fig.~\ref{fig:injection:3d-transvers}a) the $x_{1}$-$x_{2}$ and
the $x_{1}$-$x_{3}$ phase spaces are shown. As expected
there are differences between the two distributions.
In particular the width of the distribution in $x_{2}$ is 1.5 times
the width of the distribution in $x_{3}$.  

Fig.~\ref{fig:injection:3d-transvers}b) shows the $x_{1}$-$p_{2}$ and
the $x_{1}$-$p_{3}$ phase spaces. Again the width the distribution in $p_{2}$
is larger than the width in $p_{3}$. The ratio of the two values is $1.4$.
Using the values for the width of the distributions to calculate the
normalized emittance of the particle bunch in the $x_{2}$ and $x_{3}$
directions gives $\varepsilon_{n,2} = 27 \pi mm mrad$ and
$\varepsilon_{n,3} = 13 \pi mm mrad$. The value of $\varepsilon_{n,2}$ is
the same as the emittance in the 2D simulation. Another piece of
information in the table above to note is that the width of the
transverse distributions of the particles is considerably larger than the
distance of the mean values from zero (which means for $x_{2}$ and $x_{3}$ 
the distance from the center of the wakefield).

There are two main conclusions to be drawn from the results of the
3D simulation results. The first is that 2D simulations are an 
excellent
tool for studying the transverse laser injection process since they
show the same general behavior and many of the injected beam 
parameters are quantitatively the same in the 2D and 3D simulations.
The second result is that there is a $50\%$ asymmetry for the 
spotsize and the transverse momentum spread in the two transverse 
directions. These results are preliminary and more work is needed.

\section{1D Models}
  \label{sect:injection:1d-models}
  
\begin{figure}
   \begin{center}
      \epsfig{ file=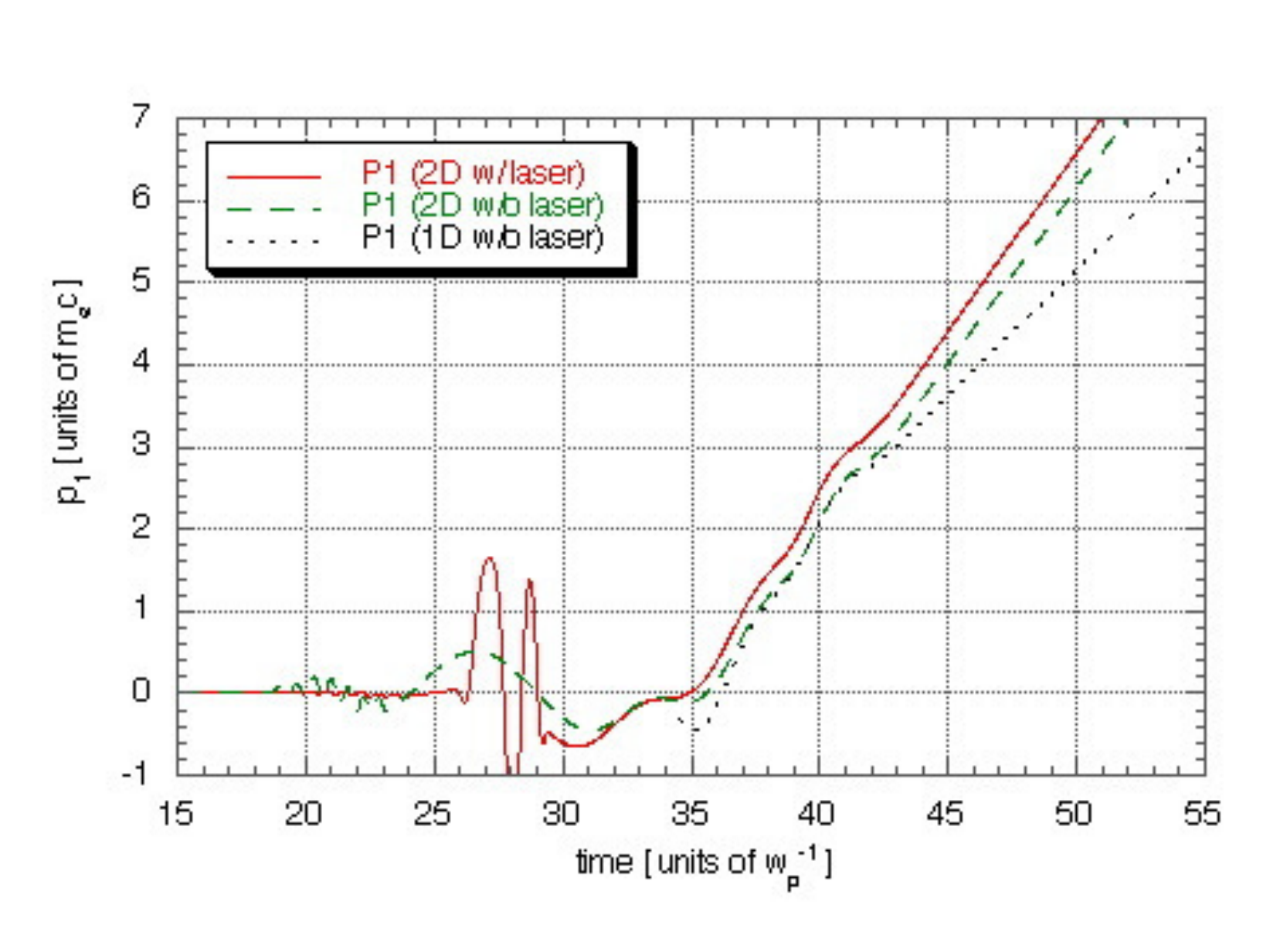, width=5.8in }
      \caption{$p_{1}$ vs. time for a particle in a 2D non-self-consistent
               simulation (solid) and for a 1D numerical calculation(dashed).
               The 1D calculation had the starting parameters $x_{0}=-0.5$,
               $w_{0}=-1.5$ and $\varphi_{0}=\frac{4}{3}\cdot\pi$.}
      \label{fig:injection:p1-of-t-1d}
   \end{center}
\end{figure}

The simulation results raise the question whether the injection is 
mostly a linear effect that arises from the superposition of the two 
plasma waves or whether it is essentially a non-linear effect arising 
from the interaction of the two plasma waves mediated by the plasma. 
To address this question, we show the results of non-self-consistent 
2D simulations and 1D numerical calculations
(Fig.~\ref{fig:injection:p1-of-t-1d}).
The 2D non-self-consistent simulations are done by turning off the field 
solver of the Pegasus code and instead calculating the fields of the 
lasers and their wakes analytically from linear theory at each 
timestep \cite{Sprangle88}. As a result it is possible to follow 
test particles in the fields caused by the linear superposition of the 
two laser pulses and their wakes. In a second non-self-consistent 2D 
simulation the injection pulse is neglected while the linear wake it 
produces is not.

The 1D numerical calculations use the following electric fields for 
the wakes to calculate the trajectory of a particle.

\noindent{Wake} from the drive pulse: 

\begin{equation}
  \label{equ:injection:1d-drive-wake} 
  E_{D} = E_{D,max} \cdot \sin \left( k_{p}x - \omega_{p}t - \varphi_{0} \right)
\end{equation}

\noindent{Wake} from the injection pulse: 

\begin{equation}
  \label{equ:injection:1d-injection-wake}
  E_{I} = E_{I,max} \cdot 2 \cdot 
  e^{1/2} \cdot \sin \left( \omega_{p} t \right) \cdot x/w_{0} \cdot 
  e^{ -2 \left (x/w_{0} \right)^{2} }
\end{equation}

\noindent{These} equations follow from the ones used for the 
non-self-consistent 2D simulations. The laser fields are also omitted 
in this 1D calculation. The initial conditions of the particle are 
given by its position in the plasma wave described by
Eq.~(\ref{equ:injection:1d-drive-wake}). 
$E_{D,max}$ and $E_{I,max}$ are taken as 0.45 and 0.35 since these 
are the values seen in the self-consistent calculation with $\psi 
=1.3\pi$ and b=1.8. For the 2D non-self-consistent simulations the 
laser amplitudes are slightly adjusted to yield those values, too. 
All other parameters of the non-self-consistent 2D simulations and 
the 1D calculations are the same as in the self-consistent simulation 
with $\psi =1.3\pi$ and b=1.8, unless stated otherwise.

The results of these idealized models can be seen in
Fig.~\ref{fig:injection:p1-of-t-1d}.
The linear superposition of two crossed plasma waves (solid line) creates 
conditions under which particles get trapped. On the other hand the 
actual development of $p_{1}$ after the injection pulse has passed 
the particle looks different from the self-consistent results, which 
suggests that the trapping process is modified by the nonlinear 
interaction between the two plasma waves. The multi-step trapping 
discussed above seems to be a result of this modification.

The result of the 2D simulation without the injection laser (dashed 
line) differ strongly from the one with the laser up to time the 
injection pulse has passed. After that time the two curves are rather 
similar and they differ mainly due to a small displacement of one 
compared to the other along the time axis. Noting that the two curves 
belong to particles with a different original position in the 
simulation, suggests that the effect of the ponderomotive force is to 
change which particles are trapped. Direct comparison of the temporal 
evolution between the 1D and 2D results is complicated because the 
same particles are not trapped. We place the 1D curve in such a way 
that it is easy to compare the trajectories once a particle is 
trapped. The similarity of this curve with the curves from the 
non-self-consistent 2D simulations indicates that the basic physics of 
the trapping can be studied by Eq.~(\ref{equ:injection:1d-drive-wake})
and Eq.~(\ref{equ:injection:1d-injection-wake}).

We close this section by commenting that determining whether the 
trapping results from a ponderomotive kick or from interfering wakes 
is important to developing simplified models to explain and extend 
the scheme investigated here. Our results suggest that the trapping 
is due to the interaction of two plasma waves rather than a plasma 
wave and a ponderomotive kick (impulse). However, this does not rule 
out the possibility that a different choice of parameters for the 
injection pulse will result in trapping due to a direct kick by the 
transverse ponderomotive force \cite{Umstadter97b}. A possible 
advantage of the mechanism found in our results here relates to
Eq.~(\ref{equ:injection:current}).
If the trapping of particles is caused by dephasing them with 
respect to the accelerating wake, as we find it here, rather than from 
directly increasing their momentum then $n_{b}/n_{0}$ could be a 
much weaker function of $\gamma_{\phi}=\omega_{0}/\omega_{p}$ 
indicating that this injection method might also be useful for larger 
$\gamma_{\phi}$.

Understanding the trapping mechanism allows one to propose and 
understand other possible geometries. A co-propagating geometry is 
the easiest to visualize \cite{Umstadter97b}. The second pulse 
should be tightly focused to interact with a single (or perhaps a few) 
bucket and it should be phased to enhance the original wake to 
amplitudes above wavebreaking. In this geometry the ponderomotive 
force and the wake are intimately connected for the first 
oscillation. However, in subsequent oscillations the interaction of 
the wakes could lead to injection.

A counter-propagating geometry is more complicated (This scheme 
differs from a recent idea of Esarey et al. \cite{Esarey97a} which 
considered a co-linear geometry with an intense pump pulse and 
two counter-streaming injection pulses). Once again a second pulse is 
focused tightly to interact with only a single bucket. In this case 
the injection pulse is phased to reinforce the electrons motion as 
they move backwards. Therefore, the wake is unequivocally essential 
in order for the electrons to be trapped as they oscillate forward. In 
simulations of this scheme we have observed an additional trapping 
mechanism at the plasma boundary. This mechanism might be of interest 
for experiments in which the plasma boundaries are sharp.

In another possible scenario, a plasma wave moving across the first 
wake (other geometries are also possible) could be gradually built up 
over time until a trapping threshold is reached. This scheme also 
clearly would rely only on the interfering wakes.

\section{Conclusion}
  \label{sect:injection:conclusion}
    
In this chapter, the 
injection scheme proposed in Ref.~\cite{Umstadter96a} was
studied using the results of 2D and 3D PIC computer simulations.
We find that the beam brightness and quality compares reasonably 
with that of electron bunches produced using conventional technologies. 
However, we find that the mechanism for the trapping of particles is 
not the transverse ponderomotive force of the injection pulse, but 
rather the interaction of the particles with the two plasma wakes.
The 2D and 3D simulations are in good agreement with each other
and give therefore confidence in the results obtained from 2D 
simulations.

These results open up a number of possibilities for future 
investigations, both to obtain analytical models and to consider other 
injection schemes and geometries. One important goal of future 
research would be to find an analytical model of the process that is 
able to predict the results seen in the simulations. This could then 
be used to determine fundamental limits on beam number and emittance, 
as well as to optimize parameters to achieve these limits. Another 
research direction would be the use of more realistic laser parameters
in 2D and 3D simulations.

\chapter{Long Wavelength Hosing of Laser Beams}
  \label{chap:hosing}

\section{Introduction}
  \label{sect:hosing:introduction} 


Understanding the evolution of short-pulse high-intensity lasers as 
they propagate through underdense plasmas is essential for the 
successfully development of some plasma accelerator \cite{Tajima79b} and 
radiation schemes \cite{Mori93}, as well as for the fast ignitor 
fusion concept \cite{Tabek94}. As a result, there has been much 
research during the past few years on short-pulse laser-plasma 
interactions. This work has resulted in the identification of 
numerous self-modulated processes, e.g., relativistic self-focusing 
\cite{Sprangle87a}, ponderomotive blowout/cavitation, and Raman 
forward scattering (RFS) instabilities \cite{Mori94, Antonsen92, Mori97}, 
including envelope self-modulation \cite{Esarey94} and hosing 
\cite{Sprangle94, Shvets94}. While the work of the last few years has 
led to the determination of the spatial-temporal growth behavior of 
the above processes \cite{Mori94, Antonsen92, Mori97, Esarey94, Sprangle94, 
Shvets94}, it is not clear which, if any, of these processes dominate 
the evolution of the laser after these processes have saturated.

In this chapter of this dissertation, we use the particle-in-cell (PIC) model PEGASUS 
\cite{Tzeng96, Tzeng98b} to investigate the final nonlinear state of short-pulse 
lasers after they have propagated through a few Rayleigh lengths of 
plasma. We find over a wide parameter space that the laser's 
evolution follows a common sequence of events. Furthermore, we find 
that the final state of the laser is dominated by a new long 
wavelength hosing instability. We present a variational principle 
analysis which provides the growth rate for the well known Raman type 
hosing instability \cite{Sprangle94, Shvets94}, but which clearly 
identifies a long wavelength hosing (LWH) regime. At higher densities, 
we find ion motion to be important. Last, we illustrate through PIC 
simulations that a consequence of LWH is for the self-trapped 
electrons \cite{Tzeng97} to be displaced sideways.

\section{Motivation}
  \label{sect:hosing:motivation} 
  
We begin by presenting results from a PEGASUS simulation in which a 
600fs/$\mu m$ laser is focused with a peak intensity of 5 $\times$ 
10$^{18}$ W/cm$^{2}$ and a spot size of 20$\mu m$ onto the edge of a 
1.4 $\times$ 10$^{19}$ plasma slab. For these parameters, 
$\omega_{0}/\omega_{p}=8.5$, $c/\omega_{p}$=1.36$\mu m$, the 
Rayleigh length is $x_{R}$=1.2 mm, the peak normalized vector 
potential $a_{0}=eA_{0}/mc^{2}=2$, and the ratio of laser power 
to the critical power for relativistic focusing is \cite{Sprangle87a} 
$P/P_{c}=a^{2}\left(k_{p}w\right)^{2}/32=27$. In the simulation 
$1.2 \times 10^{7}$ electrons are followed on a $8192 \times 256$ x-y 
cartesian grid, while the ions are modeled as a smooth neutralizing 
background.

\begin{figure}
   \begin{center}
      \epsfig{file=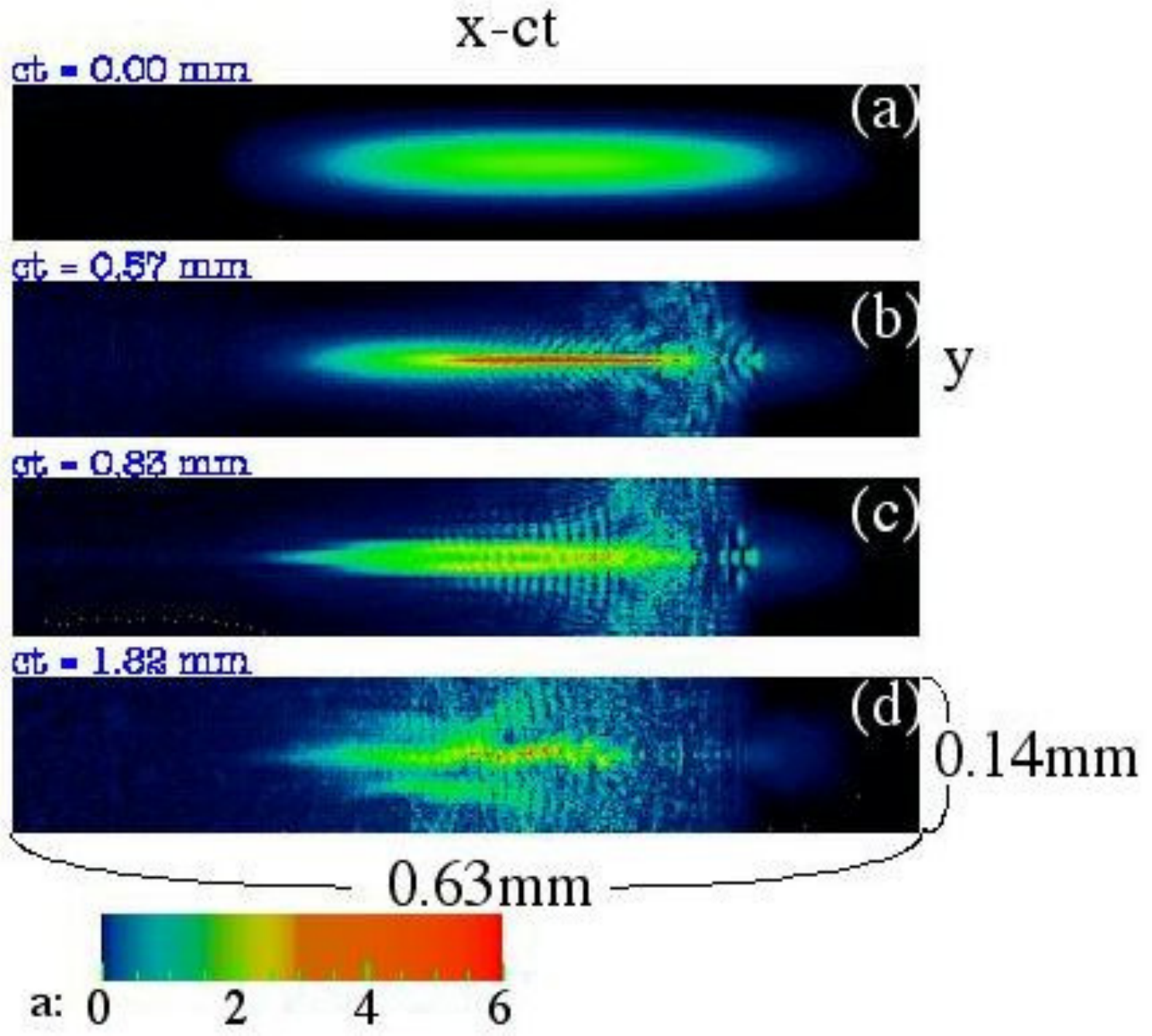, width=5.8in }
      \caption{A sequence of color contours of the laser's electric
               field in units of $eE/\left(m\omega_{0}\right)\simeq a$.
               The results are the same from the same simulation.}
      \label{fig:hosing:laser-g8.5}
   \end{center}
\end{figure}

In Fig.~\ref{fig:hosing:laser-g8.5}, we show a sequence of four color
contour plots 
of the laser's electric field with a common color map. The four 
snapshots correspond to when the laser initially impinges on the 
plasma and when the head of the laser has penetrated .57 mm, .83 mm, 
and 1.82 mm into the plasma respectively. After only .57 mm, i.e., 
.5 $x_{R}$, the head of the pulse has been depleted from Raman 
scattering while the back of the pulse has strongly self-focused. 
Details of this have been reported elsewhere \cite{Tzeng96, Tzeng98b}. 
 
Eventually , as seen in Fig.~\ref{fig:hosing:laser-g8.5}c,
the middle of the pulse is modulated 
from Raman forward scattering, while the back of the pulse expands and 
breaks up into two major filaments in which both Raman forward 
scattering and conventional hosing are occurring. However, later in 
time the pulse reaches a ``final'' nonlinear state where the back of 
the pulse has refocused into a major filament (with two weaker 
filaments surrounding it) whose average position in the y direction, 
$y_{a}$ oscillates about the original laser axis. The intensity 
contours closest to the front of the pulse alternate above and below 
the original laser axis at a wavelength of roughly twice the plasma 
wavelength, $\lambda_{p}=2\pi c/\omega_{p}$. At positions further 
back, $y_{a}$ is modulated at a longer wavelength - between 5-10 
$\lambda_{p}$. This hosing behavior at wavelengths longer than 
$\lambda_{p}$, i.e., LWH, was not discussed in the earlier 
theoretical analysis \cite{Sprangle94, Shvets94}. We emphasize that 
the nonlinear evolution of the pulse is also influenced by 
wavebreaking and intense plasma heating \cite{Tzeng98a, Pukhov98}.

\section{Theoretical Approach}
  \label{sect:hosing:theory}
  
 In order to present a possible explanation for LWH, we present here  
a variational principle approach, developed by  B.J. Duda and W.B. Mori \cite{Duda98},
to describe the evolution of short-pulse laser interacting with their
self-consistent  wakes. The standard equations for describing short-pulse lasers, 
which include the lowest order relativistic corrections and assume a 
cold plasma, are now well established to be: 

\begin{equation}
  \label{equ:hosing:std-laser-plasma-1}
  \left(\nabla_{\perp}^{2}-\frac{2}{c}\frac{\partial^{2}}{\partial\psi\partial\tau}
  -2ik_{0}\frac{\partial}{\partial\tau}\right)a
  =\frac{\omega_{p}^{2}}{c^{2}}\left(1-\phi\right)a
\end{equation}

\begin{equation}
  \label{equ:hosing:std-laser-plasma-2}
  \left(\frac{\partial^{2}}{\partial\psi^{2}}+\omega_{p}^{2}\right)\phi
    =\omega_{p}^{2}
\frac{\left|a\right|^{2}}{4}.
\end{equation}

\noindent{where} $a$ is the normalized envelope for the complex vector 
potential of the laser, $eA/mc^{2}=\left(a/2\right)exp[-i\omega_{0}\psi] + 
c.c$, $\phi$ is the scalar potential of the plasma, and $\psi=t-x/c$, 
$\tau=x/c$ are convenient variables for describing short-pulse lasers.

In the variational method a Lagrangian density, ${\mathcal{L}}$, needs 
to be found for which the Euler-Lagrange equations, obtained by 
varying the action, $S=\int dx_{\perp}d\psi d\tau {\mathcal{L}}$, 
recover Eq.~(\ref{equ:hosing:std-laser-plasma-1}) and
Eq.~(\ref{equ:hosing:std-laser-plasma-2}). We find such an ${\mathcal{L}}$ to be:

\begin{eqnarray*}
{\mathcal{L}}\left(a, a^{\ast}, \phi\right)
&=&\vec{\nabla}_{\perp}a\cdot 
\vec{\nabla}_{\perp}a^{\ast}-ik_{0}\left(a\partial 
_{\tau}a^{\ast}-a^{\ast}\partial_{\tau}a\right) \\
&&-\frac{2}{c}\left(\partial_{\psi}\phi\right)^{2}+2\frac{\omega_{p}^{2}}{c^{2}}
\phi^{2}-\frac{\omega_{p}^{2}}{c^{2}}\left(\phi -1\right)\left| 
a\right|^{2} 
\end{eqnarray*}

\noindent{where} we have dropped the so-called dispersive terms, 
i.e., those which give the mixed derivative term on the left-hand 
side of Eq.~(\ref{equ:hosing:std-laser-plasma-1}) \cite{Duda98}.
Dropping the dispersive terms leads to conservation of power, i.e.,
$\partial\tau \int dx_{\perp} \, |a|^{2} = 0$. Anderson and Bonnedal 
\cite{Anderson79} used the variational approach to study only 
self-focusing , which precludes any coupling to the plasma wave wake 
and hence their ${\mathcal{L}}$ depends upon $a$ and $a^{\ast}$ only.

In the variational method, the complexity of the system is reduced by 
substituting trial functions for $a$ and $\phi$ into the action and 
performing the $dx_{\perp}$ integration. To consider hosing, we 
assume a trial function for $a$ of the form 
$a={\mathcal{A}}e^{i\chi}e^{ik_{y}\left(y-y_{a}\right)}
e^{-2\left[\left(y-y_{a}\right)^{2}+z^{2}\right]/w^{2}}$ for 
$\phi$ of the form $\phi=\Phi 
e^{-2\left[\left(y-y_{\phi}\right)^{2}+z^{2}\right]/w^{2}}$ where 
the parameters ${\mathcal{A}}$, $\Phi$, $\chi$, $k_{y}$, $y_{a}$, and 
$y_{\phi}$ are treated as functions of $\left(\psi,\tau\right)$. The spot size, 
w, is taken to be a constant which we allow to be the same for both $a$ 
and $\phi$. The ``centroid'' variables $y_{a}$ and $y_{\phi}$ 
measure the distance that the center of the laser and its wake are 
displaced from the original axis. Performing the $dx_{\perp}$ 
integration yields a reduced action which is a functional of the 
variational parameters, i.e., $\bar{S}\left({\mathcal{A}}, \chi, 
\Phi, \alpha, k_{y}, y_{a}, y_{\phi}\right)=\int d\psi 
d\tau\bar{\mathcal{L}}$. Varying $\bar{S}$ with respect to $\chi$ 
yields the power conservation law, 
$\partial_{\tau}P=\partial_{\tau}\left({\mathcal{A}}^{2}w^{2}\right)=0$. 
Variations with respect to the functions $\alpha$ and $k_{y}$ give 
the relationships 
$\alpha=-\left(k_{0}/4\right)\partial_{\tau}\left(w^{2}\right),$ 
and $k_{y}=-k^{0}\partial_{\tau}y_{a}$, which can be substituted 
back into $\bar{\mathcal{L}}$ to yield the following reduced form of 
$\bar{\mathcal{L}}$, $\bar{\mathcal{L}}\left(\Phi ,y_{a}, y_{\phi}\right)$:

\begin{eqnarray*}
{\mathcal{L}}\left(\Phi ,y_{a},y_{\phi}\right)
&=&-\frac{k_{0}^{2}}{4}P\left(\partial_{\tau}y_{a}\right)^{2} \\
&&+\frac{k_{p}^{2}}{2}\left(w^{2}\Phi^{2}-\frac{P\Phi}{2}
e^{-\frac{\left(y_{a}-y\phi\right)^{2}}{w^{2}}}\right)
-\frac{1}{c^{2}}\left[\frac{w^{2}}{2}\left(\partial_{\psi}\Phi\right)^{2}
+\Phi^{2}\left(\partial_{\psi}y_{\phi}\right)^{2}\right].
\end{eqnarray*}

\noindent{Next} we linearize the Euler-Lagrange equations of 
$\bar{\mathcal{L}}$ about a solution in which $y_{a0}=y_{\phi 0}=0$, 
and $\Phi_{0}=a_{0}^{2}/4$, giving the coupled equations for 
$y_{a}$ and $y_{\phi}$:

\begin{equation}
  \label{equ:hosing:linear-euler-1}
  \partial_{\tau}^{2}y_{a}+c^{2}g\frac{P}{P_{c}}\frac{1}{x_{R}^{2}}y_{a}
  = c^{2}g\frac{P}{P_{c}}\frac{1}{x_{R}^{2}}y_{\phi}
\end{equation}

\begin{equation}
  \label{equ:hosing:linear-euler-2}
  \partial_{\psi}^{2}y_{\phi}+\omega_{p}^{2}y_{\phi}=\omega_{p}^{2}y_{a},
\end{equation}

\noindent{where} 
$P/P_{c}={\mathcal{A}}^{2}\left(k_{p}w\right)^{2}/32$, g is a 
geometric factor which is 1 in cylindrical and $2^{-3/2}$ in slab 
geometry (used in the simulations), and $x_{R}=k_{0}w^{2}/2$ is the 
Rayleigh length for the equilibrium laser profile. Note that these 
equations are identical in form to those which describe hosing of 
electron beams in the ion focused regime \cite{Whittum91}, and they 
reproduce Eq.~(5) in Ref. \cite{Sprangle94}. 
 
\begin{figure}
   \begin{center}
      \epsfig{file=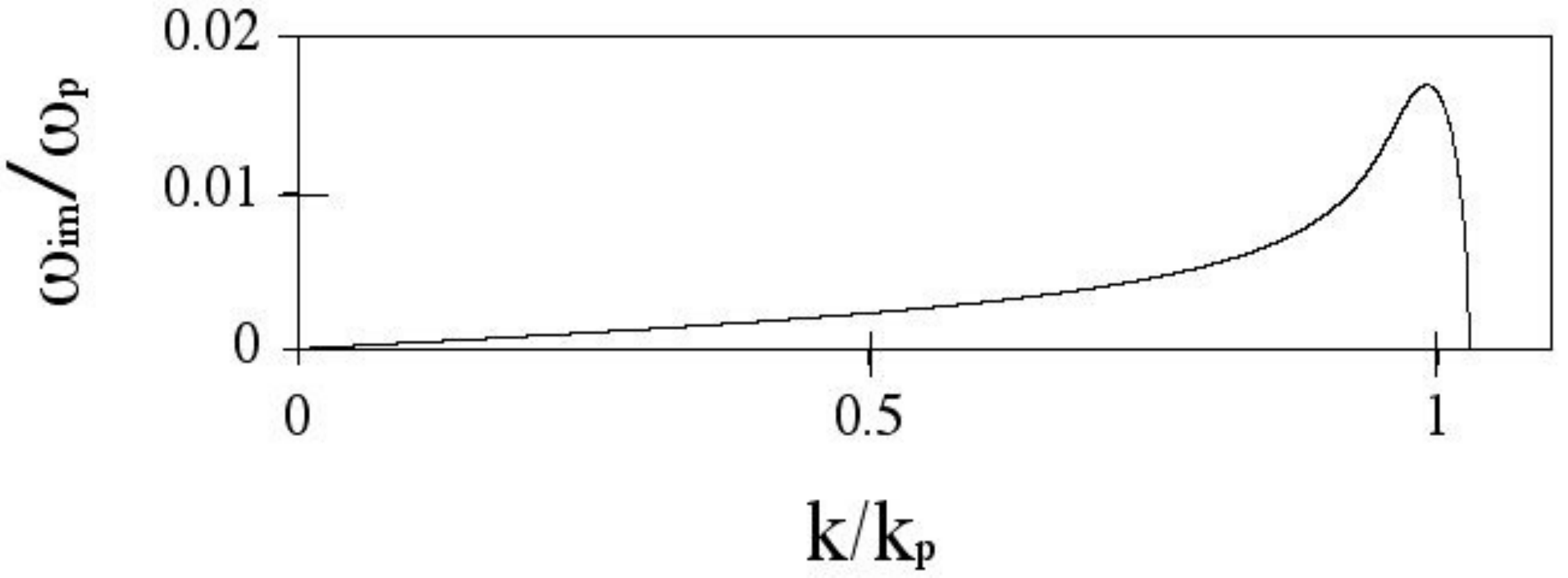, width=5.8in }
      \caption{The growth rate for hosing vs. wavenumber
               for $\tilde{x}_{R}=256$.}
      \label{fig:hosing:growthrate}
   \end{center}
\end{figure}

To discuss the growth rate and range of unstable wavelengths for 
hosing, we obtain a dispersion relation in the lab frame by using the 
transformations 
$\partial_{\tau}\rightarrow\partial_{t}+\partial_{x}$ and 
$\partial_{\psi}\rightarrow\partial_{t}$ and substituting solutions 
of the form exp(i(kx-$\omega$ t)) into
Eq.~(\ref{equ:hosing:linear-euler-1})\&(\ref{equ:hosing:linear-euler-1}),
yielding $\tilde{\omega}^{2}\left(\tilde{\omega}-\tilde{k}\right)
^{2}-\left(\tilde{\omega}^{2}g\left(P/P_{c}\right)/
\tilde{x}_{R}^{2}\right)-\left(\tilde{\omega}-\tilde{k}\right)^{2}=0$, 
where $\tilde{\omega}\equiv\omega/\omega_{p}$, $\tilde{k}\equiv 
k/k_{p}$, and $\tilde{x}_{R}\equiv k_{p}x_{R}$. In
Fig.~\ref{fig:hosing:growthrate} we plot 
the growth rate, i.e., the imaginary part of the $\tilde{\omega}$, vs. 
real $\tilde{k}$ for $P/P_{c}$=1, i.e., a matched beam. This confirms 
that the peak growth rate occurs for $\tilde{k}\sim 1$, i.e., $k\sim 
k_{p}$. This region of unstable growth is related to Raman forward 
scattering (RFS), since a plasma wave is being excited, and it is the 
regime discussed in Refs. \cite{Sprangle94, Shvets94}.

However,
Fig.~\ref{fig:hosing:growthrate} 
also makes clear that the range of unstable wave numbers extends 
continuously down to $\tilde{k}=0$. This long wavelength regime has 
heretofore never been discussed. This regime could have been obtained 
immediately if $y_{\phi}=y_{a}$ was assumed in the trial functions, 
which forces the centroids for $\phi$ and $a$ to be in phase. In this 
limit, the plasma response, $\phi$ is due almost entirely to 
relativistic mass corrections, i.e., no plasma waves are excited. 
Therefore, this long wavelength regime is the whole beam analog to 
relativistic self-phase modulation (RSPM) \cite{Max74}. LWH is 
therefore physically distinct from conventional hosing in the same way 
that RSPM is distinct from RFS.

\section{Simulation Results}
  \label{sect:hosing:simulation-results}   
 
The spatial-temporal growth for the conventional and LWH regimes also 
differ. In the RFS regime,l for $\tilde{\omega}$ near $k_{p}$ the 
asymptotic spatial-temporal growth for hosing is given by 
\cite{Sprangle94, Shvets94} $y_{a or \phi}\sim 
exp\left[\left(3^{3/2}/4\right)\left[g\left(P/P_{c}\right)
\omega_{p}\psi\right]^{1/3}\left(\tau/\tau_{R}\right)^{2/3}\right]$. 
In the LWH regime, where the inequality $\partial_{\psi}^{2}\ll\omega_{p}^{2}$ holds, 
Eq.~(\ref{equ:hosing:linear-euler-2})
leads to $y_{\phi}\cong y_{a}/(1-k^{2})$. Substituting this relationship into
Eq.~(\ref{equ:hosing:linear-euler-1})
gives the spatial-temporal growth $y_{a or \phi}\sim 
exp\left[\left(gP/P_{c}\right)^{1/2}\left(k/k_{p}\right)
^{1/2}\left(\tau/\tau_{R}\right)\right]$. These expressions are only 
valid under the ideal conditions of cold plasmas, weakly relativistic 
pumps, and matched beams.

\begin{figure}
   \begin{center}
      \epsfig{file=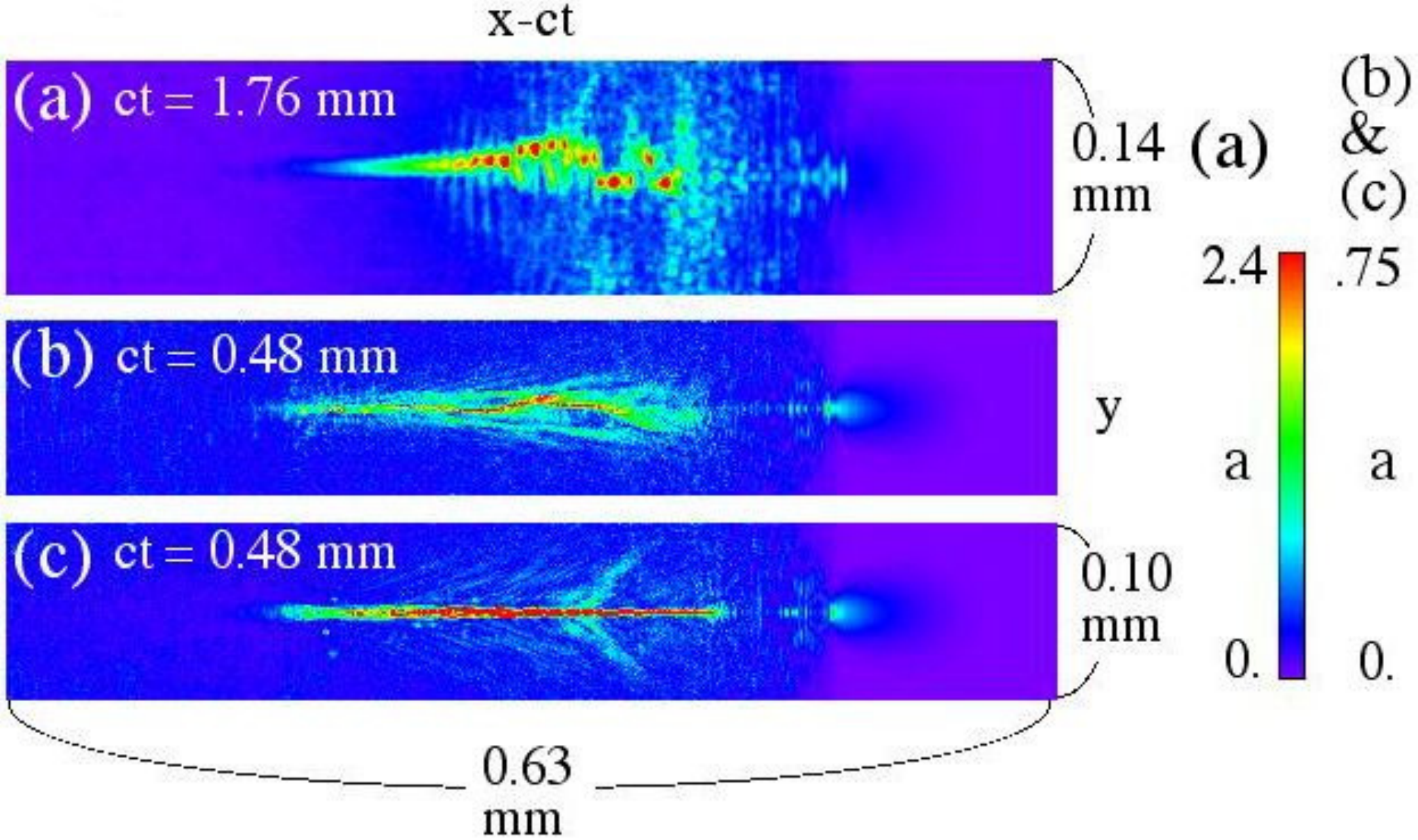, width=5.8in }
      \caption{Color contours of the laser's electric field in units of
               $eE/\left(mc\omega_{0}\right)\simeq a$ to show further
               evidence for long wavelength hosing. The results are from
               three different simulations.}
      \label{fig:hosing:laser-final}
   \end{center}
\end{figure}

However, for current experimental parameters
\cite{Modena95, Gordon98b, Umstadter96b, Wagner97, Coverdale95, Ting97} the 
conditions are far from ideal. Therefore, to accurately determine the 
relative importance of the various regimes for hosing with respect to 
other self-modulation processes, we next present additional results 
from fully nonlinear PIC simulations. In Fig.~\ref{fig:hosing:laser-final}
we show color 
contour plots of the laser's electric field in units of 
$eE/(mc\omega_{0})\approx a$ to illustrate the ``final'' nonlinear 
state of short-pulse laser from four different simulations. In each 
case 600fs laser pulse is focused to the edge of a uniform, preformed 
plasma slab and the ions are a fixed neutralizing background. It is 
clear that for each simulation the ``final'' state shows strong 
self-focusing and a dominant LWH component. In
Fig.~\ref{fig:hosing:laser-final}a, the laser's 
electric field is shown after a propagation distance of 
$1.8mm=6x_{R}$ from a simulation with parameters identical to those 
in Fig.~\ref{fig:hosing:laser-g8.5} except $w_{0}=10\mu m$ instead of $20\mu m$.
The dominant hosing wavelength is similar to that in
Fig.~\ref{fig:hosing:laser-g8.5}d
but the amplitude of 
the centroid seems larger and the instability seems to have saturated. 
The spatial-temporal theory predicts that the number of e-foldings for 
LWH scales as $1/w_{0}$ for otherwise fixed parameters. This scaling 
is consistent with the observation from Figs. \ref{fig:hosing:laser-final}a)
and Fig.~\ref{fig:hosing:laser-g8.5}d) that LWH 
is stronger when $w_{0}10\mu m$ compared to when $w_{0}=20\mu m$. 
For the parameters of this simulation, $P/P_{c}\simeq 6.75$ and 
$k/k_{0}\simeq 10$, the spatial-temporal theory predicts $\sim$9 
e-foldings of LWH growth, using the focused value of the spot size 
(w=5.6$\mu m$), and the fact that $a^{2}w$ is conserved in slab 
geometry.

The importance of LWH is further illustrated in
Fig.~\ref{fig:hosing:laser-final}b), which shows 
results from a simulation which followed $10^{8}$ particles on a 
16384 $\times$ 1024 grid. The plasma density was increased to 
$10^{20}cm^{-3}$, i.e., $\omega_{0}\omega_{p}=3.3$, the laser 
intensity was lowered to 1.25 $\times$ $10^{18}W/com^{2}$, i.e.., 
$a_{0}$, and the spot size was decreased to $6\mu m$, i.e., 
$k_{p}w_{0}=11.3$. Once again, after only a few $(480\mu m\simeq
4 x_{R})$ Rayleigh lengths of propagation the laser has strongly 
self-focused and a LWH mode is dominant. The dominant wavelength is 
$\sim$ 15-30 $k_{p}$ in this case. Using the self-focused spot size, 
the spatial-temporal theory predicts $\sim$5-8 e-foldings of LWH.

In each simulation, there is little or no evidence of the 
conventional (RFS) type of hosing, except for its presence in the 
filaments of Fig.~\ref{fig:hosing:laser-g8.5}c). However, the spatial-temporal
theory predicts 
many e-foldings of growth. Furthermore, we have independently excited 
both conventional and long wavelength hosing in smaller test case 
simulations by adding large fictitious hosing noise sources. 
Therefore, the lack of RFS hosing is due to nonlinear effects. There 
are several possible nonlinear explanations. Due to its lower initial 
noise source, hosing generally occurs after the beam has strongly 
self-modulated from RFS and self-focusing. The occurrence of RFS 
divides the beam into beamlets spaced at $\lambda_{p}$ (this is seen 
in Figs.~\ref{fig:hosing:laser-g8.5}b) and Fig.~\ref{fig:hosing:laser-g8.5}c).
When hosing occurs as seen in Fig.~\ref{fig:hosing:laser-g8.5}d), it 
appears to first displace one beamlet upward and the next beamlet 
downward. This results in a hosing wavelength of 2 $\lambda_{p}$, and 
as the laser continues to evolve, even longer wavelength modes dominate. 
Therefore, it appears that hosing behaves differently when other 
instabilities such as RFS have already grown to saturated levels. 
Another explanation for the lack of RFS hosing is that the plasma has 
been strongly heated by the time hosing occurs. RFS hosing involves 
the excitation of a plasma wave, which can be strongly damped at high 
temperatures, thereby causing a suppression of RFS hosing.

In regards to the fast ignitor, where longer pulses and higher 
densities are important (particularly for higher densities), the 
frequency of the hosing, $\omega=c k$, can be lower or on the same 
order as the ion plasma period, $\omega_{pi}=4\pi e^{2}n_{0}/m_{i}$. 
In this case, the ion dynamics cannot be ignored. In
Fig.~\ref{fig:hosing:laser-final}c), we show 
results from an identical simulation to that shown in
Fig.~\ref{fig:hosing:laser-final}b) except 
mobile hydrogen-like ions were used. The difference between the two 
cases is dramatic. The ion motion appears to stabilize the hosing (at 
least for the duration of the simulation). On the other hand, we note 
that in a simulation with ten times higher intensity, i.e., 
$a_{0}=3$, ion dynamics did not stabilize hosing. Instead, it 
appeared to cause the beam to self-focus and filament differently 
with LWH still occurring in the individual filaments. The wavelength 
for hosing was shorter than $2\pi c/\omega_{pi}$ in this case. So it 
appears that ion dynamics can stabilize hosing when 
$\lambda_{hosing}\stackrel{>}{\sim} 2\pi c/\omega_{pi}$. We also 
note that LWH can occur for densities above quarter critical where RFS 
cannot, because no plasma wave is excited. Preliminary evidence of a 
LWH effect has already been observed in simulations for the density 
regime \cite{Adam-communication}. Therefore, LWH could be important for the fast 
ignitor fusion concept.

\section{Conclusion}
  \label{sect:hosing:conclusion}
 
\begin{figure}
   \begin{center}
      \epsfig{file=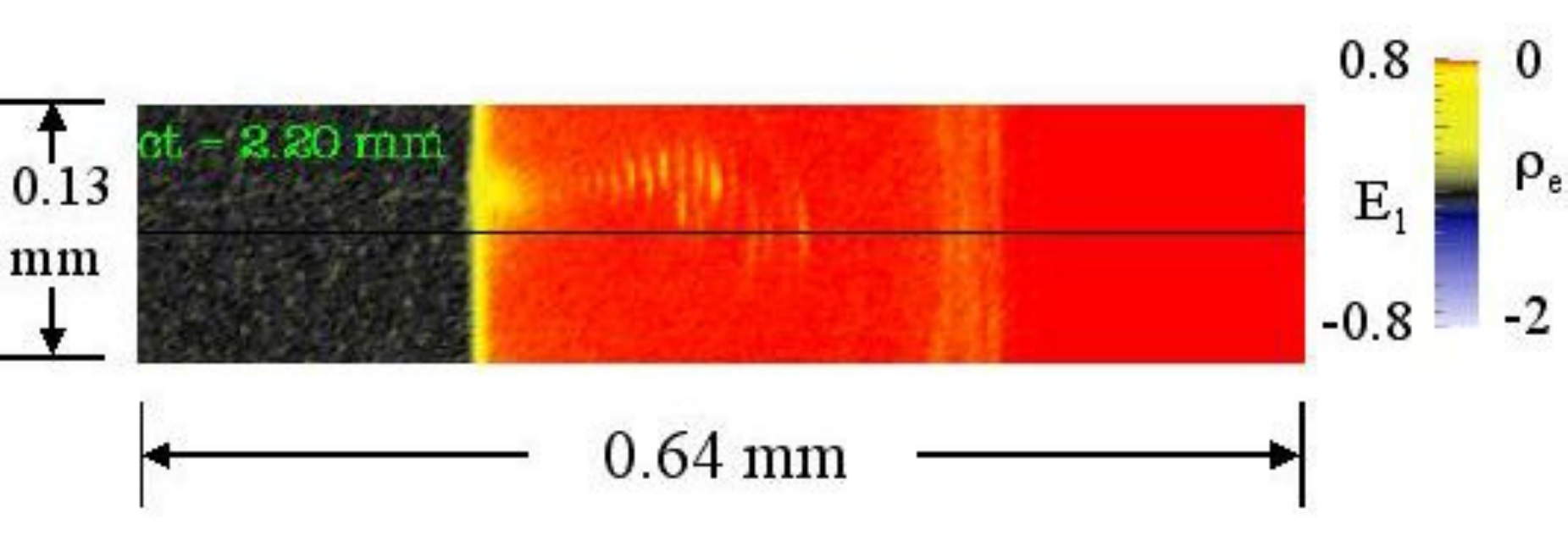, width=5.8in }
      \caption{Color contour of electron density showing self-trapped
               electrons exiting the plasma. The results are from the
               same simulation as Fig.~\ref{fig:hosing:laser-final}a).}
      \label{fig:hosing:beam-final}
   \end{center}
\end{figure}
  
We have shown analytically that a LWH regime exists and shown for the 
first time evidence of any type of hosing in self-consistent PIC
simulations. These simulations show that the LWH eventually dominates
over a wide parameter regime due to nonlinear effects.
Furthermore, we note that LWH might have important consequences for the 
electron spectra generated in self-trapped acceleration experiments 
\cite{Modena95, Gordon98b, Umstadter96b, Wagner97, Coverdale95, Ting97}.
This is illustrated in Fig.~\ref{fig:hosing:beam-final} where a color contour 
plot of the plasma density is shown as the self-trapped electrons 
exit into a vacuum region. The results are from the simulation 
corresponding to Fig.~\ref{fig:hosing:laser-final}a). The black line
is drawn in the middle for 
reference. The electrons are clearly exiting the plasma off axis by a 
distance $\sim 10 \mu m$, and their pattern corresponds to the laser 
profile in Fig.~\ref{fig:hosing:laser-final}a). When the plasma slab
was shortened to 1mm, no hosing was seen to occur, and the accelerated
electrons were not displaced \cite{Tzeng97}. In addition we believe that
LWH will be important when lasers propagate in higher density plasmas above
$n_{c}/4$. This could have important consequences to the fast ignitor
concept.


\chapter{LWFA in a Parabolic Channel}
  \label{chap:channel}

\section{Introduction}
  \label{sect:channel:intro}

The conceptually simplest plasma based acceleration concept that
uses a laser pulse as the drive beam is the laser wakefield 
accelerator(LWFA). The problem of the LWFA concept is that in its
original form it requires a very powerful laser to produce
particles with a significant energy gain.
As an example we can take a laser with a wavelength of
$\lambda_{L} = 1\mu m$ propagating in a plasma with a density of
$n_{p}= 10^{17} cm^{-3}$.  If the excited wake has an amplitude of
$eE_{max}=mc\omega_{p} \simeq 31GeV/m$ then the dephasing limited
maximum energy gain for a particle in this plasma according to 
Eq.~(\ref{equ:review-physics:energy-gain-diffraction})
is $\Delta W_{dephasing} \simeq 11GeV$. However, in order to get an
energy gain of even 1GeV we find that because of the diffraction limit
given by
Eq.~(\ref{equ:review-physics:energy-gain-diffraction})
we need a laser power of 150TW. While such lasers are technologically
feasible, they are still not readily available.



This power required by a LWFA can be reduced if the diffraction of
the laser pulse can be avoided by optically guiding the laser. Several 
possibilities of guiding have been theoretically investigated
\cite{Esarey96}. In this chapter, we present simulation results on
LWFA acceleration when the laser is guided by a parabolic plasma channel.
For a laser guided by a plasma channel the ideal cross section is
determined by channel properties. The Rayleigh length does not have
to be taken into account anymore. The cross section of the laser can
therefore be smaller than for a laser in a homogeneous plasma and the
laser will require less power. Since a full
self-consistent theory of a LWFA in a parabolic plasma channel has
so far not been developed it is a research area where computer
simulations are the best way of gaining a better understanding.
That is, the phase velocity of the wake and the laser's intensity 
will evolve as the wake is excited. In this chapter we study the 
self-consistent acceleration and excitation process for a particular 
example. 

\section{Simulation Setup}
  \label{sect:channel:setup}
  
\begin{figure}
   \begin{center}
      \epsfig{ file=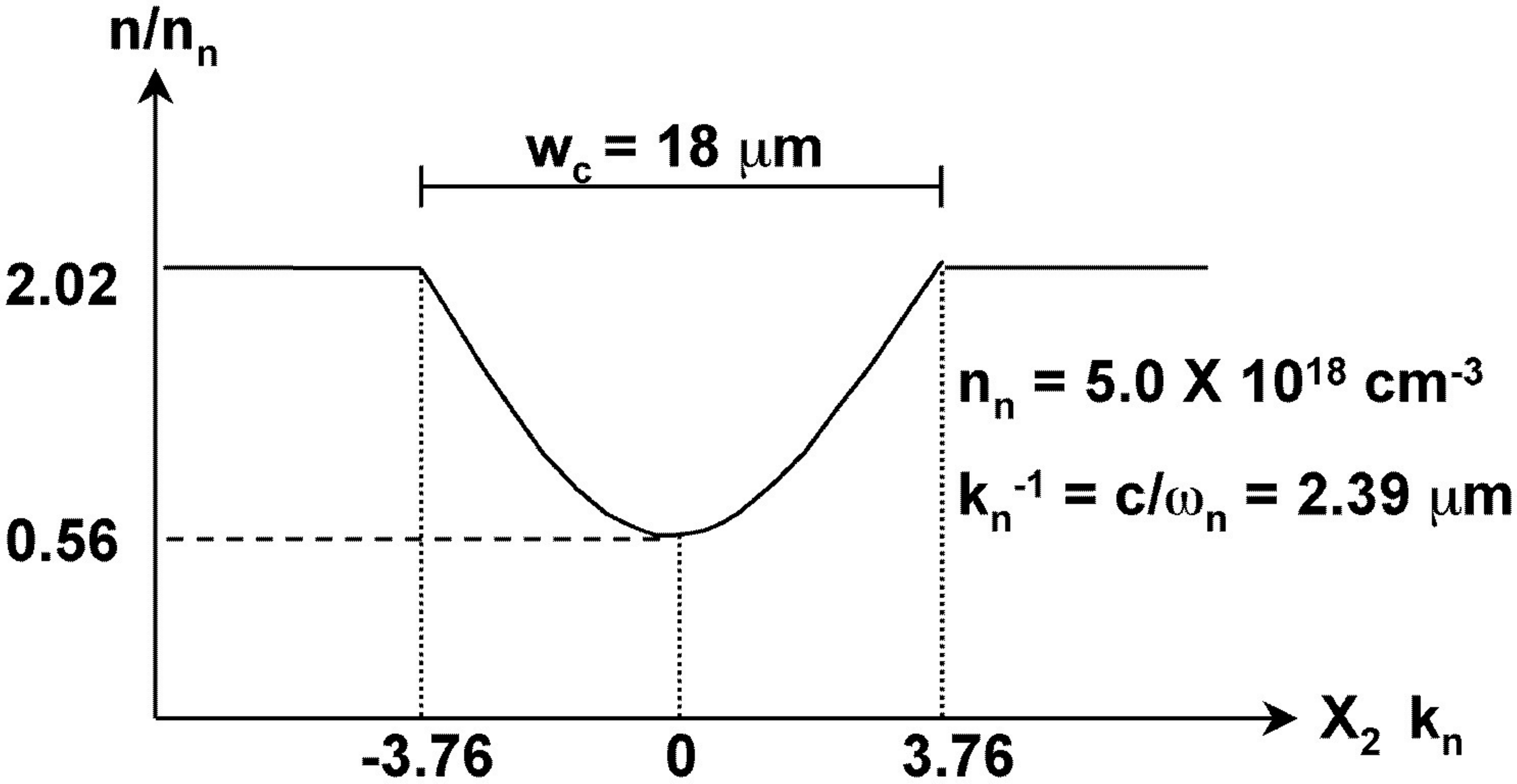, width=5.8in }
      \caption{The density profile of the plasma channel modeled in 
               the simulation.}
      \label{fig:channel:profile}
   \end{center}
\end{figure}

The simulation results presented in this chapter are based on a simulation
with parameters close to the matched beam situation explained in
section~\ref{sect:review-physics:laser-beams}. The simulation
uses a $1\mu m$ laser with a spotsize of $ w_{0}=6.4\mu m$ and a length of
$\tau_{FWHM}=19 \times 10^{-15}s$. The laser has a peak intensity of
$I_{peak}=4 \times 10^{18} W/cm^{2}$, and therefore a power of 
$P=2.6TW$ and a total energy of $E=50mJ$. The Rayleigh 
length for this laser pulse is $z_{R}\approx 128\mu m$. The channel as
described by Eq.~(\ref{equ:review-physics:channel-profile}) has the 
parameters $n_{0}=2.79 \times 10^{18} cm^{-3}$, $r_{0}=w_{0}$,
$\Delta n = 1.32$, and $\Delta n_{c} = 3.70 \times 10^{18} cm^{-3}$. The 
simulation actually uses a piecewise linear profile that deviates up to
about $10\%$ from the parabolic profile shape. The higher plasma density 
compared to the example given in the introduction of this chapter was 
used to reduce the computational requirements for the run.
A channel with these parameters should be weekly focusing for the laser 
pulse and therefore optically guide it.
The geometry of the channel is shown in Fig.~\ref{fig:channel:profile}.
The total width of the channel is $18\mu m$. In the outer areas of the
simulation the parabolic channel density profile is replaced by a 
constant density.

The simulation was a 2D cartesian simulation using the moving
simulation window and was done with the PEGASUS code. The simulation
window in normalized
units had a size along the propagation direction $x_{1}$ of
$30.72c/\omega_{n}$ and a size in the transverse direction $x_{2}$ of 
$15.36c/\omega_{n}$ with a grid of $N_{1} \times N_{2} = 1024 \times 128$. Here
c is the speed of light and $\omega_{n} = 1.26 \times 10^{14} s^{-1}$
is the plasma frequency for the normalizing density
$n_{n}=5 \times 10^{18} cm^{-3}$ used in this simulation. This 
corresponds to $c/\omega_{n} = 0.239 \mu m$ and therefore the simulation in
physical units has a size of $73\mu m \times 37\mu m$. The laser propagated
through the plasma for $50000$ timesteps of size
$dt = 0.0291 \omega_{n}^{-1}$ (corresponding to $3474\mu m$
propagation distance) for a total of $1455\omega_{n}^{-1}$
($\approx 3.5 mm \approx 27 z_{R}$).
Ten particles per cell were used for the plasma. The acceleration of 
particles by the wake was tested by uniformly placing 16128 test particles of 
negligible charge evenly over an area of $15.0\mu m \times 1.15\mu m$ in 
the center of the generated plasma wake. The plasma was initialized 
as a cold plasma while the test particles were initialized with
a momentum in $x_{1}$ of $p_{0,1}= 15 m_{e}c$ and a momentum in $x_{2}$ of
$p_{0,2}= 0.25 m_{e}c$.

\section{Simulation Results}
  \label{sect:channel:results} 

\begin{figure}
   \begin{center}
      \epsfig{ file=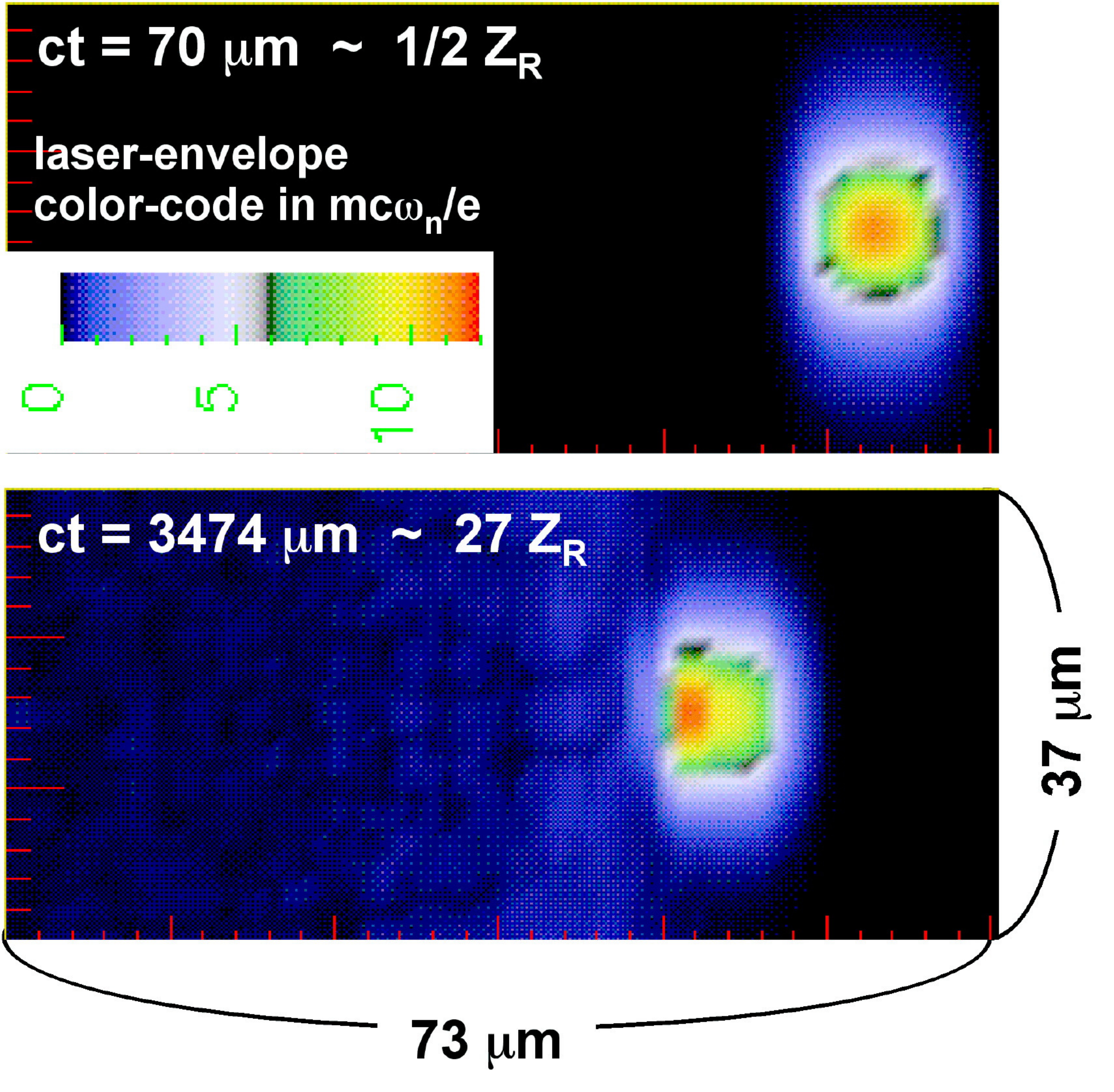, width=5.3in }
      \caption{The envelope of the matched laser beam after about
               half a Rayleigh length and after about 27 Rayleigh
               lengths of laser propagation}
      \label{fig:channel:e3-laser}
   \end{center}
\end{figure}

The electric field of the laser in the simulation is perpendicular
to the plane of the simulation. The envelope of this electric
field component, $E_{3}$, and therefore of the laser is shown in
Fig.~\ref{fig:channel:e3-laser}
at two different times. The first frame is after only about
$\frac{1}{2}$ Rayleigh
length of propagation into the channel, the second frame is after 
about $27$ Rayleigh length. The most important feature to note is that
the channel indeed prevents the diffraction of the laser pulse. 
Even though the pulse undergoes some evolution during the propagation
it is still a Gaussian beam with a spotsize that only decrease slightly
during the propagation of $\sim 26\frac{1}{2} z_{R}$. This is the 
expected result since the channel is weakly focusing with regard to 
the initial laser pulse as explained earlier. 

\begin{figure}
   \begin{center}
      \epsfig{ file=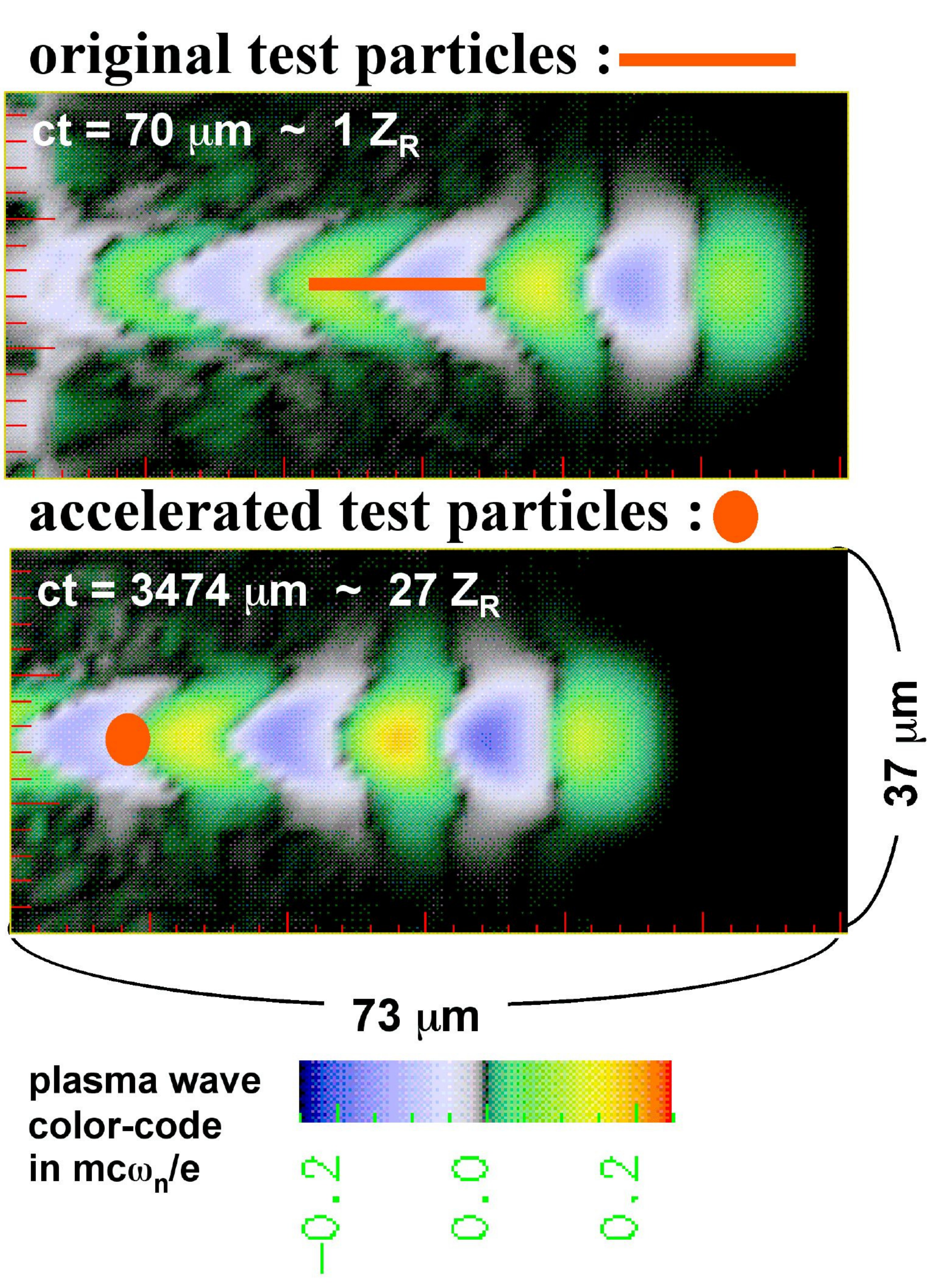, width=5.1in }
      \caption{The electric field of the plasma wake after about
               half a Rayleigh length and after about 27 Rayleigh
               lengths of laser propagation}
      \label{fig:channel:e1-wake}
   \end{center}
\end{figure}

In the propagation direction there a several effects on
the pulse. The most noticeable one is the falling back of the pulse
within the simulation window. Since the simulation window moves with
$c$ this can be used to measure the group or energy transport velocity of
the laser. The group velocity measured in this way, which is also 
roughly the phase
velocity of the wake seen in Fig.~\ref{fig:channel:e1-wake}, is
$v_{p,eff}=0.9965c$ and therefore the pulse and the wake have an effective
$\gamma$-factor of
$\gamma_{eff}=1/\sqrt{1- \left( v_{p,eff}/c \right)^{2}}=11.95$.
If we assume that this effective gamma is due to an effective density that
the laser pulse experiences while travelling through the channel
then this density is $n_{eff} = 1.58 \, n_{n}$.

Another effect is that the peak intensity of the pulse is moving 
backwards within the pulse. 
This is due to the interaction between the laser pulse and the wake
generated by the pulse \cite{Leemans96,Mori97}. The laser is focused
into a density depression of the generated plasma wave. Over the course
of the simulation the peak field of the laser first increases after entering
the plasma to about $110\%$ of its original value and then slowly falls 
off until it has a value of about $106\%$
after  propagating $27z_{R}$.

Fig.~\ref{fig:channel:e1-wake} shows the electric field component
parallel to the propagation direction, $E_{1}$, at the same times
as Fig.~\ref{fig:channel:e3-laser} shows the laser's envelope. The
frames show the plasma wake generated by the laser pulse. The wake
is essentially confined to the parabolic channel and shows a slip
relative to the simulation window. This slip is equal to the slip
seen for the laser. The peak amplitude of the plasma wave increases
over time. We concentrate on the third accelerating bucket (measured
from the front of the simulation), since this is the eventual location
of the test particles. The field evolves from its initial value of
$ E_{1,peak} =1.9 \times 10^{10}V/m$ gradually to
$E_{1,peak} = 2.6 \times 10^{10}V/m$
(an increase of $\sim 30\%$) by the
end of the simulation. This is consistent with the fact that
as the laser looses energy by generating the plasma wave it downshifts
in frequency \cite{Mori97}. This in turn causes the ponderomotive potential
of the laser
$\Phi_{L}=-e\left< \vec{E}^{2}\right>/\left(2m\omega_{L}^{2}\right)$
and therefore the wakefield amplitude, which is proportional to
it, to increase\cite{Sprangle88}. 
The wavelength of the plasma wave in the center of the channel can be
read from Fig.~\ref{fig:channel:e1-wake} to be $\lambda_{p}=17.9\mu m$.
The figure also shows the
initial placement of the test particles with regard to the plasma
wave and the position of the accelerated test particles after
$27$ Rayleigh lengths of propagation. The position of the accelerated
test particles indicates that the test particles are being accelerated
by the third accelerating bucket of the plasma wave instead of the 
second as it might have been expected from the original position of
the test particles. Since all the test particles had an initial $\gamma$
of $\sim 15$ those test particles that got eventually accelerated in the
third acceleration bucket must have first been decelerated to fall
back to the position of third bucket before eventually gaining energy.
This initial deceleration is consistent with evolution of test particle
data over time.

\begin{figure}
   \begin{center}
      \epsfig{ file=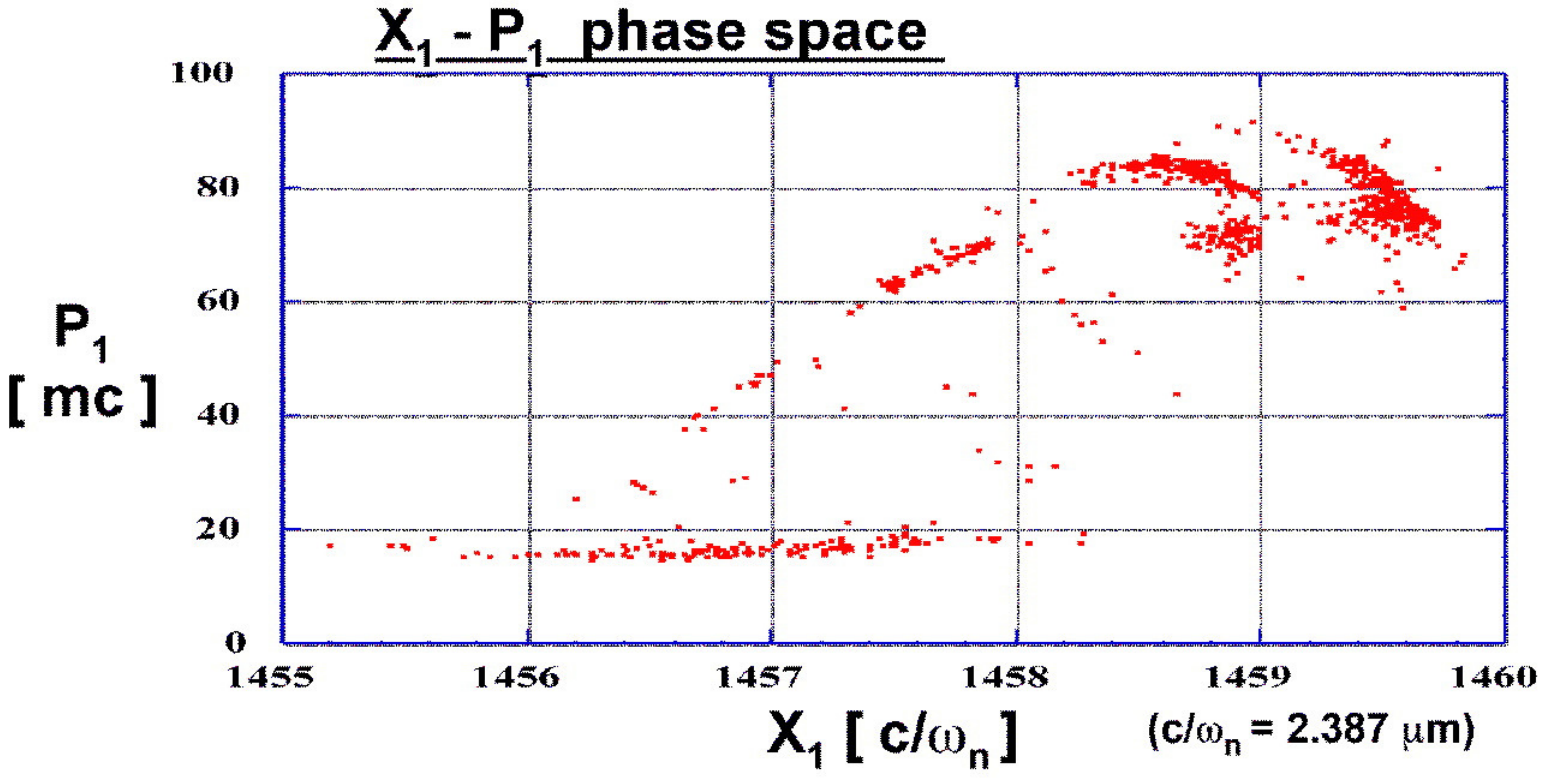, width=5.8in }
      \caption{The longitudinal, $x_{1}p_{1}$, phase space of the test 
               particles with a momentum $p_{1} \geq 15 m_{e}c$ after
               27 Rayleigh lengths of propagation. Note that the $x_{1}$
               axis of the plots only extends over about the last $1/6$-th
               of the simulation window.}
      \label{fig:channel:X1P1-1500}
   \end{center}
\end{figure}  

Fig.~\ref{fig:channel:X1P1-1500} is the longitudinal phase space of the
test particles with a momentum $p_{1} \geq 15 m_{e}c$. It shows that
some test particles exhibit behavior which is different
than expected. A certain number of test particles has essentially
the original energy, while a second group has been accelerated to an energy of
about $p_{1} = 40MeV$, and a third group has energies between these extremes.
Since the total number of particles seen in this plot is 821 out of
16128 original test particles, the remaining particles must have been lost
due to defocusing fields and deceleration. There are 570 particles with an
energy of about $40MeV$. This is about 3\% of the total number of particles
and gives a first estimate on timing precision required for injecting an
external particle bunch into the right phase of this accelerator system.
This estimates assumes that all the accelerated particles come from the same
area within the original test particle group.

\begin{figure}
   \begin{center}
      \epsfig{ file=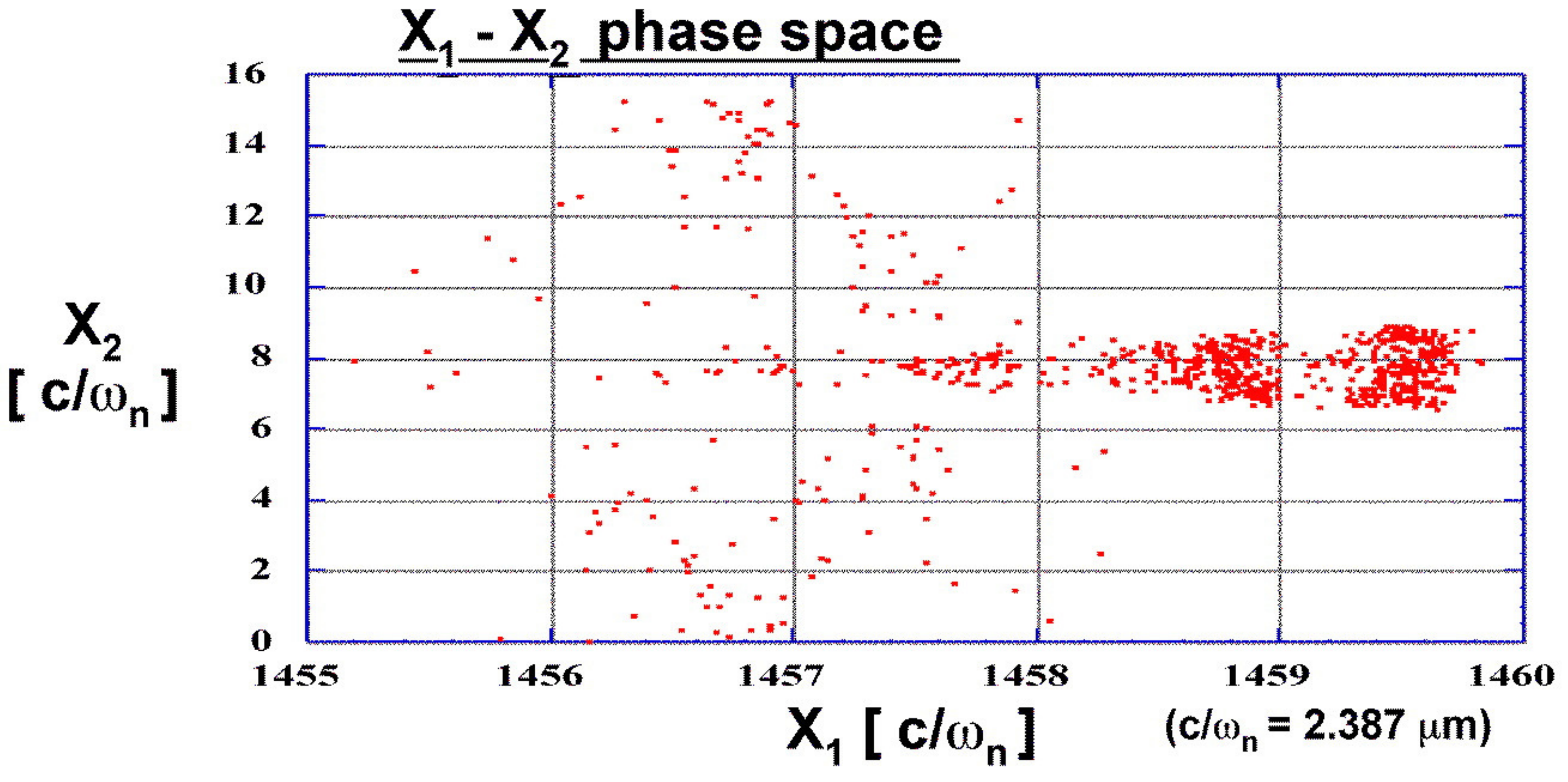, width=5.8in }
      \caption{The spatial distribution of the test 
               particles with a momentum $p_{1} \geq 15 m_{e}c$ after
               27 Rayleigh length of propagation. Note that the $x_{1}$
               axis of the plots only extends over about the last $1/6$-th
               of the simulation window.}
      \label{fig:channel:X1X2-1500}
   \end{center}
\end{figure}  

In Fig.~\ref{fig:channel:X1X2-1500} the distribution of the test particles
in real space is shown. The groups of particles with different behavior seen in
Fig.~\ref{fig:channel:X1P1-1500} can be identified with groups of particles in
this plot. All the particles with any energy gain are within a radius of about 
$4\mu$ of the central axis of the laser and the plasma wave while most of the
particles without energy gain are further away from the axis out side 
the area of the plasma channel and the wakefield.

\begin{figure}
   \begin{center}
      \epsfig{ file=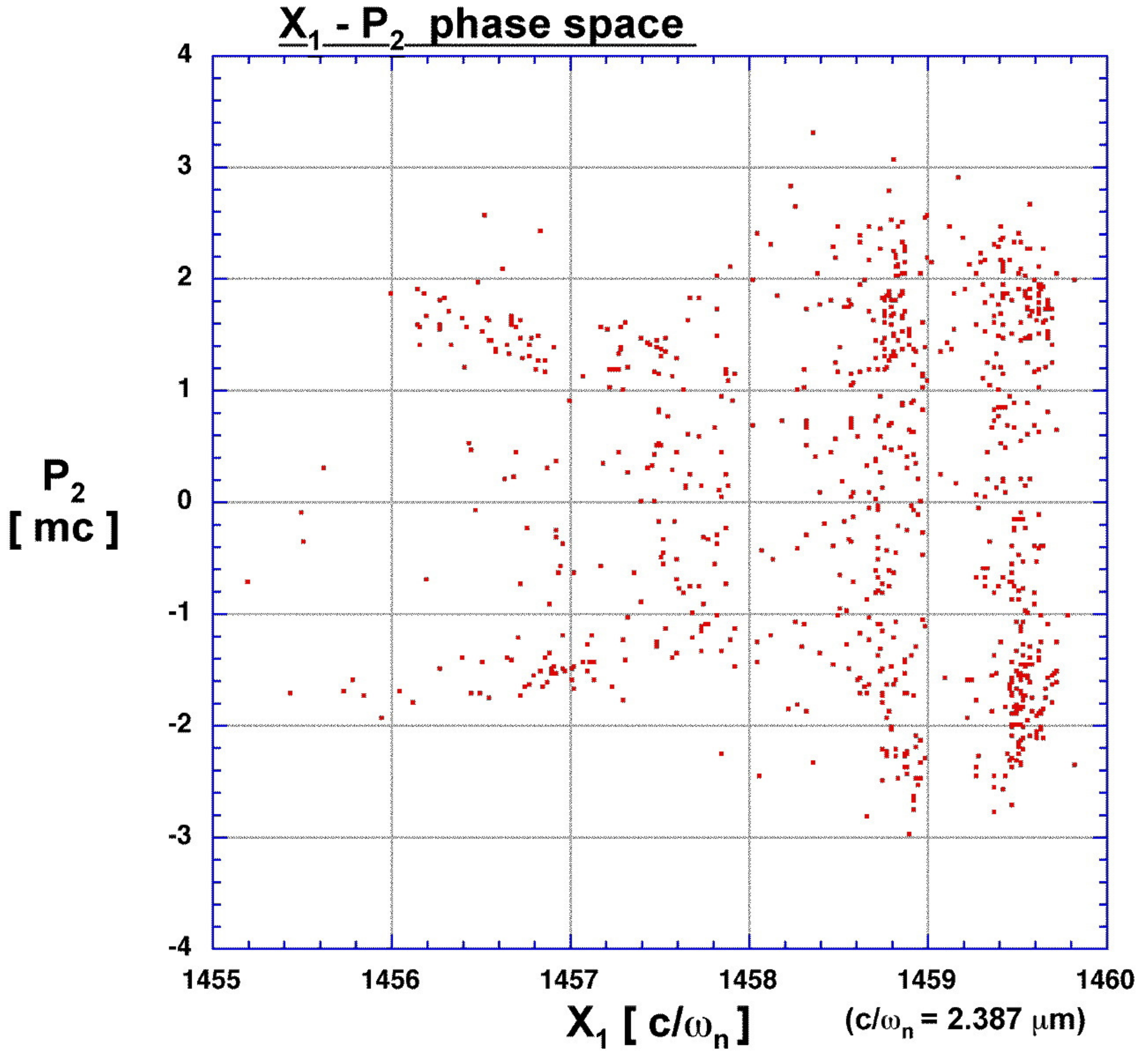, width=5.8in }
      \caption{The $x_{1}p_{2}$-phase space of the test 
               particles with a momentum $p_{1} \geq 15 m_{e}c$ after
               27 Rayleigh length of propagation. Note that the $x_{1}$
               axis of the plots only extends over about the last $1/6$-th
               of the simulation window.}
      \label{fig:channel:X1P2-1500}
   \end{center}
\end{figure}

Fig.~\ref{fig:channel:X1P2-1500} finally shows the
$x_{1}p_{2}$-phase space of the test particles. The most interesting 
fact to note about this figure is that the particles at higher $x_{1}$
which correspond to particles with higher $p_{1}$ as can be seen from
Fig.~\ref{fig:channel:X1P1-1500} have on average more transverse
momentum then the unaccelerated particles in at lower $x_{1}$.
This is consistent with the fact that the accelerated particles are
in the focusing region of the plasma wave and are undergoing
betatron oscillations. The unaccelerated particles do not undergo
any oscillations but have a net transverse momentum that moves them
out of the plasma channel as can be seen in Fig.~\ref{fig:channel:X1X2-1500}.

The particle information that was used to generate the Figs.
\ref{fig:channel:X1P1-1500}, \ref{fig:channel:X1X2-1500},and
\ref{fig:channel:X1P2-1500}
can be used to directly calculate the parameters that characterize
the bunch of 570 particles that were accelerated to about $40MeV$.
From this we obtain the following results. If not noted differently all
widths are calculated as rms-values with respect to the mean value of
a quantity. The beam
has a momentum of $p_{1}=78.3m_{e}c$ with a spread of 
$ \Delta p_{1} =5.4 m_{e}c$. The longitudinal spread of the beam is
$2 \times \Delta x_{1}= 1.94 \mu m$. The total length of the beam from the
first to the last particle is $3.59 \mu m$. For the transverse direction
the width is $2 \times \Delta x_{2}= 2.88 \mu m$ and the
momentum spread is $ \Delta p_{2} = 1.59 m_{e}c$.

These numbers result in
an energy spread of $\Delta E / E = 14\%$ and in a normalized emittance
$\varepsilon_{N} = \pi \, \Delta x_{2} \, \Delta p_{2} \, / \, m_{e}c
= 2.29 \,\pi \, mm \: mrad$.
This value can be compared with the original emittance of the test 
particles and the acceptance of the plasma wake which can be estimated
with Eq.~(\ref{equ:injection:acceptance2}). 
The initial emittance of the test particles can be calculated by 
using the total width $b$ of the initial flat distribution profile
to calculate the rms-width $\Delta x_{0,2}$ of this profile.
We get  $\Delta x_{0,2} = b / (2\sqrt{3}) = 1.15 \mu / (2\sqrt{3})$.
With this the initial normalized emittance becomes
$\varepsilon_{0,n} = \pi \, \Delta x_{0,2} \, \Delta p_{0,2} \, / \, m_{e}c
= 0.08 \, \pi \, mm \: mrad$.
In order to calculate the acceptance using Eq.~(\ref{equ:injection:acceptance2})
the normalized peak potential of the plasma wave is required. This can 
be estimated by assuming a harmonic plasma wave. In this case
$\bar{\Phi} = e\Phi/ \left( m_{e}c^{2} \right) = \bar{E}_{1,peak}/ 
\bar{k}_{p} = 0.107$ with
$\bar{E}_{1,peak}= eE_{1,peak}/ \left( m_{e} c \omega_{n} \right) = 0.105$ and
$\bar{k_{p}}= c k_{p}/ \omega_{n}=0.838$. The initial value of 
$E_{1,peak}$ is used for this estimate. Using these numbers, the 
estimated value for the acceptance is $A_{n} = 37 \,\pi \, mm \: mrad$. The 
comparison of the initial emittance, the final emittance, and the 
acceptance indicates that even though there is some emittance growth
of the beam it never reaches a matched beam equilibrium. A question that 
should be investigated is whether this emittance growth is taking
place continuously throughout the acceleration or whether it occurs
during a specific time period. 

The normalized potential calculated above can also be used to 
calculate the maximum energy gain that the test particles can 
achieve due to the linear dephasing limit. Using
Eq.~(\ref{equ:review-physics:energy-gain}) we find (applying the peak
potential at the end of the simulation $\bar{\Phi}=0.14$) that 
$\Delta W_{max} = 40.9 \, m_{e}c^{2}$. If we include the original energy
of the particles then we would expect a final energy of
$W_{final} = 28.6 \, MeV$. This means that the test particles gained
more energy than would normally be predicted for a linear 3D plasma wave.

There are two possible reasons for this. One is that the particles are 
initially not accelerated in the focusing part of the accelerating
phase of the plasma wave since acceleration by only the focusing part of 
the plasma wave was assumed in the calculation of the maximum
energy gain above. The possible energy gain over the full accelerating 
phase of the plasma wave including the defocusing part is
$42 MeV$. This is consistent with the final energy of the test particles
if we consider the fact that they were first decelerated 
before being accelerated. 
The other possibilities is that the wake generated
in the plasma channel is nonlinear enough that
Eq.~(\ref{equ:review-physics:energy-gain}) does not give a good
estimate of the maximum energy gain anymore. With the data available 
from the simulation presented here this can not be decided and
is a question for future research.

\section{Conclusion}
  \label{sect:channel:conclusion}

This chapter presented the results of a 2D simulation of a LWFA
in a parabolic plasma channel. The results suggest that the idea
of preventing diffraction of a laser pulse by propagating it through
a matched parabolic plasma channel works. Furthermore, the laser
in the channel is able to generate a plasma wave useful for 
acceleration of particles. Test particles accelerated in the plasma 
wave experience an increase in emittance. Only about $3\%$ of the
initial test particles are being accelerated which suggests that for externally
injected particles beams the length of the beam and the exact
phase will be crucial. The simulations show that the phase velocity 
of the wake and the energy gain cannot be straightforwardly obtained
from the simple 1D theory so this is an area for future research.


\chapter{Plasma Wakefield Acceleration in the Blowout Regime}
  \label{chap:pwfa}

\section{Introduction}
  \label{sect:pwfa:introduction}

The basic concept of a plasma wakefield accelerator (PWFA) is to
accelerate a low current trailing electron bunch by the wakefield 
generated by a high current driver. If the driving bunch is highly 
relativistic, then both the accelerating as well as the accelerated 
bunch are moving with about the speed of light and the accelerated 
bunch can stay in phase with the accelerating field for distances 
long enough to gain significant amounts of energy. Motivated by and 
as part of the preparations for an experiment which is currently being 
conducted at the Stanford Linear Accelerator Center (SLAC), we have 
simulated a plasma wakefield accelerator with the expected parameters 
of this experiment \cite{Assmann97}. 

\begin{figure}
   \begin{center}
      \epsfig{file=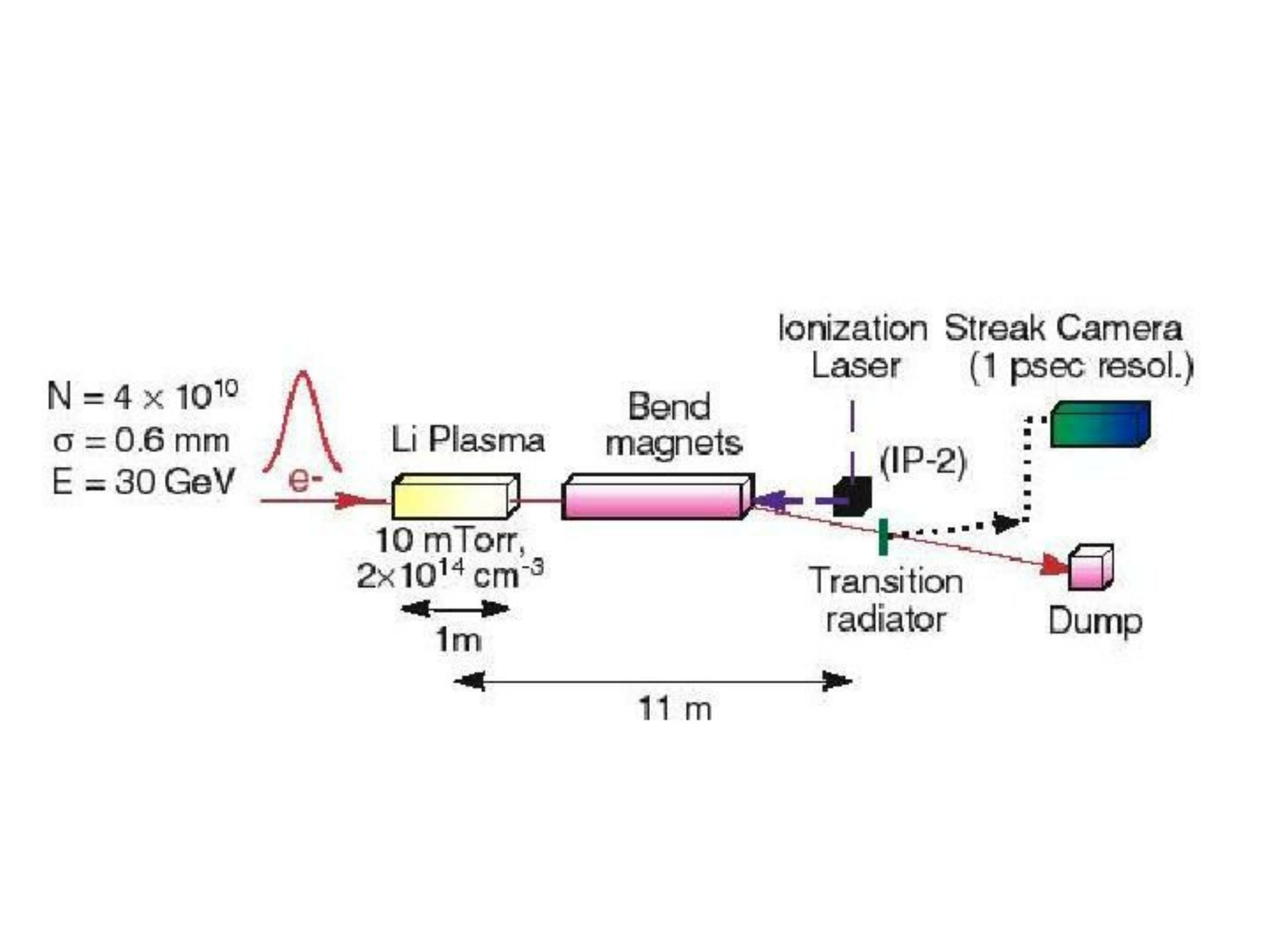, width=5.8in }
      \caption{The setup of the E-157 experiment at SLAC.}
      \label{fig:pwfa:e157-setup}
   \end{center}
\end{figure}

In this experiment a 30GeV electron beam at SLAC 
is used to excite a wake of the order 1GeV/m in a $1.4m$ long plasma 
of density $1 - 2 \times 10^{14} cm^{-3}$. In this wake the centroid energy of 
the tail of the beam is expected to increase by several hundred MeV.  
Since the beam in this experiment is typically much denser than the 
plasma (e. g., $N = 3.5 - 4 \times 10^{10}$ electrons in a
$\sigma_{z} = 0.6 mm$ bunch 
length and a spot size of $\sigma_{r} = 50 \mu m$ corresponding to a beam
density  $n_{b} = 1 \times 10^{15} cm^{-3}$), the PWFA is in the highly
non-linear or so-called 
blowout regime \cite{Rosenzweig91}. Fig.~\ref{fig:pwfa:e157-setup} shows
the experimental setup of the experiment. More details about the
setup and execution of the experiment can be found in Ref.~\cite{Hogan99}. 

The advantages that the blowout regime offers
are a high accelerating gradient, a constant accelerating structure with respect 
to the transverse dimensions, a linear focusing force, and a high 
transformer ratio. However, in this nonlinear regime neither linear theory nor 
fluid models are applicable and do not provide an accurate 
understanding of the physics. Much better insight into the physical 
processes can be gained by using Particle-In-Cell (PIC) simulations, 
which allow accurate modeling of highly non-linear processes like the 
ones occurring here. For these reasons, we conducted PIC simulations 
to investigate this regime of plasma wakefield acceleration. The 
simulations were done using both the 2D cylindrically-symmetric and 
the full 3D system packages in OSIRIS. We have also developed a 
analytic model which is a bridge between the full particle PIC models 
and the reduced description PIC models and fluid codes.

\section{2D Cylindrically-Symmetric Simulations}
  \label{sect:pwfa:2d-cyl-sym}
  
We carried out simulations for the physical parameters similar to the 
ones described above, using OSIRIS. The algorithms for the results 
presented in this section were 2D cylindrically-symmetric and used the
moving simulation window to follow the beam since this limits the
simulation domain to the beam and its immediate surroundings rather
than the whole propagation distance of the beam. The simulation window
in normalized units had a 
size along the propagation direction, z, of $25 c/\omega_{p}$ and a size in the 
radial direction, r, of $10 c/\omega_{p}$ with a grid of
$N_{z} \times N_{r} = 500 \times 200$. 
Here $c$ is the speed of light and $\omega_{p}$ is the plasma frequency for a given 
plasma density $n_{p}$. We will use a plasma density $n_{p} = 2.1 \times 10^{14}cm^{-3}$,
which corresponds to 
$c/\omega_{p} = 0.367 mm$, throughout this chapter when converting
simulation results 
back into physical units. This means the simulation window corresponds to 
a size of $9.175 mm \times 3.67 mm$. The beam propagated through the plasma for 
190000 time-steps with $dt = 0.02 \omega_{p}^{-1}$ (corresponding to 
$18.35\mu m$ of 
propagation distance per timestep) for a total of $3800 c/\omega_{p}$ 
($\sim 1.4 m$). 
Nine particles per cell were used for the background plasma and $25$ 
particles per cell for the beam. The beam�s longitudinal profile was 
fitted to the experimentally known profile of the SLAC beam \cite{Assmann97},
which 
is very close to a longitudinal Gaussian profile of length, $\sigma_{z} = 0.63 mm$, and 
a transverse Gaussian profile of width, $\sigma_{r} = 70 \mu m$.
For $3.7 \times 10^{10}$ electrons this corresponds to a peak density
of $7.56 \times 10^{14} cm^{-3}$. 

\begin{figure}
   \begin{center}
      \epsfig{file=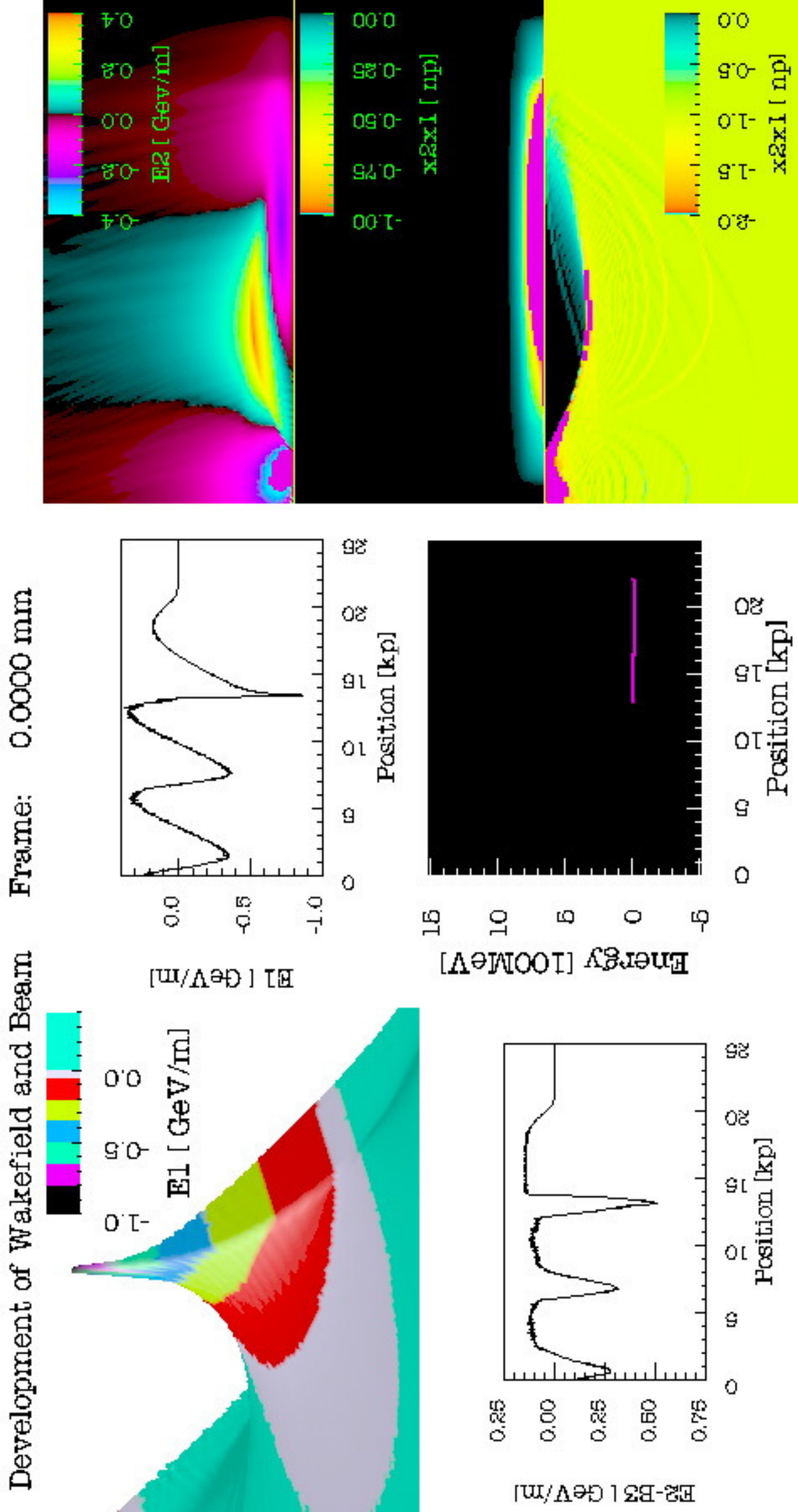, height=7.0in }
      \caption{The figure shows plots of several quantities at the beginning
               of the simulations just after the beam fully entered 
               the plasma. See the main text for a detailed explanation
               of the plotted quantities.}
      \label{fig:pwfa:movie-1002-a}
   \end{center}
\end{figure}

\begin{figure}
   \begin{center}
      \epsfig{file=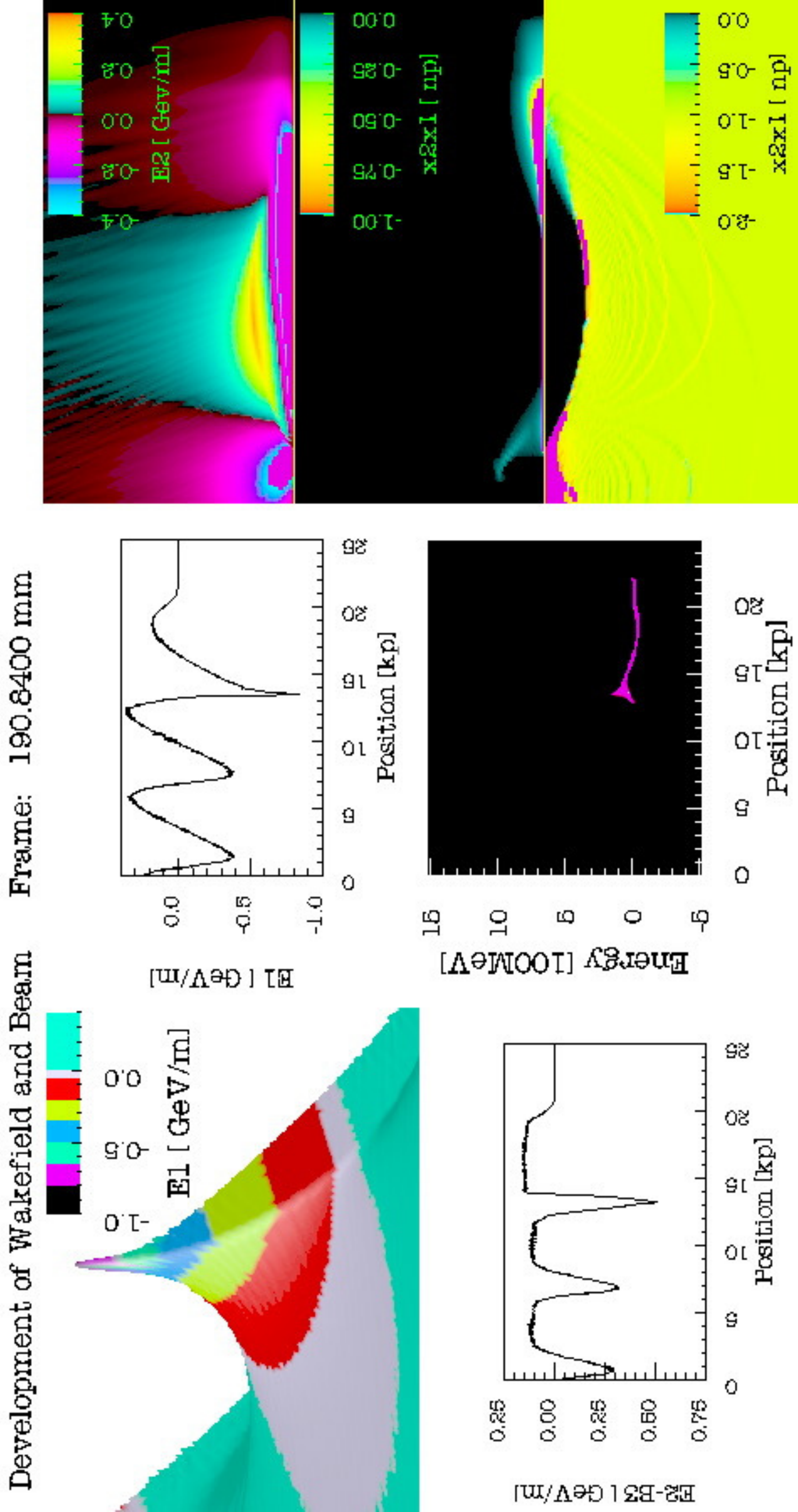, height=7.0in }
      \caption{The figure shows plots of several quantities at the first
               minimum of the betatron oscillation of the beam
               after $\sim 191 mm$ of propagation through the plasma.
               See the main text for a detailed explanation 
               of the plotted quantities.}
      \label{fig:pwfa:movie-1028-b}
   \end{center}
\end{figure}

\begin{figure}
   \begin{center}
      \epsfig{file=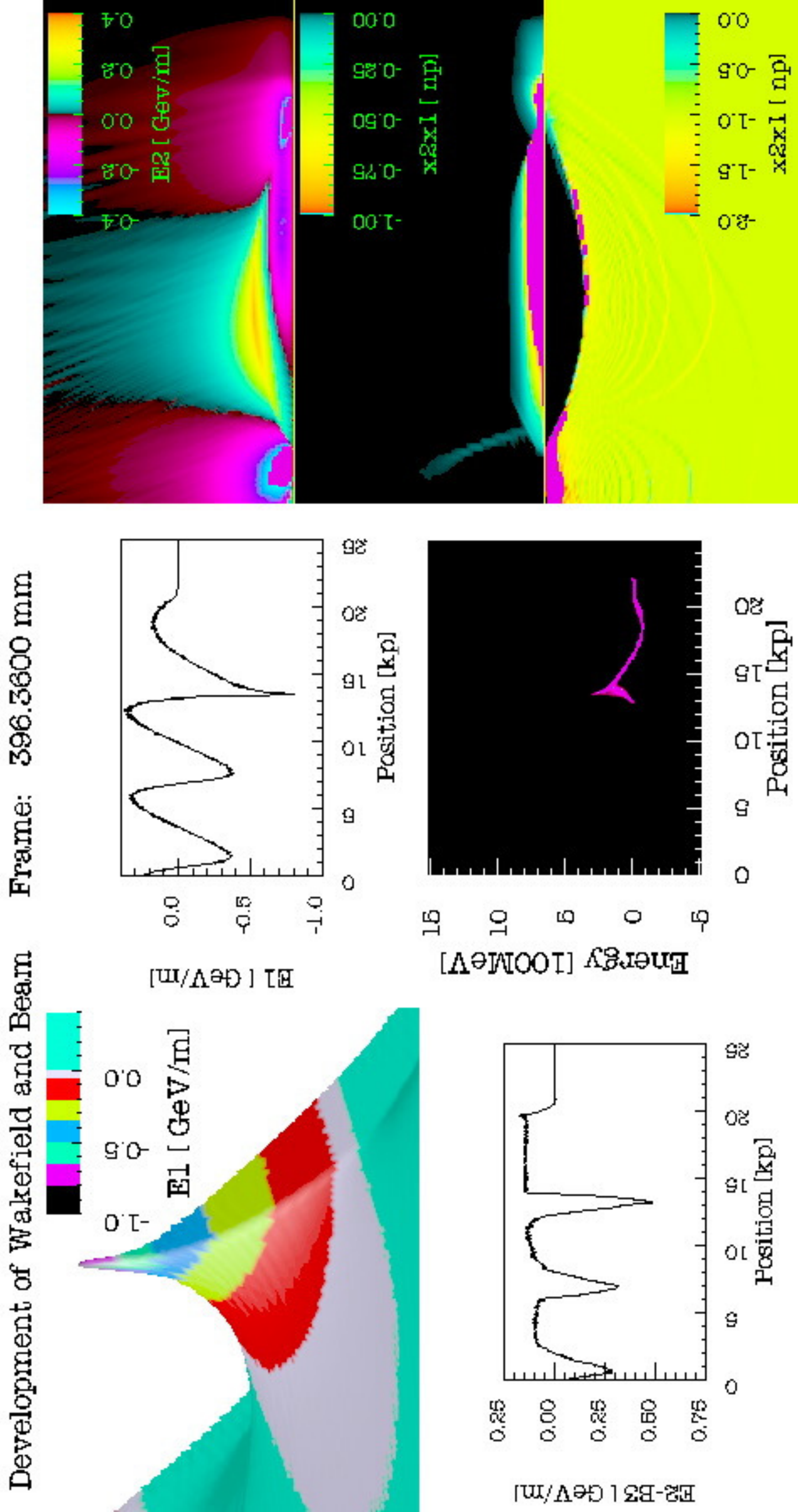, height=7.0in }
      \caption{The figure shows plots of several quantities at the first
               maximum of the betatron oscillation of the beam
               after $\sim 396 mm$ of propagation through the plasma.
               See the main text for a detailed explanation 
               of the plotted quantities.}
      \label{fig:pwfa:movie-1056-c}
   \end{center}
\end{figure}

\begin{figure}
   \begin{center}
      \epsfig{file=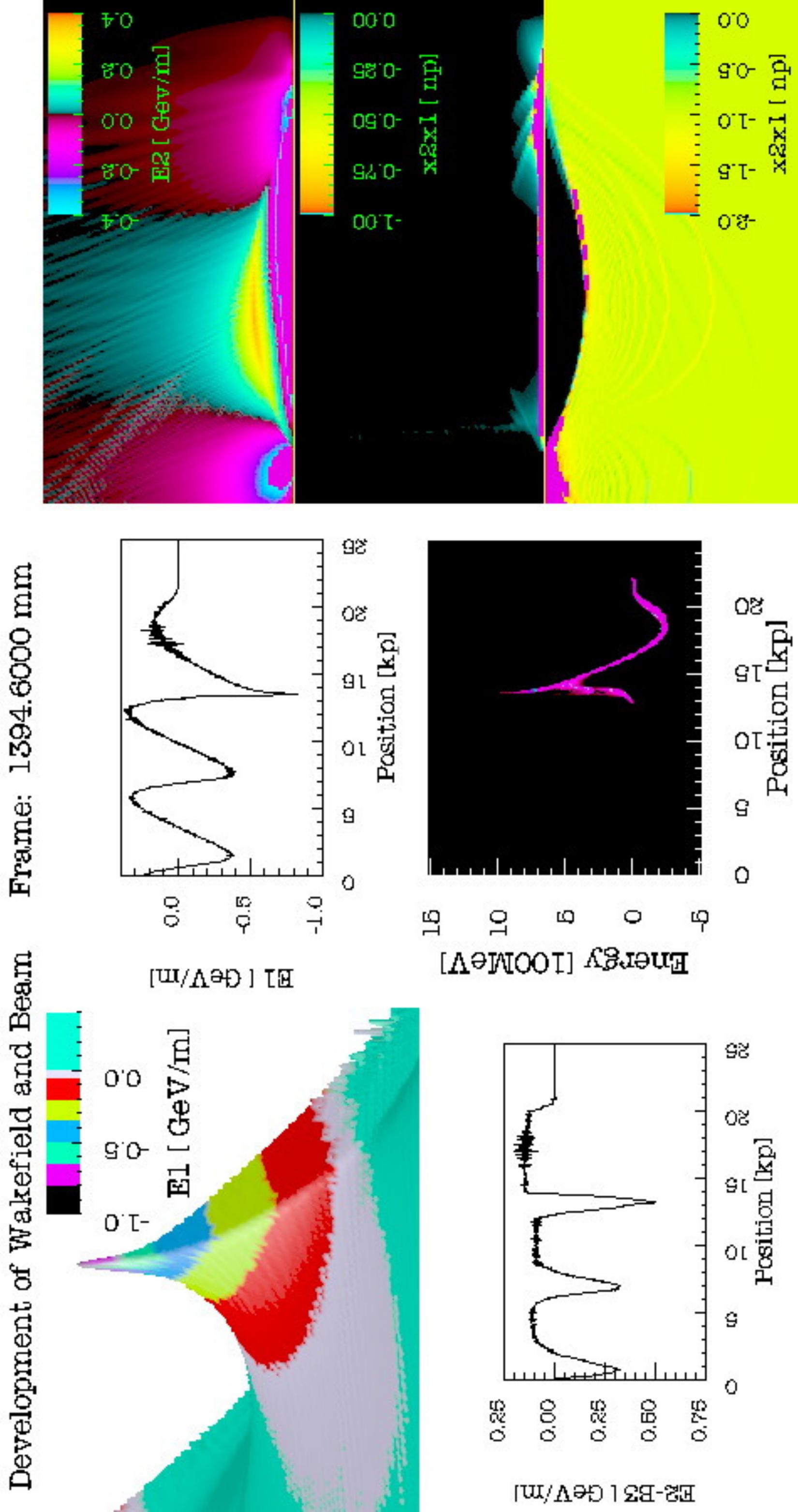, height=7.0in }
      \caption{The figure shows plots of several quantities at the end
               of the simulations
               after $\sim 1.4 m$ of propagation through the plasma.
               See the main text for a detailed explanation 
               of the plotted quantities.}
      \label{fig:pwfa:movie-1192-d}
   \end{center}
\end{figure}
  
Fig. \ref{fig:pwfa:movie-1002-a} to \ref{fig:pwfa:movie-1192-d}
give detailed information about the beam and background plasma at 
several timesteps. Together they present a picture of the development
of the beam and plasma over time. The four figure represent data after
$0 mm$, $191 mm$, $396 mm$, and $1.4m$.  The figures for the times 
$191 mm$, $396 mm$ are shown because these are the times of the first 
betatron oscillation minimum and maximum after the beam enters the
plasma.
The figure is composed of a number of plots that shows seven different
aspects of the simulation data. Please note that in these figures the indices
$1$, $2$, and $3$ are used instead of $z$, $r$, and $\Theta$ respectively.
The plot in the upper left corner is a colored, 
rubber-sheet representation of the longitudinal, accelerating electric 
field. For this visualization the elevation of a surface point as well 
as its color represent the field strengths of the electric field. Note 
that we chose a perspective for visualizing the rubber-sheet surface so 
that negative field values of the electric field would be represented by 
positive values of the surface elevation. This leads to a better 
visualization of the accelerating region. The sharp edge of the 
rubber-sheet surface going roughly from the upper left corner to the 
lower right corner is due to the axial boundary of the 2D 
cylindrically-symmetric simulation and accordingly $r$ increases starting 
from this axial boundary towards the lower left corner. Due to the chosen 
perspective the rubber-sheet does not show the data for the whole 
simulation. The upper middle plot shows the value of 
the longitudinal electric field along the axial boundary for its full 
length of $25 c/\omega_{p}$. The figure in the lower middle below the electric field 
lineout shows the energy gain and loss of the electron beam as a function 
of the axial position. The colored areas indicate the parts of this plot 
where beam electrons are present. Note that the horizontal axis of this 
plot is precisely aligned with the axis of the field lineout above. The 
plot in the lower left corner shows the focusing field experienced by the 
beam electrons, $E_{r}-B_{\Theta}$, at a position
$73.4 \mu m \cong \sigma_{r}$ off axis. The right 
column shows three color-plots in the $r-z$ plane.
The plots shown (from top to bottom) are the 
radial electric field $E_{r}$, the charge density of the beam, and the charge 
density of the background plasma. This last plot has been mirrored along 
the axis to allow for a more direct comparison of the plasma density with 
the beam density. The horizontal axis for each of the three plots goes 
from $12.5 c/\omega_{p}$ to $22.5 c/\omega_{p}$ of the simulation window
and the vertical 
axis shows $0$ to $5 c/\omega_{p}$ along the radial direction. The field
and density 
values are given by the colorbars in each of the plots. Note that the 
areas of the plots colored in magenta are areas in which the field or 
density values are outside the respective color-scales. Since the 
color-code of the beam charge density plot reaches from $0$ to $�1$, 
which is the normalized density of the background plasma, the 
magenta-colored areas in this plot indicate densities above the 
background plasma density.

The first fact to note is the lack of 
change over time in the evolution of the accelerating electric field, 
and the focusing field. With the exception of the peak accelerating 
field which fluctuates slightly by about $\pm$0.05GeV/m around a value of 
about 0.75GeV/m ($\sim \pm 7\%$) and some slight variation in the level 
of numerical noise, the accelerating electric field essentially does 
not change over time. This is in strong contrast with the dynamic 
development of the beam radius (middle plot in the right column) and 
energy (lower plot in the center column), and the radial electric 
field (upper plot in the right column). The energy plot shows that 
every part of the beam except the front part and the very tail gains 
or loses energy linearly as a function of time. This is consistent 
with the constant longitudinal field since at an initial energy of 
about 30GeV the beam electrons experience no significant phase slippage 
over the time of the simulation.

Two other effects can also be observed. First there is a slight 
broadening of the front part of the decelerated area of the beam along 
the energy axis, which means that not all electrons at a given z 
experience exactly the same decelerating field. Secondly there is a 
large energy spread of the very back of the beam tail, which splits 
up into two parts. The first observation can be understood when looking 
at the background plasma charge density. The plasma charge density plot 
shows that in the front part of the beam the area of total electron 
blowout is smaller than in the later parts of the beam, and therefore 
the radius up to which the focusing force $F_{r}$ is independent of $z$ is 
smaller. According to the Panofsky-Wenzel theorem,
$\partial F_{r}/\partial z = \partial F_{z}/\partial r$, 
this implies an acceleration gradient that varies along the radial 
position beyond a small value of $r$\cite{Rosenzweig91}.
This can also be noticed for the region of decelerating 
field that is visible in the lower right corner of the $E_{z}$-rubbersheet 
plot. The radially flat area increases slightly in width towards the 
back. The broadening of the front part of the deceleration area of the 
beam is a result of this non-uniform accelerating field. The energy 
spread of the tail of the beam can be understood by looking at the 
narrowing of the accelerating and focusing field profile near the 
peak-accelerating field. It shows that a part of the tail of the 
beam, in contrast to the rest of the beam, experiences a strong 
defocusing force that pushes it radially out of the accelerating 
field. The blowout of some of the tail-electrons of the beam can 
also be seen in the development of the beam charge density.

The evolution of the main part of the beam, as seen in the beam 
charge density plot, is clearly dominated by the betatron oscillation 
of the beam in the focusing field. The focusing field is mainly due 
to the ions left in the plasma blowout area, as seen in the plasma 
charge density plot, since the effects of electric and magnetic 
fields of the relativistic beam on itself cancel each other almost 
completely. The linear focusing force in the blowout area results 
in the same oscillation frequency for all beam electrons in that 
area. The beam propagates while undergoing betatron oscillations 
with a wavelength for the spotsize 

\begin{equation}
  \label{equ:pwfa:lambda-spotsize}
  \lambda_{spotsize} = \lambda_{\beta}/2
                     = \pi \sqrt{\frac{\gamma m c^{2}}{2 \pi e^{2} n_{0}}} 
\end{equation}

where $\lambda_{\beta}$ is the betatron wavelength of a single particle.
This wavelength follows directly from
Eq.~(\ref{equ:review-physics:beam-k-beta-in-ion-column}).
Measuring this wavelength using the minima of the oscillation of the beam 
density gives a wavelength $\lambda_{spotsize} = 40cm$ as predicted by
Eq.~(\ref{equ:pwfa:lambda-spotsize}) 
for the density of the simulation \cite{Assmann97}.

The dynamics of the front part of the beam is more complex because 
the blowout area there is not as wide. This leads to non-harmonic 
oscillations or so-called aberrations in the focusing force, which 
leads to phase mixing of the electrons. The oscillation frequency of 
the beam electrons decreases towards the front. Even though this is
not to clearly visible from the figures above the full data set of the
simulation shows clearly 
that after the main part of the beam reaches an oscillation 
minimum this minimum moves forward towards the front of the beam as 
the electrons there execute betatron oscillations with lower 
frequencies. This happens while at the same time the main part of the 
beam starts to expand again. This dynamics at the front of the beam
leads to a subtle point. 
Namely, the focusing field for the beam, $E_{r}-B_{\Theta}$,
shows an unexpected 
behavior with time. Initially the focusing force rises slowly over 
the first one-quarter of the beam, but once the head of the beam 
begins to pinch the rise becomes steeper. The unexpected behavior 
results because the transverse profile never relaxes back to the 
original one. Instead, there is always an axial slice of the beam 
at the head of the beam that is near a pinch. So on average, the 
beam density at the front of the beam is always larger than it was 
at $t=0$. As a result the occurrence of complete blowout is earlier 
in the beam and the region of blowout is wider leading to more of 
the beam undergoing the uniform betatron oscillations than might have 
been expected.
Unlike the beam, the plasma electrons respond predominantly to only 
$E_{r}$. Thus, the blowout of the plasma electrons and their oscillation 
back onto the axis in the back of the pulse is caused by the total 
radial electric field that they experience. The figures show that the 
radial field has two distinct regions. The front, where the plasma 
electrons are not blown out yet, is dominated by the electric field of 
the beam; and the back, where the plasma electrons are blown out, is 
dominated by the radial electric field of the remaining ion charge. 
The plasma charge density plot shows the effect of this. In the 
moving window frame, i.e., in the $z-ct$ coordinate, the plasma 
electrons stream backward past the stationary drive beam. After the 
radial field force deflects the electrons outward most of them 
coalesce in a narrow, high density surface layer that lies at the 
edge of the blowout region. The radius of the blowout region and 
therefore the radial position of the layer is roughly 
$0.77 c/\omega_{p} \approx 280 \mu m$. This is consistent with
the rough theoretical estimate \cite{Lee99}.

\begin{equation}
  \label{equ:pwfa:blowout-estimate}
  r_{blowout} = 2 \sigma_{r} \sqrt{\frac{n_{b}}{n_{p}}}
\end{equation}

where $n_{b}$ is the peak density of a beam. Note that for a long 
pulse for which the electrons are blown out adiabatically, 
$r_{blowout} = \sigma_{r} \sqrt{n_{b}/n_{p}}$ \cite{Whittum91}.

\begin{figure}
   \begin{center}
      \epsfig{file=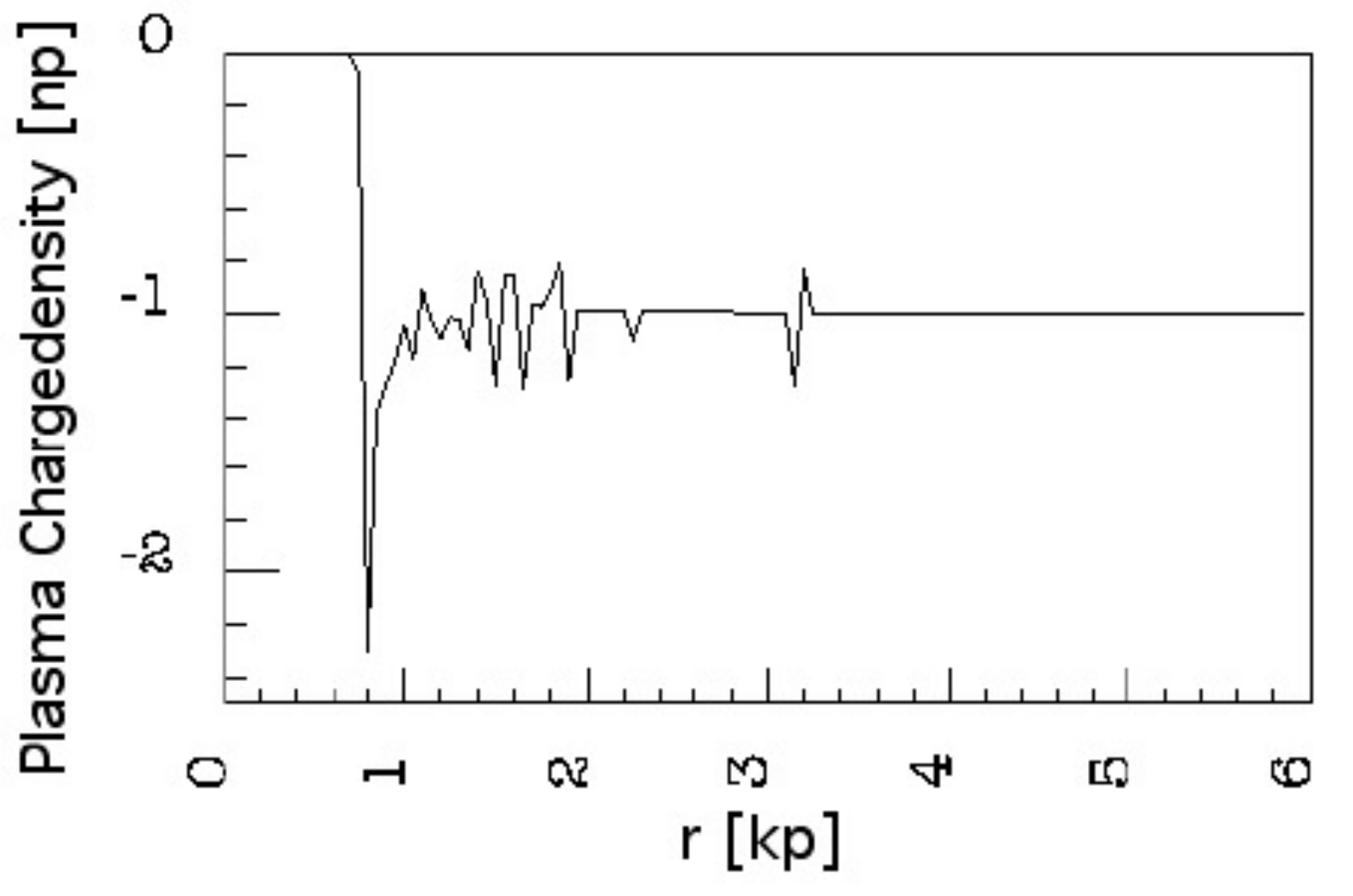, width=5.8in }
      \caption{A radial lineout of the plasma charge density at the
               center of the beam after $1.4$ meters of propagation.}
      \label{fig:pwfa:np-1192-x2+175}
   \end{center}
\end{figure}

The electrons 
stream backward within this narrow surface layer and converge on the 
axis creating a very dense spike and therefore a sharp peak in the 
accelerating field. (Note that in the lab frame individual electrons 
are blown out and then return while remaining near their initial $z$ 
value, but we will use the moving window point of view for it�s 
convenience of description). The insensitivity of the accelerating 
wake field to the dynamic beam development is a consequence of the 
beam being narrow when compared to the radius at which the surface 
layer is located. For most of the plasma oscillation, all of the plasma 
electrons are outside of the beam so that from Gauss� law the 
electrostatic field affecting them is independent of the radius of the 
charge inside. Thus the betatron pinching of the beam has little effect 
on the plasma electrons and hence the wake. The slower evolution in the 
front of the beam does not have any significant effect either since the 
slight variations in the initial trajectories of electrons become 
insignificant after the blown out electrons reach the surface layer. 
The surface layer is shown in Fig.~\ref{fig:pwfa:np-1192-x2+175},
where a radial lineout of the 
plasma charge density at the center of the beam is plotted after $1.4$ 
meters of propagation. The plasma blowout as well as the surface layer 
are clearly visible.

%

\begin{figure}
   \begin{center}
      \epsfig{file=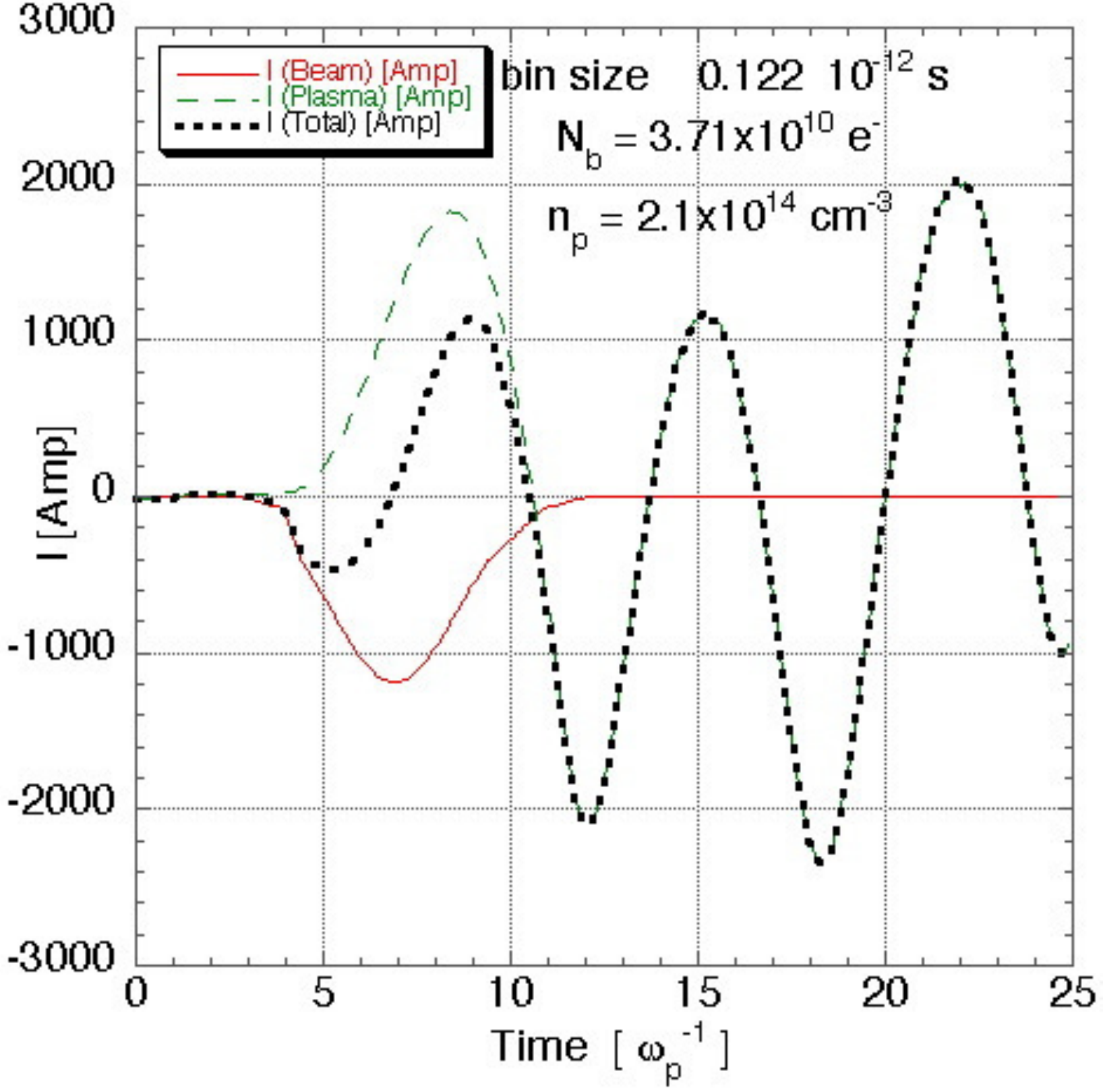, width=5.8in }
      \caption{The beam current, the plasma current, and the total 
               current for each 0.12 pico-second bin at the beginning of 
               the simulation just after the beam fully entered the 
               plasma.}
      \label{fig:pwfa:currents}
   \end{center}
\end{figure}

Another useful quantity to illustrate the plasma response is the return current
carried by the plasma electrons. Fig.~\ref{fig:pwfa:currents} shows the 
plasma current,
the beam current, and the sum of both. They are calculated from phase 
space data for each 0.12 pico-second bin of the moving simulation window.
In the first half of the beam the plasma current is smaller than the 
beam current but increases strongly after an initial delay compared to
the beam current. At the center of the beam the beam current and the plasma
current roughly balance. After this point the plasma current
dominates. The main result here is that in the later half of the beam the
plasma current completely shields the plasma further away from the beam
from the magnetic field generated by the beam. This is consistent the
observation above that the most of the plasma response is confined to a
narrow layer outside the blowout region.

\begin{figure}
   \begin{center}
      \epsfig{file=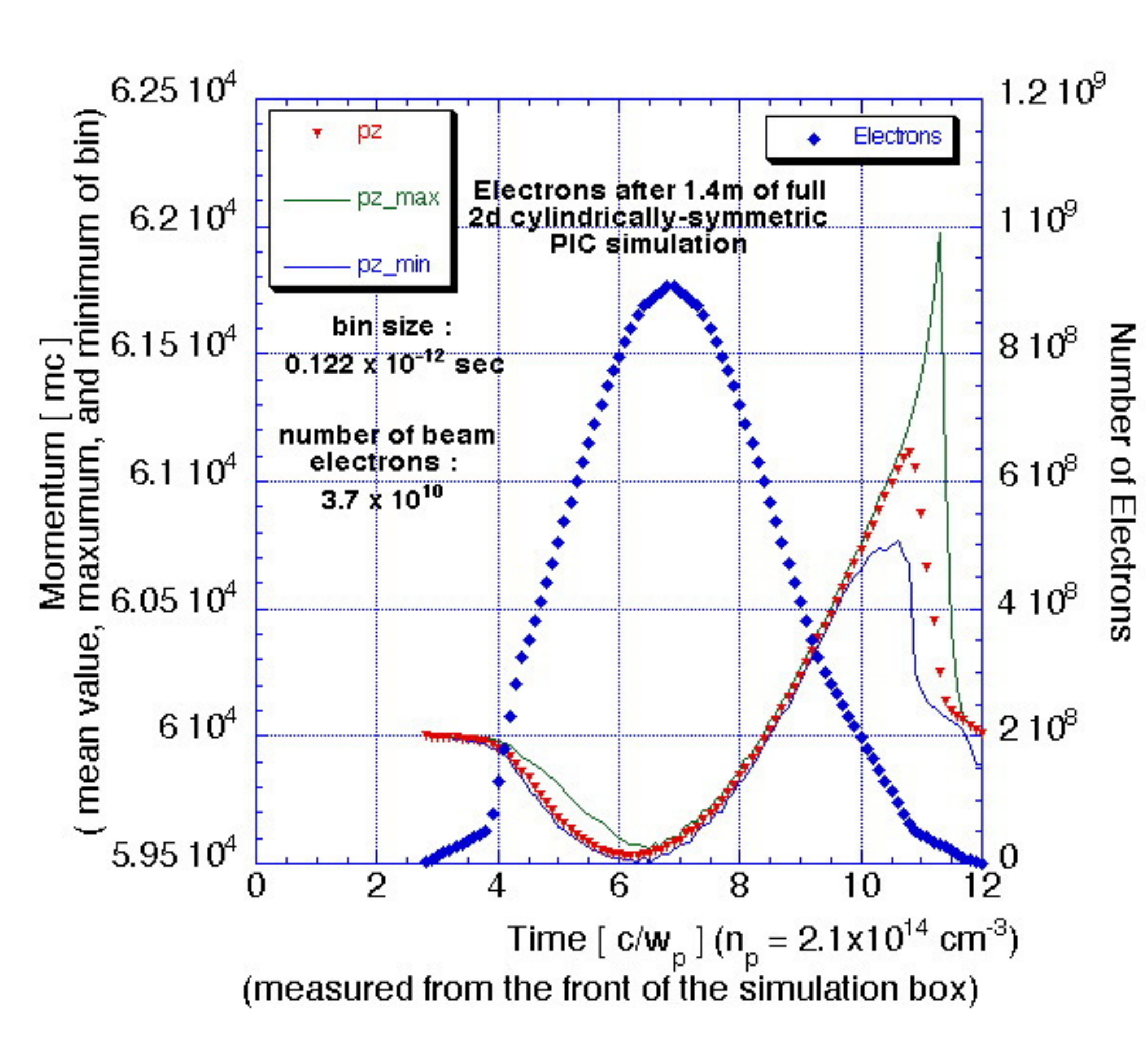, width=5.8in }
      \caption{The mean, maximum, and minimum momentum in the propagation 
               direction, $p_{z}$, as well as the
               number of electrons for each 0.12 pico-seconds bin
               after 1.4 meters of propagation using the full PIC
               simulation to propagate the beam.}
      \label{fig:pwfa:par-a-1192}
   \end{center}
\end{figure}

\begin{figure}
   \begin{center}
      \epsfig{file=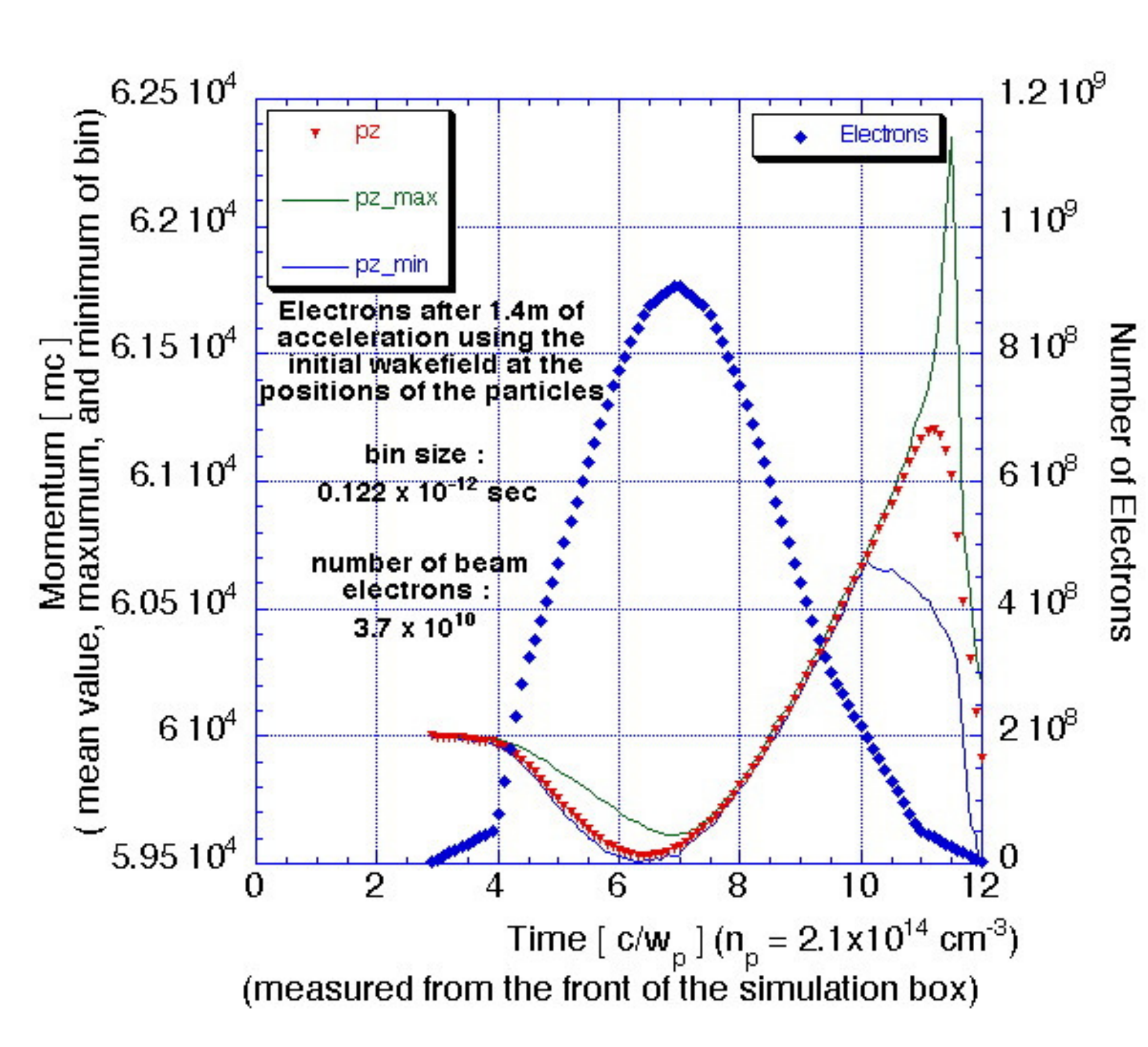, width=5.8in }
      \caption{The mean, maximum, and minimum momentum in the propagation 
               direction, $p_{z}$, as well as the
               number of electrons for each 0.12 pico-seconds bin
               after 1.4 meters of propagation using the initial fields
               at the initial positions of the particles to propagate
               the beam.}
      \label{fig:pwfa:par-b-1004-192}
   \end{center}
\end{figure}

Due to the invariance of the accelerating field, the expected energy 
gain can be predicted with confidence for a specified beam charge and 
profile. The longitudinal momentum $p_{z}(\cong \gamma)$ vs. $ct$
phase space is shown 
in Fig.~\ref{fig:pwfa:par-a-1192} and Fig.~\ref{fig:pwfa:par-b-1004-192}
to illustrate the expected acceleration of the beam after 
$1.4 m$ of propagation. The mean, maximum, and minimum energy of the beam 
are plotted in $0.12$ pico-second bins along the length of the beam 
(Fig.~\ref{fig:pwfa:par-a-1192}). This is done in figure 
Fig.~\ref{fig:pwfa:par-a-1192} for the actual simulation particle 
data after $1.4m$. Fig.~\ref{fig:pwfa:par-b-1004-192} by contrast was
generated by using the initial 
particle data propagated for $1.4m$ using the initial fields at the initial 
positions of the particles. This makes the assumption of a non-evolving 
field and neglects the betatron oscillation of each particle. The mean, 
maximum, and minimum energies resulting from these two graphs are very 
similar for most of the beam. The results only differ at the very end 
of the beam where Fig.~\ref{fig:pwfa:par-b-1004-192} shows larger
average and maximum energies 
and lower minimum energies than Fig.~\ref{fig:pwfa:par-b-1004-192}.
The similarity between 
the two figures for the main part of the beam is consistent with our 
assumption of non-evolving wakefield if the accelerating field has a 
constant value within the radial range of the betatron oscillation for 
each particle. The differences in the tail are due to the fact that the 
particles in the tail at larger radii do not experience a constant 
accelerating field during their radial motion. For the full simulation 
this leads to an averaging out of the different accelerations experienced 
by each particle due to its transverse motion. For the particles 
accelerated with the initial field this averaging does not happen and the 
maximum and minimum energies in the beam tail of
Fig.~\ref{fig:pwfa:par-b-1004-192} are therefore 
a measure of the maximum and minimum accelerating field in that part of 
the wake. Based on these figures we can say that the maximum field is 
about 0.85GeV/m but that the maximum energy gain by a particle after 
1.4 m will be about 1GeV. The maximum mean energy for a 
0.12 pico-second bin is 550MeV with about $7 \times 10^{7}$ electrons
in this maximum energy bin. This is again consistent with the information in 
Fig.~\ref{fig:pwfa:movie-1192-d} for these numbers. The conclusion from
Fig.~\ref{fig:pwfa:par-a-1192} and Fig.~\ref{fig:pwfa:par-b-1004-192} is 
that the betatron oscillations do not have a significant influence on 
the acceleration of the beam.

\begin{figure}
   \begin{center}
      \epsfig{file=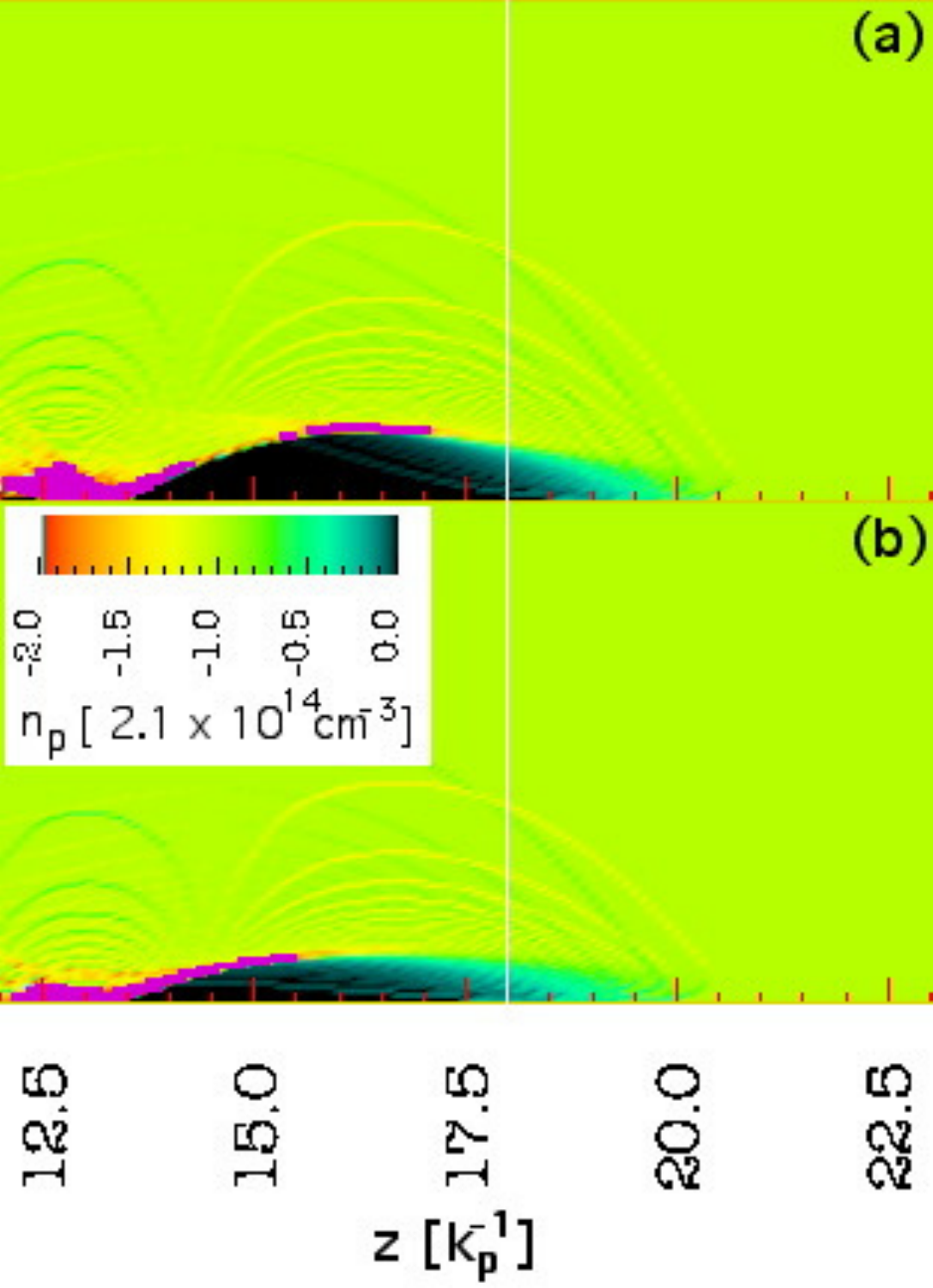, width=4.0in }
      \caption{The initial response of plasma in simulations with
               (a) $N_{1}=3.7  \times 10^{10}$ beam electrons and
               (b) $N_{2}=1.85 \times 10^{10}$ beam electrons.}
      \label{fig:pwfa:blowout-comp}
   \end{center}
\end{figure}

There are a number of parameters in the experiment E-157 that can vary.
The number of particles in the electron beam is one of these parameters
and in order to investigate the effects of a change in the number of
electrons a simulations with only half the number of beam electrons 
as before was done. All other parameters of the this simulation were
kept the same. Fig.~\ref{fig:pwfa:blowout-comp} shows the initial response
of plasma for the previous simulation with $N_{1}=3.7 \times 10^{10}$ as
well as the plasma response for the simulation with only have the number
of beam electrons, $N_{2}=1.85 \times 10^{10}$. The overall structure of
plasma response is the same for both beams. The main change is as would
be expected from Eq.~(\ref{equ:pwfa:blowout-estimate}) a decrease of the
blowout radius by a factor of roughly $\sqrt{2}$. In addition the begin
of the total blowout area moves further to the back relative to the
center of the electron beam (indicated in Fig.~\ref{fig:pwfa:blowout-comp}
by the vertical white line through the figure).

\begin{figure}
   \begin{center}
      \epsfig{file=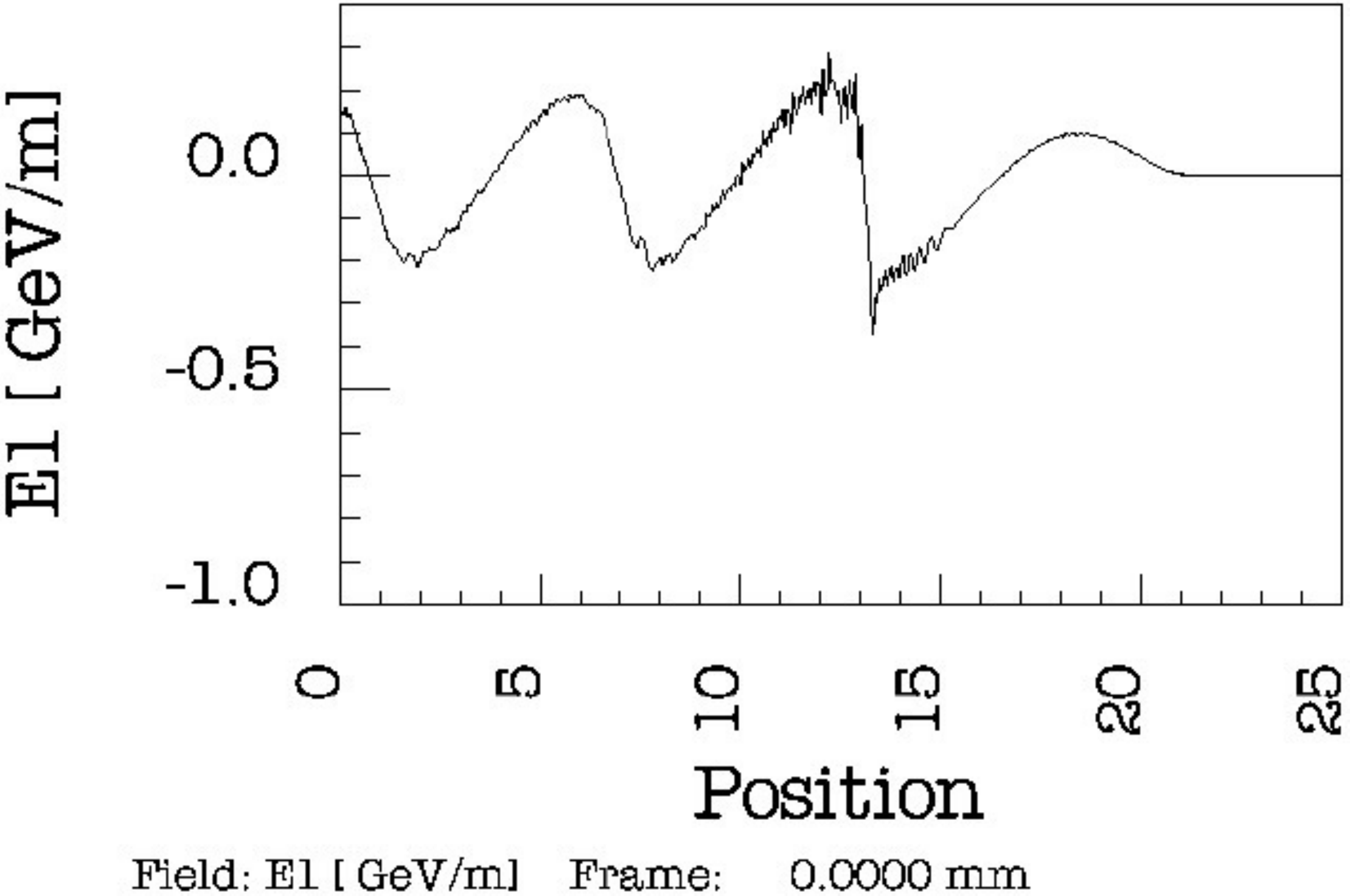, width=5.8in }
      \caption{ The lineout of the accelerating field along the axis
                for a simulation with $N_{1}=1.85 \times 10^{10}$.}
      \label{fig:pwfa:n1.85-E1-lineout}
   \end{center}
\end{figure}

Fig.~\ref{fig:pwfa:n1.85-E1-lineout} which is a lineout of the accelerating
field for the simulation with the smaller number of beam electrons shows
an effect of the decreased blowout. The peak accelerating field decreases
to $\sim$0.35GeV/m. This is less than half the value seen in the other
simulation. 

\begin{figure}
   \begin{center}
      \epsfig{file=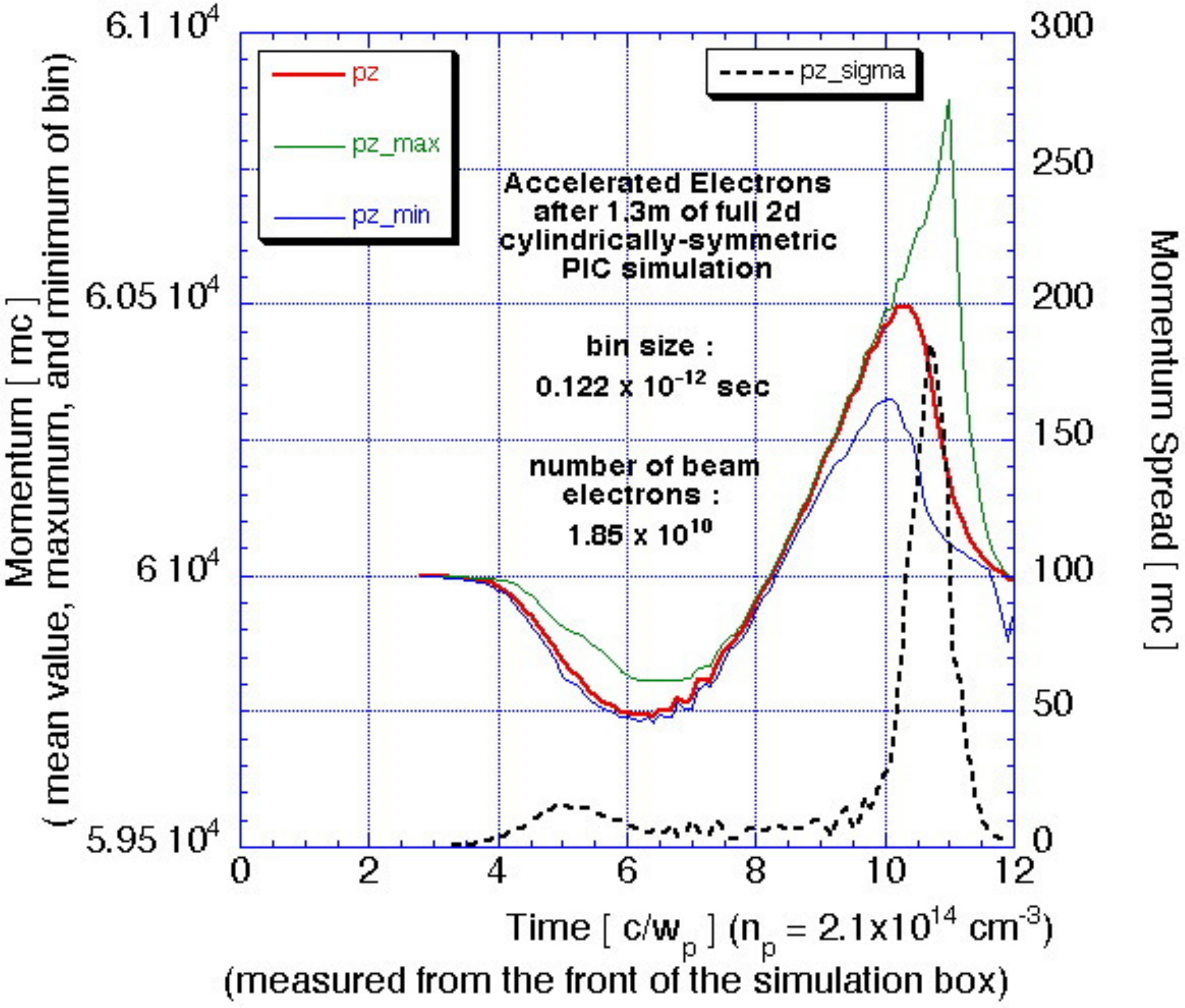, width=5.8in }
      \caption{The mean, maximum, and minimum momentum in the propagation 
               direction, $p_{z}$, as well as the width of the 
               distribution of $p_{z}$, $\sigma_{p_{z}}$, 
               for each 0.12 pico-seconds bin
               after 1.3 meters of propagation using the full
               2D cylindrically-symmetric PIC
               simulation to propagate the beam.}
      \label{fig:pwfa:n1.85-par-1176}
   \end{center}
\end{figure}

The resulting decrease in the energy gain of the accelerated tail of the
beam is shown in Fig.~\ref{fig:pwfa:n1.85-par-1176}. The figure shows
the mean, maximum, and minimum value of $p_{z}$ as well as $\sigma_{p_{z}}$
for 0.122ps-bins after 1.3 meters of propagation. The maximum mean energy
gain is only about 250MeV.

\section{3D Cartesian Simulations}
  \label{sect:pwfa:3d-simulations}
  
\begin{figure}
   \begin{center}
      \epsfig{file=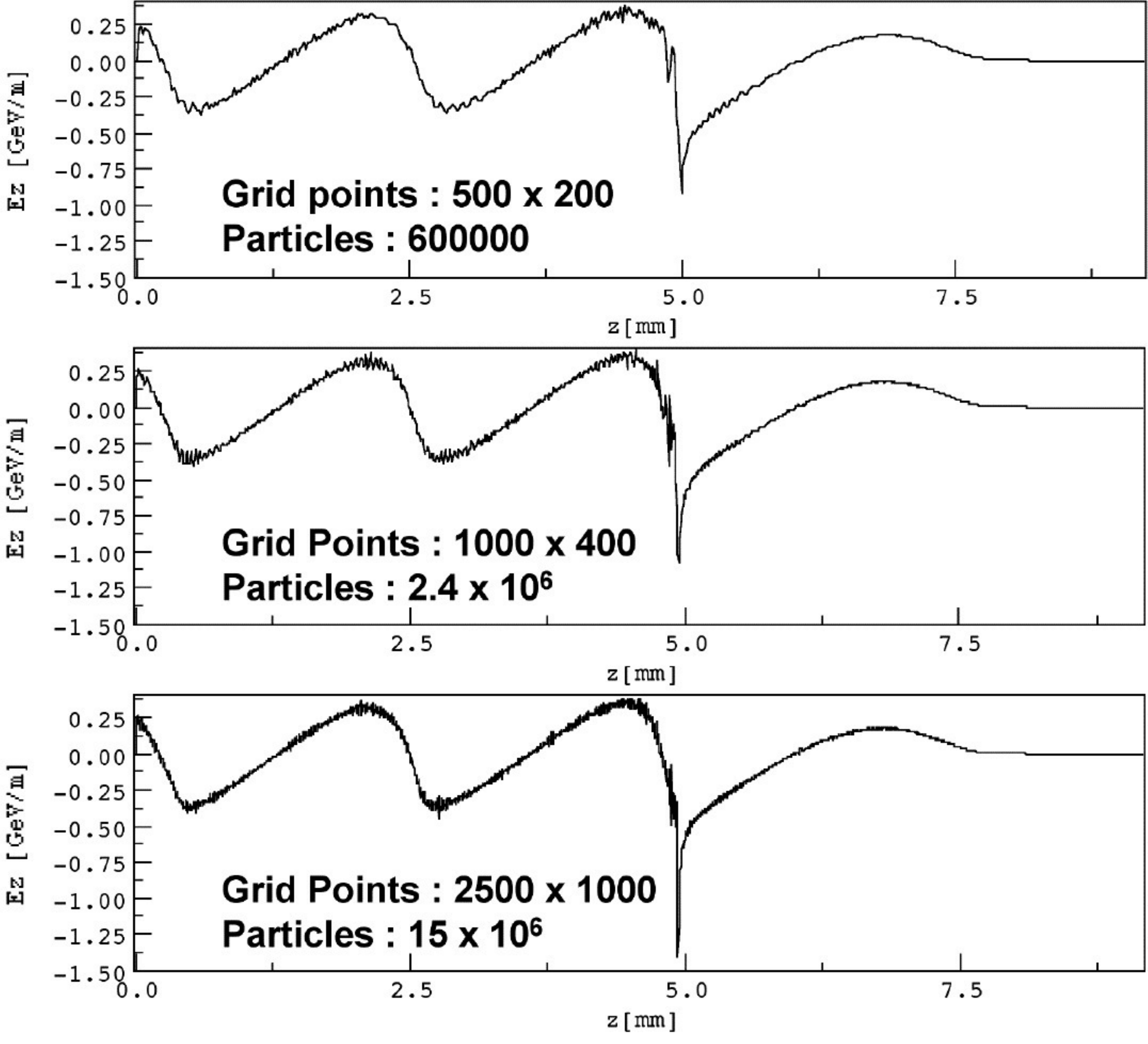, width=5.8in }
      \caption{The axial lineout of the accelerating field of 
               simulations of the PWFA using three different
               grid resolutions. The number of particles per cell
               is the same for all three simulations.}
      \label{fig:pwfa:resolutions}
   \end{center}
\end{figure}

A concern regarding the accuracy of the results presented in 
section~\ref{sect:pwfa:2d-cyl-sym} is the degree of numerical resolution 
needed to resolve the spike in the accelerating electric field and the use
of 2D cylindrically-symmetric algorithms.
In order to address the first point simulations with the beam only propagating
for a short distance were done for several cell sizes. The results are
summarized in Fig.~\ref{fig:pwfa:resolutions}. The figure shows the lineouts
of the axial accelerating fields for the resolution used above, for twice this
resolution, and for five times this resolution, i.e.,
the number of grid cells in each direction was increased by a 
factor of two and five, respectively, while the number of particles per
cell was kept the same. The most important effect of the increased resolution
is that the peak accelerating field becomes larger. Since this affects only
a very narrow spike it means that the electric field has a very high value
in a very small spatial area. This high field value is not properly resolved
by the original simulation, but the rest of the beam plasma interaction
is modeled accurately. The question is whether the insufficient resolution in
this small area actually matters considering the purpose of the simulations.
Since the higher field values will only affect the acceleration of a view
simulation particles which will only contribute very little to the mean
momentum in a bin of about 1 pico-second (this is the time resolution for measurements
of the E-157 experiment\cite{Hogan99}) it will not be of particular interest
for predicting the energy gain in the experiment or the dynamics of the rest
of the beam and the plasma. Furthermore, this large spike may not persist 
as the beam executes betatron oscillations.

In order to avoid the limitations of 2D cylindrically-symmetric 
simulations full 3D simulation were done as well. The simulations
were propagated only over short distances that were just long enough
for the beam to completely enter in to the plasma. 
This was done because of the much larger computational cost of
full 3D simulations compared to 2D cylindrically-symmetric simulations.

In order to study both the possible numeric inaccuracies near the 
axis for a cylindrically-symmetric code and the effects of asymmetric 
drive pulses full 3D simulations were performed. The 3D simulations 
only modeled the initial excitation of the wake. Presently, it is not 
computationally feasible to perform full 3D simulations which model 
$\sim$meter propagation distances.
The moving simulation window was also shortened
so that only the first oscillation of the plasma wake fits into the simulations 
box. The size of the grid cells was kept
$dx_{1}=dx_{2}=dx_{3} = 0.05 c/\omega_{p}$. The number of particles per
cell for the plasma as well as the beam was four. The total 
size of the simulation was 14 million grid cells and about 56 million particles.

\begin{figure}
   \begin{center}
      \epsfig{file=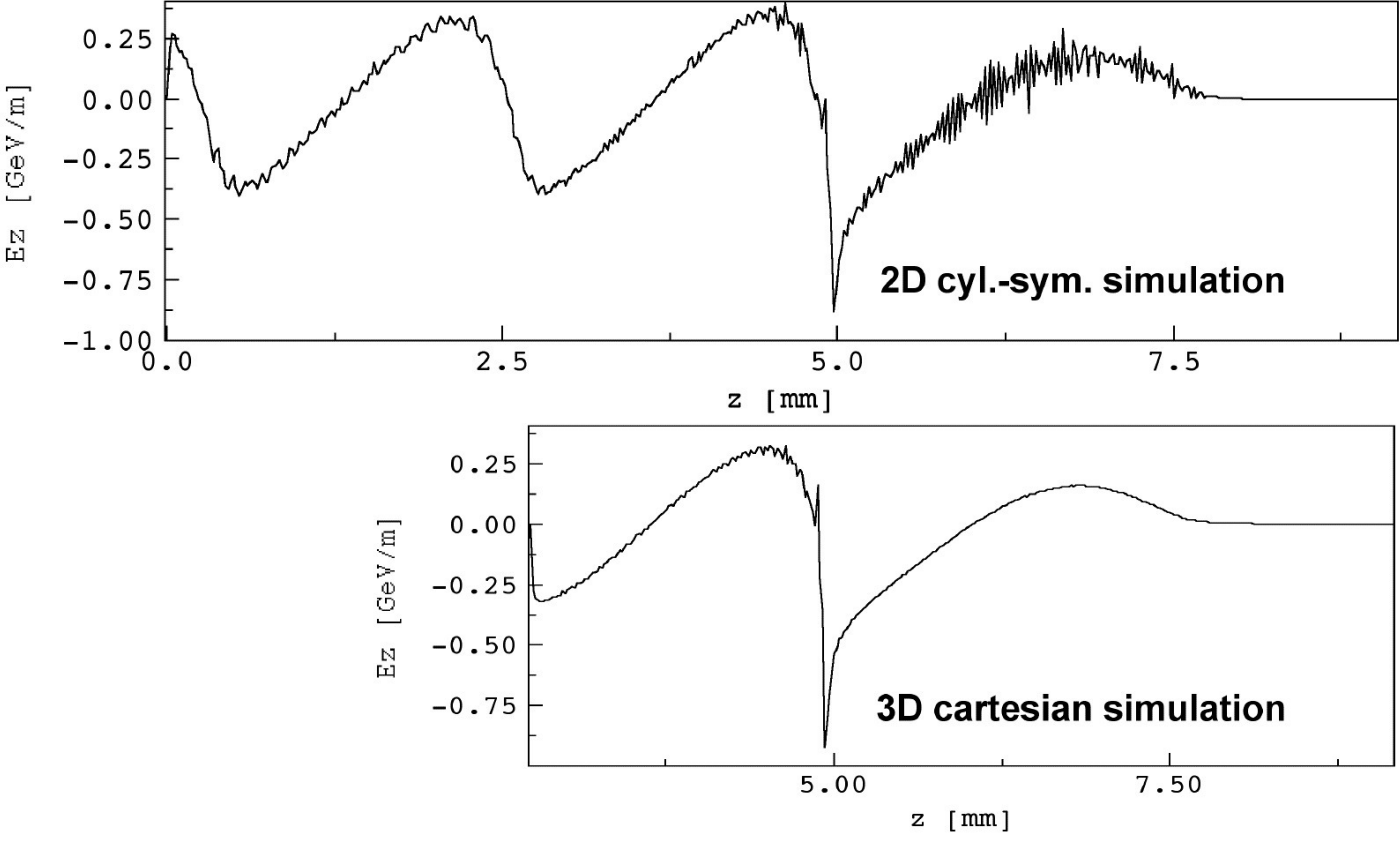, width=5.8in }
      \caption{The lineouts of the accelerating field along the
               axis of the beam for a 2D
               cylindrically-symmetric and a 3D Cartesian simulation
               with the same beam and plasma parameters.}
      \label{fig:pwfa:lineout-2d-3d}
   \end{center}
\end{figure}

Fig.~\ref{fig:pwfa:lineout-2d-3d}
shows a comparison between the central lineouts of the
accelerating fields of the 2D cylindrically-symmetric simulation and 
of the full 3D simulation. The figure shows that the wake on axis
for both simulations has the same structure and the amplitude agrees 
to within a few percent. Perhaps the most significant agreement is the 
amplitude and the structure of the spike in the accelerating field.
As a result we have confidence that the peak value for the
spike does not depend on which algorithms was used in the PIC code.
The cell size dependence of the
peak accelerating field value in Fig.~\ref{fig:pwfa:resolutions}
could have been a artifact of the 2D cylindrically-symmetric
algorithms on axis but the 3D simulations do not have the same
kind of problems on axis. Furthermore, since the 2D and 3D algorithms 
are very different, we have confidence that both are correct.

\begin{figure}
   \begin{center}
      \epsfig{file=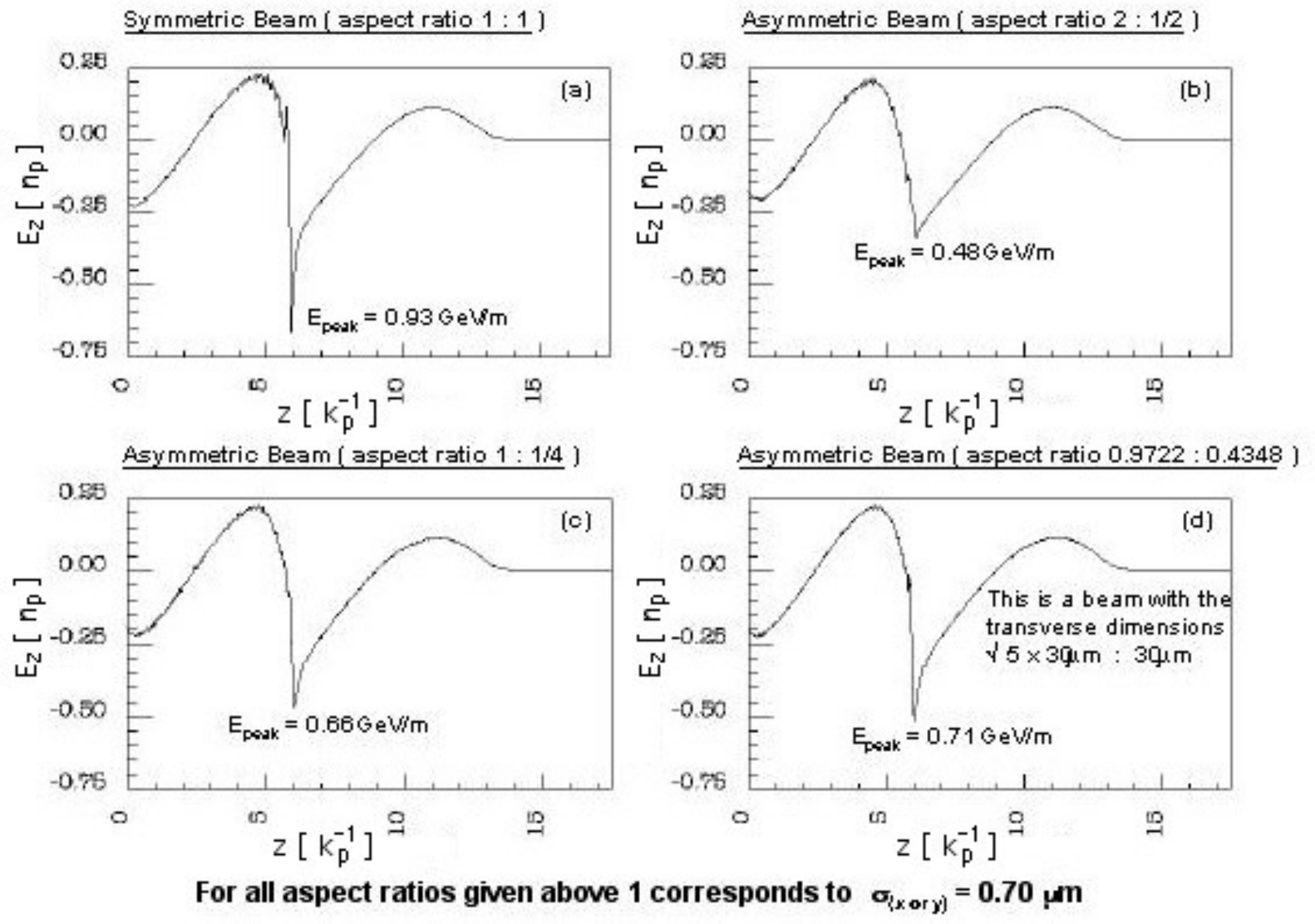, width=5.8in }
      \caption{The lineouts of the accelerating electric field
               along the axis of the beam in the propagation direction $z$
               for different aspect ratios of the transverse beam 
               spotsizes: (a) $1$:$1$ (b) $2$:$\frac{1}{2}$
               (c) $1$:$\frac{1}{4}$ (d) $0.9722$:$0.4348$}
      \label{fig:pwfa:lineouts-asym}
   \end{center}
\end{figure}

Now that we have confidence with the 3D algorithm, we can
investigate the effects of drive beam asymmetries
on the wake generation. Fig.~\ref{fig:pwfa:lineouts-asym}
shows lineouts of the accelerating electric field along the
axis of the beam in propagation direction $z$
for different aspect ratios of the transverse beam spotsizes,
$\sigma_{x} : \sigma_{y}$. Fig.~\ref{fig:pwfa:lineouts-asym}a)
shows again the lineout for the symmetric case.
Fig.~\ref{fig:pwfa:lineouts-asym}b) shows the lineout for a simulation
were the beam spotsize was increased by a factor of 2 in one
transverse direction
and decreased by a factor of 2 in the other transverse direction.
For this $2$:$\frac{1}{2}$ aspect ratio the magnitude of the peak
accelerating field decreases
to about half of its magnitude in the symmetric case. This decrease of
the peak field amplitude is consistent with the idea given earlier
that the sharp peak in the accelerating field is due to the return of
the blown-out plasma electrons to precisely the same point on axis. If the
spotsize of the beam in the different transverse directions in not the
same then electrons which are blown out on different transverse trajectories
will return to the axis at different points. This means the density spike on
axis will be smoothed out over a larger area causing the peak of 
the accelerating field to be smaller than for a symmetric beam.

Fig.~\ref{fig:pwfa:lineouts-asym}c) shows the lineout for a simulation
were the beam spotsize was kept the same in one transverse direction
and decreased by a factor of 4 in the other transverse direction. This
$1$:$\frac{1}{4}$ aspect ratio is different than the 
$2$:$\frac{1}{2}$ because in this case the peak density of the beam is
larger by a factor of 4. For this aspect ratio the peak accelerating field 
still decreases, but only by about $70\%$ of the $1$:$1$ case in 
Fig.~\ref{fig:pwfa:lineouts-asym}a). This case is important because 
it shows that any asymmetry in the beam will smear out the spike 
even if the blowout radius is much larger than either transverse 
dimension.


Fig.~\ref{fig:pwfa:lineouts-asym}d) we show the lineout for a
simulation were the beam spotsize was decreased only slightly in one
transverse direction and decreased by a factor of roughly 2 in the other
transverse direction. This $0.9722$:$0.4348$ aspect ratio corresponds 
to that of a typical SLAC beam at energies higher than the 30GeVs used
in the E-157 experiment\cite{Mori-communication}. This case looks most similar to 
the $1$:$\frac{1}{4}$ case. However, although the peak density is 
smaller the peak wake amplitude is still higher. This clearly 
demonstrates that round beams are desirable.


\begin{figure}
   \begin{center}
      \epsfig{file=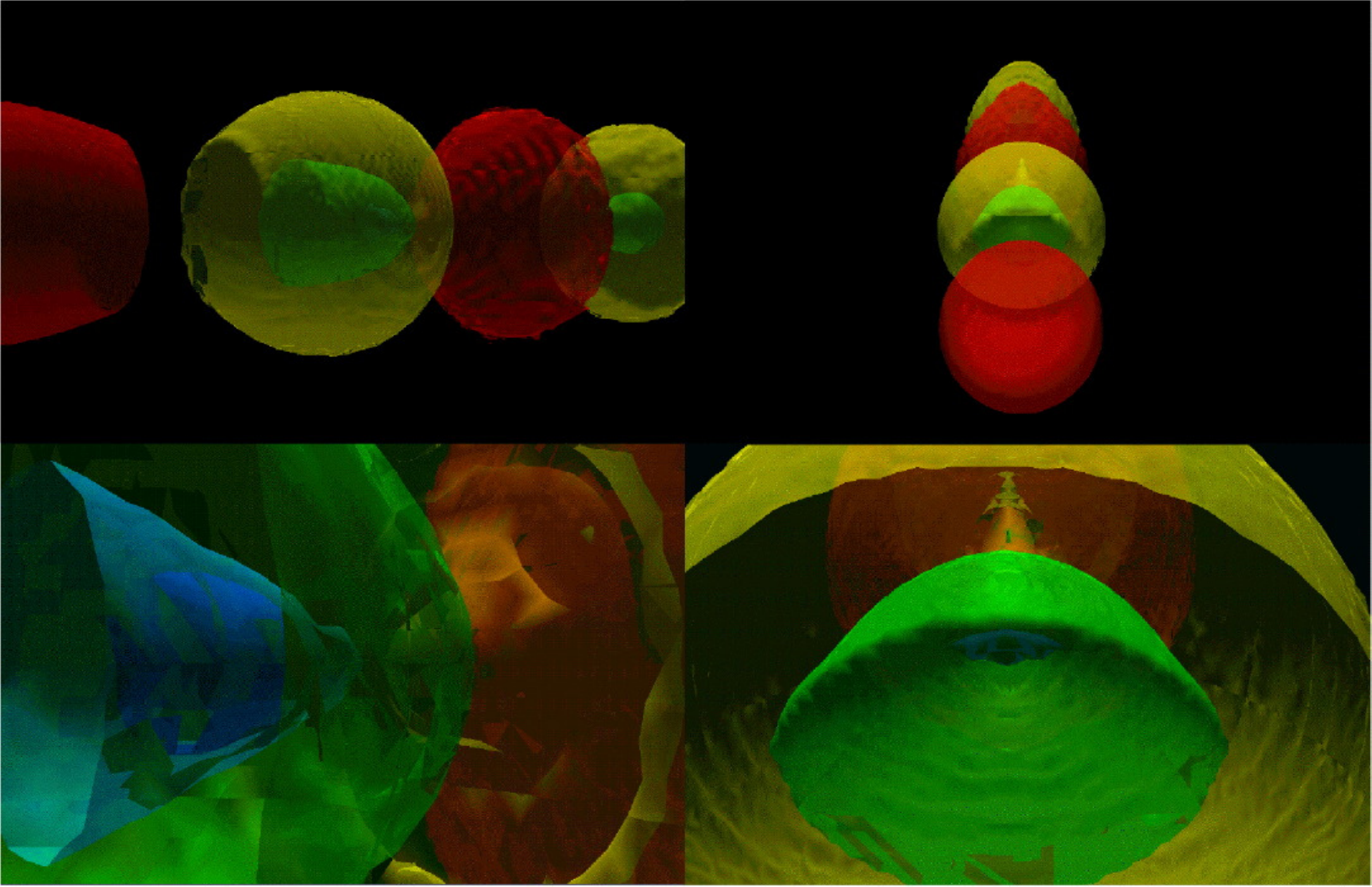, width=5.8in }
      \caption{Selected isosurfaces of the accelerating field.
               The dark blue, light blue, green,
               and yellow surfaces corresponds to an acceleration gradients
               of 0.5 , 0.4, 0.2, and 0.1 GeV/m while the red surfaces
               correspond to a decelerating gradient of 0.1 GeV/m.
               The fields shown in the left column and right column are
               from the simulation with an aspect ratio of $1$:$1$ and
               $2$:$\frac{1}{2}$ respectively. This figure has been 
               made possible by the help of the Office of Academic
               Computing at UCLA.}
      \label{fig:pwfa:isosurfaces-asym}
   \end{center}
\end{figure}

In Fig.~\ref{fig:pwfa:isosurfaces-asym} selected isosurfaces of the accelerating
field are shown for two of the 3D simulations. The dark blue, light blue,
green, and yellow surfaces corresponds to an acceleration gradients of 0.5 ,
0.4, 0.2, and 0.1 GeV/m while the red surfaces correspond to a decelerating
gradient of 0.1 GeV/m. The fields shown in the left column and right column
are from the simulation with an aspect ratio of $1$:$1$ and 
$2$:$\frac{1}{2}$
respectively. The viewpoints and perspectives were chosen to best 
visualize information about the size and shape of the accelerating areas
of the wake. The interesting fact to note in this figure is that the peak accelerating
volume for the asymmetric beam is small compared to the peak accelerating
volume for the symmetric beam. Beam asymmetry therefore does not just decrease
amplitude of the peak accelerating field but also causes the accelerating volume
to decrease. This is at this time only a qualitative statement and it will require
further research and data analysis to be able to quantify this statement more
precisely.

\section{An Analytical Model}
  \label{sect:pwfa:theory}
  
The simulation results presented above serve
the purpose of giving an understanding of the wake excitation process
and of the dependence of blowout on the beam parameters. This can now
be used to guide the development
of an analytical model that gives a deeper understanding
of the physics. In this section we will develop a model that within certain
limits predicts the motion of individual plasma electrons, and hence the
plasma response to the beam.

The analytical approach combines Whittum's frozen field formalism 
\cite{Whittum91} and 
a cylindrical sheet model used by Mori et al.\cite{Mori90}.
We start with the wave equations for the vector and scalar potential
in the Lorentz gauge \cite{Jackson}.

\begin{equation}
  \label{equ:pwfa:wave-lorentz-a}
  \left( \frac{1}{c^{2}} \frac{\partial^{2}}{\partial t^{2}}
        - \vec{\nabla}^{2} \right) \vec{A} = \frac{4\pi}{c} \vec{j}
\end{equation}

\begin{equation}
  \label{equ:pwfa:wave-lorentz-phi}
  \left( \frac{1}{c^{2}} \frac{\partial^{2}}{\partial t^{2}}
        - \vec{\nabla}^{2} \right) \Phi = 4\pi \rho
\end{equation}

\noindent{The} mathematical transformation to the speed-of-light variables
$\xi = z - ct$ and $s=z$ means that

\begin{equation}
  \label{equ:pwfa:dz}
  \frac{\partial}{\partial z} = 
  \frac{\partial}{\partial \xi} + \frac{\partial}{\partial s}
\end{equation}

\noindent{and}

\begin{equation}
  \label{equ:pwfa:dt}
  \frac{\partial}{\partial t} = - c \frac{\partial}{\partial \xi} 
\end{equation}

\noindent{From} the simulations we know that the plasma response is essentially
stationary in the moving frame and we can make the frozen-field
approximation \cite{Whittum91}
$\partial / \partial s \rightarrow 0$. With this
Eq.~(\ref{equ:pwfa:wave-lorentz-a}) and
Eq.~(\ref{equ:pwfa:wave-lorentz-phi}) become:

\begin{equation}
  \label{equ:pwfa:wave-soflv-a}
        - \vec{\nabla}_{\perp}^{2} \vec{A} = \frac{4\pi}{c} \vec{j}
\end{equation}

\begin{equation}
  \label{equ:pwfa:wave-soflv-phi}
        - \vec{\nabla}_{\perp}^{2} \Phi = 4\pi \rho
\end{equation}

\noindent{As} noted noted by Whittum, this has essentially reduced this problem
to a 2D Poisson problem. We next assume cylindrical symmetry to
reduce this even further to a one dimensional problem that can be
solved analytically under certain conditions.

Since we know for the plasma electrons that $|v_{z}| \ll c$
where $v_{z}$ is the axial velocity of the plasma,
we will henceforth assume $v_{z} = 0$. This approximation can be 
verified a posteriori. When considering the Lorentz force
[Eq.~(\ref{equ:review-codes:lorentz-force})] on the plasma electrons
this means that the magnetic field can be neglected. The force
is on the plasma electrons is therefore \cite{Jackson}:

\begin{equation}
  \label{equ:pwfa:force-plasma}
   \vec{F} = \left(
     \begin{array}{l}
     \vec{F}_{z} \\
     \vec{F}_{\perp} 
     \end{array}
   \right) = q \left(
     \begin{array}{l}
     -\frac{\partial}{\partial z} \Phi
     -\frac{1}{c} \frac{\partial}{\partial t} A_{z} \\
     -\vec{\nabla}_{\perp} \Phi
     -\frac{1}{c} \frac{\partial}{\partial t} \vec{A}_{\perp}
     \end{array}
   \right) = q \left(
     \begin{array}{l}
     -\frac{\partial}{\partial \xi} \left( \Phi- A_{z} \right) \\
     -\vec{\nabla}_{\perp} \Phi
     +\frac{\partial}{\partial \xi} \vec{A}_{\perp}
     \end{array}
   \right)
\end{equation}

\noindent{Since} we are neglecting the axial motion of the plasma electrons we will only
consider the perpendicular component of 
Eq.~(\ref{equ:pwfa:force-plasma}).
The source term of $\Phi$, i.e., $\rho$, has three contributions: the
charge density due to the beam $\rho_{b} = -e n_{b}$, the charge
density due to the ion background $\rho_{i} = e n_{p}$, and the
charge density due to the plasma electrons $\rho_{e} = -e n_{e}$.
If we neglect the ion motion and the radial motion of the beam
electrons due to the betatron oscillations then
the source term $\frac{1}{c}\vec{j}_{\perp}$ of $\vec{A}_{\perp}$ is given
by $-e v_{r} n_{e} / c$. Since we know from the simulation and 
analytical estimates that
$v_{r}/c$ is small compared to 1 we can to lowest order neglect this term and 
therefore $\vec{A}_{\perp}$ as well.

In case of a cylindrically-symmetric problem the perpendicular component
of Eq.~(\ref{equ:pwfa:wave-soflv-phi}) reduces to 

\begin{equation}
  \label{equ:pwfa:wave-soflv-cs-phi}
        - \frac{1}{r} \frac{\partial}{\partial r}
                   r  \frac{\partial}{\partial r} \bar{\Phi} 
        =  4 \pi \rho \: \frac{e}{mc^{2}}
        = - k_{p}^{2} \frac{n}{n_{p}}
\end{equation}

\noindent{where} $\bar{\Phi}$ is the normalized potential 
$e\Phi/\left(mc^{2}\right)$, $k_{p}$ the plasma wave vector, and $n$ the
density of charged particles. Note that $n$ has to be negative in order
to make this notation work for positive ions.

The potential due to a given charge distribution can be can be obtained using
the Green's function for Eq.~(\ref{equ:pwfa:wave-soflv-cs-phi}) which is

\begin{equation}
  \label{equ:pwfa:greens-function}
  G\left(r,r'\right) = \frac{ln\left(r/r'\right)}{2\pi} \; H\left(r - r'\right)
\end{equation}

\noindent{Here} $H\left(r - r'\right)$ is the Heaviside step function which is 1 for 
$r \geq r'$ and $0$ otherwise.
For a Gaussian drive beam $\rho_{b}$ is:

\begin{equation}
  \label{equ:pwfa:rho-beam}
  \rho_{b} = -e N_{b}
  \times \frac{1}{\sqrt{2\pi}\sigma_{z}}
  \; e^{-\frac{\left( \xi - \xi_{0}\right)^{2}}{2 \sigma_{z}^{2}}} 
  \times \frac{1}{2\pi\sigma_{r}^{2}}
  \; e^{-\frac{r^{2}}{2 \sigma_{r}^{2}}}
\end{equation}

\noindent{where} $N_{b}$ is the number of electrons in the beam, $\sigma_{z}$
the width of the Gaussian distribution in propagation direction, $\sigma_{r}$
the radial spotsize of the beam, and $\xi_{0}$ the position of the
beam center using speed-of-light variables. 
With this we get

\begin{eqnarray}
  \nonumber
  \bar{\Phi}_{b}\left( r, \xi \right)
  & = & 2 \pi \int_{0}^{r} \; \frac{ln\left(r/r'\right)}{2\pi}
        \left( \;
        \left( 2\pi \right)^{-3/2} \; 
        \frac{N_{b}}{\sigma_{r}^{2}\sigma_{z}} \;
        \frac{k_{p}^{2}}{n_{p}}\;
         e^{-\frac{\left( \xi - \xi_{0}\right)^{2}}{2 \sigma_{z}^{2}}} \;
         e^{-\frac{r'^{2}}{2 \sigma_{r}^{2}}}
        \;\right) \, r' \, dr' 
  \\ \nonumber
  \\
  \label{equ:pwfa:phi-beam-1}
  & = & -2 \; \left( 2\pi \right)^{-3/2} \; \frac{N_{b}}{\sigma_{z}} \;
        \frac{k_{p}^{2}}{n_{p}} \;
         e^{-\frac{\left( \xi - \xi_{0}\right)^{2}}{2 \sigma_{z}^{2}}} 
        \\ \nonumber
  & \times & \left( 
             \; \int_{0}^{\beta} \;
             \alpha \; ln\left(\alpha\right) \; e^{-\alpha^{2}}
             \, d\alpha
             - \; ln\left(\beta\right)
             \; \int_{0}^{\beta} \;
             \alpha \; e^{-\alpha^{2}}
             \, d\alpha
             \right)
\end{eqnarray}

\noindent{with} $\beta = r/(\sqrt{2}\sigma_{r})$.
The two integrals can be solved, leaving:

\begin{eqnarray}
  \label{equ:pwfa:phi-beam-2}
  \bar{\Phi}_{b}\left( r, \xi \right)
  & = &  -2 \; \left( 2\pi \right)^{-3/2} \; \frac{N_{b}}{\sigma_{z}} \;
        \frac{k_{p}^{2}}{n_{p}} \;
         e^{-\frac{\left( \xi - \xi_{0}\right)^{2}}{2\sigma_{z}^{2}}} \;
         \\ \nonumber
  & \times & \left( \; \frac{1}{4} \;
         \left( -\gamma 
          -\Gamma\left(0,\left(\frac{r}{\sqrt{2}\sigma_{r}}\right)^{2}\right)
         \right) \; 
         -  \frac{1}{2} \,
         ln\left( \frac{r}{\sqrt{2}\sigma_{r}} \right) \;
         \right)
\end{eqnarray}

\noindent{where} $\gamma \simeq 0.577216$ is Euler's constant and 
$\Gamma\left(a,z\right)= \int_{z}^{\infty} t^{a-1} e^{-t} dt$ is the
incomplete Euler-gamma function.

For the constant background of fixed ions, we have $\rho_{i}= -e \; (-n_{p})$.
For such a  $\rho$ it is straight forward to integrate
Eq.~(\ref{equ:pwfa:greens-function}) giving

\begin{equation}
  \label{equ:pwfa:phi-ion}
  \phi_{i}\left(r\right) = - \frac{k_{p}^{2}}{4} \; r^{2}
\end{equation}

Calculating the potential due to the plasma electron density $n_{e}$ is more
difficult since $n_{e}$ is not explicitly known. However, the simulations 
show a largely laminar flow of electrons passing the beam (in the beam's
moving frame). Based on this observation we can make an approximation that
will allow us to calculate 
a potential that a given electron experiences. We assume that there
is no ring crossing of cylindrical charge rings around the axis of 
the beam. That is, for any given ring of charge the number of electrons 
inside and outside its radial position at any time does not change.
We can therefore obtain the potential and fields experienced by a 
particular ring by simple considerations.
From Gauss' law the field at the position of a particular ring at 
position r is given by the field due to the amount of electron charge 
enclosed by the ring. As long as no rings cross each other then the enclosed electron 
charge is simply that which was enclosed when the electron ring was at
its original position $r_{i}$.
The potential due to the electrons is then given by

\begin{equation}
  \label{equ:pwfa:phi-cylinder}
  \Phi_{\lambda}\left(r\right) = -2 \; \lambda \; ln\left(r/r_{0}\right)
\end{equation}

\noindent{where} $\lambda$ is the charge per unit length for $r<r_{i}$ and
$r_{0}$ provides an arbitrary integration constant.
\noindent{An} initial uniform charge distribution $\lambda=\pi r_{i^{2}} n_{p}$ 
gives

\begin{equation}
  \label{equ:pwfa:phi-electrons}
  \bar{\Phi}_{e}\left(r,\xi\right) =
  \bar{\Phi}_{e}\left(r\left(r_{i},\xi\right)\right) =
  \frac{k_{p}^{2}}{2} \; r_{i}^{2}
  \; ln\left(\frac{r\left(r_{i},\xi\right)}{\sqrt{2}\sigma_{r}}\right)
\end{equation}

\noindent{where} for simplicity 
$r_{0}$ in Eq.~(\ref{equ:pwfa:phi-cylinder}) has been set to
$\sqrt{2}\sigma_{r}$.

The total $\bar{\Phi}$ is therefore given by

\begin{eqnarray}
  \nonumber
  \bar{\Phi}\left( r, \xi \right) \; 
   & = & \; \bar{\Phi}_{b} \; + \; \bar{\Phi}_{i} \; + \; \bar{\Phi}_{e}
  \\
  \label{equ:pwfa:phi-total}
   & = & -2 \; \left( 2\pi \right)^{-3/2} \; \frac{N_{b}}{\sigma_{z}} \;
        \frac{k_{p}^{2}}{n_{p}} \;
         e^{-\frac{\left( \xi - \xi_{0}\right)^{2}}{2\sigma_{z}^{2}}} \;
  \\
  \nonumber
   & & \times  \left( \; \frac{1}{4} \;
         \left( -\gamma 
          -\Gamma\left(0,\left(\frac{r}{\sqrt{2}\sigma_{r}}\right)^{2}\right)
         \right) \; 
         -  \frac{1}{2} \,
         ln\left( \frac{r}{\sqrt{2}\sigma_{r}} \right) \;
         \right)
  \\
  \nonumber
   & & - \; \frac{k_{p}^{2}}{4} \; r^{2} \; + \; \frac{k_{p}^{2}}{2} \; r_{i}^{2}
       \; ln\left(\frac{r}{\sqrt{2}\sigma_{r}}\right)
\end{eqnarray}

\noindent{The} equation of motion for an electron ring is therefore given by

\begin{equation}
  \label{equ:pwfa:force-plasma-radial}
  \frac{1}{c} \frac{d}{dt} \bar{p}_{r} \; = \;
  \frac{\partial}{\partial\xi} \; \gamma\;
  \frac{\partial}{\partial\xi} \; r = \;
  \frac{\partial}{\partial r} \; \bar{\Phi}
\end{equation}

\noindent{where} $\bar{p}_{r}  \, \equiv \, p_{r}/(mc)$.
Using the fact that the motion of the plasma is assumed to be purely 
radial this can be rewritten as

\begin{equation}
  \label{equ:pwfa:plasma-equation}
  \gamma^{3} \;
  \frac{\partial^{2}}{\partial\xi^{2}} \; r = \;
  \left( 1\;-\; \left( \; \frac{\partial}{\partial\xi}
  \; r \right)^{2} \right)^{-\,\frac{3}{2}} 
  \frac{\partial^{2}}{\partial\xi^{2}} \; r = \;
  \frac{\partial}{\partial r} \; \bar{\Phi}
\end{equation}

\noindent{This} is a differential equation for the radial motion of
a particle with an initial position $r_{i}$. The solution will give
us $r$ as function of $\xi$ for a given initial $r_{i}$. 
While much analytical work can still be done with
Eq.~(\ref{equ:pwfa:plasma-equation}), we next solve it numerically.

\begin{figure}
   \begin{center}
      \epsfig{file=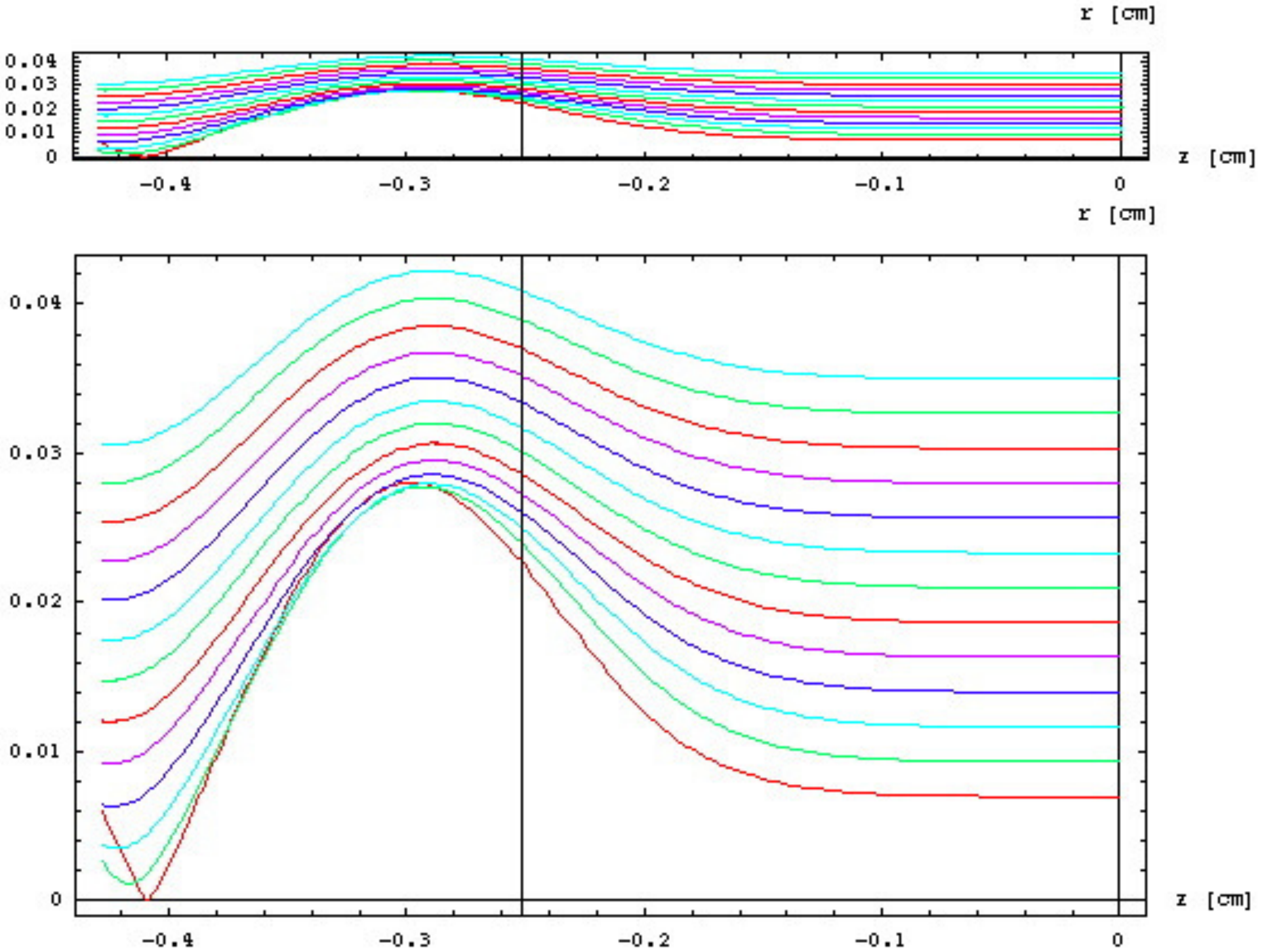, width=5.8in }
      \caption{Solutions for the electron trajectories for 13 different
               initial radii starting at $r_{i} = \sigma_{r}$ and then
               increasing in equal steps of $\sigma_{r}/3$. The beam
               has $N_{b} = 3.7 \times 10^{10}$. The upper part of the
               picture shows the results using the same scaling for both
               axes. The lower part of the figure shows the same results
               but with a blown up radial axis in order to show more details.
               The vertical line in the center of the figure indicates 
               the center of the electron beam.}
      \label{fig:pwfa:traject-3.7-1}
   \end{center}
\end{figure}

Fig.~\ref{fig:pwfa:traject-3.7-1} shows the solutions of
Eq.~(\ref{equ:pwfa:plasma-equation}) for 13 different initial radii 
starting at $r_{i} = \sigma_{r}$ and then increasing in equal steps
of $\sigma_{r}/3$. The beam and plasma parameters used are the same
as the ones used for the PIC simulation presented above with
$N_{b} = 3.7 \times 10^{10}$. The upper part of the picture shows the
results using the same scaling for both axes. The lower part of the figure
shows the same results but with a blown up radial axis in order to
show more detail. The are two interesting things to note in this 
figure. First, particles with an initial radius of about $\sigma_{r}$
have trajectories that cross other trajectories and therefore the 
question is raised as to how valid the model is for particles with radii
of $\sigma_{r}$ or smaller. However, the maximum blowout radius seen in
this Fig.~\ref{fig:pwfa:traject-3.7-1} is about $280\mu m$, which is
within a few percent of the blowout radius seen in the full PIC simulation.
This means that at least up to the point of maximum blowout the
model is quite accurate.

\begin{figure}
   \begin{center}
      \epsfig{file=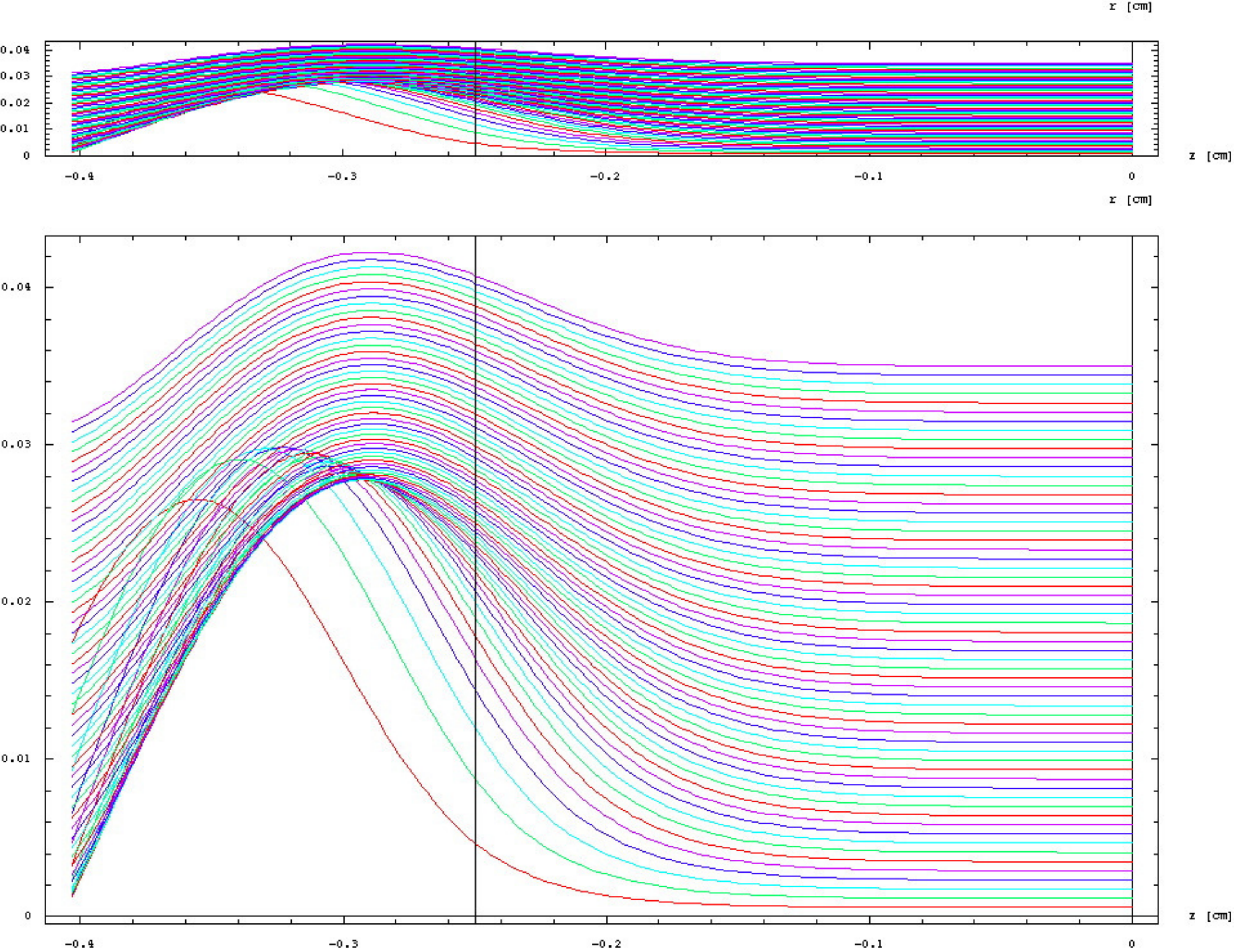, width=5.8in }
      \caption{Solutions for the electron trajectories for 60 different
               initial radii starting at $r_{i} = \sigma_{r}/12$ and then
               increasing in equal steps of $\sigma_{r}/12$. The beam
               has $N_{b} = 3.7 \times 10^{10}$. The upper part of the
               picture shows the results using the same scaling for both
               axes. The lower part of the figure shows the same results
               but with a blown up radial axis in order to show more details.
               The vertical line in the center of the figure indicates 
               the center of the electron beam.}
      \label{fig:pwfa:traject-3.7-2}
   \end{center}
\end{figure}

\begin{figure}
   \begin{center}
      \epsfig{file=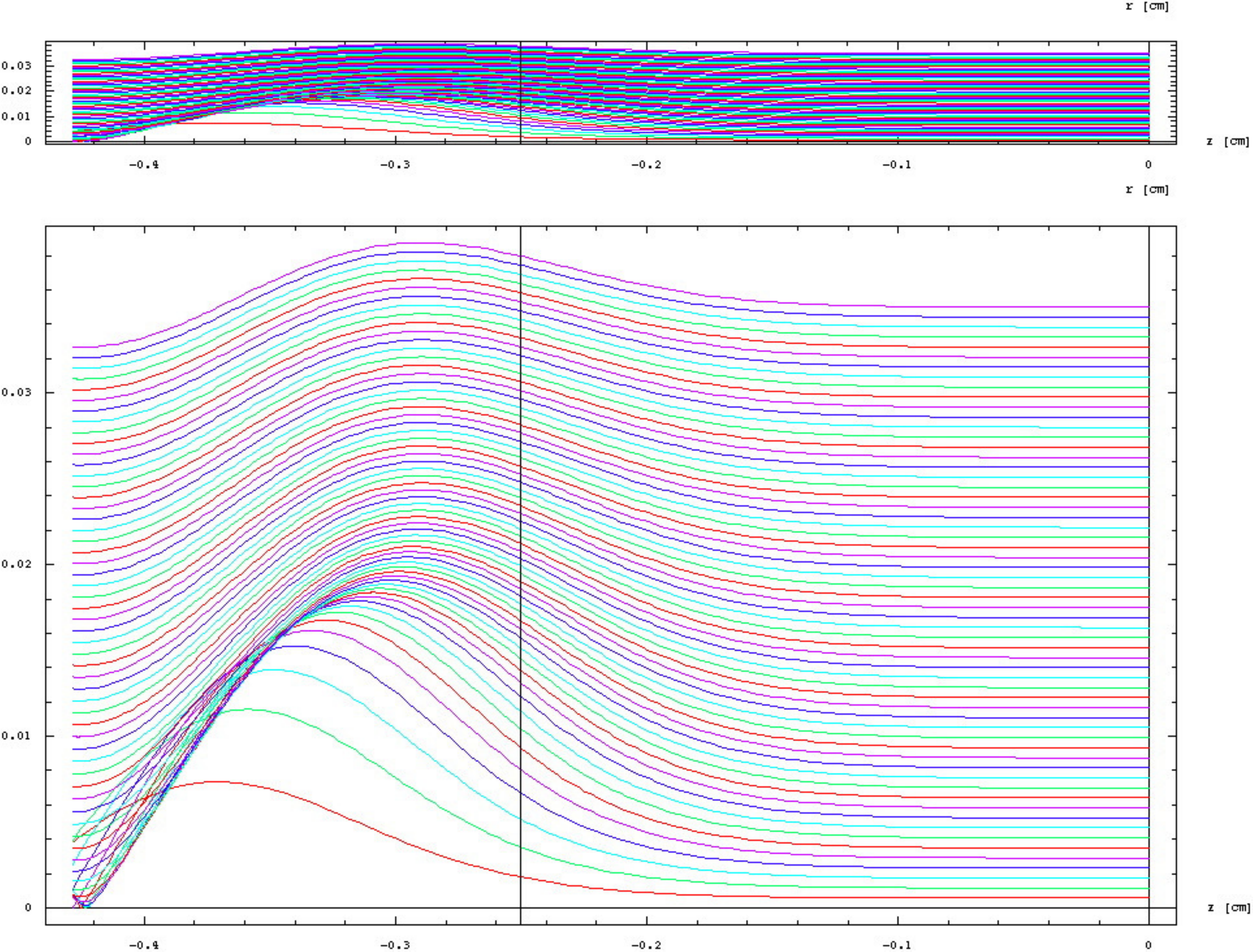, width=5.8in }
      \caption{Solutions for the electron trajectories for 60 different
               initial radii starting at $r_{i} = \sigma_{r}/12$ and then
               increasing in equal steps of $\sigma_{r}/12$. The beam
               has $N_{b} = 1.85 \times 10^{10}$. The upper part of the
               picture shows the results using the same scaling for both
               axes. The lower part of the figure shows the same results
               but with a blown up radial axis in order to show more details.
               The vertical line in the center of the figure indicates 
               the center of the electron beam.}
      \label{fig:pwfa:traject-1.85-1}
   \end{center}
\end{figure}

Fig.~\ref{fig:pwfa:traject-3.7-2} shows the solution for 60 different
initial radii starting at $\sigma_{r}/12$ and then increasing in 
equal steps up to $5\, \sigma_{r}$. This ``density'' of 
trajectories corresponds to the same transverse density of particles
as used in the PIC simulation. The range of $\xi$ over which the
trajectories are shown is shorter than in Fig.~\ref{fig:pwfa:traject-3.7-1}
because some of the particles with $r_{i}<\sigma_{r}$ will cross the
axis, $r=0$, for $\xi<0.4cm$ so that the model breaks down.
Fig.~\ref{fig:pwfa:traject-1.85-1} shows trajectories for the same 
initial radii as in Fig.~\ref{fig:pwfa:traject-3.7-2} but for a beam
with only half the number of electrons. The range of $\xi$ over which the
trajectories are shown in Fig.~\ref{fig:pwfa:traject-1.85-1}
is the same as in Fig.~\ref{fig:pwfa:traject-3.7-1} because no 
problems with particles crossing the axis appeared. In
Fig.~\ref{fig:pwfa:traject-3.7-2} as well as in Fig.~\ref{fig:pwfa:traject-1.85-1}
the upper part shows the trajectories using the same scaling
for both axes while the lower part again blows up the radial axis
to make certain details more visible.

Fig.~\ref{fig:pwfa:traject-3.7-2} and Fig.~\ref{fig:pwfa:traject-1.85-1}
should be compared with Fig.~\ref{fig:pwfa:blowout-comp}, which shows the
plasma response for the PIC simulations with the same parameters. The
similarity of the results is good enough that even certain particle
trajectories can be identified which each other.
This indicates that the model developed here is quite good for 
estimating the blowout radius despite the fact that at least some 
sheet crossing does take place and the longitudinal motion was 
neglected. For the two different beams shown in
Fig.~\ref{fig:pwfa:traject-3.7-1} to
Fig.~\ref{fig:pwfa:traject-1.85-1} the PIC simulations and the
analytical model agree within about $5\%$ for the values of the
blowout radii.

\begin{figure}
   \begin{center}
      \epsfig{file=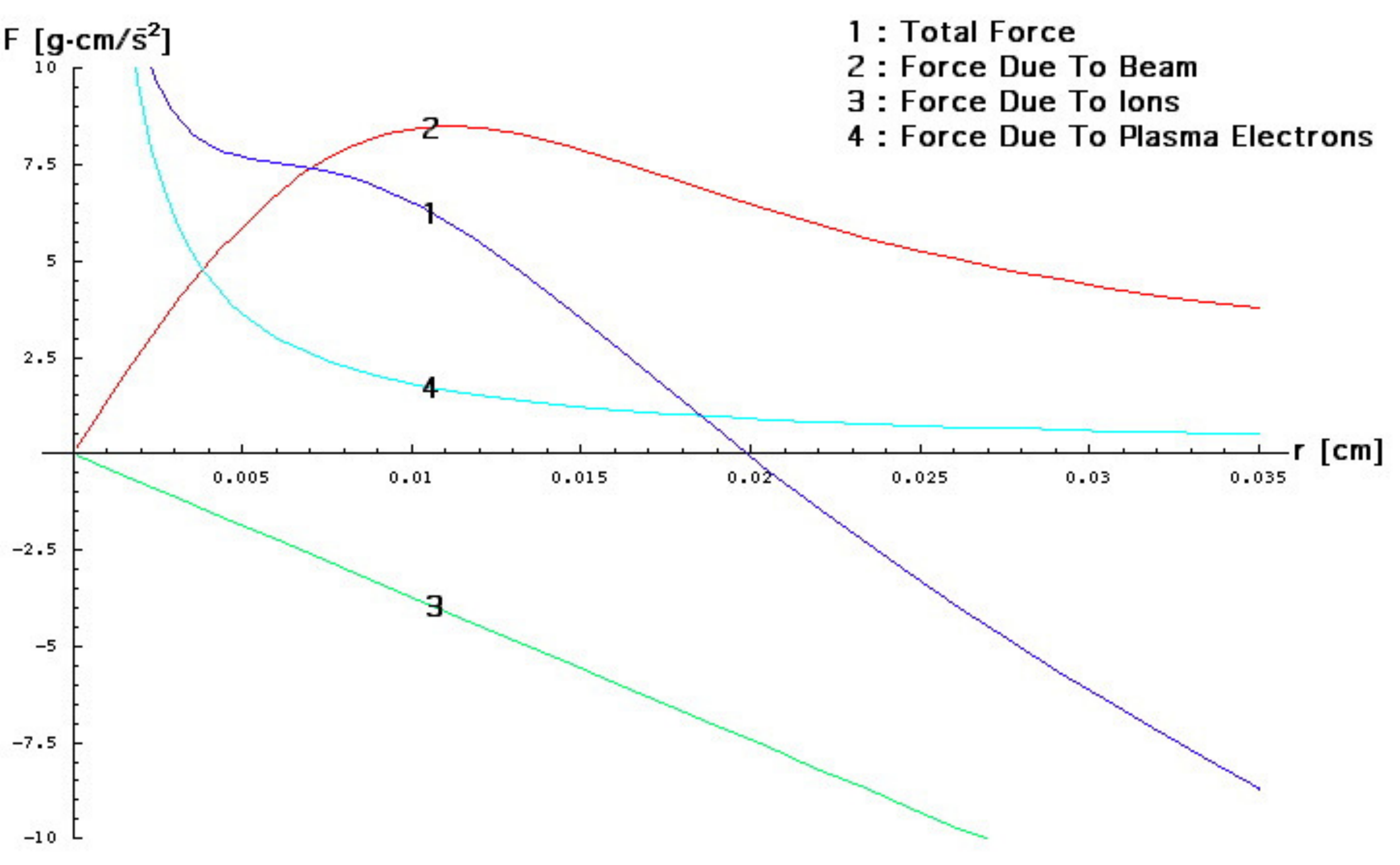, width=5.8in }
      \caption{The forces acting on an electron starting at an
               initial radius $r_{i}=\sigma_{r}$.}
      \label{fig:pwfa:forces}
   \end{center}
\end{figure}  

To make the agreement more understandable we plot in
Fig.~\ref{fig:pwfa:forces} the forces acting
on an electron with an initial radius $r_{i}=\sigma_{r}$. It
shows that the contribution of the plasma electrons to the total force
is relatively small for a radius of the order of the blowout radius.


\section{Conclusion}
  \label{sect:pwfa:conclusion}

The main result of the analysis of the beam and wakefield dynamics is 
that the wakefield is rather insensitive to the betatron oscillation 
dynamics of the beam and therefore essentially constant over time. In 
addition the acceleration and deceleration of the beam electrons is 
not affected by the betatron oscillation either. On the other hand
the magnitude of the peak accelerating field is decreased significantly
by a decrease in the number of beam electrons and by asymmetries in the
beam spotsize. These are therefore parameters that have to be controlled 
carefully in PWFA experiments. The analytical model we developed 
predicts the blowout radii and trajectories seen in the PIC simulations
well. It can therefore be used as a starting point for further
analytical work.

The results in this chapter indicate that the blowout regime
provides stable and robust plasma wakefield acceleration. 
Energy gains on the order of a GeV should be achievable in this blowout 
regime if the physical parameters of the simulations can be realized in 
an experiment. Far higher gradients and energy gains may be possible 
with shorter bunches and longer denser plasmas\cite{Assmann97}. Such
beams would undergo 
hundreds rather than a few betatron oscillations. Although it maybe 
desirable to match the beam emittance to the plasma focusing strength to 
avoid betatron oscillations as discussed in Ref.\cite{Assmann97},
the analysis here 
suggests that the presence of the oscillation is not necessarily 
detrimental. A major issue for the scaling of plasma wakefield
acceleration to the 10�s and 100�s of GeV, is the possibility of a hosing
instability of the beam, which might reduce the achievable energy gain 
and lead to emittance growth of beam. Hosing is inherently a 
three-dimensional instability and is therefore absent in the 2D 
simulations. We are currently carrying out 3D simulations over long
propagation distances to study the importance of hosing and other 3D
effects and  this is an area of future work.


\chapter{Summary}
  \label{chap:summary}

\section{Important Results}
  \label{sect:summary:important}
  
The complex nonlinear interactions between the driver, the plasma wake,
and the accelerated electron bunch in plasma-based accelerators make 
PIC modeling the only method of getting a complete, integrated picture
of the evolution of such a system. In this dissertation, we have 
described the development of a new object-oriented code, OSIRIS, 
which is ideally suited for modeling laser plasma
and beam plasma interactions.

The object-oriented design of OSIRIS made it possible to create a currently 
unique combination of advanced algorithms. OSIRIS now contains
algorithms for 1D, 2D, and 3D simulations in Cartesian coordinates and for 2D
simulations in cylindrically symmetric coordinates. For all of these algorithms
the code is fully relativistic and presently uses a charge-conserving current
deposition algorithm. It allows for a moving simulation window and arbitrary
domain decomposition for any number of dimensions. This combination of 
algorithms makes OSIRIS a useful tool for many different research problems
besides plasma-based accelerators, with
the possibility to be extended much further by adding new modules.

We applied the PIC modeling technique to investigate a number of 
problems in plasma-based accelerator research. The main results
of our research are:

\begin{itemize}

\item  Using 2D and 3D PIC
simulations we studied the injection scheme proposed in
Ref.~\cite{Umstadter96a}. We find that the beam brightness and quality
compares reasonably with that of electron bunches produced using conventional
technologies. We believe that the mechanism for the trapping of particles in our 
simulation is the interaction of the particles with the two plasma wakes.
The 3D simulation is the only one to date on this problem.

\item High resolution simulations of long laser pulses indicated the existence
of a hosing instability with a wavelength longer than the plasma wavelength.
The simulations motivated the development of a theory by Duda and Mori 
that explains the long wavelength hosing seen in the simulations. It is
found that this effect might significantly increase the emittance of electron
beams produced in the SMLFWA and it may be important when 
high-intensity lasers propagate in densities above the quarter 
critical density as in the fast ignitor concept. 
      
\item 2D simulations of a LWFA in a parabolic plasma channel indicate that
the idea of preventing diffraction of a laser pulse by propagating it through
a matched parabolic plasma channel works. Only about $3\%$ of the length of
one plasma wave oscillation accelerates externally injected particles. The
test particles in the simulation gain more energy than the linear
scaling laws for a uniform plasma would predict.

\item 2D and 3D simulations of a PWFA in the blowout regime showed
that the wakefield is rather insensitive to the betatron oscillation 
dynamics of the beam or the acceleration and deceleration of the beam
electrons and is therefore essentially constant over time. On the other hand
the magnitude of the peak accelerating field is decreased significantly
by a decrease in the number of beam electrons and by asymmetries in the
beam spotsize. An analytical model was developed that predicts the
blowout radii and trajectories seen in the PIC simulations well.

\end{itemize}

\section{Future Challenges}
  \label{sect:summary:challenges}

Research on plasma-based acceleration has made significant 
progress in the areas of experiment, theory, and simulations, but much
work remains to done.   
Of the research topics in this dissertation the LWFA in a channel and 
the PWFA in the blowout regime are the ones that will strongly benefit
from more simulation research. 3D simulations of the LWFA will require
considerable computational resources but would allow a much more 
detailed understanding of the whole system. One of the current 
questions for PWFA in the blowout regime is the significance of
the tail hosing instability.  Hosing is inherently a 
three-dimensional instability and is therefore absent in 2D 
simulations. Therefore, 3D simulations over long propagation
distances are required and are currently being done to study the importance
of hosing and other 3D effects.

An important challenge for the future are simulations of accelerators
over the full distance of acceleration. This problem is a motivation
for a number of possible extensions to OSIRIS. The challenge lies in
the fact that there are in principle three length scales that have be
considered, the scale of the driver evolution (laser or particle beam),
the scale of plasma wave length which is needed to resolve the
evolution of the accelerating wake, and finally the total length of
the particle acceleration. The problem of simulating plasma based
acceleration over the distances comparable to real experiments requires very 
efficient use of the available computational resources. Mechanisms 
like dynamic load balancing, adaptive mesh refinement \cite{Berger89, Bryan99},
and more sophisticated boundary conditions which reduce the required 
number of simulation cells are ways to strongly improve the code performance 
while maintaining the accuracy of simulation results. For laser 
drivers, explicit PIC algorithms lead to the problem 
of having to resolve the laser wavelength, $\lambda_{L}$, which is typically 
several 
orders of magnitude smaller that the plasma wavelength, $\lambda_{p}$, and therefore 
the required computational effort is the square of 
$\lambda_{p}/\lambda_{L}$ times larger than if just the  plasma 
wavelength would be resolved.
This increase could be avoided if the laser evolution 
is not described by a fully explicit PIC algorithm but by other means as for 
example following a laser envelope equation \cite{Mora97}. 
For particle beam drivers (and even laser drivers) the timescales for 
the evolution of the driver are orders of magnitude smaller than the 
plasma frequency. As a result, algorithms which evolve the drive beam 
separately would be extremely useful. Such codes already exist. Fully 
explicit codes will however always be needed to benchmark any reduced 
description code.
Furthermore, OSIRIS with its object-oriented design should make the implementation
and use of these feature easier and less time consuming while providing a framework for 
automatic parallelization and providing sophisticated diagnostic
possibilities.

Another problem for the future is the management of a large multi-purpose,
multi-user, and multi-author code like OSIRIS. Our current way of management
is that there is one code manager who needs to be quite familiar with the
details of the code and who manages the ``master copy'' of the code.
This means this code manager has the most updated version of the code. He
updates this version with the changes that other authors have made to their
versions of the code and then redistributes the updated version to all
users. This model has been workable so far but might become more difficult
over time as the code grows. The use of professional code development
management software might become necessary in the future to solve this
difficulty.

A final point to make is that OSIRIS was designed for distributed 
computing and more specifically for the currently pervasive
type of supercomputers, multiple-instruction-multiple-data parallel
computers with local memory. This design can be expected to remain
dominant for at least another 5 to 10 years. After that time it is an open
question. Depending on how strongly the design for supercomputers
will change over time and on how flexible OSIRIS turns out to be
the code might either become outdated or it might be able to be adapted
to new architectures and remain a helpful tool for a long time to come.

\appendix


\chapter{OSIRIS - A Brief User's Guide}
  \label{chap:os-input}

\section{General Information}
  \label{sect:os-input:general} 
  
This appendix will describe how to use OSIRIS and in particular the
input file. In addition to the actual input file "os-stdin", which
describes the simulation, the code requires five files
to be in the same directory as the executable of the code.
They contain path-strings which provide the code with direct
information on where to find certain files. The path-strings in these
files have to conform to the conventions of the computer system the
code is running on. The files are:

\begin{itemize}

\item{path.bin:} The path-file that contains the path to the directory 
with executable. It is also used to find the "os-stdin" as well as the other
path.* files.

\item{path.mass:} The path-file that contains the path to the mass-storage directory.
This mass-storage directory is used for all diagnostic data dumps.

\item{path.rest:} The path-file that contains the path to the restart-file directory.
The restart file directory is the directory the restart files are written
to or read from.

\item{path.home:} The path in this file is currently not used by the code.
This means the file can be empty but it has to exist in order to prevent
an accidental crash.

\item{path.work:} The path in this file is also currently not used but 
the file still needs to exists.

\end{itemize}

The code also assumes a certain subdirectory structure in the directory given
by the path.mass file. The structure that is required is the following one
(this list is using Unix notation):

\begin{itemize}

\item For the writing of full field data into mass storage: FLD, FLD/B1, FLD/B2, FLD/B3,
 FLD/E1, FLD/E2, FLD/E3, FLD/J1, FLD/J2, FLD/J3, FLD/RH

\item For the writing of averaged data into mas storage: AVE, AVE/b1, AVE/b2, AVE/b3,
AVE/e1, AVE/e2, AV/eE3, AVE/j1, AVE/j2, AVE/j3, AVE/rh

\item For the writing of particle data: PAR

\item For the writing of phase space data for each particle species:
PHA, PHA/x2x1, PHA/x3x1, PHA/p1x1, PHA/p2x1, PHA/p3x1,
PHA/x3x2, PHA/p1x2, \\
PHA/p2x2, PHA/p3x2, 
PHA/p1x3, PHA/p2x3, PHA/p3x3, 
PHA/p2p1, \\
PHA/p3p1, 
PHA/p3p2. 
Each of the directories for a specific phase space again needs to have
subdirectories for each species. For example:
PHA/x2x1/01, PHA/x2x1/02, PHA/x2x1/03, \ldots 
PHA/x2x1/[last-species]
 
\end{itemize}

Another point that should be mentioned is that even though most 
parts of OSIRIS have been written for runtime polymorphism some parts
have not. Those parts have been written to be compiletime polymorph. This 
specifically means that the code has to be recompiled when switching
from 3D to 2D simulations or when switching between the different
deposition algorithms for 2D simulations. In the first case the 
parameter "p\_x\_dim" in the file "os-param.f" has to set to
2 or 3, depending on which dimensionality is wanted, before 
recompiling. In the other case an interface in the 
file "os-spec.f" has to be specified in the following way

\small{ \begin{verbatim}

        interface getjr
          module procedure getjr_2d_quadratic   ! ISIS algorithm
!          module procedure getjr_2d             ! TRISTAN algorithm
          module procedure getjr_3d
        end interface

\end{verbatim} } 

\noindent{in} order to use the ISIS method.
In order to use the TRISTAN method it has to be
changed slightly to

\small{ \begin{verbatim}

        interface getjr
!          module procedure getjr_2d_quadratic   ! ISIS algorithm
          module procedure getjr_2d             ! TRISTAN algorithm
          module procedure getjr_3d
        end interface

\end{verbatim} } 

\noindent{This} compiletime polymorphism can be easily changed to 
runtime polymorphism and it will be in a future version
of the code.

The actual input file that describes the physics of the simulation, the
diagnostic data dumps, and the parallel node-configuration of the simulation
is divided into sections corresponding to each object for which data needs to
be read in from the input file. Each of those sections starts and ends
with the character ``/''. (Note that this is different from the rules
for a standard Fortran namelist file.) Between the different sections
any kind of comment can be written as long as is doesn't contain the
character "/".

One of advantages of this type of structured input file is that
comments can be scattered throughout the file at the  places where they
are relevant. The next section is therefore an actual input-file for a 3D
run related to the research results presented in chapter \ref{chap:injection}
with comments on the meaning of the different input variables. The comments
are written in such a way that they could also be in an actual input file.

All quantities in the input file are
given in dimensionless units which are obtained by normalization with
regard to a normalizing frequency, mass, and charge (usually the plasma
frequency, the electron mass, and the electron charge), and by 
normalization with respect to the speed of light. A consistent
normalization for all quantities can be derived in this way.

\section{An Input File Example}
  \label{sect:os-input:example}

\small{ \begin{verbatim}
  
-------------------- OSIRIS INPUT DEC --------------------
 This input file is structured into blocks according to
 the data structures in the program. Each class has its
 own routine to read in data from this file.
 SLASHES are reserved for structuring this file and
 can not be used for any other purpose.


-------------------------COMMENTS-------------------------

 RUN: run655.3d     date: 11-01-99

 MOTIVATION AND EXPLANATION
 This is a 3D version of run651.2d.

 It has been suggested that an injection pulse propagating
 transversely to the pump pulse of a LWFA and passing by
 behind it should cause a large number of electrons to be
 trapped in the plasma wave following the pump pulse.
 This run is set up to investigate this possibility.

 CHANGES COMPARED TO PREVIOUS RUNS
 In contrast to run651.2d this run is using a larger
 grid cell size and timestep size in order to cut down on the
 computational time required in 3D. The grid size is now such
 that the laser wavelength corresponds to 14 gridpoints.
 dx(i) = 0.08975 for all i.

 PHYSICS
 A laser pump pulse is propagating through a plasma of 4% of critical
 density. It has a length which is 2*Pi*c:wp in order to create a
 plasma wake wave.
 A second pulse , the injection pulse, is launched in the direction
 transverse to the first pulse and in such a way as to pass by
 directly behind the first pulse. 
 The polarization of both pulses is in the x1-x2 plane of the simulation.
 The pump pulse propagates 4-5 Rayleigh lengths during the simulation.

 RESULTS

----------------------- INPUT DATA -----------------------


--------the node configuration for this simulation--------
/ &nl_node_conf
node_number(1:3) = 16, 2, 2,
if_periodic(1:3) = .false., .true., .true.,
/
"node_number" is the number of nodes in each direction of the simulation.
The example above has 16 nodes in x1, 2 nodes in x2, and 2 nodes in x3.
The total number of nodes in this simulation is therefore 16x2x2=64.
In this case a full 3D decomposition is used.
The decomposition could be turned into a 2D decomposition by only
requesting 1 node in a certain direction
(e.g., node_number(1:3) = 16, 2, 1,) or into a 1D decomposition by
requesting only 1 node in two directions
(e.g., node_number(1:3) = 16, 1, 1,). If all three numbers
are 1 the simulation is done on a single node. 
"if_periodic" is a switch for each of the three directions to turn on
periodic boundary conditions. If ".true." is specified here for a certain
direction it will override any other boundary condition specified later
on in this input file. OSIRIS treats periodic boundary conditions as a
aspect of the node configuration because these boundary conditions
specify in general that the boundary of one node will communicate with
the boundary of another node. Only under specific circumstances the node
will have to  "communicate" with its own opposite boundary.
Note that for 2D simulations "node_number" and "if_periodic" have only 2
components each. For example "node_number would be given by
"node_number(1:2) = 16, 2,".

----------spatial grid----------
/ &nl_grid
  nx_p(1:3) = 400, 280, 280,
  coordinates = "cartesian",
/
"nx_p" gives the number of grid cell for the global grid in each 
direction. The local grid for each node is calculated by 
nx_p(i) divided node_number(i) for direction i. If for a
given direction i nx_p(i) is not a multiple of node_number(i)
but nx_p(i) = u x node_number(i) + m then the first m nodes
will have u+1 grid cells and the remaining node_number(i) - m
nodes will have u grid cells.
"coordinates" is a character string that determines the coordinate
system used in a simulation. The currently valid values are
"cartesian" and "cylindrical".

----------time step and global data dump timestep number----------
/ &nl_time_step
  dt     =   0.0513d0,
  ndump  =    136,
/
"dt" is the length of a timestep in the simulation.
"ndump" is the number timesteps after which the code checks
for all objects that require writing of data into a file whether
data should be written. In this way it provides a basic measure of
time for diagnostic and restart dumps. The filenames of all dumped
files have a numerical string appended that is based on after how
many multiples of "ndump" timesteps the file was written. 

----------restart information----------
/ &nl_restart
  ndump_fac = 1,  file_name  = ' ',
  if_restart=.false.,
/
"ndump_fac" gives the code the information after how
many multiples of "ndump" to write restart files. In this
example restart files are written every 136 timesteps. The
names of the successive files would be "rst-1001", "rst-1002",
"rst-1003", and so on. If "ndump_fac = 2" then a restart file
would be written every 272 timesteps and the sequence of
successive files would be "rst-1002", "rst-1004", "rst-1006",
and so on. "ndump_fac = 0" turns restart file writing off.
"if_restart" is a switch that turns on the reading of a restart
file in beginning of the simulation. if  "if_restart=.true.,"
then the code expects to find restart files in the directory
given by the file "path.rest" that was discussed in the
beginning of this chapter.
"file_name" provides the possibility to  attach a prefix to
filenames of the restart files.

----------spatial limits of the simulations----------
(note that this includes information about
 the motion of the simulation box)
/ &nl_space
  xmin(1:3) =   0.0000d0 ,  0.000d0 ,  0.000d0 ,
  xmax(1:3) =  35.9000d0 , 25.130d0 , 25.130d0 ,
  if_move= .true., .false., .false.,
/
"xmin" and "xmax" give the upper and lower boundary
of the global simulation space at the beginning of the
simulation.
"if_move" is a switch for the motion of the space in any of
the three directions. If one of the components is ".true." the
simulation window will move in that direction with the speed
of light. Setting "if_move(i)" to ".true." will override any
other boundary condition specified for these boundaries in 
direction i with the exception of the periodic boundary
conditions that are specified in the node-configuration.
For 2D simulation all variables should have one component less.

----------time limits ----------
/ &nl_time
  tmin = 0.0d0, tmax  = 104.652d0,
/
"tmin" and "tmax" are the initial and final time of the
simulation.

----------field solver set up----------
/ &nl_el_mag_fld
  b0(1:3)= 0.0d0, 0.0d0, 0.0d0,
  e0(1:3)= 0.0d0, 0.0d0, 0.0d0,
/
"b0" and "e0" are constant external fields that are added to the
electromagnetic field.

----------boundary conditions for em-fields ----------
/ &nl_emf_bound
  type(1:2,1) =   5,  5,
  type(1:2,2) =   5,  5,
  type(1:2,3) =   5,  5,
/
"type" defines here the type of boundary for the electromagnetic field.
The following boundary conditions are currently implemented.
   1 : boundary moving into the simulation box with c
   2 : boundary moving outward from the simulation box with c
   5 : conducting boundary with particle absorption
  20 : axial b.c. for 2D cylindrically-symmetric coordinates
  30 : Lindman open-space boundary - limited implementation
Each of the lines above defines boundary conditions for the lower
and upper boundaries in one direction. "type(1,3)" for example means
the lower boundary in direction 3.
The boundary conditions specified here can be overwritten if
periodic or moving boundaries are specified above. For a 2D simulation
the last line "type(1:2,3) =   5,  5," would be removed.
For a 2D cylindrically-symmetric simulation "type(1,2)"
needs to be set to 20.

----------diagnostic for electromagnetic fields----------
/ &nl_diag_emf
  ndump_fac_all = 1,  file_name_all = ' ',
  ndump_fac_ave = 1,  file_name_ave = ' ',
  n_ave(1:3)      = 16, 10, 10,
  ifdmp_efl(1:3) = .true. , .true. , .true. ,
  ifenv_efl(1:3) = .false. , .false. , .false. ,
  ifdmp_bfl(1:3) = .true. , .true. , .true. ,
  ifenv_bfl(1:3) = .false. , .false. , .true. ,
/
This section defines the diagnostic for the electric and magnetic 
field. There are two different diagnostics. One that writes the full
field data of a given field component on each node. The data from
this diagnostic has to be merged after the simulation to get the
full data set in one file that can then be post-processed further.
The second diagnostic averages the data or takes their envelope for
a given number of grid cells and then merges the resulting reduced
field data at runtime into one field that is written into mass
storage. 
"ndump_fac_all" determines the times at which the full field data are
written and "ndump_fac_ave" determines the times at which the averaged
field data are written. If "ndump_fac_all" or "ndump_fac_ave" are 0
then that particular diagnostic is turned off.
"file_name_all" provides the possibility to  attach a prefix to
filenames of the full data files. "file_name_ave"  does the same for
the filenames of the averaged data.
"n_ave" gives the number of grid cells that the averaging diagnostic
averages over in each direction.
"ifdmp_efl" are switches that turn on both diagnostics for the
different components of the electric field if set to ".true.".
"ifdmp_bfl" does the same for the magnetic field components.
"ifenv_efl" decides for each component of the electric field whether
the averaging diagnostic really does take the average (.false.) or 
whether it takes the the maximum absolute value (.true.) of the field 
values in the grid cells determined by "n_ave". "ifenv_bfl" does
the same for the magnetic field components.

----------number of particle species----------
/ &nl_particles   num_species = 2,   /

"num_species" determines the number of species in the simulation.
This section has to be followed by the appropriate kind and number of
of sections for each species. In this case the data for exactly
two species have to be provided below. The code will crash if this is
not the case.

----------diagnostics for all particles----------
/ &nl_diag_particles
  ndump_fac = 1,  file_name = ' ',
  if_particles_all = .true.,
  gamma_limit = 6.0d0,
  particle_fraction = 1.0d0,
/
This section defines a common diagnostic for all particles species.
The particle data are written into a text file. For each particle
the number of its species, its position, its momentum, and its 
simulation charge are written.
"ndump_fac" and "file_name" have the same functions as in the sections
where they appeared before but this time they effect the writing
of particle data dumps.
"if_particles_all" is a switch that can turn this diagnostic on and off.
"gamma_limit" is a filter for this diagnostic. Only particles with a
gamma above "gamma_limit" are written into the dump file. 
"particle_fraction" determines which fraction of the particles that
are above "gamma_limit" are actually written. The particles that are
written are randomly selected from the particles that could be written.

----------information for species 1----------
/ &nl_species
  num_par_max = 1600000,
  rqm=-1.0,
  num_par_x(1:3) = 1, 2, 2,
  vth(1:3) = 0.0d0 , 0.0d0 , 0.0d0 ,
  vfl(1:3) = 0.0d0 , 0.0d0 , 0.0d0 ,
  den_min = 1.d-5,
  if_unneutralized = .false.,
  num_dgam = 0,
  dgam = 0.0,
/
This section defines the basic variables for the first
particle species.
"num_par_max" sets the maximum number of particle on each node
for this species.
"rqm" is the mass to charge ratio of the species in normalized units.
For electrons this is -1.
"num_par_x(1:3)" gives the number of particles in each direction in a 
grid cell. The total number of particles per cell is the product of the
numbers in the different directions. The current example has a total
of 1x2x2=4 particles per cell. The number of components of "num_par_x"
depends on the dimensionality of the run. For example for a 2D simulation
a correct setup of 4 particles per cell would be "num_par_x(1:2) = 2, 2,".
"vth(1:3)" defines the thermo-velocity of this particle species in each
direction.
"vfl(1:3)" defines a global fluid velocity for this particle species.
This variable is currently not fully implemented.
"den_min" defines an approximate minimum density up to which particles
are still initialized in a grid cell. This is necessary in order to
avoid the initialization of particle with zero charge. "den_min" should
be chosen well below the minimum density of interest  for this species.
"if_unneutralized" is currently not fully implemented.
The current algorithms in OSIRIS automatically assume an neutralizing
background for newly initialized particles. In future version of OSIRIS
"if_unneutralized = .true.," will turn on calculation of the initial
electric and magnetic fields due to a the species that is not neutralized.
"dgam" and "num_dgam" are used to specify the acceleration of a 
beam particle species due to an assumed external field in the x1
direction. "dgam" is here the increase in the gamma of all particles
of the species at each timestep of the simulation. "num_dgam" is
the number of timesteps for which the gamma of the beam particle
species is increased by "dgam". Note that during the timesteps for which
this species is being accelerated the transverse momentum of it is not
updated.


----------density profile for this species----------
  number of points in profile along each direction
/ &nl_num_x  num_x = 6,   /

"num_x" is the number of points at which the functions
that are read in from the next section are defined.
 
actual profile data 
/ &nl_profile
  fx(1:6,1) = 1., 1., 0., 0., 1., 1.,
   x(1:6,1) = 0., 0.1001, 0.1002, 35.9001, 35.9002, 10000.,
  fx(1:6,2) = 0., 0., 1., 1., 0., 0.,
   x(1:6,2) = 0., 5.065, 5.0651, 20.13, 20.1301, 25.13,
  fx(1:6,3) = 0., 0., 1., 1., 0., 0.,
   x(1:6,3) = 0., 5.065, 5.0651, 20.13, 20.1301, 25.13,
/
This section defines the initial density function of this species
anywhere in space. The density function is specified as the product
of 3 piecewise linear functions. Each of the functions gives the
behavior of the density along one of the axes and is given by
a number of function values at certain positions. " x(1:6,1) = ..."
gives the positions along the x1 direction and "fx(1:6,1) = ..."
gives the function values at these positions. The number of points
where the function is given for each direction is determined by
the variable "num_x" in the previous section. For 2D simulations only
functions for 2 directions have to be defined.

----------boundary conditions for this species----------
/ &nl_spe_bound
  type(1:2,1) =   5,  5,
  type(1:2,2) =   5,  5,
  type(1:2,3) =   5,  5,
/
"type" defines here the type of boundary for this particle species.
The following boundary conditions are currently implemented.
   1 : boundary moving into the simulation box with c
   2 : boundary moving outward from the simulation box with c
   5 : conducting boundary with particle absorption
  20 : axial b.c. for 2D cylindrically-symmetric coordinates
  30 : Lindman open-space boundary - limited implementation
The specification of boundary conditions for specific boundaries
works in exactly the same way as described above for the boundary
conditions of the electromagnetic field.

----------diagnostic for this species----------
/ &nl_diag_species
  ndump_fac_pha = 1,  file_name = ' ',
  ps_xmin(1:3) = 0.0,  0.0,   0.0,  ps_pmin(1:3) =  -5.0, -10.0, -10.0,
  ps_xmax(1:3) = 0.0, 25.13, 25.13, ps_pmax(1:3) =  25.0,  10.0,  10.0,
  ps_nx(1:3)   = 400,  280,   280,  ps_np(1:3)   =   300,   100,   100,
  if_x2x1    = .true.,
  if_x3x1    = .true.,
  if_p1x1    = .true.,
  if_p2x1    = .true.,
  if_p3x1    = .true.,
  if_x3x2    = .true.,
  if_p1x2    = .true.,
  if_p2x2    = .true.,
  if_p3x2    = .true.,
  if_p1x3    = .true.,
  if_p2x3    = .true.,
  if_p3x3    = .true.,
  if_p2p1    = .true.,
  if_p3p1    = .true.,
  if_p3p2    = .true.,
/
This section specifies the diagnostic data dumps for this species.
"ndump_fac" and "file_name" work in the same way as described earlier
for other diagnostics. The logical variables "if_x2x1", "if_x3x1", ...
"if_p3p2" are switches that turn on the writing of a specific
phase space if they are set to ".true.". 
"ps_xmin(1:3) = ..." and "ps_max(1:3) = ..." are defining the lower
and upper boundary of the ranges of interest for each of the
directions in position space. "ps_nx(1:3) = ..."gives the number of
points that each direction should be resolved by.
"ps_pmin(1:3) = ...", "ps_pmax(1:3) = ...", and "ps_np(1:3) = ..."
specify the same information for the momentum axes of the phase spaces.
For example, with the information specified in this input file 
the simulation will generate phase space data that show
the projection of the full phase space of this particle species onto
the x2-p1 plane in the area from 0 to 25.13 in x2 and -5 to 25 in p1.
This data would be written as a 2D array with a resolution of
280 in x2 and 300 in p1. Note that the projection means that for a
phase space plot in a given plane the other position and momentum space
directions are integrated over. The numerical values of the array 
elements are the density of charge with regard to the plane of the
given phase space in normalized units. 
The upper and lower boundary for x1 are given as zero above. This will
not be used directly by the code but triggers the code to use the
instantaneous boundaries of the simulation space as the boundaries
of the phase space. This can be done for any of the position space
axes.


The following information is the information for the second particle
species in this simulation. These sections with the information for the
second particle species have exactly the same structure as the ones for
the first species and require therefore no additional comments.

----------information for species 2----------
/ &nl_species
  num_par_max = 600000,
  rqm=-1.0,
  num_par_x(1:3) = 1, 1, 1,
  vth(1:3) = 0.0d0 , 0.0d0 , 0.0d0 ,
  vfl(1:3) = 0.0d0 , 0.0d0 , 0.0d0 ,
  den_min = 1.d-12,
  if_unneutralized = .false.,
  num_dgam = 0,
  dgam = 0.0,
/
----------density profile for this species----------
  number of points in profile along each direction
/ &nl_num_x  num_x = 6,   /
  actual profile
/ &nl_profile
  fx(1:6,1) = 0., 0., 1.0d0, 1.0d0, 0., 0.,
   x(1:6,1) = 0., 46.30000, 46.30001, 46.70000, 46.70001, 1000.0,
  fx(1:6,2) = 0., 0., 1.0d-4, 1.0d-4, 0., 0.,
   x(1:6,2) = 0., 12.36500, 12.36501, 12.83000, 12.83001,  25.13,
  fx(1:6,3) = 0., 0., 1.0d-4, 1.0d-4, 0., 0.,
   x(1:6,3) = 0., 12.36500, 12.36501, 12.83000, 12.83001,  25.13,
/
----------boundary conditions for this species----------
/ &nl_spe_bound
  type(1:2,1) =   5,  5,
  type(1:2,2) =   5,  5,
  type(1:2,3) =   5,  5,
/
----------diagnostic for this species----------
/ &nl_diag_species
  ndump_fac_pha = 1,  file_name = ' ',
  ps_xmin(1:3) = 0.0,  0.0,   0.0,  ps_pmin(1:3) =  -5.0, -10.0, -10.0,
  ps_xmax(1:3) = 0.0, 25.13, 25.13, ps_pmax(1:3) =  25.0,  10.0,  10.0,
  ps_nx(1:3)   = 400,  280,   280,  ps_np(1:3)   =   300,   100,   100,
  if_x2x1    = .true.,
  if_x3x1    = .true.,
  if_p1x1    = .true.,
  if_p2x1    = .true.,
  if_p3x1    = .true.,
  if_x3x2    = .true.,
  if_p1x2    = .true.,
  if_p2x2    = .true.,
  if_p3x2    = .true.,
  if_p1x3    = .true.,
  if_p2x3    = .true.,
  if_p3x3    = .true.,
  if_p2p1    = .true.,
  if_p3p1    = .true.,
  if_p3p2    = .true.,
/


----------number of pulses----------
/ &nl_pulse_sequence   num_pulses = 2,   /

"num_pulses" determines the number of laser pulses launched in the
simulation. This section has to be followed by the appropriate number of
of sections which means one for each laser pulse. In this case the data
for exactly two pulses have to specified below. The code will crash if
this is not the case.

----------information for pulse 1----------
/ &nl_pulse
  iflaunch = .true.,
  wavetype=1,
  w0=3.0d0,     rise=3.14,      fall=3.14,      length=0.0,
  vosc=1.0d0,   rkkp=5.0,       pol=0.0,        phase = 0.0d0,
  start=6.29d0, focus=-30.0,    offset(1:2)=0.0, 0.0, time=0.0,
/
The following information for a laser pulse is specified in this
section:
"iflaunch" is a switch that can turn the laser pulse on or off.
"wavetype" allows to choose between different kinds of laser pulses.
Currently pulses propagating in x1 (wavetype=1) and a type propagating
in x2 a (wavetype=2) are implemented. In both cases the pulse shape
is approximately Gaussian in the transverse directions.
"w0" specifies the spotsize of the laser in the focal plane.
"rise" gives the distance over which the pulse rises from 0 at the 
front of the pulse to the peak intensity. The shape of the rise is
approximately Gaussian.
"length" defines the length over which the pulse has its peak intensity
after reaching it at the end of the rise at the front.
"fall" gives the distance over which the pulse falls off from the peak
intensity to 0 at the back of the pulse. The shape of the fall off is
approximately Gaussian.
"vosc" specifies the maximum vector potential of the laser pulse at 
the time it is initialized.
"rkkp" is the wavenumber of the laser pulse in normalized units.
"pol" specifies the plane of polarization of the laser. "pol=90" 
corresponds to a laser polarized in x3. "pol=0" is laser with
a polarization changed by 90 degrees from x3.
"phase" specifies an overall phase change to the laser. This can be
used to generate circularly polarized lasers by superposing two laser
with different polarization and phase.
"start" gives the position of the front of the laser with respect to
the side of the box that the pulse is moving towards. In the example
above this means that the front of the pulse is at an x1 position of
29.61 since the simulation window has a length of 35.9 in x1.
"focus" determines the position of the focal plane of the laser in the
same way as "start" specifies the position of the front of the laser.
Note that in the example above this means that the focal plane is outside
the initial box.
"offset(1:2)" specifies an offset of the pulse transverse to the
propagation direction from the center of the simulation. If
"offset(1:2)=0.0, 0.0," then the pulse is centered in the simulation
box with respect to the transverse coordinates. The components 1 and 
2 are referring to x2 and x3 for a  pulse propagating in x1 and to
x1 and x3 for a pulse propagating in x2.
In a 2D simulation there is only one transverse coordinate and
the statement becomes "offset(1:1)=0.0,".
"time" specifies the time at which the pulse is initialized in the 
simulation. 

----------information for pulse 2----------
/ &nl_pulse
  iflaunch = .true.,
  wavetype=2,
  w0=3.0d0,     rise=1.57,      fall=1.57,      length=0.0,
  vosc=1.8,     rkkp=5.0,       pol=0.0,        phase = 0.0d0,
  start=21.065, focus=12.565,   offset(1:2)=10.95,0.0, time=19.90,
/
The section for the second pulse has exactly the same structure as 
the section for the first one and requires no additional comments.


----------smoothing for currents----------
/ &nl_smooth

  ifsmooth(1)     = .false.,
  smooth_level(1) = 3,
  swfj(1:3,1,1)   = 1,2,1,
  swfj(1:3,2,1)   = 1,2,1,
  swfj(1:3,3,1)   = 1,2,1,

  ifsmooth(2)     = .false.,
  smooth_level(2) = 3,
  swfj(1:3,1,2)   = 1,2,1,
  swfj(1:3,2,2)   = 1,2,1,
  swfj(1:3,3,2)   = 1,2,1,

  ifsmooth(3)     = .false.,
  smooth_level(3) = 3,
  swfj(1:3,1,3)   = 1,2,1,
  swfj(1:3,2,3)   = 1,2,1,
  swfj(1:3,3,3)   = 1,2,1,
/
This section specifies the smoothing of the current density. It does 
this for each direction separately. The smoothing is done in position
space using weighting factors.
For each direction i:
"ifsmooth(i) = .true." switches the smoothing on.
smooth_level(i) = ..., gives the number of times the smoothing
is iteratively done over the nearest neighbors. "swfj(1:3,1,i)" gives
the weighting factors for the first smoothing iteration,
"swfj(1:3,2,i)" gives the weighting factors for the second
smoothing iteration, and so on. For more details see Ref.\end{verbatim}
\cite{BirdsallLangdon}.\begin{verbatim}


----------diagnostic for currents----------
/ &nl_diag_phy_field
  ndump_fac_all = 1,  file_name_all = ' ',
  ndump_fac_ave = 0,  file_name_ave = ' ',
  n_ave(1:3)    = 1, 1, 1,
  ifdmp_phy_field(1:3) = .true. , .true. , .true. ,
/
This section specifies the diagnostic data dumps for the current.
"ndump_fac_all" and "file_name_all" work in the same way as described
earlier for the electromagnetic field diagnostics. "ndump_fac_ave",
"file_name_ave", and "n_ave" are currently not implemented but are
going to have the same functionality in future versions of the
code as described for the electromagnetic field.
"ifdmp_phy_field(1:3)" are switches for turning the diagnostic
on and off for the different components of the current.


The following two sections define the smoothing and the diagnostic for 
the charge density in the same way as described above for the current.
The only difference is that the variable "ifdmp_phy_field(1:1)" only
needs one component to work as a switch for the charge density diagnostic
since the charge density is a scalar.
-----------smoothing for charge-----------
/ &nl_smooth

  ifsmooth(1)     = .false.,
  smooth_level(1) = 3,
  swfj(1:3,1,1)   = 1,2,1,
  swfj(1:3,2,1)   = 1,2,1,
  swfj(1:3,3,1)   = 1,2,1,

  ifsmooth(2)     = .false.,
  smooth_level(2) = 3,
  swfj(1:3,1,2)   = 1,2,1,
  swfj(1:3,2,2)   = 1,2,1,
  swfj(1:3,3,2)   = 1,2,1,

  ifsmooth(3)     = .false.,
  smooth_level(3) = 3,
  swfj(1:3,1,3)   = 1,2,1,
  swfj(1:3,2,3)   = 1,2,1,
  swfj(1:3,3,3)   = 1,2,1,
/
-----------diagnostic for charge-----------
/ &nl_diag_phy_field
  ndump_fac_all = 1,  file_name_all = ' ',
  ndump_fac_ave = 0,  file_name_ave = ' ',
  n_ave(1:3)    = 1, 1, 1,
  ifdmp_phy_field(1:1) = .true.,
/
 

\end{verbatim} } 


\pagebreak
\addcontentsline{toc}{chapter}{Bibliography}
\begin{singlespace}
    \bibliography{Back/dissertation}
    \bibliographystyle{unsrt}
\end{singlespace}

\end{document}